%% file: thesis.tex
\documentclass[11pt]{report}
\pdfoutput=1
\usepackage{sty/isuthesis}
\usepackage{sty/traditional}
\usepackage{amssymb}
\usepackage{amsmath}
\usepackage{multirow}
\usepackage{graphics}
\usepackage{graphicx}
\usepackage{hyperref}
\chaptertitle
\alternate
\usepackage{rotating}
\usepackage{sty/natbib}
\newcommand{\ii}{{\rm i}}
\begin{document}

\include{chapters/title}
\tableofcontents 
\addtocontents{toc}{\def\protect\@chapapp{}}
\cleardoublepage 
\addcontentsline{toc}{chapter}{LIST OF TABLES}
\listoftables 
\cleardoublepage 
\addcontentsline{toc}{chapter}{LIST OF FIGURES} 
\listoffigures
\addtocontents{toc}{\def\protect\@chapapp{CHAPTER\ }}
\newpage
\pagenumbering{arabic}
\include{chapters/overview/overview}
\include{chapters/QCD_intro/QCD_intro}
\include{chapters/proton_intro/proton_intro}
\include{chapters/transverse_protons/transverse_protons}
\include{chapters/jpsi_intro/jpsi_intro}
\include{chapters/detector/detector}
\include{chapters/jpsi_AN/jpsi_AN}
\include{chapters/jpsi_pol/jpsi_pol}
\include{chapters/conclusions/conclusions}

\renewcommand{\bibname}{\centerline{BIBLIOGRAPHY}}
\unappendixtitle
\interlinepenalty=300
\addcontentsline{toc}{chapter}{BIBLIOGRAPHY}
\bibliography{bib/thesis}
 
\appendixtitle
\appendix       
\include{chapters/bg_contributions/bg_contributions}
\include{chapters/angular_acceptance/angular_acceptance}
\include{chapters/data_tables/data_tables}

\end{document}

%% file: chapters/title.tex
\title{Understanding the $J/\psi$ Production Mechanism at PHENIX}
\author{Todd Kempel}
\degree{DOCTOR OF PHILOSOPHY} 
\major{Nuclear Physics}
\level{doctoral} 
\mprof{John G. Lajoie} 
\format{dissertation}
\committee{4} 
\members{
Kevin L De Laplante \\
S\"oren A. Prell \\
J\"org Schmalian \\
Kirill Tuchin
} \notice
\maketitle

%% file: chapters/overview/overview.tex
\chapter{Overview}\label{ch:overview}

The $J/\psi$ meson, a bound state of a charm and an anti-charm quark, was discovered in 1974, yet the mechanism for the binding of the two quarks is still largely unknown.  Several models have been proposed for the so-called production mechanism, but none of them satisfactorily describe all available data.  Because the $J/\psi$ is composed of two heavy quarks, the production of the quarks should have a different energy scale from their relative motion, and production may be understood by separating the two scales using non-relativistic QCD (quantum chromodynamics).  Such an understanding would shed light on one of the basic components of QCD: hadronization.  In this document I will present two novel measurements sensitive to the $J/\psi$ production mechanism using the PHENIX detector.  

The first measurement utilizes polarized proton beams at RHIC for a transverse single spin asymmetry of $J/\psi$ mesons, sensitive to both the internal structure of the colliding protons and to the production mechanism of the $J/\psi$.  This is the first time such a measurement has been made, and the resulting asymmetry is inconsistent with zero, a fact which has strong implications for understanding of both the proton and the $J/\psi$.

The second measurement is a determination of the angular distribution of decay leptons in the $J/\psi$ rest frame.  Measurements of such distributions have been made in the past, but very few have determined all relevant coefficients or considered the coordinate system used in the $J/\psi$ rest frame.  Using various coordinate systems and measuring all relevant coefficients will provide a much richer understanding of the $J/\psi$ at a phenomenological level and should provide guidance to theorists in pursuit of the production mechanism.

The organization of this document will be as follows:  In Chapter~\ref{ch:QCD_intro}, I will give a short review of QCD, both to give the reader a bit of background on the subject and to introduce terminology which will be used throughout the document.  Chapter~\ref{ch:proton_intro} will discuss the structure of unpolarized protons, and Chapter~\ref{ch:trans_protons} will present the current theoretical and experimental understanding of transversely polarized protons.  In Chapter~\ref{ch:jpsi_intro}, I will give an overview of the $J/\psi$ production mechanism as it is currently understood.  Chapter~\ref{ch:detector_overview} will present a description of the RHIC complex and the detectors making up the PHENIX experiment, detectors which are used in the analyses discussed in subsequent chapters.  In Chapter~\ref{ch:jpsi_AN}, I will discuss the methods for measuring the transverse single spin asymmetry of $J/\psi$ mesons, present the results of such a measurement, and give a short discussion of the results.  Chapter~\ref{ch:jpsi_pol} will present a similar discussion of the angular decay coefficients for $J/\psi$ mesons.  Finally, Chapter~\ref{ch:conclusions} will present conclusions which can be drawn from the two measurements and a possible path to an even richer understanding. 

%% file: chapters/QCD_intro/QCD_intro.tex
\chapter{Quantum Chromodynamics}\label{ch:QCD_intro}

The prediction and discovery of the $J/\psi$ meson and other bound states of charm and anti-charm quarks are the culmination of a vast theoretical foundation: the Standard Model of Particle Physics, Quantum Chromodynamics (QCD), and the parton model.   In order to understand the relevance of the $J/\psi$ meson in enriching our understanding of QCD, we need to get our footing in the theoretical framework which predicted its discovery.   

\section{The Standard Model}

The Standard Model of Particle Physics is arguably the most successful theoretical model in contemporary physics.  It was first proposed by Glashow~\cite{Glashow:1961tr} as an attempt to describe the hundreds of new particles being observed at moderate and high-energy particle colliders in the 1950s and 1960s and was later modified and improved by Weinberg and Salam~\cite{Weinberg:1967tq,Salam:1968rm}.  The model is based on the idea that the majority of the matter we observe is composed of a relatively small number of common constituents called leptons, quarks, and gauge Bosons~\cite{GellMann:1964xy,Feynman:1969wa}.  The Standard Model describes the interactions between the constituent particles in a single elegant expression, the Standard Model Lagrangian, derived from fundamental symmetries using group theory, and it has provided some of the most precise predictions in the history of theoretical physics.  While it is not in the scope of this document to present a detailed discussion of the predictions and experimental status of the Standard Model (I will instead provide a brief introduction), a comprehensive guide can be found in the standard reference of particle physics~\cite{PDG}. 

The integer spin mediators of fundamental interactions in the Standard Model are known as gauge Bosons (Table~\ref{tab:gauge-bosons}), and the fundamental spin-$\frac{1}{2}$ particles are the leptons and quarks (Table~\ref{tab:leptons_quarks}).  Particles composed of two and three quarks are named mesons and baryons respectively, collectively known as hadrons.  



\begin{table}[h!tb]
\isucaption[Properties of the Standard Model gauge Bosons.]{\label{tab:gauge-bosons} Properties of the Standard Model gauge Bosons.  Dependence of the strong force on distance will be discussed in Section~\ref{sec:asymp_confine}.  From~\cite{PDG}.}
\begin{tabular}{|cc|c|c|c|c|}
\hline
\multicolumn{2}{|c|}{\small Interaction Mediated}                   & \multicolumn{3}{|c|}{\small Electro-Weak}                      & \small Strong \\ \cline{3-5}
                   &                                                &  \small Electromagnetic &   \multicolumn{2}{|c|}{\small Weak}  &        \\ \hline
\multicolumn{2}{|c|}{\small gauge Boson}                            & \small $\gamma$ (Photon)& \small $W^{+/-}$    & \small $Z^{0}$     & \small gluon  \\ \hline
\multicolumn{2}{|c|}{\small mass (GeV/c$^{2}$)}                     &   \small 0              &  \small 80.398$\pm$0.025   & \small 91.1876$\pm$0.0021 &  \small 0    \\ \hline
\multicolumn{2}{|c|}{\small Electric Charge}                        &   \small 0              &   \small +1/-1      &   \small 0         & \small 0   \\ \hline
\scriptsize Strength Relative to EM            & \small $10^{-18}$m         &   \small 1       &   \small 0.8        &   \small 0.8       & \small 25  \\ \cline{3-6}
\scriptsize coupling for two up quarks at     & \small $3\times 10^{-17}$m &   \small 1       &   \small $10^{-4}$  &   \small $10^{-4}$ & \small 60  \\ \hline
\end{tabular}
\end{table}

\begin{table}[h!tb]
\isucaption[Properties of the Standard Model quarks and leptons]{\label{tab:leptons_quarks} Properties of the Standard Model quarks and leptons. From~\cite{PDG}.}
\begin{tabular}{|c|c|c||c|c|c|}
\hline
\multicolumn{3}{|c||}{\small Leptons}                                  &                 \multicolumn{3}{|c|}{\small Quarks}                \\ \hline
\small Flavor     & \small mass (GeV/c$^2$)        & \small Electric Charge & \small Flavor & \small mass (GeV/c$^2$) & \small Electric Charge \\ \hline 
$\begin{array}{c}\text{\small electron} \\ \text{\small neutrino}\end{array}$ & \small $<7\times 10^{-9}$ & 0  & \small up & \small $2.55 \begin{array}{c}+0.75\\-1.05\end{array} \times 10^{-3}$  &  $\frac{2}{3}$ \\ \hline
\small electron   & \small $0.54858\times 10^{-6}$ & -1 & \small down & \small $5.04 \begin{array}{c}+0.96\\-1.54\end{array}\times 10^{-3}$  &  $-\frac{1}{3}$ \\ \hline
$\begin{array}{c}\text{\small muon} \\ \text{\small neutrino}\end{array}$ & \small $<0.0003$ & 0 & \small charm & \small $1.27 \begin{array}{c}+0.07\\-0.11\end{array}$  &  $\frac{2}{3}$ \\ \hline
\small muon   & \small $0.10566\times 10^{-3}$ & -1 & \small strange & \small $105 \begin{array}{c}+25\\-35\end{array}\times 10^{-3}$  &  $-\frac{1}{3}$ \\ \hline
$\begin{array}{c}\text{\small tau} \\ \text{\small neutrino}\end{array}$ & \small $<0.03$ & 0  & \small top & \small $171.3 \begin{array}{c}+1.1\\-1.2\end{array}$  &  $\frac{2}{3}$ \\ \hline
\small tau   & \small $1.77684 \pm 0.17$ & -1 & \small bottom & \small $4.20 \begin{array}{c}+0.17\\-0.07\end{array}$ & $-\frac{1}{3}$ \\ \hline
\end{tabular}
\end{table}



The overriding symmetries of the Standard Model are given by the group SU(3)$\otimes$SU(2)$\otimes$U(1).  The SU(3) group provides the symmetries for the strong interaction of quantum chromodynamics (QCD), SU(2) for the weak interaction, and U(1) for the electromagnetic interaction of quantum electrodynamics (QED).  The number of gauge Bosons for a given interaction corresponds to the number of dimensions of the gauge group: $N^{2}-1$ for SU($N$).  This means that there is one gauge Boson (the photon) for the electromagnetic interaction, 3 gauge Bosons (the $W^{+}$, $W^{-}$, and $Z^{0}$) for the weak interaction, and 8 gauge Bosons (8 types of gluons) for the strong interaction.

\section{Quarks and Gluons}

In QED, stable particles have only three possible charges: positive, negative, or zero.  In QCD, however, there are 6 charges and 6 anti-charges, and all stable particles which can be observed in isolation have a net charge of zero. 
 The charges are called red, blue, and green with anti-charges anti-red, anti-blue, and anti-green.\footnote{To have a net color charge of zero, a particle can either have an equal number of colors and anti-colors, like red and anti-red, or it can have one of each color or anti-color, red, blue, and green or anti-red, anti-blue, and anti-green.}  They are collectively referred to as `colors' and `anti-colors,' and stable particles are said to be `colorless,' while charged particles are `colored.'  All gluons in QCD carry an equal number of colors and anti-colors, and it is these charges which determine the type of gluon.  The eight gluon types are:
\begin{equation*}
\mathbf{g}_{i} = \mathbf{\lambda}_{i}
\left( 
\begin{array}{ccc}
r\bar{r} & r\bar{b} & r\bar{g} \\
b\bar{r} & b\bar{b} & b\bar{g} \\
g\bar{r} & g\bar{b} & g\bar{g}
\end{array}
\right)
\end{equation*}
for $i=1,2, ... 8$, where the $\mathbf{\lambda}_{i}$ are the Gell-Mann matrices, which are generators of the SU(3) group, and $r$, $b$, and $g$ are red, blue, and green respectively (the bar denotes anti-color).  The eight color configurations for gluons are collectively known as color-octet states.  All gluons are colored so that none inhabit the color-singlet state\footnote{The definitions of a color-octet and color-singlet state will be particularly useful when we talk about $J/\psi$ meson production in Chapter \ref{ch:jpsi_intro}.}
\begin{equation*}
\frac{1}{\sqrt{2}}(r\bar{r}+b\bar{b}+g\bar{g}).
\end{equation*}
Quarks carry a single color ($r$, $b$, or $g$), while anti-quarks carry a single anti-color ($\bar{r}$, $\bar{b}$, or $\bar{g}$).  Leptons are colorless and do not experience the strong interaction.  

The fact that gluons carry color charge makes the strong interaction considerably more complicated than electromagnetism, where the photon has zero charge.  An interaction wherein the gauge Bosons have zero charge, like electromagnetism, is called Abelian.  An interaction wherein the gauge Bosons are charged, like the strong or weak interactions, is called non-Abelian.  The non-Abelian nature of QCD implies that gluons interact not only with quarks but also with other gluons.  The interaction between gluons leads to a theory wherein quarks and gluons cannot be found in isolation but are confined to bound states (confinement) by a force which grows stronger as they move away from other quarks and gluons and weaker as they approach one another (asymptotic freedom).

\section{Asymptotic Freedom and Confinement}\label{sec:asymp_confine}

The coupling constants of the Standard Model determine the strength of a given interaction.  They are not `constants,' however, as they depend on the mass of the exchanged gauge Boson. If a gauge Boson is short-lived (virtual), the Heisenberg uncertainty principle allows for its mass to fluctuate away from the nominal value, and even a virtual photon or gluon is allowed to have some mass. The mass of the gauge Boson, real or virtual, sets the energy scale $Q$ of the interaction, defined as $Q^{2} \equiv -\text{(mass)}^{2}$.  

If we know the QED coupling $\alpha_{em}$ at a scale $Q^{2}=m^{2}$, the coupling can be calculated at a much larger scale $q^{2} \gg m^{2}$ by
\begin{equation}
\alpha_{em}(q^{2}) = \alpha_{em}(m^{2}) [1 + \frac{\alpha_{em}(m^{2})}{12\pi}\log{\frac{q^{2}}{m^{2}}} + \mathcal{O}(\alpha_{em}^{2})].
\end{equation}
For QCD we have
\begin{equation}
\alpha_{S}(q^{2}) = \alpha_{S}(m^{2}) [1 + \frac{\alpha_{S}(m^{2})}{12\pi}\log{\frac{q^{2}}{m^{2}}}(2n_{f}-11N) + \mathcal{O}(\alpha_{S}^{2})],
\label{eq:QCD_coupling}
\end{equation} 
where $n_{f}$ is the number of (active) quark flavors, which depends on the energy scale,\footnote{Since the quarks have very different masses, a flavor of quarks is only relevant at an energy scale greater than or approximately equal to the rest mass energy for that flavor of quark.} and $N$ is the number of colors.  There are 3 colors and no more than 6 active flavors so that $\alpha_{S}$ decreases with increasing $Q^{2}$, while $\alpha_{em}$ increases.  To put things a bit more simply: $Q$ is inversely proportional to distance $d \sim \frac{\hbar c}{Q}$, and Eq.~\ref{eq:QCD_coupling} implies that the strong interaction is weaker for two quarks which are close together than it is for two quarks which are far apart.  This surprising discovery, known as asymptotic freedom~\cite{Politzer:1973fx,Gross:1973id}, leads to the confinement of quarks into bound state hadrons.

A non-relativistic approximation for the interaction potential between a quark and anti-quark can be written~\cite{Gupta:1982kp} in terms of $r$, the radial distance between quark and anti-quark, as 
\begin{equation}
V_{q\bar{q}}(r) = -\frac{f\alpha_{S}}{r} \left(1-\frac{3\alpha_{S}}{2\pi}+\frac{\alpha_{S}}{6\pi}(33 - 2 n_{f})[\ln(\mu r) + \gamma_{E}]\right) + \frac{r}{4\pi\epsilon_{0}}
\label{eq:singlet_octet_potential}
\end{equation}
where $\alpha_{S}$ is the strong coupling constant, $n_{f}$ is the number of active flavors, $\epsilon_{0}$ is the permittivity of free space, $\mu$ and $\gamma_{E}$ depend on the renormalization scheme.\footnote{The theory of QCD contains several divergences, and a renormalization scheme is a method to handle those divergences in a rigorous way.  Renomalization in QCD was first introduced in the 1970s by 't Hooft~\cite{'tHooft:1972fi}}  The color factor $f$ is $\frac{4}{3}$ for a color-singlet configuration and $-\frac{1}{6}$ for a color-octet configuration. The divergence of the potential in Eq.~\ref{eq:singlet_octet_potential} for $r\rightarrow \infty$, as well as the (related) increase of $\alpha_{S}$ at large distances, mean that colored objects (quarks, gluons, and colored hadrons) cannot be found in isolation.  The difference in sign between the color factors for the color-singlet and color-octet configurations implies that the quarks making up a stable hadron can only be in a color-singlet configuration.

The fact that quarks and gluons must be confined to colorless hadrons in QCD implies that a single quark or gluon can never leave a collision in isolation.  Instead, it is energetically favorable for a parton leaving a collision to continue pulling colorless pairs of quarks and gluons from the vacuum until a totally colorless state can be reached.  This process is known as fragmentation, and the binding of partons into colorless objects is called hadronization.  Fragmentation cannot be calculated from first principles in QCD.  Instead, the properties of fragmenting quarks and gluons are parameterized by fragmentation functions determined from experimental data.  The parameterized measurements use groups of particles, called jets, observed in a particle detector every time a quark has fragmented, and the parameterization depends heavily on the experimental definition of a jet.

%% file: chapters/proton_intro/proton_intro.tex
\chapter{The Proton}\label{ch:proton_intro}

The particle collider is one of the most common tools for studying elementary particles.  Collider experiments measure the flavor of outgoing particles and the directions in which they are produced from a collision.  These measurements can then be compared to theoretical predictions in order to develop a combined understanding.  

Many contemporary experiments study particle production by colliding two beams of protons.  As we will show, the proton is a composite object, made of quarks and gluons.  To develop an understanding of the particles produced in proton collisions, like the $J/\psi$ meson, we need to have a thorough understanding of the structure of the proton itself.  Otherwise, quantities like cross-sections from proton collisions cannot be predicted, leaving little theoretical guidance for experiment.

\section{Cross-Sections and Luminosities}\label{sec:cross_sec_lumi}

In scattering experiments, it is necessary to quantify the number of particles measured so that results can be compared between separate experiments.  This is done in part with differential cross-sections $\frac{d\sigma}{d\Omega}$ where $d\Omega$ is a unit of solid angle and 
\begin{equation}
\sigma =  \displaystyle\int d\Omega \frac{d\sigma}{d\Omega} 
\end{equation}
is the total cross section.  The cross-section depends on the luminosity $\mathcal{L}$ as $\sigma = \frac{\text{rate}}{\mathcal{L}}$, and for a beam of particles in a circular collider
\begin{equation}
\mathcal{L} = f n \frac{N_{1} N_{2}}{A}
\end{equation}
where $f$ is the crossing frequency and $n$ the number of bunches of colliding particles in one beam,\footnote{Beams of particles in circular accelerators are not continuous but are composed of a number of equally spaced `bunches.'} $N_{1}$ is the number of particles per bunch in one beam, while $N_{2}$ is the number of particles per bunch in the other beam, and $A$ is the cross-sectional colliding area of the beams.

The differential cross-section can be measured with respect to any physical quantity, but one of the most basic is the invariant cross section with respect to the energy $E$ and three-momentum $\vec{p}$.  Through a change of variables, this quantity can be rewritten as
\begin{equation}
E \frac{d^{3}\sigma}{dp^{3}} = \frac{1}{2\pi p_{T}} \frac{d^{2}\sigma}{dy dp_{T}}
\end{equation}
where $y$ is the rapidity and $p_{T}$ the transverse momentum of the particle.  
 
For measurements of particle production, rapidity is preferable to $\theta$, the angle relative to the beam axis, because it is additive while maintaining Lorentz invariance.  Rapidity is defined as
\begin{equation}
y = \frac{1}{2}\frac{E+p_{z}}{E-p_{z}},
\end{equation}
where $E$ is the total energy of the particle and $p_{z}$ its momentum along the beam axis.  It is often convenient to use the pseudo-rapidity, a quantity which is equivalent to rapidity if the particle is massless but depends only on the outgoing angle of the particle:
\begin{eqnarray}
\eta &=& \frac{1}{2}\ln \left(\frac{|\vec{p}|+p_{z}}{|\vec{p}|-p_{z}}\right) \nonumber \\
     &=& -\ln\left(\tan \frac{\theta}{2}\right).
\end{eqnarray}

Experimentally, both the luminosity and the detector efficiency $\varepsilon$ must be known in order to determine a cross section.  For example, the differential cross-section as a function of transverse momentum $p_{T}$ can be written as
\begin{equation}
\frac{d\sigma}{d p_{T}} = \frac{1}{\mathcal{L}\varepsilon}\frac{dN}{d p_{T}}
\end{equation}
where $dN$ is the number of particles measured in a transverse momentum range $d p_{T}$, and $\mathcal{L}$ is the provided luminosity.

\section{Deep-Inelastic Scattering}\label{sec:DIS}

\begin{figure}[h!tb]
 \centering\includegraphics*[width=0.65\columnwidth]{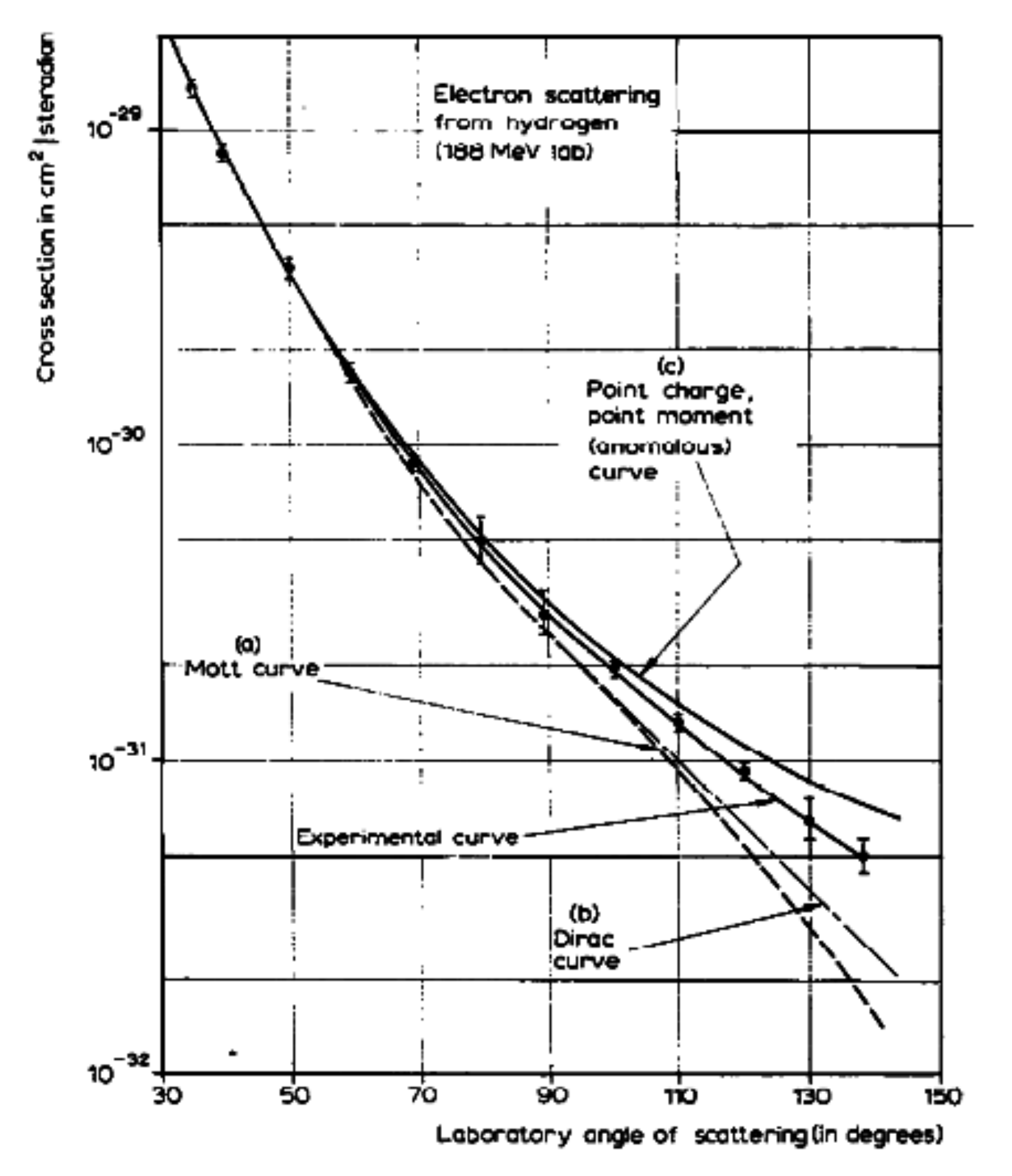} 
\isucaption[Scattering of a 188~MeV electron beam on a hydrogen target]{Scattering of a 188~MeV electron beam on a hydrogen target showing that the proton is a composite object.  From the body of work that won Hofstadter and Schiff the 1961 Nobel Prize~\cite{Hofstadter:1961nl}.}\label{fig:proton_structure}
\end{figure}

Scattering of leptons on protons has long provided precise information on the substructure of the proton.  The charged point-like structure of leptons makes them powerful tools for probing the electronic structure of the much larger proton.   Early measurements compared the total scattering cross-section for $ep \rightarrow eX$ with calculations from several ansatz for a point-like charged proton structure: 
\begin{itemize}
\item
The `Mott cross-section' for scattering of a spin-1/2 electron on a classical potential
\item
The `Dirac cross-section' for scattering of a spin-1/2 electron on a spin-1/2 point charge
\item
A modified Dirac cross-section, which takes into account the anomalous proton magnetic moment.
\end{itemize}
The measured cross-section does not agree with any of these point-like ansatz (see Fig.~\ref{fig:proton_structure}), implying that the proton is not a point-like object.\footnote{Admittedly, the anomalous magnetic moment of the proton already provides evidence that it is not a point-like particle, but the verification of this conclusion with electron scattering was quite important historically.}

The quantum numbers of the proton (charge and isospin) indicate that its dominant constituents in the quark model ought to be two up quarks and one down quark, called the valence quarks.  While this is a good approximation for collisions which are barely inelastic, high energy collisions begin to probe the QCD vacuum structure of the proton, and a `quark sea,' consisting of all possible flavors of quark and anti-quark, is observed along with a very large density of gluons.  Since the details of this structure cannot be calculated from first principles, the probability distributions for finding a gluon or a specific flavor of quark or anti-quark are determined experimentally with deep-inelastic scattering (DIS).

\begin{figure}[h!tb]
 \centering\includegraphics*[width=0.25\columnwidth]{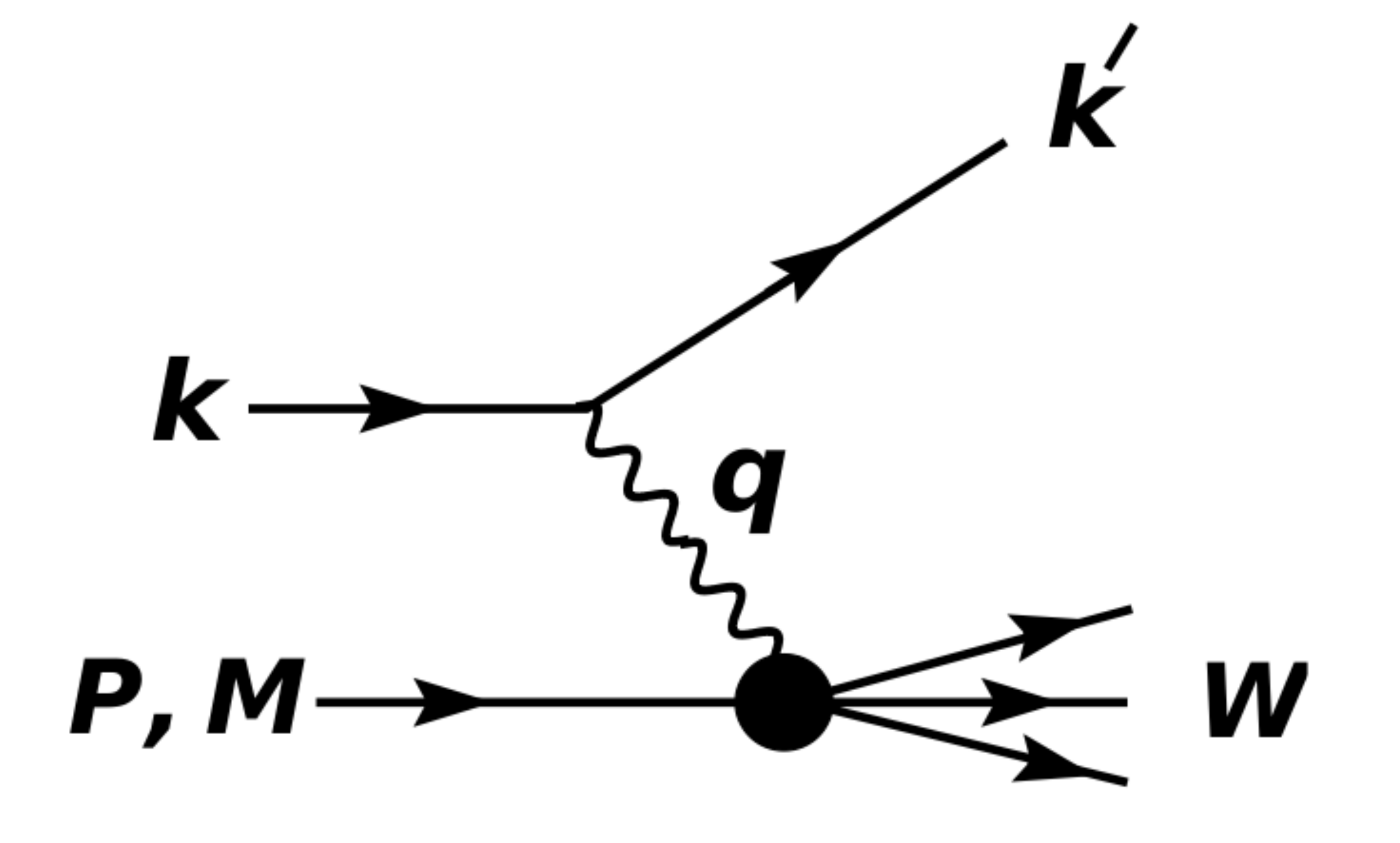} 
\isucaption[Commonly used variables in DIS]{Illustration of commonly used variables in DIS.}\label{fig:DIS}
\end{figure}

In DIS, a lepton is scattered on a proton with enough energy to probe its internal structure, and the outgoing lepton is observed.\footnote{Semi-inclusive deep-inelastic scattering (SIDIS) is equivalent to DIS except that other outgoing particles are measured in addition to the scattered lepton.}  The scattering resolution is $\sim$$\frac{\hbar c}{Q}$ where $Q^{2}=-q^{\mu}q_{\mu}$ and $\vec{q}=\vec{k}^{\prime}-\vec{k}$ with $\vec{k}$ the 4-momenta of the incoming lepton and $\vec{k}^{\prime}$ the 4-momenta of the same lepton after scattering (Fig.~\ref{fig:DIS}).  Because the proton is approximately 1 Fermi in diameter, quarks become the relevant degrees of freedom at $Q$$\sim$$0.5$~GeV.  

The scattering cross-section of a lepton on a proton with 4-momentum $\vec{P}$ is $d\sigma \propto L_{\mu\nu}W^{\mu\nu}$ where
\begin{eqnarray*}
L_{\mu\nu} &\propto& \displaystyle\sum_{\lambda,\lambda^{\prime}}\left(\bar{u}(k^{\prime},\lambda^{\prime})\gamma_{\mu}u(k,\lambda)\right)
\left(\bar{u}(k^{\prime},\lambda^{\prime})\gamma_{\nu}(0)u(k,\lambda)\right)^{*}, \\
W^{\mu\nu} &\propto& \displaystyle\sum_{\sigma}\int \frac{d^{4} \xi}{2\pi} e^{\ii q\cdot\xi}
\left<\vec{P},\sigma|[J_{\mu}^{\dagger}(\xi),J_{\nu}(0)]|\vec{P},\sigma\right>
\end{eqnarray*}
are the leptonic and hadronic tensors respectively, and $J$ is a matrix describing the effective vertex between the virtual photon and the proton.\footnote{An effective vertex uses a single coupling to describe a much more complicated interaction.}  The helicities (scalar product between the spin and momentum vector) of the incoming and outgoing lepton are $\lambda$ and $\lambda^{'}$ respectively, while $\sigma$ is the helicity of the incoming proton.  $W_{\mu\nu}$ is considered universal, meaning that the same $W_{\mu\nu}$ can be applied for scattering of a proton on any particle, not just a lepton.  The total scattering cross-section for a proton-proton collision, for example, is $d\sigma \propto W_{\mu\nu}W^{\mu\nu}$.

\section{Structure Functions and Bjorken Scaling} \label{sec:struct_func}

For unpolarized DIS using electrons or muons, the hadronic tensor can be parameterized as
\begin{equation}
W_{\mu\nu} = \widetilde{g}_{\mu\nu} W_{1}(x_{B},Q^{2}) + \widetilde{P}_{\mu}\widetilde{P}_{\nu} W_{2}(x_{B},Q^{2}) 
\label{eq:DIS_param}
\end{equation}
where $\widetilde{P}_{\mu} = (P_{\mu} - \frac{P \cdot q}{q^{2}}q_{\mu})/M$, $x_{B}=\frac{Q^{2}}{2M q^{0}}$, $M$ is the mass of the proton, and $\widetilde{g}$ the inverse metric tensor.  

The quantity $x_B$ (called the Bjorken-$x$) is especially convenient, because for large proton momenta it can be interpreted as the fraction of the hadron momentum carried by a quark.  An additional convenience is that at large energy transfer $q^{0}\rightarrow\infty$ and fixed $x_{B}$, the structure functions $W_{1}(x_B,Q^{2})$ and $W_{2}(x_B,Q^{2})$ depend only on $x_B$ and not $Q^{2}$ so that $M W_{1}(x_{B},Q^{2}) = F_{1}(x_{B})$ and $\frac{Q^{2}}{2 M x_{B}} W_{2}(x_{B},Q^{2}) = M F_{2}(x_{B})$~\cite{Bjorken:1968dy}. 

$F_{1}(x_B)$ and $F_{2}(x_B)$ are called the `unpolarized proton structure functions,' and they can be related to the probability for the scattering of a polarized virtual photon on a quark in the hadron.  In the parton model, the longitudinal structure function $F_{L} = F_{2} - 2x_{B}F_{1}$ (corresponding to a longitudinally polarized photon) can be written as $F_{L} = \frac{Q^{2}}{4\pi^{2}\alpha_{em}}\sigma_{L}(x_{B},Q^{2})$, where $\sigma_{L}$ is the cross section for interaction between a quark and a longitudinally polarized photon.  Similarly, the transverse structure function $F_{T} = x_{B} F_{1}$ (corresponding to a transversely polarized photon) can be written as $F_{T} = \frac{Q^{2}}{4\pi^{2}\alpha_{em}}\sigma_{T}(x_{B},Q^{2})$, where $\sigma_{T}$ is the cross section for interaction between a quark and a transversely polarized photon.  Callan and Gross have shown that $\sigma_{L}/\sigma_{T}$ scale as $1/Q^{2}$, meaning that $F_{L} \rightarrow 0$ as $q^{0} \rightarrow \infty$~\cite{Callan:1969uq} so that $F_1$ and $F_2$ are related at large momentum transfer.

In proton-proton collisions, the quantities $x_1$ and $x_2$ are defined~\cite{0201503972} to be approximately equivalent to $x_{B}$ at leading order in $\alpha_{S}$ as
\begin{equation}
x_{1} = \frac{M}{\sqrt{s}}e^{y}, \quad x_{2}=\frac{M}{\sqrt{s}}e^{-y} 
\end{equation}
for partons from the proton travelling in the positive and negative $\hat{z}$ direction respectively, where $M$ is the particle's mass, $y$ its rapidity, and $\sqrt{s}$ the center of mass collision energy.  If the measured particle is sufficiently massless, $M$ can be replaced with the transverse momentum $p_{T}$ and $y$ with the pseudo-rapidity $\eta$.

The parameterization of Eq.~\ref{eq:DIS_param} along with Bjorken's observation that the structure functions do not depend on $Q^{2}$ as $q^{0} \rightarrow \infty$ are very powerful due to the fact that they can be easily interpreted in term of quarks.  At significantly large energies in electron-proton collisions,\footnote{The actual approximation is that the perturbative expansion is taken only to leading order in $\alpha_{em}$, the electromagnetic coupling constant.} the electron probes the structure function $F_{2}$ of a proton as
\begin{equation}
F_{2}(x) = x \displaystyle\sum_{q,\bar{q}} e^{2} [q(x) + \bar{q}(x)],
\label{eq:parton_model_F2}
\end{equation}
where $q(x)$ ($\bar{q}(x)$) is the probability to find a quark (anti-quark) of flavor $q$ at momentum fraction $x$.  Similar expressions can be determined for DIS using other lepton probes (i.e. $\nu P \rightarrow \nu X$ and $e^{+}P \rightarrow e^{+}X$), and these measurements together determine both $q(x)$ and $\bar{q}(x)$.  

\section{Altarelli-Parisi Evolution}\label{sec:altarelli-parisi}

Part of the reason that Eq.~\ref{eq:parton_model_F2} and the Callan-Gross relation work as $q^{0}$$\rightarrow$$\infty$ is that gluons have very little effect on high energy quarks.\footnote{Another reason is that the momenta of most quarks is preferentially along the direction of the proton's motion, meaning that the effects of transverse motion in the proton are negligible.  Transverse partonic motion inside the proton will be discussed at length in Chapter~\ref{ch:trans_protons}.}  At lower $q^{0}$, gluons become much more important, and Bjorken scaling no longer applies.  The contribution of gluons to the structure function $F_{2}$ follows Altarelli-Parisi evolution, a set of differential equations relating the quark distributions to gluon distributions:
\begin{eqnarray*}
\frac{d g(x,Q)}{d \log{Q}} &=& \frac{\alpha_{s}(Q^{2})}{\pi} \displaystyle\int_{x}^{1} \frac{dz}{z} \{P_{g \leftarrow q}(z)
\displaystyle\sum_{q}[q(\frac{x}{z},Q) + \bar{q}(\frac{x}{z},Q)] + P_{g \leftarrow g}(z)g(\frac{x}{z},Q)\}, \\
\frac{d q(x,Q)}{d \log{Q}} &=& \frac{\alpha_{s}(Q^{2})}{\pi} \displaystyle\int_{x}^{1} \frac{dz}{z} \{P_{q \leftarrow q}(z)q(\frac{x}{z},Q) + P_{q \leftarrow g}(z)  g(\frac{x}{z},Q) \}, \\
\frac{d \bar{q}(x,Q)}{d \log{Q}} &=& \frac{\alpha_{s}(Q^{2})}{\pi} \displaystyle\int_{x}^{1} \frac{dz}{z} \{ P_{q \leftarrow q}(z) \bar{q}(\frac{x}{z},Q) + P_{q \leftarrow g}(z) g(\frac{x}{z},Q) \}
\end{eqnarray*}
where the $g(x,Q)$, $q(x,Q)$, and $\bar{q}(x,Q)$ denote the probability for the scattering of a gluon, quark, and anti-quark respectively~\cite{Altarelli:1977zs}. 
The splitting functions $P_{j \leftarrow i}$ are 
\begin{eqnarray*}
P_{q \leftarrow q}(z) &=& \frac{4}{3} \left(\frac{1+z^{2}}{[1-z]_{+}} + \frac{3}{2}\delta(1-z) \right), \\
P_{g \leftarrow q}(z) &=& \frac{4}{3} \left(\frac{1+(1-z)^{2}}{z}\right), \\
P_{q \leftarrow g}(z) &=& \frac{1}{2} \left(z^{2}+(1-z)^{2}\right), \\
P_{g \leftarrow g}(z) &=& 6 \left(\frac{1-z}{z} + \frac{z}{[1-z]_{+}} + z (1-z) + \left(\frac{11}{12} - \frac{n_{f}}{18}\right)\delta(1-z) \right)
\end{eqnarray*}
where $n_f$ is the number of (active) quark flavors, and $\left[F(z)\right]_{+}$ is defined as
\begin{equation*}
\displaystyle\int_{0}^{1}dz f(z) [F(z)]_{+} = \displaystyle\int_{0}^{1}dz (f(z) - f(1))F(z).
\end{equation*}
The Altarelli-Parisi equations can be used along with the measured $q(x)$ and $\bar{q}(x)$ distributions from Eq.~\ref{eq:parton_model_F2} to determine 
both the quark and gluon parton distribution functions~(PDFs). 

\begin{figure}[h!tb]
  \begin{center} 
    \includegraphics*[width=0.65\columnwidth]{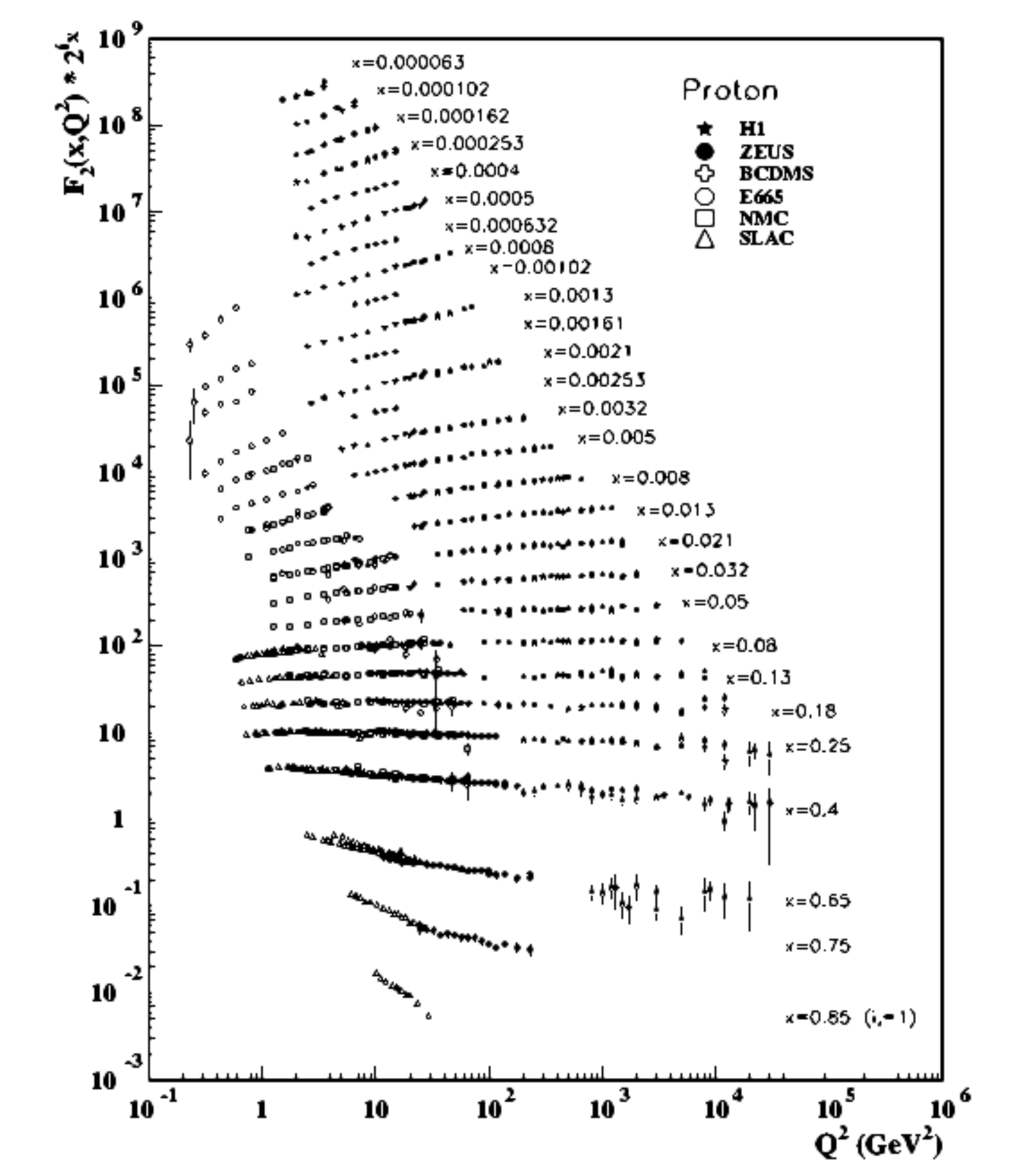} 
  \end{center}
  \isucaption[Compilation of measurements of the $F_{2}$ structure functions]{Compilation of measurements of the $F_{2}$ structure function (from~\cite{PDG}).  Deviations from the flat line (which is indicative of Bjorken-scaling) are due to Altarelli-Parisi evolution at small-x and elastic scattering contributions at large-x.}\label{fig:F2_data}
\end{figure}

\begin{figure}[h!tb]
 \begin{center}
   \includegraphics*[width=0.7\columnwidth]{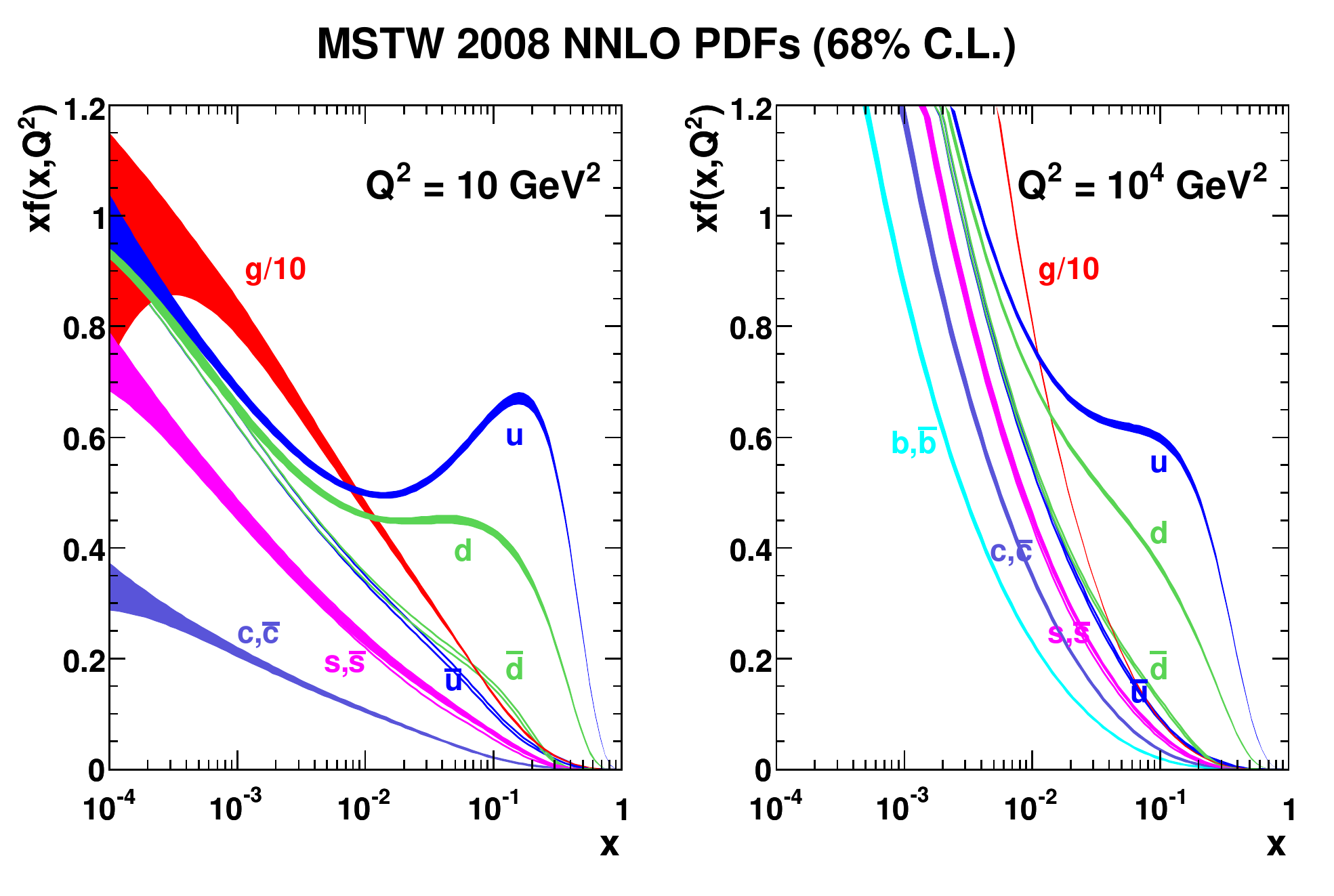} 
 \end{center}
\isucaption[Parton Distribution Functions from the MSTW collaboration calculated at next-to-leading order in $\alpha_{s}$]{Parton Distribution Functions from the MSTW collaboration calculated at next-to-next-to-leading order in $\alpha_{s}$ for $Q^{2} = 10$~GeV$^{2}$ on the left and $Q^{2}=10^{4}$~GeV$^{2}$ on the right (from~\cite{Martin:2009iq}).}\label{fig:MSTW2008_PDFs}
\end{figure}

The structure function $F_{2}$ has been extensively measured, and a compilation of results can be found in Fig.~\ref{fig:F2_data}.  In order to determine the PDFs from these measured structure functions, a global analysis must be performed, taking into account information from several scattering processes at various energies.  Global analyses are continually updated in order to include the most recent structure function data and to improve on methods for the parton model fits~\cite{Pumplin:2002vw,Ball:2008by,Martin:2009iq}.  An example of a recent fit from the MSTW collaboration~\cite{Martin:2009iq} is shown in Fig.~\ref{fig:MSTW2008_PDFs} where it is clear that the composition of the proton is well determined for a large range in $x$.

%% file: chapters/transverse_protons/transverse_protons.tex
\chapter{Transversely Polarized Protons}\label{ch:trans_protons}

The discussions of Chapter~\ref{ch:proton_intro} focused on unpolarized protons, but an even richer understanding of the proton structure can be developed by studying collisions of polarized protons.   The structure of a transversely polarized proton is different from that of a longitudinally polarized proton, because partons are highly relativistic so that their boosts and rotations do not commute.  In this work, I will only discuss the structure of a proton with spin oriented transverse to its momentum (transverse polarization), but a great deal of understanding has also been developed from the study of protons polarized along their momentum direction~\cite{Gluck:1995yr,Gluck:2000dy,deFlorian:2008mr,deFlorian:2009vb}. 

Early experiments with transversely polarized protons yielded surprising results, and the theoretical understanding continues to develop.  Before we discuss the measurements and the associated theory in detail, it is necessary to define the basic observable: a transverse single spin asymmetry (SSA).  Put simply, transverse SSAs are an imbalance of the production cross-section to one side of the proton spin relative to the other.

For proton-proton collisions in a collider, the asymmetry is measured using particle yields in the left and right hemispheres from opposing spin orientations of a single polarized beam.  The term `left' is defined as the axis which forms a right-handed coordinate system between the polarized beam direction and one of the proton spin orientations, denoted as $\uparrow$.  If the momentum vector of the outgoing particle is $\vec{p}$ and the colliding proton momentum is $\vec{P}$ with spin $\vec{S}$, left is defined as $\vec{p}\cdot\left(\vec{S}\times\vec{P}\right)>0$, and the transverse SSA on the left is
\begin{equation}
A_{N} = \frac{f}{\mathcal{P}} \frac{\sigma_{L}^{\uparrow} - \sigma_{L}^{\downarrow}} {\sigma_{L}^{\uparrow} + \sigma_{L}^{\downarrow}}
\label{eq:AN_sigma}
\end{equation}
where $\sigma_{L}^{\uparrow}$ ($\sigma_{L}^{\downarrow}$) represents the production cross-section in the left hemisphere with beam polarized in the $\uparrow$ ($\downarrow$) direction, and $\mathcal{P}$ is the beam polarization.  An overall minus sign is required for $A_{N}$ on the right.

The geometric scale factor $f$ takes into account the convolution of an azimuthal asymmetry with an incomplete detector acceptance.  We assume that the modulation of $A_N$ is sinusoidal in azimuth (a shape which will be justified in the following sections) and ensure that we are measuring the amplitude of such a modulation by defining
\begin{equation}
f \equiv \left( \frac{\displaystyle\int_{0}^{\pi} \epsilon(\phi) \sin\phi d\phi} {\displaystyle\int_{0}^{\pi}d \phi} \right)^{-1}
\label{eq:f_factor}
\end{equation}
for $A_{N}$ on the left, where $\epsilon(\phi)$ is the efficiency for measuring a particle with azimuthal angle $\phi$ between $\vec{p}$ and $\vec{S}$.  The limits of integration correspond to the hemisphere in which the measurement is made.

Measurements of $A_N$ (Eq.~\ref{eq:AN_sigma}) for various event topographies currently provide most of the basis for understanding the structure of a transversely polarized proton.

\section{Experimental Observations}\label{sec:SSA_measurements}

QCD na\"ively predicts that $A_{N}$ should scale as $m_{q} / \sqrt{s}$, where $m_{q}$ is the mass of the scattered quark from the polarized proton~\cite{Kane:1978nd}.  Because a large fraction of the proton is composed of light up and down quarks, the asymmetry should be $\mathcal{O}(10^{-4})$ for $\sqrt{s}=$20~GeV collisions, but experimentally measured asymmetries have consistently been $\mathcal{O}(10^{-1})$.  This contradiction has led to a great deal of theoretical and further experimental activity, and the theoretical understanding of large measured asymmetries is still far from complete.

\begin{figure}[h!tb]
 \centering\includegraphics*[width=0.3\columnwidth]{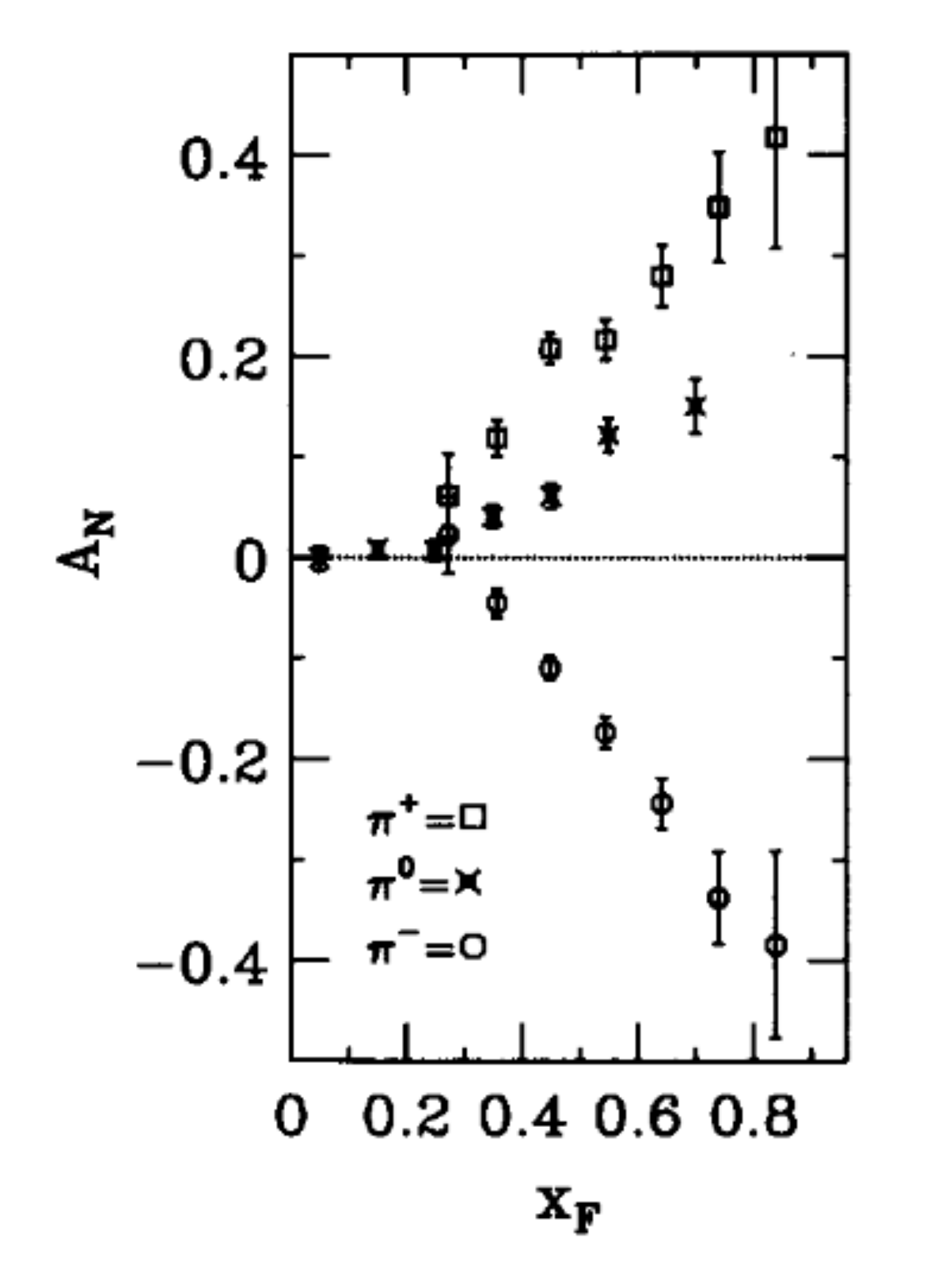} 
\isucaption{Fermilab E704 transverse SSAs for $\pi^{+}$, $\pi^{0}$ and $\pi^{-}$ from $p$+$p$ collisions at $\sqrt{s}=20$~GeV as a function of Feynman-x (from~\cite{Adams:1991cs}).}\label{fig:E704_AN}
\end{figure}

\begin{figure}[h!tb]
 \centering\includegraphics*[width=0.9\columnwidth]{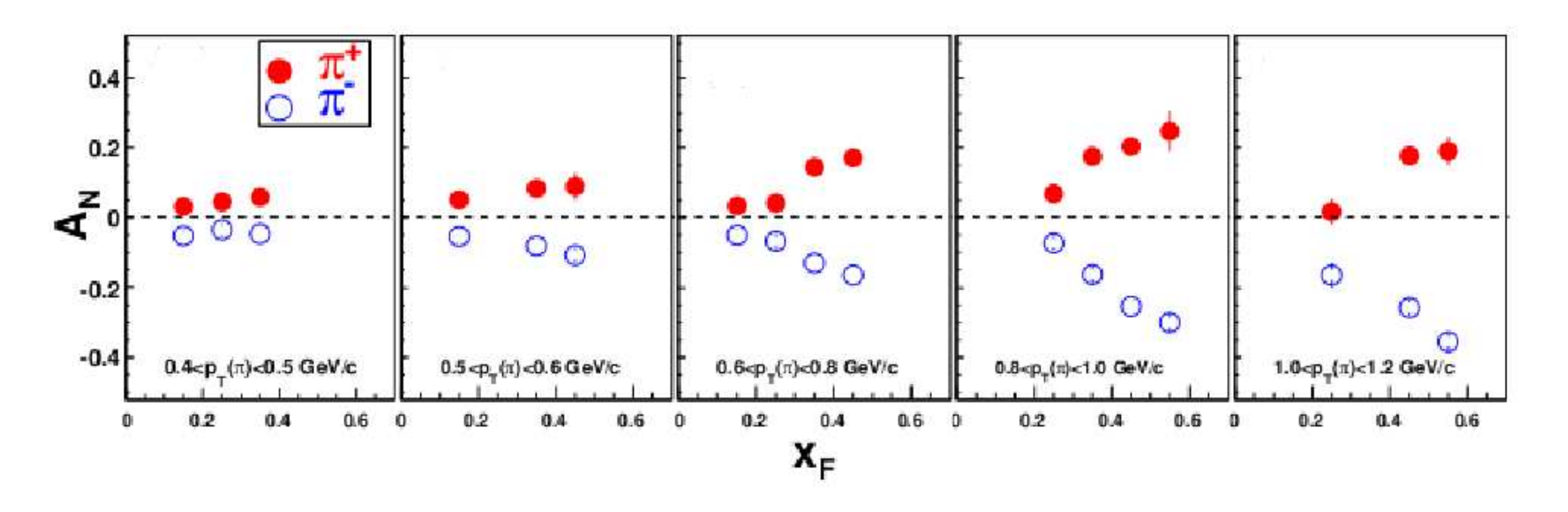} 
\isucaption{Brahms transverse SSAs for $\pi^{+}$ and $\pi^{-}$ from $p$+$p$ collisions at $\sqrt{s}=62.4$~GeV as a function of Feynman-x for various ranges of transverse momenta (from~\cite{Arsene:2008mi}).}\label{fig:Brahms_AN}
\end{figure}

\begin{figure}[h!tb]
 \centering\includegraphics*[width=0.6\columnwidth]{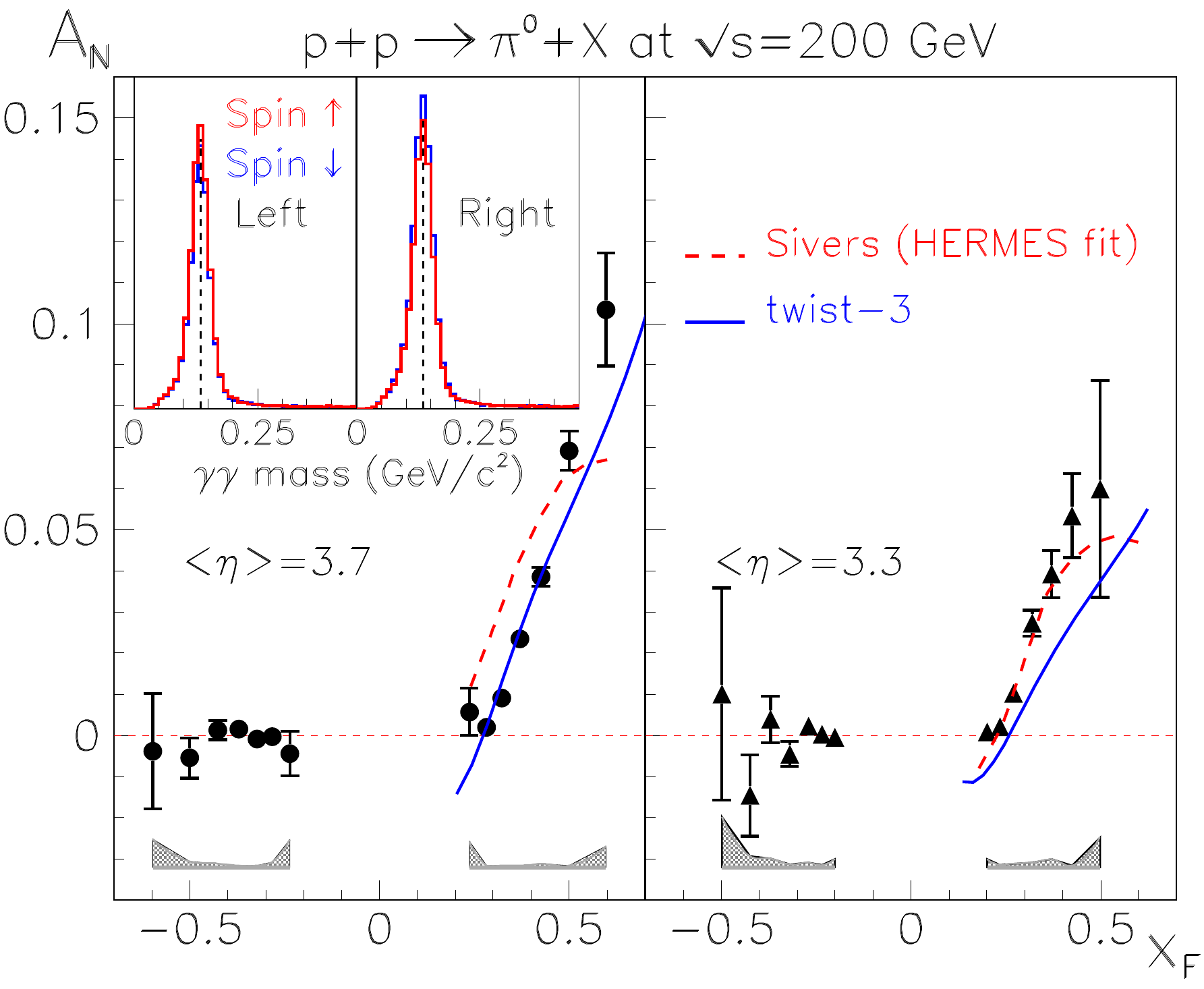} 
\isucaption{STAR transverse SSAs for $\pi^{0}$ from $p$+$p$ collisions at $\sqrt{s}=200$~GeV as a function of Feynman-x (from~\cite{Abelev:2008qb}).}\label{fig:STAR_AN}
\end{figure}

Unexpectedly large transverse SSAs of up to $\sim 40$\% were first observed in pion production from a $p^{\uparrow}$ beam\footnote{The $\uparrow$ denotes a transversely polarized proton} incident on a liquid hydrogen target in 1976 at the Argonne zero-gradient synchrotron (ZGS)~\cite{Klem:1976ui} and subsequently in a number of experiments using hadronic collisions over a range of energies extending up to $\sqrt{s}=200$~GeV.  Several examples of such measurements are:

\begin{itemize}
\item
$\pi^{+/-}$ production from a 22~GeV/$c$ polarized proton beam incident on C at the Brookhaven AGS~\cite{Allgower:2002qi}
\item
$\pi^{0}$ production from a 24~GeV/$c$ proton beam incident on a $p^{\uparrow}$ target at the CERN experiment PS141~\cite{Antille:1980th}
\item
$\pi^{+/-}$ production from a 200~GeV/$c$ polarized proton beam incident on a liquid Hydrogen target at the E704 experiment~\cite{Adams:1991cs} (Fig.~\ref{fig:E704_AN})
\item
$\pi^{0}$  production from a 200~GeV/$c$ $\bar{p}$ beam incident on a $p^{\uparrow}$ target at the Fermilab E581 experiment~\cite{Adams:1991rw} 
\item
$\pi^{+/-}$ and $K^{+/-}$ production from $p^{\uparrow}$+$p$ collisions with $\sqrt{s}=$62.4~GeV at the BRAHMS experiment at BNL~\cite{Arsene:2008mi} (Fig.~\ref{fig:Brahms_AN})
\item
$\pi^{0}$ production from $p^{\uparrow}$+$p$ collisions with $\sqrt{s}=$200~GeV at the STAR experiment at BNL~\cite{Adams:2003fx,Abelev:2008qb} (Fig.~\ref{fig:STAR_AN})
\item
$\pi^{0}$ production from $p^{\uparrow}$+$p$ collisions with $\sqrt{s}=$200~GeV and $\sqrt{s}=$62.4~GeV at the PHENIX experiment at BNL~\cite{Adler:2005in,Aidala:2008qj} (Fig.~\ref{fig:PHENIX_pi0})
\end{itemize}
The asymmetries have also been observed in SIDIS using $p^{\uparrow}$~\cite{Airapetian:2004tw,Diefenthaler:2007rj} and polarized deuteron \cite{Alexakhin:2005iw,Martin:2007au,Alekseev:2008dn} targets.

Transverse SSAs are typically measured as a function of Feynman-x, $x_{F} = x_{1} - x_{2}$, where $x_{1}$ is the $x$ of the scattering parton from the polarized and $x_{2}$ from the unpolarized proton.  The asymmetries become non-zero around $x_{F}$=0.2 and are zero for $x_{F}<0$, statements that hold true for collisions at $\sqrt{s}$=19.4~GeV, (Fig.~\ref{fig:E704_AN}), $\sqrt{s}$=62.4~GeV (Fig.~\ref{fig:Brahms_AN}), and $\sqrt{s}$=200~GeV (Fig.~\ref{fig:STAR_AN}).  Because the up and down valence quarks dominate for $x\gtrsim0.1$, we can infer that the transverse SSA is dominated by effects from valence quarks in the polarized proton.

The theoretical explanations of large transverse SSAs typically suggest additional parton distribution functions which describe correlations between the polarization of the proton and the partons which make up that proton:
\begin{description}
\item[The Sivers Function.]
  Distribution of unpolarized quarks in a transversely polarized proton.
\item[The Boer-Mulders Function.]
  Distribution of transversely polarized quarks in an unpolarized proton.
\item[Transversity.]
  Distribution of transversely polarized quarks in a transversely polarized proton.
\end{description}
In addition to these distributions, transverse SSAs can be produced by the Collins effect (the correlation between the spin of a polarized quark and hadrons in a fragmenting jet).

While these distributions each describe some aspect of the hadronic structure, the mechanisms for creating asymmetries are quite different.  In reality, it is likely that no single effect causes the transverse SSA but that effects from all of these additional parton distribution functions contribute.

\section{The Sivers Effect}\label{sec:Sivers}

In 1989, Dennis Sivers proposed that partonic motion transverse to the proton's momentum, $\vec{k_{T}}$, ought to be taken into account when calculating PDFs for transversely polarized protons~\cite{Sivers:1989cc,Sivers:1990fh}.  He suggested that this could be done by creating transverse momentum dependent parton distribution functions $f_{q/P^{\uparrow}}(x,\vec{k_{T}},\vec{S_{T}})$ (TMDs) as opposed to the usual $f_{1}^{q}(x)$ used for unpolarized protons, where $q$ corresponds to the flavor of quark, anti-quark, or gluon. The new TMD includes a so-called Sivers function, $\Delta^{N} f_{q/P^{\uparrow}}(x,\vec{k_{T}})$: 
\begin{equation}
 \displaystyle f_{q/P^{\uparrow}}(x,\vec{k_{T}},\vec{S_{T}})=f_{1}^{q}(x,\vec{k_{T}}) + \frac{1}{2} \Delta^{N} f_{q/P^{\uparrow}}(x,\vec{k_{T}}) \hat{S_{T}}\cdot\left(\hat{P}\times\hat{k_{T}}\right)
\label{eq:Sivers_TMD}
\end{equation}
for a parton with transverse momentum $\vec{k_{T}}$ and spin component $\vec{S_{T}}$ transverse to the three-momentum $\vec{P}$ of the proton.  Alternatively, the TMD can be written as
\begin{equation}
 \displaystyle f_{q/P^{\uparrow}}(x,\vec{k_{T}},\vec{S_{T}})=f_{1}^{q}(x,\vec{k_{T}}) + \frac{1}{2} \Delta^{N} f_{q/P^{\uparrow}}(x,\vec{k_{T}}) \sin(\phi_{S}-\varphi)
\end{equation}
where $\phi_{S}$ is the azimuthal angle of the proton spin and $\varphi$ the angle of the partonic $\vec{k_{T}}$.  Because the Sivers function is weighted by a sinusoidal modulation, it can be observed as a difference between TMDs, corresponding to the numerator of Eq.~\ref{eq:AN_sigma}, or directly as a sinusoidal modulation in particle production.

The Sivers effect is not universal.  Instead, it should have the same magnitude and opposite sign in SIDIS relative to Drell-Yan production (quark/anti-quark annihilation into a pair of leptons) from $p$+$p$ collisions~\cite{Collins:2002kn}.  There has not yet been an experimental verification of this sign-change, but such a verification will be quite important in proving the existence of a Sivers effect.

Unfortunately, Eq.~\ref{eq:Sivers_TMD} and the associated formalism is not valid for hadron production from hadron-hadron collisions.  It was shown in 2007 that factorization is violated for the TMD approach to the Sivers function in the case of back-to-back hadron production from hadron-hadron collisions~\cite{Collins:2007nk}.  Furthermore, in 2010, such violations were shown to occur for any hadronic production from hadron-hadron collisions~\cite{Rogers:2010dm}. 

An alternative approach to calculating the transverse SSA caused by unpolarized quarks in a transversely polarized hadron was developed by Qiu and Sterman, wherein the transverse SSA is generated by an incoming or outgoing parton which exchanges a gluon with a parton in the initial state hadron remnant~\cite{Qiu:1991pp}.  Kanazawa and Koike found that the same effect could occur for the exchange of a gluon with a parton in the final state remnant~\cite{Kanazawa:2000hz}.  The collinear approach, which does not violate factorization, orders the perturbation theory in powers of $\left(\frac{2 P\cdot q}{Q^{2}}\right)^{s}\left(\frac{1}{Q}\right)^{d-s-2}$, where $P$ is the four momentum of the polarized hadron, $Q^{2}$ the momentum transfer, $q$ the momentum of the quark in the polarized hadron, $d$ the dimension and $s$ the spin of the operator in a given term of the expansion.  The size of the contributions decrease with the twist, $d-s$, of the contributing operators, and it is the twist-3 terms, called multi-parton correlation functions, which cause the transverse SSA.  For processes involving only gluons we will refer to these twist-3 terms as trigluon correlation functions.

Because the collinear approach expands about $\frac{1}{Q}$, the perturbation theory is valid for collisions with transverse momentum transfer $q_{\perp}$ and $Q \gg \Lambda_{QCD}$,\footnote{$\Lambda_{QCD}$ depends on the renormalization scheme and does not have a clear physical interpretation, but it is roughly the energy scale at which quarks bind into hadrons.} while the TMD approach requires a separation between scales of the transverse and longitudinal momentum transfer so that $q_{\perp}\ll Q$.  The two formalisms have been shown to be equivalent in their region of overlap $\Lambda_{QCD} \ll q_{\perp} \ll Q$~\cite{Ji:2006ub}.\footnote{Throughout the rest of the document we will use the term `Sivers effect' when discussing the TMD and collinear approaches collectively.  This should not be confused with term `Sivers function,' which is only applicable to the TMD approach.}

\section{Transversity}\label{sec:Transversity}

The transversity distribution describes the kinematics of transversely polarized quarks in a transversely polarized hadron~\cite{Ralston:1979ys,Jaffe:1991ra}.  Transversity cannot be responsible on its own for the large transverse SSAs discussed in Section~\ref{sec:SSA_measurements}, because quarks from an unpolarized hadron in hadron-hadron collisions will na\"\i vely scatter with an equal probability in all polarizations.  Likewise, in SIDIS, the virtual photon interacting with quarks in the polarized hadron will not have a preferred polarization direction.  Instead, the transversity distribution must couple with another distribution to create a transverse SSA.  The distribution coupled with transversity can either be a property of the initial state of the unpolarized hadron, such as the Boer-Mulders function (Section~\ref{sec:Boer-Mulders}), or it can be a final state distribution, such as the Collins function (Section~\ref{sec:Collins}).  

Soffer has shown that the transversity distribution is bounded by measurements of the unpolarized and longitudinally polarized PDFs~\cite{Soffer:1994ww}.  The model-independent `Soffer bound' requires that 
\begin{equation*}
\left|\Delta_{T} q(x)\right| \leq \frac{1}{2}(q(x) + \Delta q(x))
\end{equation*}
where $\Delta_{T} q(x)$ is the transversity distribution for a quark of flavor $q$, $q(x)$ is the unpolarized PDF, and $\Delta q(x)$ the longitudinally polarized PDF. Because gluons are massless, they must have a definite helicity, which means that the transversity distribution for gluons must be zero.

\section{The Boer-Mulders Effect}\label{sec:Boer-Mulders}

Expanding on the idea of the Sivers effect, Boer and Mulders suggested that transverse SSAs could also be generated by polarized quarks in an unpolarized hadron~\cite{Boer:1997nt}.  The Boer-Mulders function, $h_{1q}^{\perp}(x,\vec{k_{T}})$, corresponds to a modification of the unpolarized PDFs which takes into account the polarization of the constituent quarks in an unpolarized hadron.  The parton distribution function for a polarized quark in an unpolarized hadron becomes
\begin{equation}
f_{q^{\uparrow}/P}(x,\vec{k_{T}},\vec{S_{T}})=\frac{1}{2}\left(f_{1}^{q}(x,\vec{k_{T}}) - h_{1q}^{\perp}(x,\vec{k_{T}}) \frac{\hat{S_{T}}\cdot\left(\hat{P}\times\hat{k_{T}}\right)}{M}\right)
\end{equation} 
where $M$ is the mass of the hadron~\cite{Bacchetta:2004jz}.  A non-zero Boer-Mulders function in an unpolarized hadron coupled with transversity in a polarized hadron could potentially lead to transverse SSAs.  Such a distribution should also create modulations in the azimuthal distribution of particle yields from \textit{un}polarized hadron collisions.

\section{The Collins Effect}\label{sec:Collins}

In an alternative explanation of large transverse SSAs, Collins suggested that a transversely polarized quark should fragment differently than an unpolarized quark, causing an azimuthal asymmetry about the jet axis~\cite{Collins:1992kk}.  If an outgoing quark has spin $\vec{S_{q}}$, the unpolarized fragmentation function $D_{1}^{q}(z,\vec{p_{T}})$ is modified by a Collins fragmentation function $\Delta^{N}D_{h/q^{\uparrow}}(z,\vec{p_{T}})$ for a hadron $h$ with mass $M_{h}$ and transverse momentum $p_{T}$, carrying a fraction $z$ of the quark's momentum $\vec{k_{q}}$:
\begin{equation}
 \displaystyle D_{h/q^{\uparrow}}(z,\vec{p_{T}},\vec{S_{q}})=D_{1}^{q}(z,\vec{p_{T}})+2 |p_{T}| \Delta^{N}D_{h/q^{\uparrow}}(z,\vec{p_{T}}) \hat{S_{q}}\cdot\left(\hat{p_{T}}\times\hat{k_{q}}\right).
\end{equation} 
If there is a non-zero transversity distribution in the polarized hadron, a quark going into the hard scattering will have some polarization, and it is likely that an outgoing quark will maintain that polarization.  The polarization of the quark causes its fragmentation function to be modified in such a way that there is a sinusoidal modulation in hadron yields about the jet axis.

The Collins function can not be larger than the unpolarized fragmentation function it modifies, a constraint called the positivity bound.

\section{PDF Extractions and Global Fits}\label{sec:trans_global}

The kinematics of polarized SIDIS are such that the Sivers and Collins functions create different angular distributions and can be separated experimentally.  Instead of the one dimensional distribution given in  Eq.~\ref{eq:AN_sigma}, SIDIS experiments take the target polarization direction into account and measure the two dimensional distribution
\begin{equation}
A^{h}_{UT}(\phi,\phi_{S}) = \frac{1}{\mathcal{P}}\frac{N_{h}^{\uparrow}(\phi,\phi_{S})-N_{h}^{\downarrow}(\phi,\phi_{S})}{N_{h}^{\uparrow}(\phi,\phi_{S})+N_{h}^{\downarrow}(\phi,\phi_{S})}
\end{equation}
for a hadron of type $h$.  $\mathcal{P}$ is the target polarization, $\uparrow$ and $\downarrow$ the spin direction, $\phi_{S}$ the azimuthal angle between the interaction plane and the target spin, and $\phi$ the azimuthal angle between the interaction plane and the outgoing hadron momentum.  The Sivers moment $\left<sin(\phi - \phi_{S})\right>^{h}_{UT}$ and Collins moment $\left<sin(\phi + \phi_{S})\right>^{h}_{UT}$ can then be extracted by fitting the two dimensional distribution with
\begin{equation}
\frac{A^{h}_{UT}}{2} =\left<sin(\phi - \phi_{S})\right>^{h}_{UT} sin(\phi-\phi_{S}) + \left<sin(\phi + \phi_{S})\right>^{h}_{UT} \frac{1-\left<y\right>}{A(\left<x_{B}\right>,\left<y\right>)} sin(\phi+\phi_{S}),
\end{equation}
where $y=\frac{\vec{P} \cdot \vec{q}}{\vec{P} \cdot \vec{k}}$ (with $\vec{P}$, $\vec{q}$, and $\vec{k}$ defined in Section~\ref{sec:DIS}) and
\begin{equation*} 
A(x,y) \equiv \frac{y^{2}}{2} + \frac{(1-y)(1+R(x,y))}{1+\frac{2 M x_{B}}{E y}}
\end{equation*}
with $R(x,y)$ the ratio of the longitudinal to transverse structure functions discussed in Section~\ref{sec:struct_func}.

The Sivers moment is sensitive only to the Sivers function and can be extracted directly.  The Collins moment, on the other hand, is sensitive to a convolution of the Collins effect and transversity, and another measurement is necessary to extract both distributions.   

\begin{figure}[h!tb]
 \centering\includegraphics*[width=0.75\columnwidth]{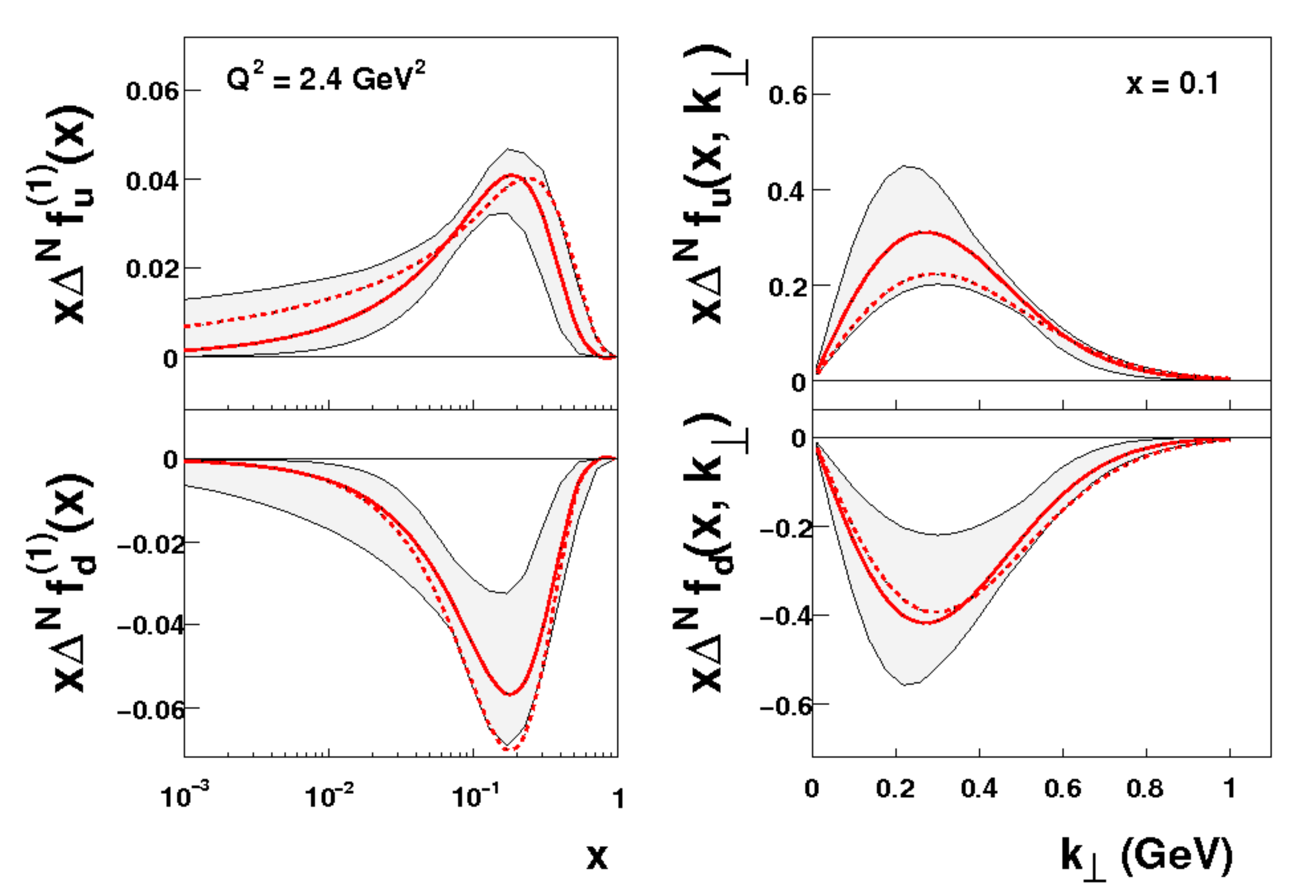} 
\isucaption[Sivers function from a global fit to SIDIS data.]{Sivers function from a global fit to SIDIS data.  The solid line is the central value of the fit from~\cite{Anselmino:2008sga} and the dashed line from~\cite{Anselmino:2005ea} (Figure from~\cite{Anselmino:2008sga}).}\label{fig:Anselmino_Sivers}
\end{figure}

A global fit to extract the Sivers function~\cite{Anselmino:2008sga} was performed on SIDIS data with a hydrogen target from HERMES~\cite{Airapetian:2004tw,Diefenthaler:2007rj} and a deuteron target from COMPASS~\cite{Alexakhin:2005iw,Martin:2007au,Alekseev:2008dn}.  The first moment in $k_{T}$ of the Sivers function
\begin{equation}
\Delta^{N}f^{(1)}_{q/P^{\uparrow}}(x) \equiv \displaystyle\int d^{2}k_{T} \frac{k_{T}}{4 M} \Delta^{N}f_{q/P^{\uparrow}}(x,k_{T})
\end{equation}
from the fit is shown in Fig.~\ref{fig:Anselmino_Sivers}.  The data favors a non-zero first moment in SIDIS which is positive for up quarks, negative and nearly identical in magnitude for down quarks.\footnote{Recall that the Sivers function should have an opposite sign for hadron-hadron collisions.} 

\begin{figure}[h!tb]
 \centering\includegraphics*[width=\columnwidth]{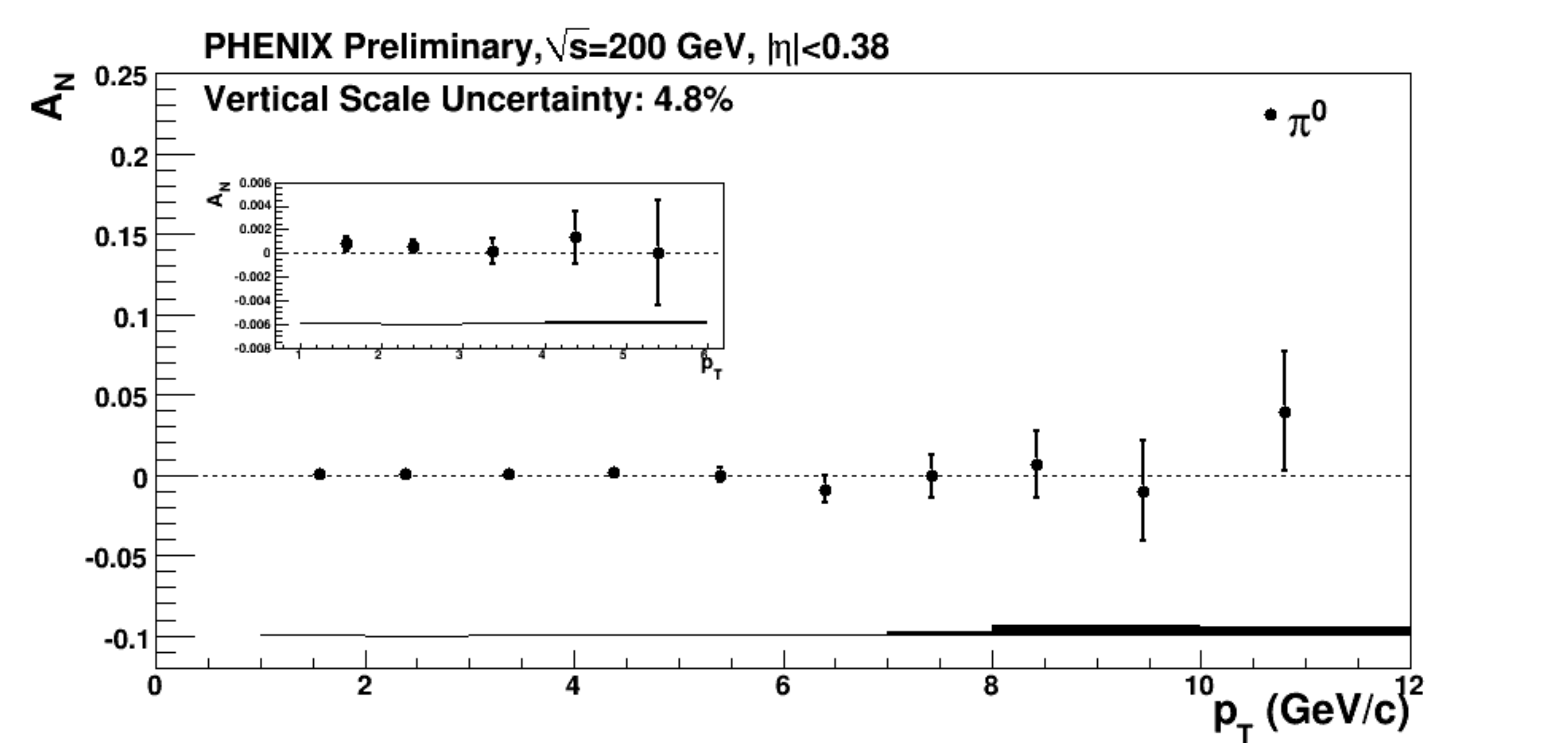} 
\isucaption{Preliminary PHENIX transverse SSAs for $\pi^{0}$ from $p$+$p$ collisions at $\sqrt{s}=200$~GeV as a function of $p_{T}$.}\label{fig:PHENIX_pi0}
\end{figure}

\begin{figure}[h!tb]
 \centering\includegraphics*[width=0.55\columnwidth]{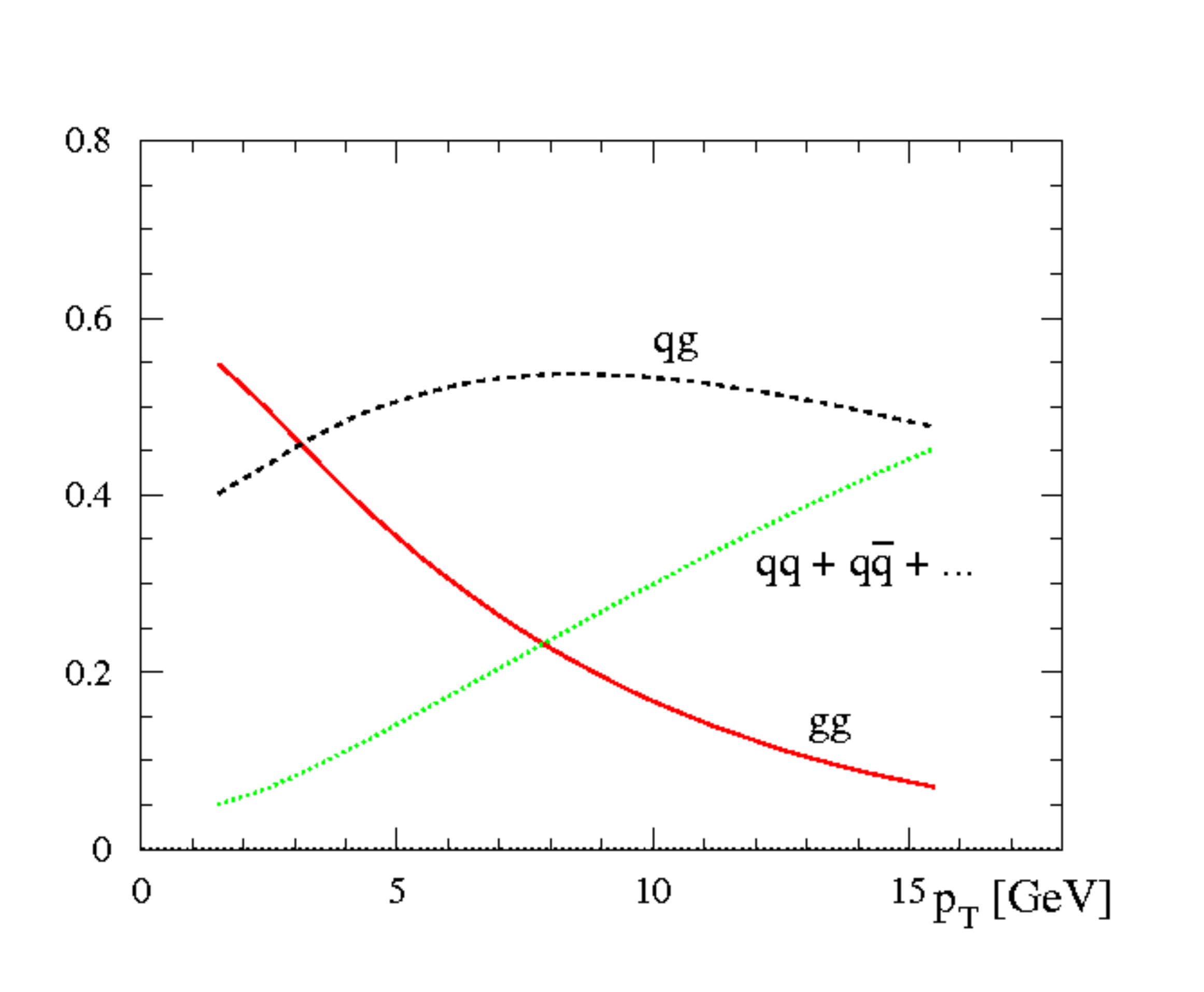} 
\isucaption[Incoming partons from processes contributing to $\pi^{0}$ production at mid-rapidity from $p$+$p$ collisions at $\sqrt{s}$=200~GeV.]{Incoming partons from processes contributing to $\pi^{0}$ production at mid-rapidity from $p$+$p$ collisions at $\sqrt{s}$=200~GeV (from~\cite{Stratmann:2007hp}).}\label{fig:pi0_processes}
\end{figure}

Because an electromagnetic probe is used, the gluon Sivers and trigluon correlation functions are not well constrained by SIDIS measurements.  Measurements of the transverse SSA for $\pi^{0}$s at PHENIX~\cite{Adler:2005in} seem to prefer a small gluon Sivers function at mid-rapidity~\cite{Anselmino:2006fq}, and recent preliminary results (Fig.~\ref{fig:PHENIX_pi0}) confirm a small transverse SSA out to large $p_{T}$.  A numerical constraint from the data has not been made, and such constraints are complicated by the process dependence of $\pi^{0}$ production.  At low $p_{T}$, $\pi^{0}$ production from $p$+$p$ collisions at $\sqrt{s}$=200~GeV is dominated by gluon-gluon collisions, but quark-gluon collisions become more important at moderate $p_{T}$, and quark-quark collisions become dominant at sufficiently large $p_{T}$~\cite{Stratmann:2007hp} (see Fig.~\ref{fig:pi0_processes}).

In order to extract the Collins and Transversity distributions from SIDIS, those measurements are taken together with measurements of the Collins distribution from BELLE~\cite{Abe:2005zx}.  The BELLE detector is located at the KEKB accelerator, which collides unpolarized $e^{+}$ and $e^{-}$, and the Collins function appears as a modulation $\cos(\phi_{1}+\phi_{2})$ in the azimuthal angles $\phi_{1}$ and $\phi_{2}$ of outgoing hadrons from back-to-back jets with respect to the reaction plane.  Because there are no hadrons in the initial state, the measurement from BELLE is only sensitive to the Collins function and not to transversity.  This means that the BELLE data can be used in a global fit with data from HERMES and COMPASS to extract both the Collins and Transversity distributions~\cite{Anselmino:2008jk}.  

\begin{figure}[h!tb]
 \centering\includegraphics*[width=0.75\columnwidth]{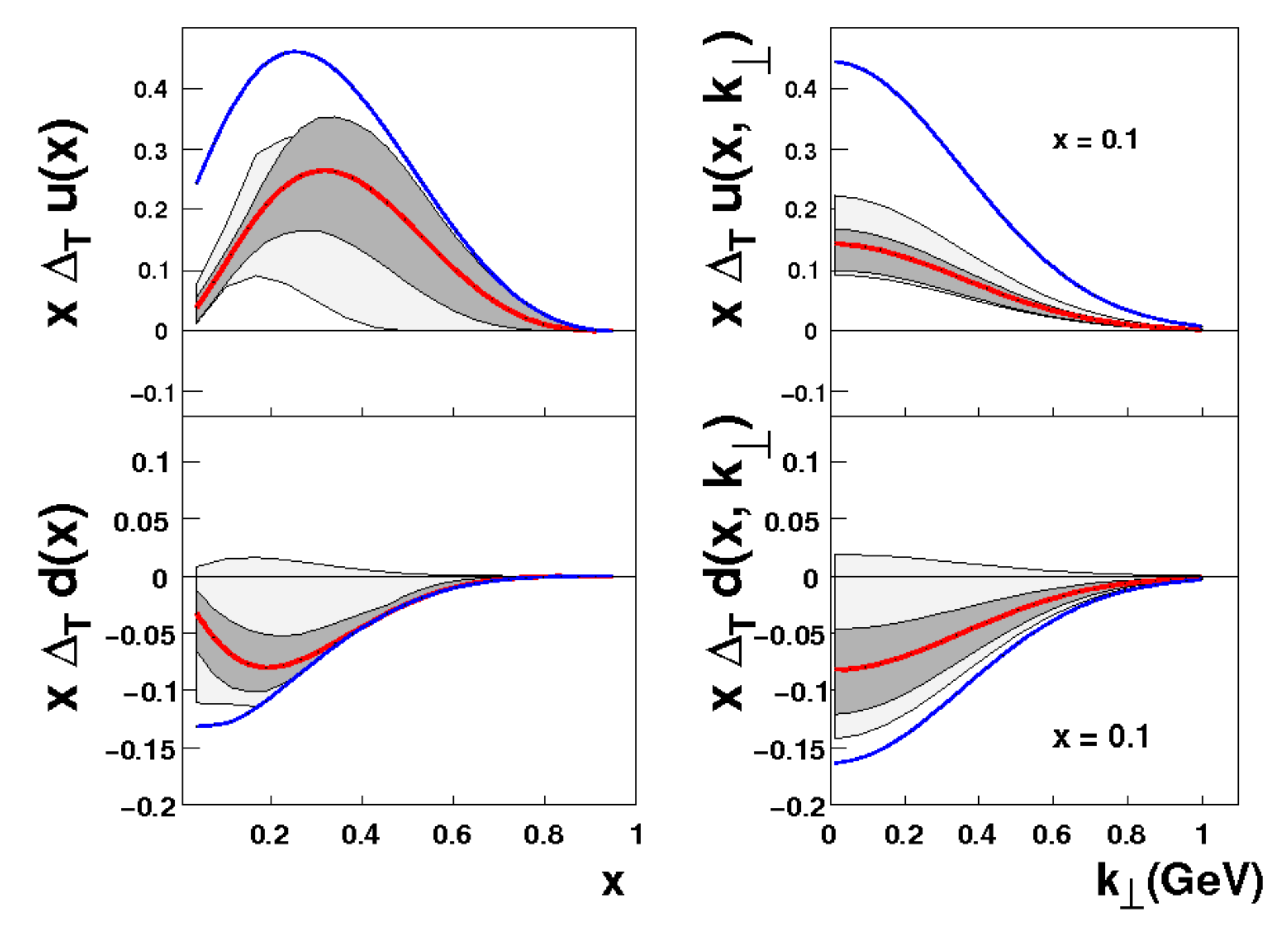} 
\isucaption[Transversity distribution from a global fit to SIDIS and $e^{+}e^{-}$ data.]{Transversity distribution from a global fit to SIDIS and $e^{+}e^{-}$ data.  The blue line represents the Soffer bound.  Light uncertainty bands are those from~\cite{Anselmino:2007fs}, while dark bands are from~\cite{Anselmino:2008jk} (Figure from~\cite{Anselmino:2008jk}).}\label{fig:Anselmino_Transversity}
\end{figure}

Transversity distributions from the global fit (Fig.~\ref{fig:Anselmino_Transversity}) are non-zero and smaller than the Soffer bound and again appear to have an opposite sign for down quarks relative to up quarks.

\begin{figure}[h!tb]
 \centering\includegraphics*[width=0.75\columnwidth]{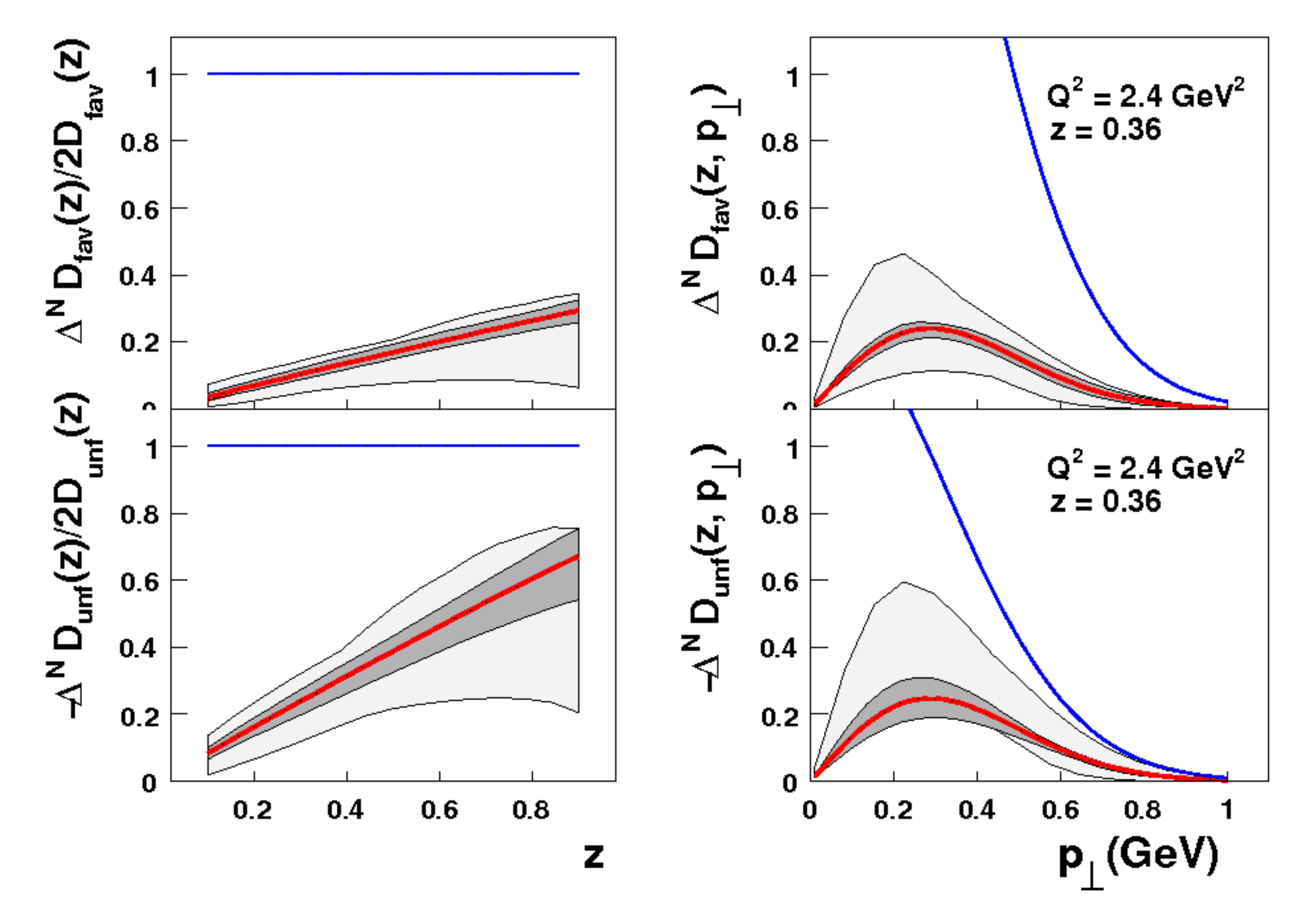} 
\isucaption[Collins function from a global fit to SIDIS and $e^{+}e^{-}$ data.]{Collins function from a global fit to SIDIS and $e^{+}e^{-}$ data.   The dark blue line represents the positivity constraint.  Light uncertainty bands are those from~\cite{Anselmino:2007fs}, while dark bands are from~\cite{Anselmino:2008jk} (Figure from~\cite{Anselmino:2008jk}).}\label{fig:Anselmino_Collins}
\end{figure}

Collins distributions are extracted separately for favored fragmentation, where the observed hadron contains a valence quark in the flavor of the struck quark, and unfavored fragmentation, where it does not.  The Collins fragmentation functions extracted from the global fit (Figure~\ref{fig:Anselmino_Collins}) show an increase of the Collins effect with respect to $z$ as we would expect.\footnote{If a single hadron carries most of the momentum of a jet, the azimuthal distribution created by the quark polarization will not be diluted by extensive fragmentation.}

The Boer-Mulders function is a property of unpolarized protons and can, in principle, be extracted from unpolarized SIDIS or Drell-Yan.  Recent measurements at COMPASS~\cite{Kafer:2008ud,Bressan:2009eu} and HERMES~\cite{Giordano:2009hi} of the $\cos2\phi$ modulation in azimuthal angle $\phi$ of hadrons with respect to the production plane from unpolarized SIDIS were used in a proof of principle global fit to extract Boer-Mulders functions~\cite{Barone:2009hw}.

\begin{figure}[h!tb]
 \centering\includegraphics*[width=0.85\columnwidth]{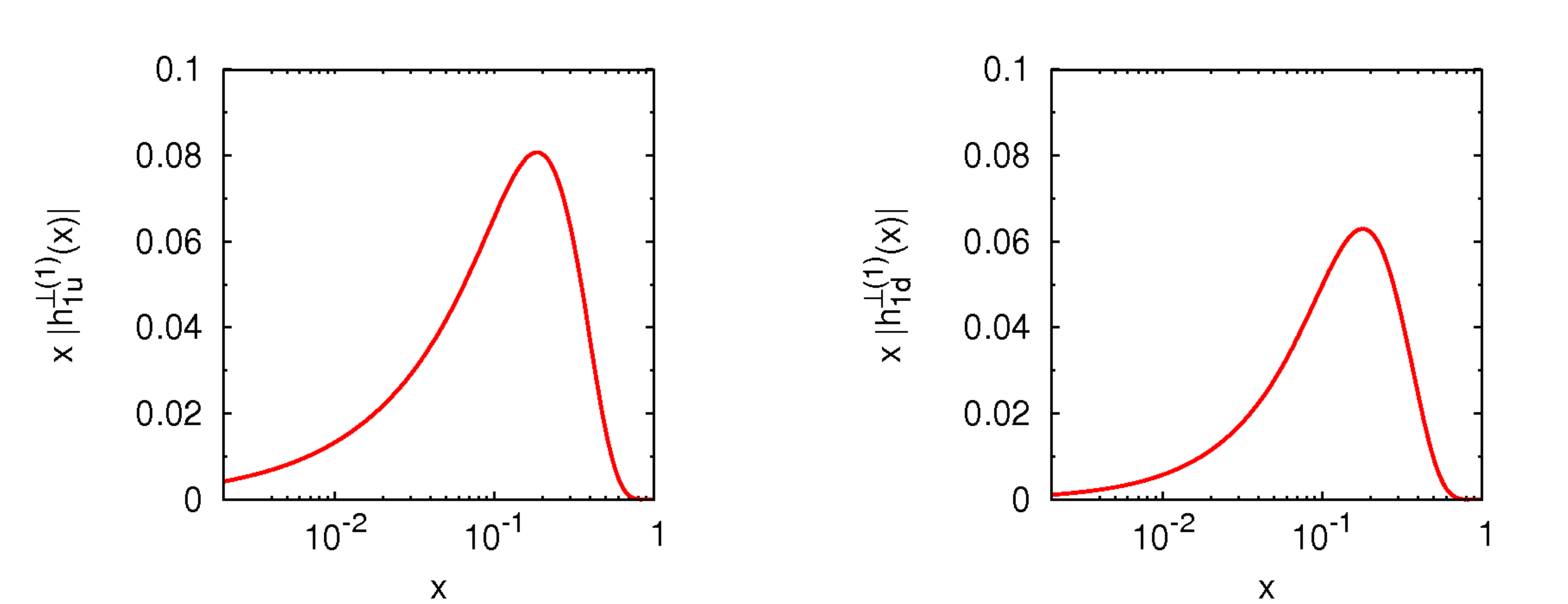} 
\isucaption[Boer-Mulders function from a proof-of-principle global fit to SIDIS data assuming that the Boer-Mulders function is proportional to the Sivers function.]{Boer-Mulders function from a proof-of-principle global fit to SIDIS data assuming that the Boer-Mulders function is proportional to the Sivers function from the global fit in~\cite{Anselmino:2008sga} (Figure from~\cite{Barone:2009hw}).}\label{fig:Prokudin_BoerMulders}
\end{figure}

One complication to the extraction of the Boer-Mulders function is the Cahn effect~\cite{Cahn:1978se,Cahn:1989yf}, a $\cos2\phi$ modulation in hadron production due to non-zero $k_{T}$ in the proton.  Fortunately, the Cahn effect has a different kinematic dependence than the Boer-Mulders effect in $p_T$, $k_T$, and $Q^{2}$, and the two can be separated in the fit.  While the available data do not allow a full extraction of the Boer-Mulders effect, a fit can be performed assuming that the Boer-Mulders function is proportional to the Sivers function extracted in~\cite{Anselmino:2008sga}.  Results of a fit which makes such an assumption to extract the first moment in $k_{T}$ of the Boer-Mulders function
\begin{equation}
h_{1q}^{\perp(1)}(x) \equiv \displaystyle\int d^{2}k_{T} \frac{k^{2}_{T}}{2 M^{2}} h_{1q}^{\perp}(x,k^{2}_{T})
\end{equation}
are shown in Fig.~\ref{fig:Prokudin_BoerMulders}.  The fit prefers a Boer-Mulders function which has approximately the same magnitude for up and down quarks, twice as large as the Sivers function for the up quark.\footnote{The plots shown in this document are a bit confusing on this last point, because the Sivers function shown in Fig.~\ref{fig:Anselmino_Sivers} is scaled differently than the Boer-Mulders function in Fig.~\ref{fig:Prokudin_BoerMulders}.  A thorough discussion of the notation used for the various transverse PDFs and the relationship between different notations is given in~\cite{Bacchetta:2004jz}.}

The global fits performed for the Sivers, transversity, Collins, and Boer-Mulders distributions all prefer magnitudes which are inconsistent with zero, implying that all of these distributions play some part in the large $A_N$ seen by experiments.  These fits are still in their early stages, however, and a great deal more data is required to sort out which effects dominate.  The gluon Sivers effect, for instance, is still poorly constrained, and further measurements from hadron-hadron collisions, like the measurement presented in Chapter~\ref{ch:jpsi_AN}, will be necessary to fully explain the large measured transverse SSAs.

%% file: chapters/jpsi_intro/jpsi_intro.tex
\chapter{The $J/\psi$ Meson}\label{ch:jpsi_intro}

\section{Charmonium}

\begin{figure}[h!tb]
        \centering\includegraphics*[width=0.75\columnwidth]{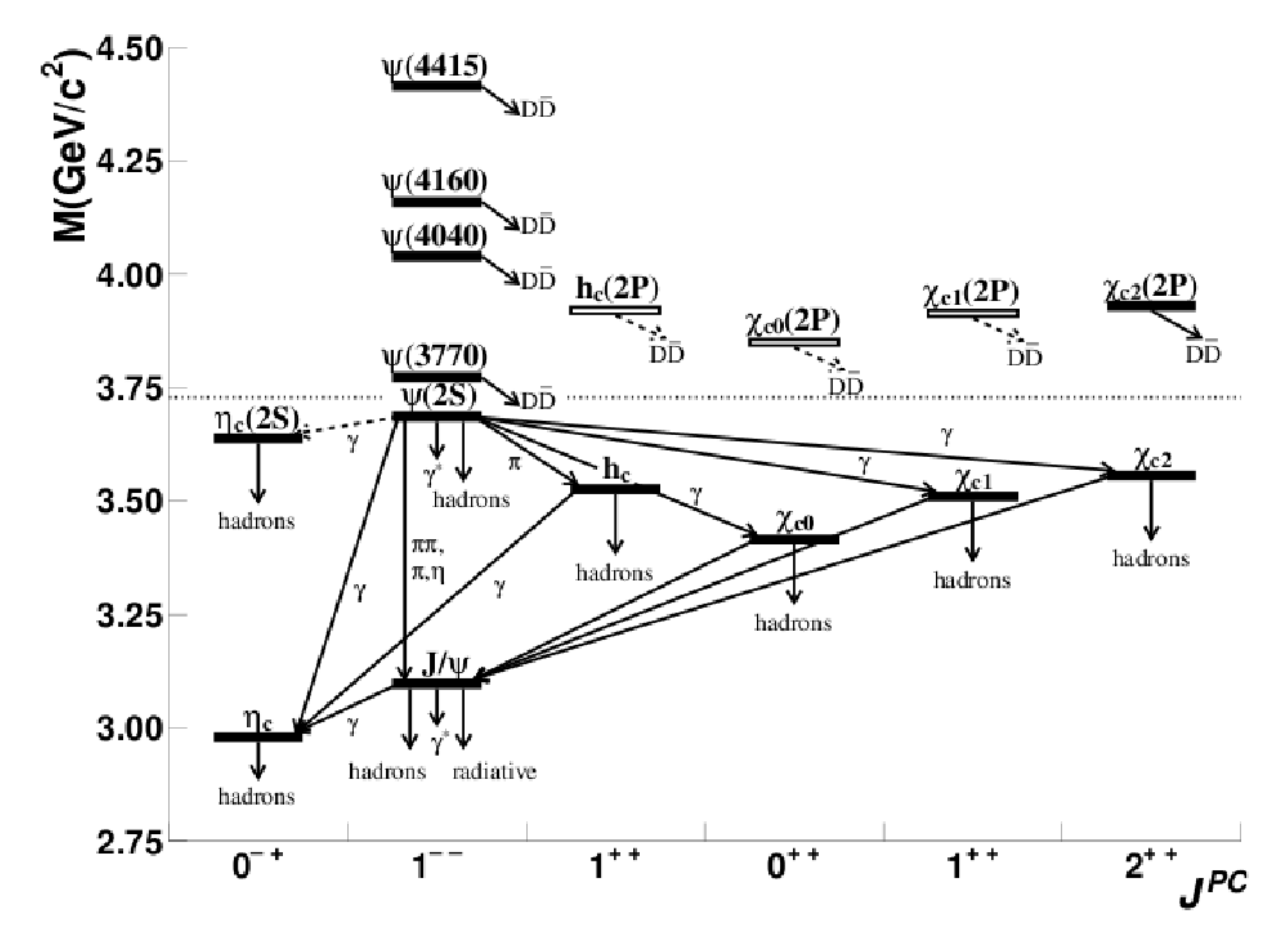} 
\isucaption[Various charmonium resonances and their decay channels.]{Various charmonium resonances and their decay channels (from~\cite{PDG}).  Three $\chi_{c}$ states, as well as the $\psi(2\text{S})$, contribute to $J/\psi$ production in addition to those directly produced.}\label{fig:charmonium_states}
\end{figure}

Analogous to the term `positronium' (a bound state of $e^{+}e^{-}$), the term `charmonium' refers to any bound state of a charm and anti-charm quark.  The various charmonium states differ in their total angular momentum, charge conjugation, parity, and principal quantum number.  It was proposed soon after the discovery of such bound states that a spectroscopic arrangement of the various resonances would be beneficial~\cite{De_Rujula:1976qd}.  Such an arrangement can be seen in Fig.~\ref{fig:charmonium_states}, and continues to be used in lattice QCD to successfully predict the masses of and transitions between states~\cite{Davies:1995db,Dudek:2007wv}.  

\begin{figure}[h!tb]
        \centering\includegraphics*[width=\columnwidth]{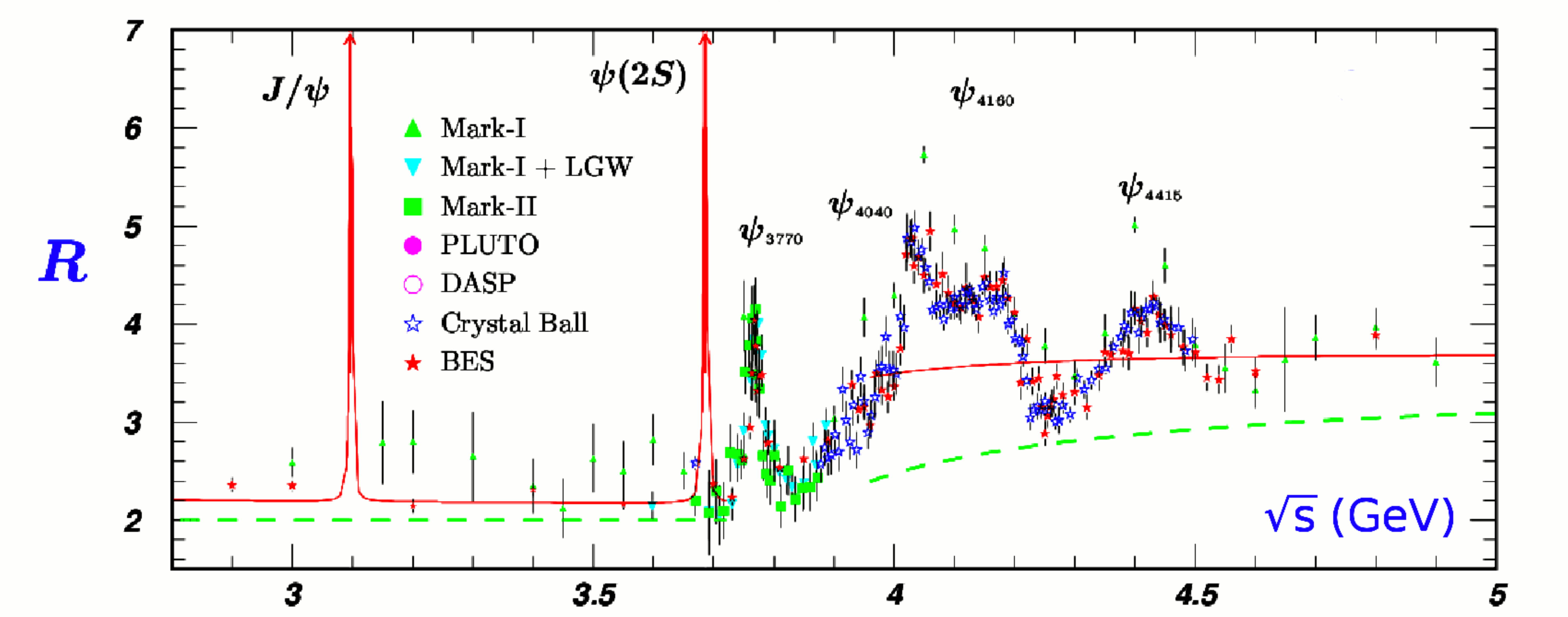} 
\isucaption[Charmonium resonances confirmed in data from various experiments.]{Various charmonium resonances confirmed in data from several experiments.  The $\hat{y}$-axis is the R value, defined as the ratio of the cross section of oppositely charged hadrons from a colliding $e^{+}$ and $e^{-}$ at center of mass energy $\sqrt{s}$  to the cross section for producing a pair of oppositely charged muons (figure from~\cite{PDG}).}\label{fig:charmonium_states_data}
\end{figure}

The most copiously produced charmonium state is the $J/\psi$, a spin-1, parity odd, charge-0 meson in an s-wave orbital state.  The $J/\psi$ meson was simultaneously discovered at BNL~\cite{Aubert:1974js} and SLAC~\cite{Augustin:1974xw}, where it was named the `$J$' and `$\psi$' respectively.  A composite name has remained, although early papers differ on the order.  Since the discovery of the $J/\psi$, many other charmonium resonances have been confirmed experimentally, a number of which can be seen in Fig.~\ref{fig:charmonium_states_data}.  It is important to note that many of the $J/\psi$ mesons observed at colliders are not directly produced from collisions but are the result of decays from other states in the charmonium spectroscopy or from decays of B mesons or heavier charmonium states.\footnote{B mesons are composed of a bottom or anti-bottom quark along with an up, down, charm, or strange quark or anti-quark.}  Recent models estimate that approximately 30$\pm$10\% of $J/\psi$ mesons produced at RHIC come from $\chi_{c}$ decays, and overall only 59$\pm$10\% of the $J/\psi$ mesons are directly produced~\cite{Brodsky:2009cf}.

\section{The OZI Rule}

\begin{figure}[h!tb]
 \centering\includegraphics*[width=0.4\columnwidth]{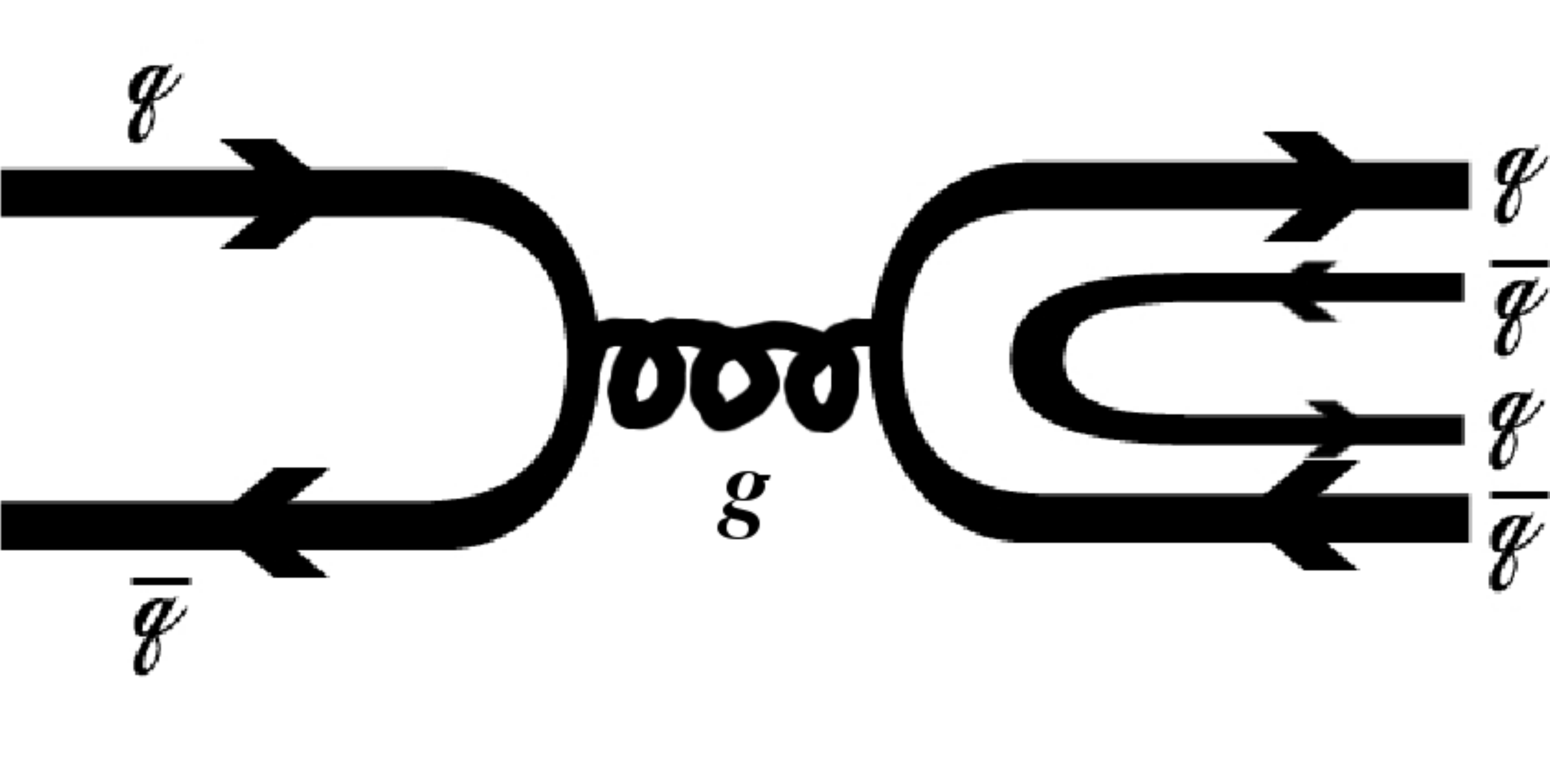} 
\isucaption[Example of a decay process suppressed by the OZI rule.]{Example of a decay process suppressed by the OZI rule.}\label{fig:OZI}
\end{figure}

Because gluons are colored and hadrons are not, interactions with a single gluon connecting the initial and final states ought to be suppressed.  The OZI rule, posited by Zweig~\cite{Zweig:1964jf} and elaborated by Iizuka~\cite{Iizuka:1966wu} and Okubo~\cite{Okubo:1963fa,Okubo:1977rk}, generalizes this principle to show that there is a supression of any interaction that can be split into separate initial and final states by cutting through gluon lines.  This simple observation has striking consequences for decaying bound states like charmonium, as it explains the suppresion of decays to hadrons relative to leptons and the associated narrow decay width (93.2$\pm$2.1~keV~\cite{PDG}) of the $J/\psi$.  While we might expect diagrams for charmonium decays to hadrons like that in Fig.~\ref{fig:OZI} to make up the entire decay width, the observed decays to leptons make up a significant fraction (Table \ref{tab:JPsi_decay_modes}).

\begin{table}[htbp]
\centering
\isucaption[Dominant $J/\psi$ decay modes.]{\label{tab:JPsi_decay_modes}Dominant $J/\psi$ decay modes.  Branching ratios are from~\cite{PDG}.}
\begin{tabular}{cc}
\hline
Decay Channel                     &  Percentage of all decays  \\ \hline
hadrons                           & 87.7$\pm$0.5\%           \\ \hline
$\gamma^{*}\rightarrow$ hadrons   & 13.50$\pm$0.30\%         \\ \hline
$e^{+}e^{-}$                      &  5.94$\pm$0.06\%         \\ \hline
$\mu^{+}\mu^{-}$                  &  5.93$\pm$0.06\%         \\ \hline
\end{tabular}
\end{table}

\section{Angular Distributions of Decay Products}\label{sec:angular_dists}

Charmonium systems are well described by hydrogen-like potentials having quantum numbers $n$, $l$, and $m$ corresponding to the principal quantum number, eignvalue for the total angular momentum $J$ squared, and magnetic quantum number respectively.  The $J/\psi$ is in an s-wave orbital state ($l=0$), meaning that it can't be `polarized' in the sense that $\left<J_{z}\right>\ne 0$, because the quantum number $m$ must be zero.\footnote{The real-space wavefunction which solves a hydrogen-like Hamiltonian contains a radial piece multiplied by Legendre polynomials, which are zero for $|m|>l$.  Likewise, the Wigner D-matrices $D^{l}_{m\lambda}$ in Eq.~\ref{eq:Wigner_D_expand} are zero for $|m|>l$.  This is not coincidental, as the $d$-functions of Eq.~\ref{eq:Big_d_Small_d} are related to the Legendre polynomials by $d^{l}_{0,0}=P_{l}(\cos\beta)$.} It is possible, though, for the spin to be preferentially aligned along some axis with an equal probability to be pointing either along or against that axis, and this alignment will cause a non-uniform angular distribituion for leptons from the decay.  Here we present a derivation of the angular distribution of leptons from $J/\psi$ decays, borrowing heavily from the much more thorough treatment of angular decay distributions in~\cite{Jackson:1965lh}.

Suppose we have a particle $\gamma$ in eigenstates $j$ and $m$ of total angular momentum and magnetic sub-state respectively.  The $\gamma$ decays into two particles $\alpha$ and $\beta$ with momenta $\vec{p}$ and $-\vec{p}$ in the $\gamma$ rest frame and they have corresponding helicities $\lambda_{\alpha}$ and $\lambda_{\beta}$.\footnote{Helicity is defined as the projection of a particle's spin along its direction of motion, $\lambda=\vec{S}\cdot\vec{P}$.}  The decayed state is characterized by a ket
$|\hat{p}\lambda_{\alpha}\lambda_{\beta}\rangle$ where $\hat{p}$ is a unit vector in the direction of $\vec{p}$.   This state can be expanded into angular momentum eigenstates of the parent
\begin{equation}\label{eq:Wigner_D_expand}
\displaystyle|\hat{p}\lambda_{\alpha}\lambda_{\beta}\rangle = \sum_{j,m} |j m \lambda_{\alpha}\lambda_{\beta}\rangle \sqrt{\frac{2j+1}{4\pi}} D^{j \; *}_{m\lambda}(\varphi,\vartheta,\psi)
\end{equation}
where $\lambda = \lambda_{\alpha} - \lambda_{\beta}$ and $D^{j}_{m\lambda}(\varphi,\vartheta,\psi)$ are the so-called `Wigner $D$-matrices' corresponding to matrix elements $\langle j m'|R(\varphi,\vartheta,\psi)|j m\rangle$ of the rotation operator $R(\varphi,\vartheta,\psi)$.  The three Euler angles $\varphi$, $\vartheta$, and $\psi$ specify the direction of $\hat{p}$, and the Wigner $D$-matrices can be written in terms of the more convenient real-valued $d$-functions
\begin{equation}\label{eq:Big_d_Small_d}
D^{j}_{m \lambda} (\varphi,\vartheta,\psi) = e^{-i m \varphi} d^{j}_{m \lambda}(\vartheta) e^{-i \lambda \psi},
\end{equation}
some of which can be found in~\cite{PDG}, as well as many other sources. 

The amplitude of the decay $\gamma \rightarrow \alpha + \beta$ is 
\begin{equation}
A_{m}(\hat{p},\lambda_{\alpha},\lambda_{\beta}) = \langle \hat{p} \lambda_{\alpha} \lambda_{\beta} | \hat{U} | j m \rangle 
\end{equation}
where $\hat{U}$ is an operator invariant under rotations and reflections.  Using Eq.~\ref{eq:Wigner_D_expand} this can rewritten as 
\begin{equation}
A_{m}(\hat{p},\lambda_{\alpha},\lambda_{\beta}) = \sqrt{\frac{2j+1}{4\pi}} M(\lambda_{\alpha},\lambda_{\beta}) D^{j}_{m\lambda}(\varphi,\vartheta,\psi)
\end{equation}
where we have, for convenience, defined the matrix element
$M(\lambda_{\alpha},\lambda_{\beta}) = \langle j m \lambda_{\alpha} \lambda_{\beta} | \hat{U} | j m\rangle$ which is independent of $j$ and $m$.

If the production mechanism of $\gamma$ is uncertain, the pure quantum state is unknown, and we can not calculate a scattering probability directly from this amplitude.  Instead, we will use a density matrix to represent a more generalized quantum state.  For any system of total angular momentum $j$ and magnetic sub-state $m$, we can define a density matrix in the ($j$,$m$) basis as
\begin{equation}
\displaystyle \mathbf{\rho} = \sum_{m,m'} |j m\rangle \rho_{m,m'} \langle j m'|.
\end{equation}
The density matrix is Hermitian and has a trace of 1, meaning that it can be characterized by 4$j$($j$+1) independent real numbers.  If we force the $\hat{x}$ and $\hat{z}$-axes to be in the production plane (the plane formed by the two incoming hadron momenta boosted into the $\gamma$ rest frame), we can further restrict the matrix with
\begin{equation}
\rho_{m,m'}=(-1)^{m-m'}\rho_{-m,-m'},
\end{equation}
which reduces the number of independent parameters to $4j$.  

Using the density matrix we can now write the probability for $\gamma \rightarrow \alpha + \beta$ with $\hat{p}$ specified by Euler angles $\varphi$, $\vartheta$, and $\psi$ in the $\gamma$ rest frame:
\begin{eqnarray}
\displaystyle \frac{d\sigma}{d^{4}q d\Omega} &=& \frac{2j+1}{4\pi} \sum_{m,m'} \sum_{\lambda_{\alpha}\lambda_{\beta}} A_{m}(\hat{p}\lambda_{\alpha}\lambda_{\beta})\rho_{m m'} A^{\; *}_{m'}(\hat{p}\lambda_{\alpha}\lambda_{\beta}) \nonumber \\
                                             &=& \frac{2j+1}{4\pi} \sum_{m,m'} \sum_{\lambda_{\alpha}\lambda_{\beta}} \left|M(\lambda_{\alpha},\lambda_{\beta})\right|^{2} D^{j \; *}_{m \lambda}(\varphi,\vartheta,\psi) D^{j}_{m' \lambda}(\varphi,\vartheta,\psi) \rho_{m m'}.
\end{eqnarray}
For the specific case of $\gamma$ as a $J/\psi$, we have a spin-1 particle with $m$=0, and helicity conservation requires $\lambda_{\alpha} \neq \lambda_{\beta}$.  For decay leptons, $\lambda_{\alpha}$ and $\lambda_{\beta}$ can be either $\frac{1}{2}$ or $-\frac{1}{2}$ so that the sum simplifies to
\small
\begin{equation}
\displaystyle \frac{d\sigma}{d^{4}q d\Omega} = \frac{3}{4\pi} \left|M(\frac{1}{2},-\frac{1}{2})\right|^{2} \sum_{m,m'}\rho_{m,m'}e^{(m-m')\varphi}(d^{1}_{m 1}(\vartheta) d^{1}_{m' 1}(\vartheta) + d^{1}_{m -1}(\vartheta) d^{1}_{m' -1}(\vartheta)).
\end{equation}
\normalsize 
After requiring the $\hat{x}$ and $\hat{z}$-axis to be in the production plane, the density matrix can be simplified in terms of 4 real-valued parameters
\footnotesize
\begin{equation*}
\mathbf{\rho}
= 
\left(
\begin{array}{ccc}
\rho_{1 1}  &  \rho_{1 0}   &  \rho_{1 -1}   \\
\rho_{0 1}  &  \rho_{0 0}   &  \rho_{0 -1}   \\
\rho_{-1 1} &  \rho_{-1 0}  &  \rho_{-1 -1} 
\end{array}
\right)
=
\left(
\begin{array}{ccc}
            \rho_{1 1}               &  Re(\rho_{1 0}) + \ii Im(\rho_{1 0})   &            \rho_{1 -1}                \\
Re(\rho_{1 0}) - \ii Im(\rho_{1 0})  &              1 - 2 \rho_{1 1}          &  -Re(\rho_{1 0})+\ii Im(\rho_{1 0})   \\
            \rho_{1 -1}              &  -Re(\rho_{1 0}) - \ii Im(\rho_{1 0})  &            \rho_{1 1}
\end{array}
\right)
\end{equation*}
\normalsize
and we can use the property $d^{j}_{m' m} = (-1)^{m-m'}d^{j}_{m m'} = d^{j}_{-m -m'}$ along with the contributing $d$-functions
\begin{equation*}
d^{1}_{1 1}(\vartheta)  = \frac{1+\cos(\vartheta)}{2} \qquad d^{1}_{1 0}(\vartheta) = \frac{-sin(\vartheta)}{\sqrt{2}} \qquad d^{1}_{1 -1}(\vartheta) = \frac{1-\cos(\vartheta)}{2}
\end{equation*}
to write out the full amplitude
\begin{eqnarray}
\frac{d\sigma}{d^{4}q d\Omega} =  \frac{3}{4\pi} \left|M(\frac{1}{2},-\frac{1}{2})\right|^{2} \{ \rho_{1 1} (1 + \cos^{2}\vartheta) + (1 - 2\rho_{1 1})(1 - \cos^{2}\vartheta) \nonumber \\
 + Re(\rho_{1 0}) \sqrt{2} sin2\vartheta \cos\varphi + \rho_{1 -1} sin^{2}\vartheta \cos2\varphi  \}. 
\end{eqnarray}
It should be noted that the angular distribution has no dependence on $Im(\rho_{1 0})$, implying that a measurement of each angular decay coefficient still allows for an undetermined phase in the density matrix.

\begin{figure}[h!tb]
        \centering\includegraphics*[width=0.25\columnwidth]{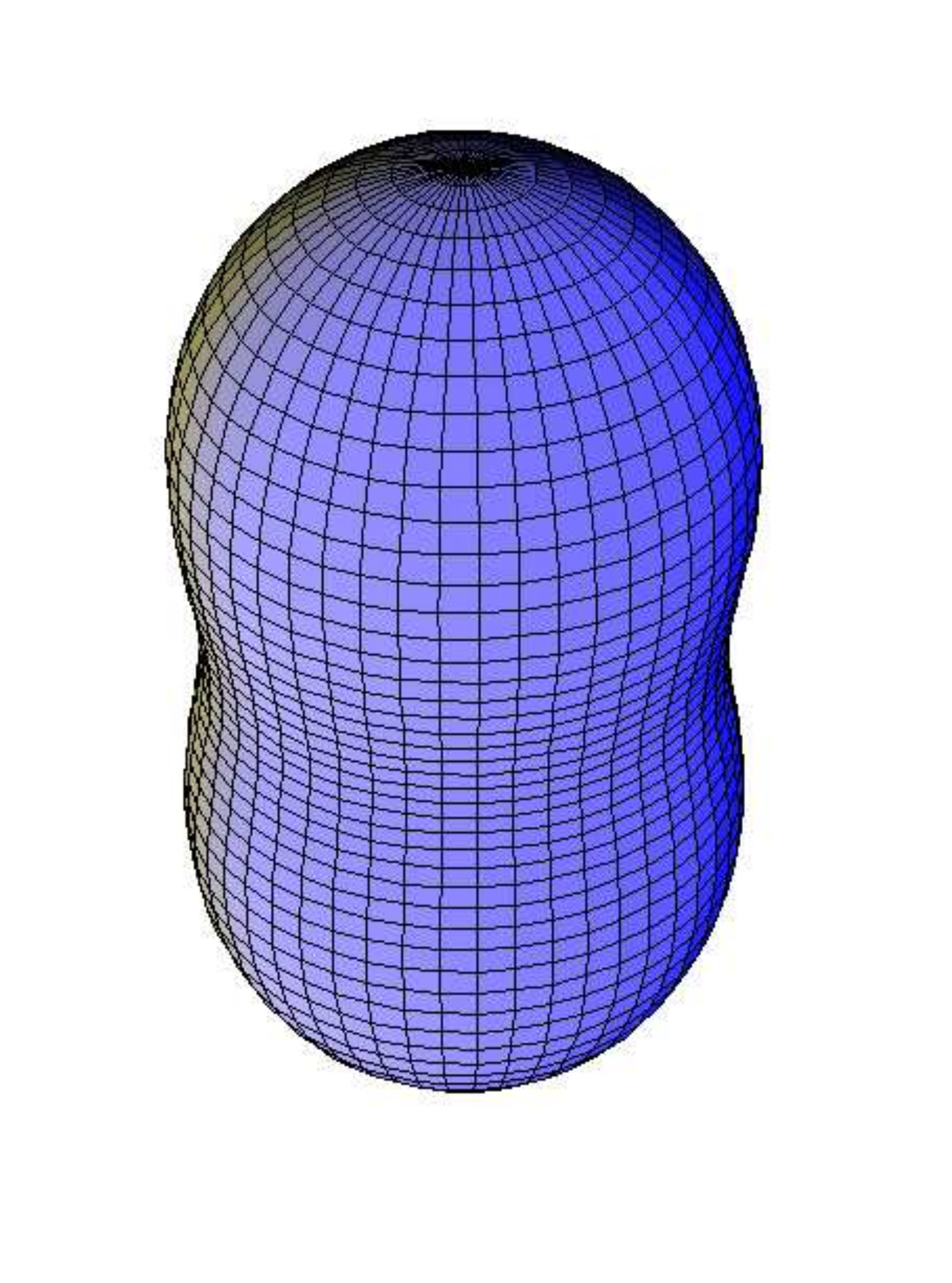} 
        \centering\includegraphics*[width=0.35\columnwidth]{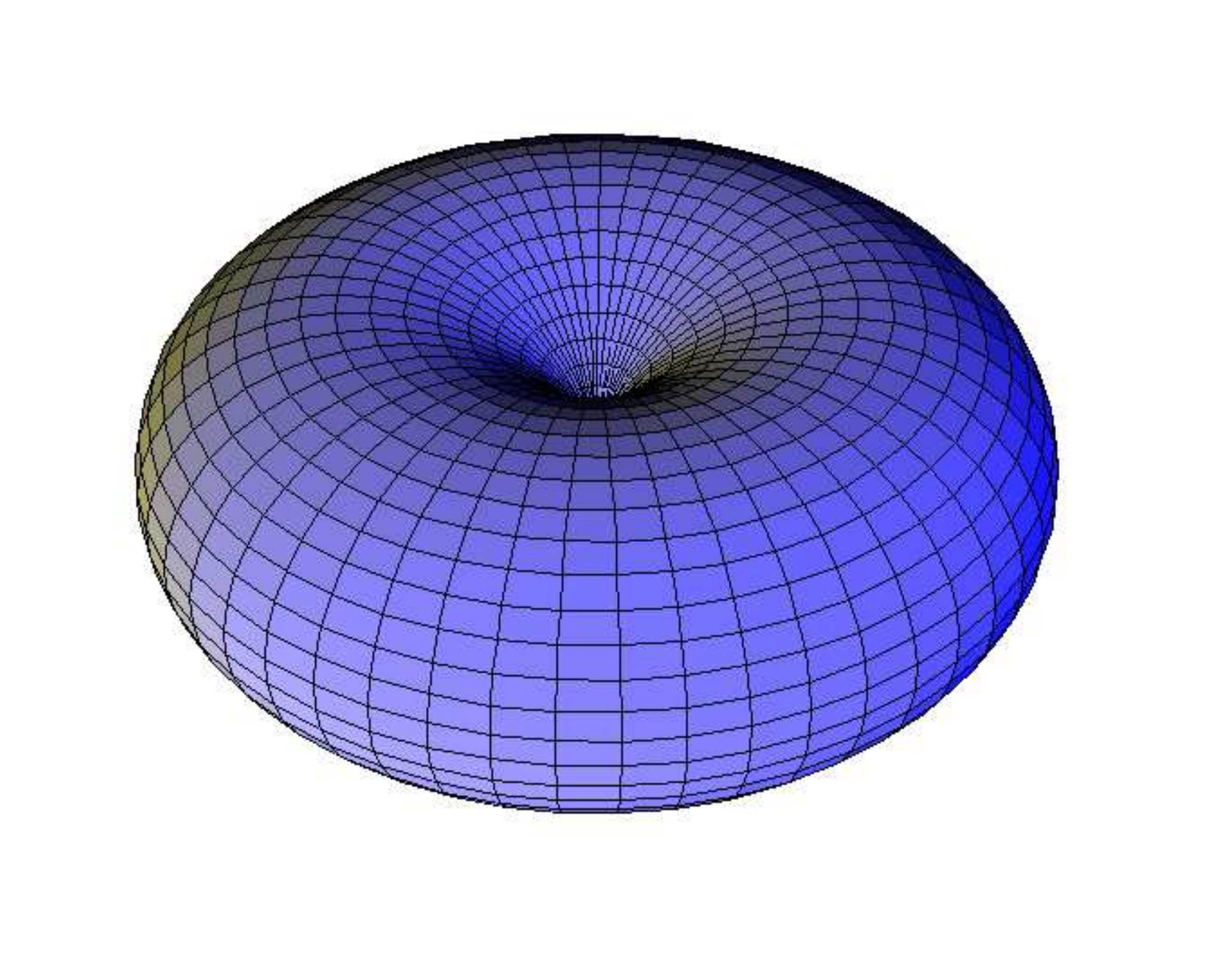}
\isucaption[Angular distributions of decay products from a spin-1 boson decaying into two spin-1/2 fermions.]{Angular distributions of decay products from a spin-1 boson decaying into two spin-1/2 fermions for `transverse' spin-alignment on the left ($\lambda_{\vartheta}$=1, $\lambda_{\varphi}$=$\lambda_{\vartheta\varphi}$=0) and `longitudinal' spin-alignment on the right ($\lambda_{\vartheta}$=-1, $\lambda_{\varphi}$=$\lambda_{\vartheta\varphi}$=0).  The $\hat{z}$-axis points towards the top of the page.}\label{fig:spin1_pol_dists}
\end{figure}

It is customary~\cite{Collins:1977iv,Lam:1978pu} to parameterize the cross-section as
\begin{eqnarray}
\frac{d\sigma}{d^{4}q d\Omega} = \frac{3}{4\pi} \left|M(\frac{1}{2},-\frac{1}{2})\right|^{2} \{ W_{T} (1 + \cos^{2}\vartheta) + W_{L} (1 - \cos^{2}\vartheta) \nonumber \\
 + W_{\triangle} \sqrt{2} sin2\vartheta \cos\varphi +  W_{\triangle\triangle} sin^{2}\vartheta \cos2\varphi \}
\end{eqnarray}
where $W_{T}$ and $W_{L}$ are called the `transverse' and `longitudinal' components respectively.  The reason for the `transverse' and `longitudinal' labels is that $W_{T}$ corresponds only to the $\rho_{11}$ and $\rho_{-1-1}$ elements of the density matrix (spin transverse to the $\hat{z}$-axis), while $W_{L}$ corresponds only to the $\rho_{00}$ element (spin along the $\hat{z}$-axis).  Plots of the decay distributions for purely longitudinal and purely transverse spin-alignment can be found in Fig.~\ref{fig:spin1_pol_dists}.   $W_{\triangle}$ and $W_{\triangle\triangle}$ are called the `single-spin flip' and `double-spin flip' components. 

The angular distribution of the decay leptons is often rewritten as
\begin{equation}
\frac{d\sigma}{d(\cos\vartheta)d\varphi} \propto 1 + \lambda_{\vartheta} \cos^{2}\vartheta + \lambda_{\vartheta\varphi} \sin2 \vartheta \cos\varphi + \lambda_{\varphi} sin^{2}\vartheta \cos2\varphi
\label{eq:angle_dist}
\end{equation}
where $\lambda_{\vartheta} \equiv \frac{W_{T} - W_{L}}{W_{T} + W_{L}} = -\frac{1-3\rho_{1 1}}{1-\rho_{1 1}}$, $\lambda_{\vartheta\varphi} \equiv \frac{\sqrt{2} W_{\triangle}}{W_{T}+W_{L}} = \frac{\sqrt{2} Re(\rho_{1 0})}{1-\rho_{1 1}}$, and $\lambda_{\varphi} \equiv \frac{2 W_{\triangle\triangle}}{W_{T}+W_{L}} = \frac{2 \rho_{1 -1}}{1-\rho_{1 1}}$.   Experimentally, only the $\lambda_{\vartheta}$ parameters have typically been measured~\cite{Affolder:2000nn,Chang:2003rz,Abulencia:2007us,Adare:2009js}, but this is clearly not adequate to fully describe the decay. Several recent experiments have also measured $\lambda_{\vartheta\varphi}$ and $\lambda_{\varphi}$~\cite{Abt:2009nu}, and in Chapter~\ref{ch:jpsi_pol}, a new measurement of $\lambda_{\vartheta}$, $\lambda_{\vartheta\varphi}$, and $\lambda_{\varphi}$ will be presented from PHENIX data.

The coefficients $\lambda_{\vartheta}$, $\lambda_{\varphi}$, and $\lambda_{\vartheta\varphi}$ are clearly correlated.  In particular, a rotation of the reference frame in the production plane by an angle
\begin{equation}
  \delta = \frac{1}{2}\arctan\left(\frac{2\lambda_{\vartheta\varphi}}{\lambda_{\varphi}-\lambda_{\vartheta}}\right)
\label{eq:delta_angle}
\end{equation}
(90$^{\circ}$ when $\lambda_{\vartheta}$=$\lambda_{\varphi}$) corresponds to a rotation into the frame in which $\lambda_{\vartheta\varphi}$ is zero, while $\lambda_{\vartheta}$ and $\lambda_{\varphi}$ describe the entire angular distribution~\cite{Faccioli:2008dx}.

In hadronic collisions, intrinsic partonic transverse momenta allows asymmetries in the spin alignment to exist with respect to the production plane on an event-by-event basis.  Such asymmetries modify Eq.~\ref{eq:angle_dist} to include two additional coefficients.  For a subprocess~$i$,
\begin{eqnarray}
\frac{d\sigma^{(i)}}{d(\cos\vartheta)d\varphi} \propto 1 + \lambda^{(i)}_{\vartheta} \cos^{2}\vartheta + \lambda^{(i)}_{\vartheta\varphi} \sin(2 \vartheta) \cos\varphi + \lambda^{(i)}_{\varphi} sin^{2}\vartheta \cos2\varphi \nonumber \\ 
+ \lambda^{\perp(i)}_{\vartheta\varphi} \sin(2 \vartheta) \sin\varphi + \lambda^{\perp(i)}_{\varphi} sin^{2}\vartheta \sin2\varphi.
\label{eq:prod_plan_angle_dist}
\end{eqnarray}
For inclusive measurements, however, these additional coefficients must vanish~\cite{Faccioli:2010ej}, leaving the expression in Eq.~\ref{eq:angle_dist}.

\section{Reference Frames}\label{sec:reference_frames}

\begin{figure}[h!tb]
 \centering\includegraphics[width=0.35\columnwidth]{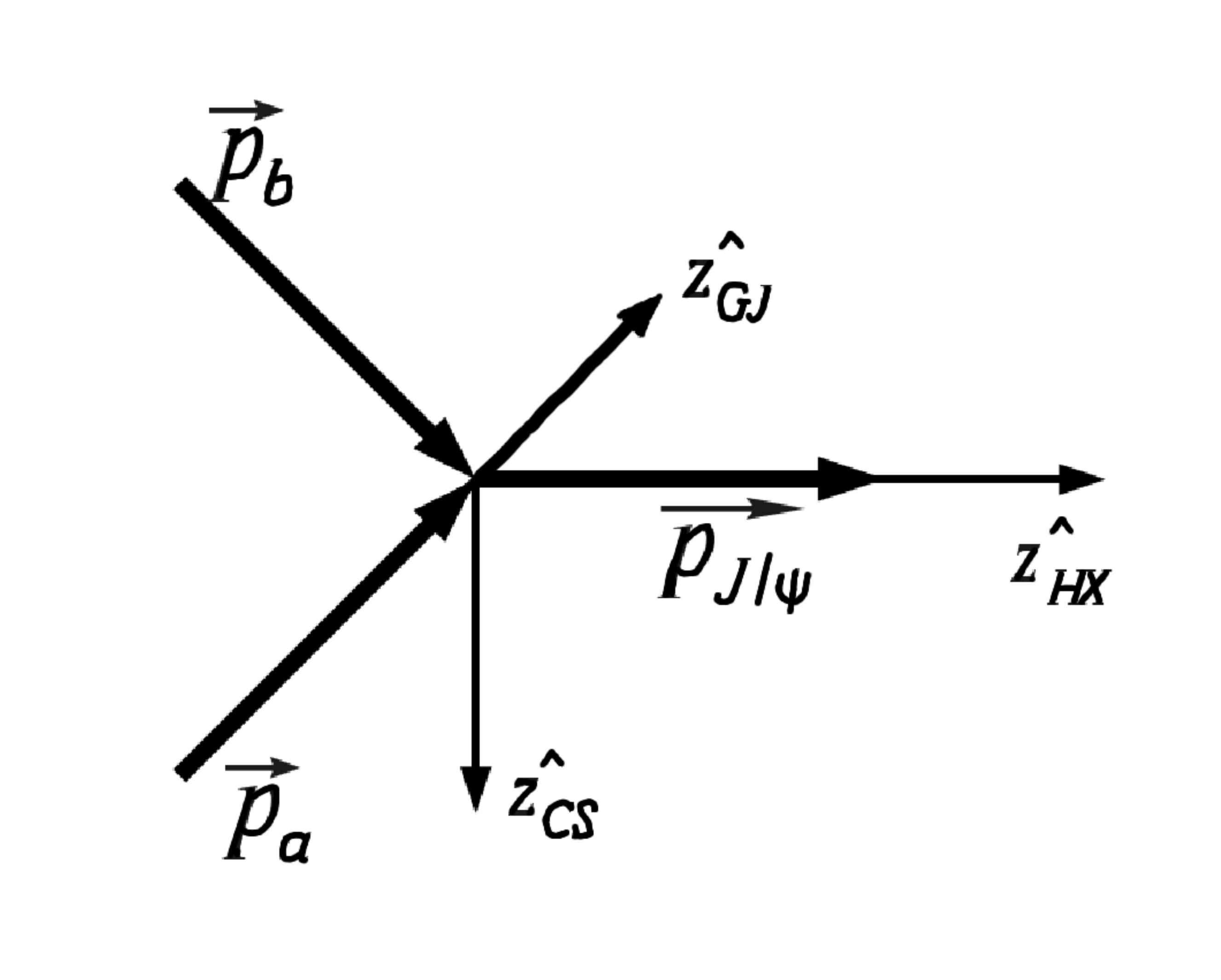}
 \centering\includegraphics[width=0.35\columnwidth]{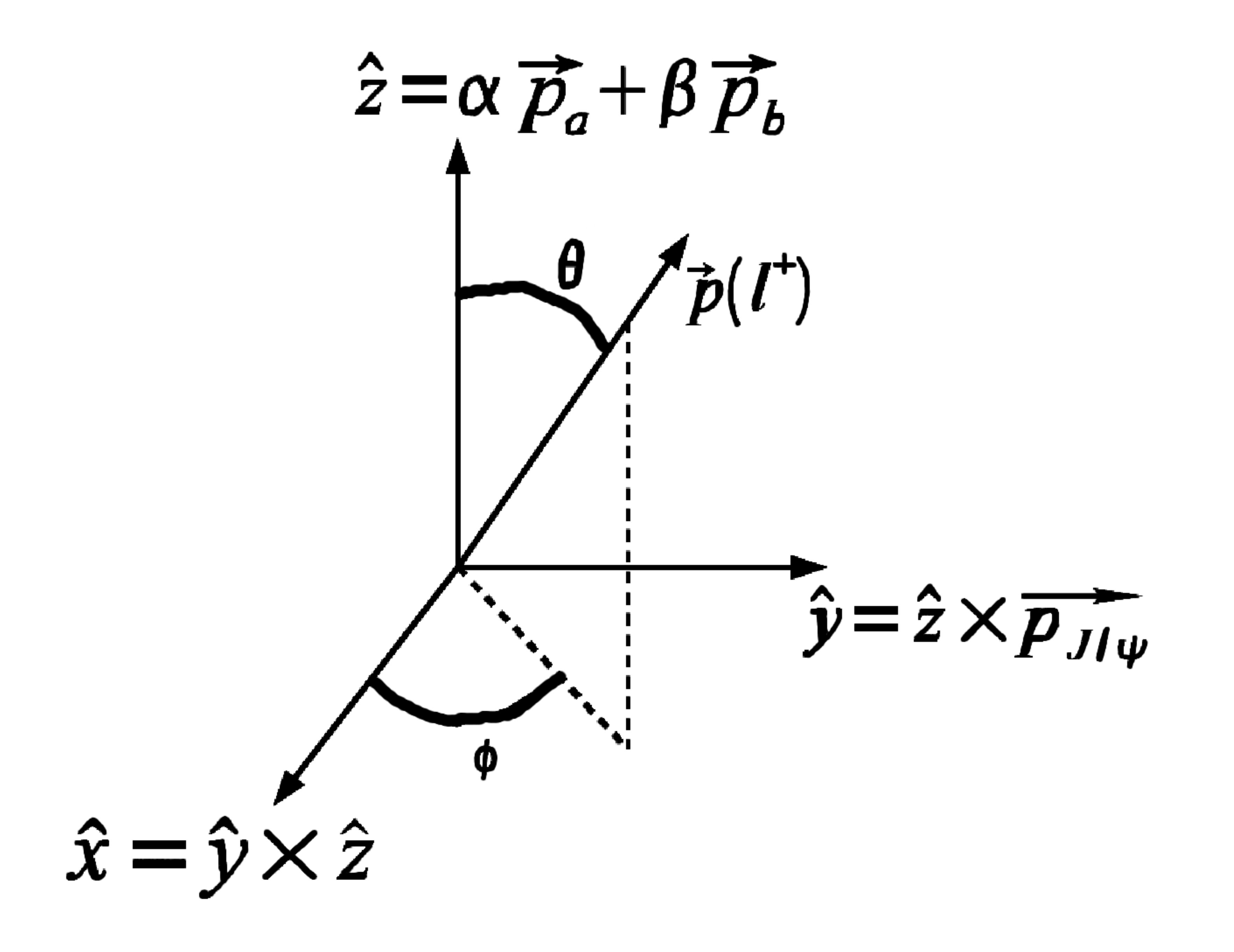}
 \isucaption[Illustration of $\hat{z}$-axes, $\vartheta$, and $\varphi$ in several reference frames]{Illustration of $\hat{z}$-axes for several reference frames (left) and for the angles $\vartheta$ and $\varphi$ as defined in those frames (right).}\label{fig:ref_frames}
\end{figure}

The derivation of Section~\ref{sec:angular_dists}, leading to Eq.~\ref{eq:angle_dist}, gave angular distributions in terms of $\vartheta$ and $\varphi$ with the $\hat{z}$ and $\hat{x}$ axes required to lie in the production plane.  However, there are an infinite number of possible reference frames, each leading to an alternate definition of $\vartheta$ and $\varphi$.  Three of these reference frames are typically chosen for measurements of angular decay distributions:
\begin{enumerate}
\item
  The Jacob-Wick Helicity Frame~\cite{Jacob:1959at} \\
  The $\hat{z}$-axis is chosen as the $J/\psi$ velocity direction in the lab frame.
\item
  The Gottfried-Jackson Frame~\cite{Gottfried:1964nc} \\
  The $\hat{z}$-axis is chosen as the 3-momentum of one beam boosted into $J/\psi$ rest frame.
\item
  The Collins-Soper Frame~\cite{Collins:1977iv} \\
  $\hat{z}$ $= \frac{\vec{P}_{b}}{|\vec{P_{b}}|} - \frac{\vec{P}_{a}}{|\vec{P}_{a}|}$ \\
  where $\vec{P}_{a}$ and $\vec{P}_{b}$ are the 3-momenta of each beam boosted into $J/\psi$ rest frame.
\end{enumerate}
After the $\hat{z}$-axis is chosen, the $\hat{y}$-axis is taken as normal to the production plane, $\hat{y} = \hat{z} \times \vec{P}_{b}$, and the $\hat{x}$-axis is $\hat{x}=\hat{y} \times \hat{z}$.  A cartoon of the various frames and the angles associated with each frame can be found in Fig.~\ref{fig:ref_frames}.  The Helicity frame has typically been used by collider experiments, while the Gottfried-Jackson and Collins-Soper frames have been used by fixed target experiments.  The Collins-Soper frame was initially motivated by Drell-Yan production and has been widely used for measurements of that process.

The orientation of the $\hat{y}$-axis is especially important for the determination of $\lambda_{\vartheta\varphi}$.  While the $\lambda_{\vartheta}$ and $\lambda_{\varphi}$ coefficients do not depend on the orientation of the $\hat{y}$-axis, the sign of $\lambda_{\vartheta\varphi}$ does.  To make the situation even more complicated, the sign of $\lambda_{\vartheta\varphi}$ changes depending on the sign of the rapidity~(see Fig.~\ref{fig:frame_alignment}).  In order to keep $\lambda_{\vartheta\varphi}$ from becoming zero when integrating over rapidity, the direction of the $\hat{z}$ and $\hat{y}$-axes should be flipped when going from positive to negative rapidity~\cite{Faccioli:2010kd}.

One additional subtlety is that the Gottfried-Jackson frame is ambiguous in a collider environment, because either beam can be boosted into the $J/\psi$ rest frame to define the $\hat{z}$-axis.  This ambiguity is especially troublesome at forward rapidities where the detector acceptance is quite different depending on the beam used in the frame definition.  To avoid ambiguity, we define two separate Gottfried-Jackson frames: the Gottfried-Jackson forward frame, using the beam circulating in the same direction as the $J/\psi$ momentum, and the Gottfried-Jackson backward frame, using the beam circulating in the opposite direction.

\begin{figure}[h!tb]
 \centering\includegraphics[width=0.55\columnwidth]{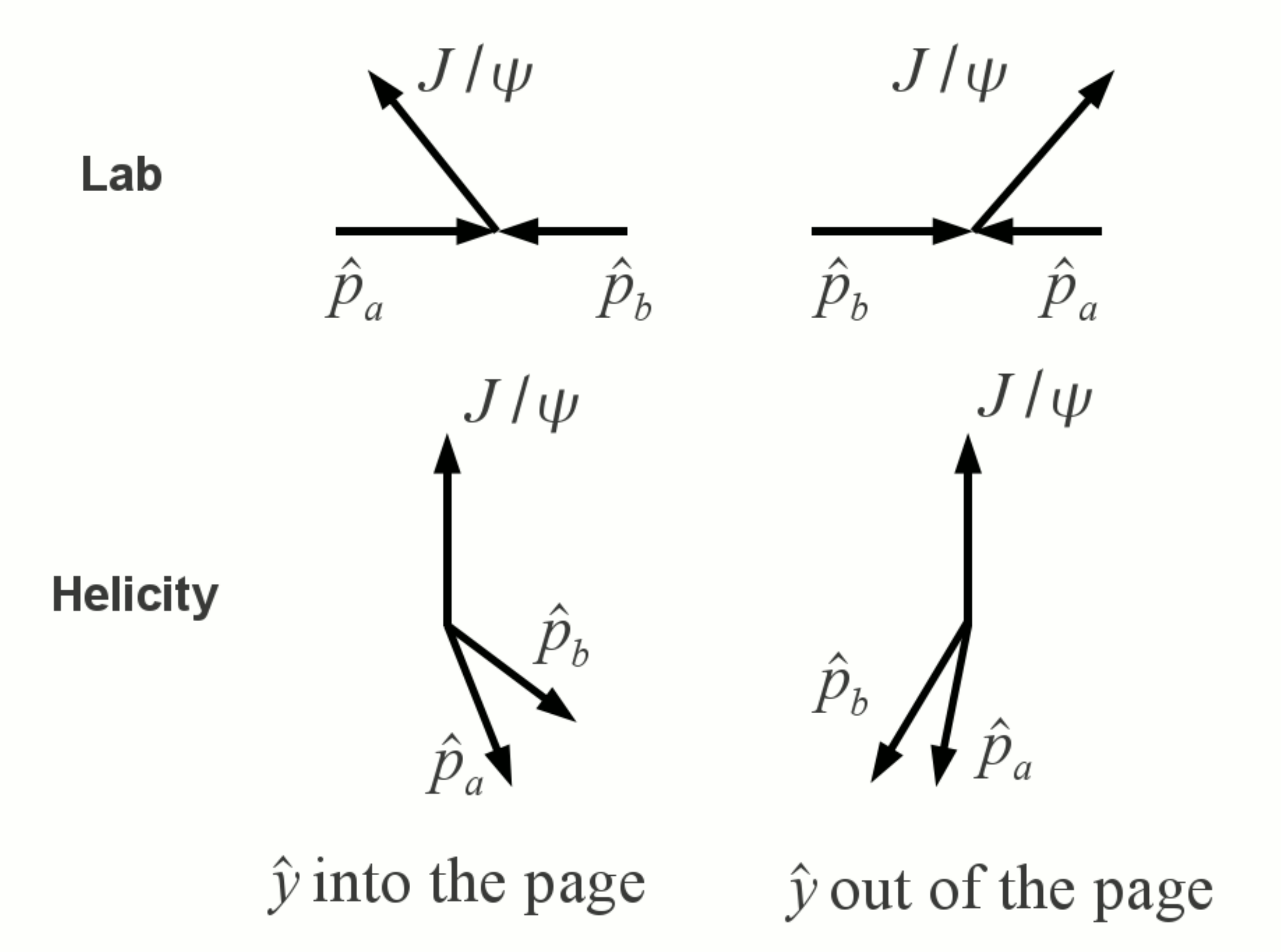}
 \isucaption[Orientation of the $\hat{y}$-axis depending on the direction of the boost.]{Orientation of the $\hat{y}$-axis depending on the direction of the Lorentz boost.}
\end{figure}\label{fig:frame_alignment}

Once a reference frame is defined and the various amplitudes are measured, it is possible to make calculations for other frames by rotating the density matrix~\cite{Argyres:1982kg}.  The fact that the real-space distribution of decays must be invariant under such rotations, along with the property that $|\lambda_{\vartheta}|<1$ (by definition) lead  to the inequalities
\begin{equation}
|\lambda_{\varphi}|\leq\frac{1}{2}(1+\lambda_{\vartheta})\quad,\quad |\lambda_{\vartheta\varphi}|\leq\frac{1}{2}(1-\lambda_{\varphi}),
\end{equation}
implying additional bounds $|\lambda_{\varphi}|\leq 1$ and $|\lambda_{\vartheta\varphi}|\leq 1$~\cite{Faccioli:2010ji,Faccioli:2010ej,Faccioli:2010kd}.  These inequalities also allow for a family of frame-invariant quantities,\footnote{The derivation of these quantities uses the property $d^{1}_{1,m}(\vartheta) + d^{1}_{-1,m} = \delta_{|m|,1}$ and assumes that a reference frame exists wherein only the $\rho_{11}$ and $\rho_{-1-1}$ density matrix elements are non-zero.} one of which is
\begin{equation}
\widetilde{\lambda} = \frac{\lambda_{\vartheta}+3\lambda_{\varphi}}{1-\lambda_{\varphi}},
\label{eq:lambda_tilde}
\end{equation}
which corresponds to the weighted sum of $\lambda_{\vartheta}$ over all $n$ contributing subprocesses
\begin{equation}
\widetilde{\lambda} = \frac{\displaystyle\sum_{i=1}^{n}\frac{f^{(i)}}{3+\lambda_{\vartheta}^{(i)}}\lambda_{\vartheta}^{(i)}}{\displaystyle\sum_{i=1}^{n}\frac{f^{(i)}}{3+\lambda_{\vartheta}^{(i)}}},
\label{eq:lambda_tilde_sum}
\end{equation}
where $f^{(i)}$ is the fraction of measured particles produced by process $i$.

It should be noted that the entire formalism developed in Sections~\ref{sec:angular_dists} and~\ref{sec:reference_frames} is valid for any spin-1, $m$=0, particle decaying to two leptons, including both the $J/\psi$ and photon.  A na\"\i ve model for Drell-Yan production (the decay into two leptons of a virtual photon from the annihilation of a quark and anti-quark) would expect $\lambda_{\vartheta} = 1$ and $\lambda_{\vartheta\varphi}$=$\lambda_{\varphi}$=0, because real photons are transversely polarized.  Virtual photons are likely to maintain much of that transverse polarization and transfer it to the decay leptons.\footnote{There is an unfortunate ambiguity in the use of the words `transverse' and `longitudinal' when describing polarization throughout the literature.  Typically, when the polarization of a gauge Boson is being discussed, as we are discussing the photon here, the labels correspond to the orientation of the field vectors (electric and magnetic in this case) with respect to the momentum of the gauge boson.  For hadrons, however, the same labels usually correspond to the orientation of the hadron spin (which is perpendicular to the field vectors) with respect to the momentum of the hadron.  When we discussed transversely polarized protons in Chapter~\ref{ch:trans_protons}, for example, the `transverse' label meant that the proton spin, not the field vectors, was transverse to its momentum.  To avoid this ambiguity as much as possible, we will not speak in this document of the `$J/\psi$ polarization' as the spin-alignment is called elsewhere in the literature.}  However, non-zero total transverse momentum of the decay products can lead to $\lambda_{\vartheta} \neq 1$~\cite{Chiappetta:1986yg}, and non-zero transverse momentum of the colliding partons can lead to $\lambda_{\vartheta\varphi} \neq 0$ and $\lambda_{\varphi} \neq 0$~\cite{Cleymans:1981zw}.\footnote{The $\lambda_{\varphi}$ coefficient in Drell-Yan production is also sensitive to the Boer-Mulders effect discussed in Section~\ref{sec:Boer-Mulders}~\cite{Barone:2009hw}}

Lam and Tung have shown, however, that the spin-1/2 nature of the quarks leads to the constraint that $1 - \lambda_{\vartheta} - 4\lambda_{\varphi} = 0$, meaning that $W_{L}=2 W_{\triangle\triangle}$~\cite{Lam:1978pu,Lam:1980uc}.\footnote{The Lam-Tung relation is analogous to the Callan-Gross relation in DIS, a consequence of the interaction between a photon probe and half-integer spin quarks, resulting in the condition $F_{L}=F_{2}-2xF_{1}$, discussed in Section~\ref{sec:DIS}.}  While pion induced Drell-Yan experiments on deuterium and tungsten show clear violations of the Lam-Tung relation~\cite{Guanziroli:1987rp,Heinrich:1991zm}, measurements of proton induced Drell-Yan on a deuterium target show no such violation~\cite{Zhu:2008sj}.  The expression in Eq.~\ref{eq:lambda_tilde_sum} is formally equivalent to the Lam-Tung relation when $\widetilde{\lambda}$=1, and any violation of the Lam-Tung relation could be described by a suitably modified frame-invariant expression~\cite{Faccioli:2010ej}.

\section{Production Mechanism}\label{sec:production_mechanism}

Since the first observation of the $J/\psi$ meson, theorists have been attempting to model the mechanism for its production.  Among the most prominent of these models have been the Color Evaporation Model, the Color Singlet Model, and the Color Octet Model.  Each model makes distinct predictions for the production cross section and most also predict the angular decay coefficients of the produced charmonium states.

If the mass of the charm quark is taken to be large compared to $\Lambda_{QCD}$, then a system of two hadronizing charm quarks is approximately non-relativistic.\footnote{In the $\overline{MS}$ factorization scheme, $\Lambda_{QCD}$ is approximately 206-231~MeV for 5 active flavors of quarks~\cite{Bethke:2006ac}, whereas the charm quark rest mass energy is approximately 1.27~GeV~\cite{PDG}.}  All three models attempt to factorize $J/\psi$ production into a relativistic part describing the production of the charm and anti-charm quark, $d\sigma_{c\bar{c}[n]+X}$, and a non-relativistic part describing the bound state of the two quarks, $F_{c\bar{c}[n]}(\Lambda)$: 
\begin{equation}
d\sigma(J/\psi+X) = \displaystyle\sum_{n}\int d\Lambda \frac{d\sigma_{c\bar{c}[n]+X}}{d\Lambda} F_{c\bar{c}[n]}(\Lambda)
\label{eq:Production_Model}
\end{equation}
where the $[n]$ denotes the quantum state of the $c\bar{c}$ pair and $\Lambda$ the energy scale.

\subsection{Color Evaporation Model}\label{sec:Color_Evaporation}

The Color Evaporation Model (CEM)~\cite{Fritzsch:1977ay} was the earliest attempt to calculate the charmonium cross section and has only been applied to hadronic collisions.  The non-relativistic part is assumed to be non-zero and constant between $4m_{c}^{2}$ and $4 m_{\text{D}}^{2}$ and zero for all other energies, where $m_{c}$ and $m_{\text{D}}$ are the mass of the charm quark and D meson respectively.  The model does not take the quantum state of the $c\bar{c}$ pair into account explicitly, but instead sums over all quantum states and multiplies by a factor of $\frac{1}{9}$, the statistical probability for the $c\bar{c}$ pair to be in the color-singlet state.  The differential cross section is
\begin{equation}
d\sigma(J/\psi +X) = \frac{F_{c\bar{c}[J/\psi]}}{9} \sum_{n} \int_{2m_{c}}^{2m_{\text{D}}} dM \frac{d\sigma_{c\bar{c}[n]+X}}{dM},
\label{eq:CEM_equation}
\end{equation}
which has a single constant parameter $F_{c\bar{c}[J/\psi]}$ to be determined from a fit to data.

\begin{figure}[h!tb]
  \centering\includegraphics*[width=0.75\columnwidth]{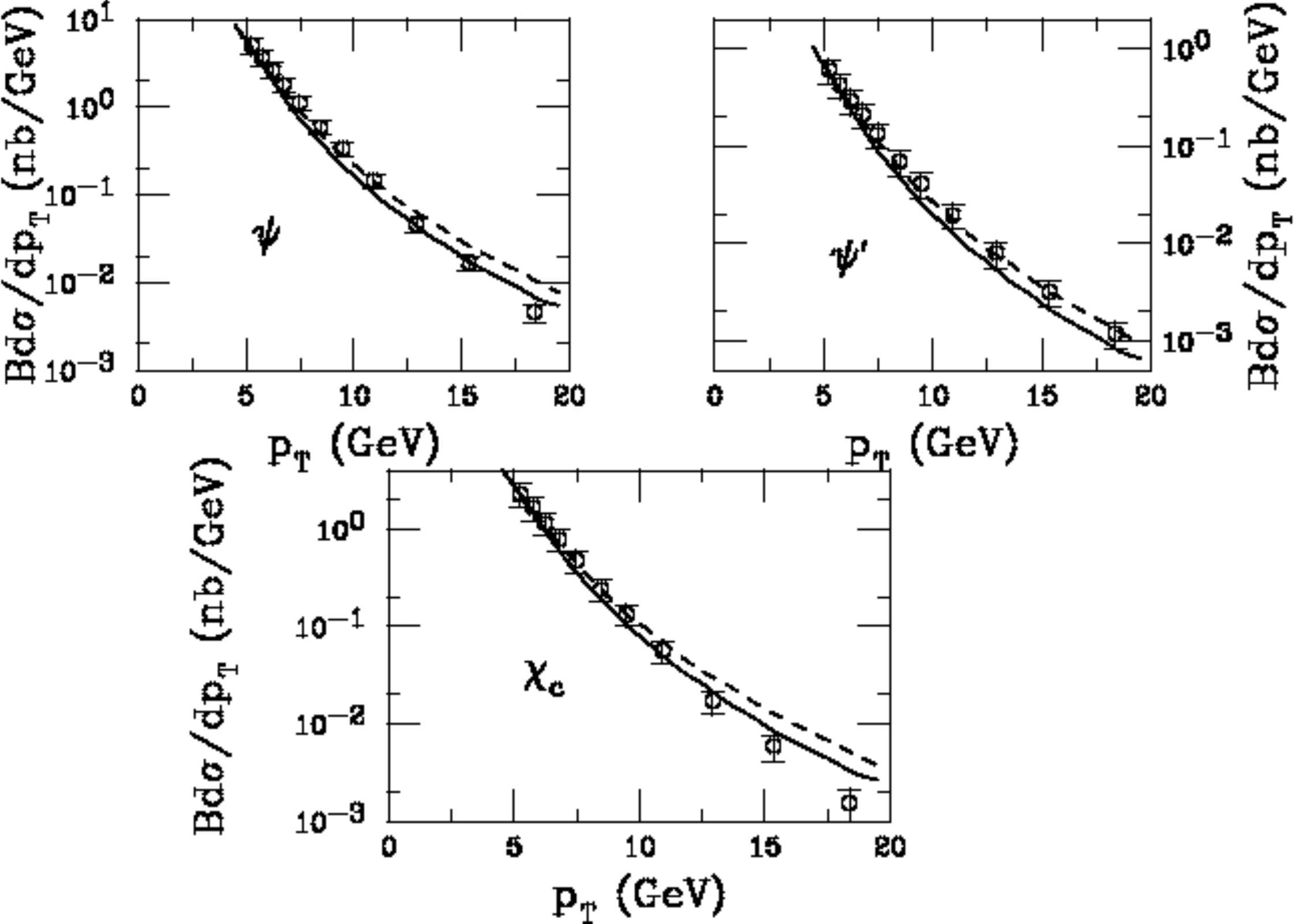} 
  \isucaption[Comparison of the Color Evaporation Model at NLO in $\alpha_{S}$ with $J/\psi$ cross-sections from CDF.]{Comparison of Color Evaporation Model predictions to measured cross-sections for production of direct $J/\psi$ (top left), prompt $J/\psi$ from decays of $\psi^{\prime}$ (top right), and prompt $J/\psi$ from decays of $\chi_{c}$ (bottom) at CDF as a function of $p_{T}$. The dashed and solid curves use the MRST98 HO (higher order)~\cite{Martin:1998np,Martin:1998ab} and GRV98 HO~\cite{Gluck:1998xa} PDF sets with $m_{c}$=1.2 and 1.3~GeV/$c^{2}$ respectively (from~\cite{Brambilla:2004wf}).}\label{fig:CEM_NLO}
\end{figure}

The CEM is actually more of a parameterization than a model, but it has predictive power for the shape $J/\psi$ cross-section to the extent that all $d\sigma_{c\bar{c}[n]+X}$ can be calculated.   There is no clear prediction for the $J/\psi$ spin-alignment from the CEM, but it has been suggested that multiple soft gluon exchanges destroy the spin-alignment of the $c\bar{c}$ pair~\cite{Amundson:1996qr}.  A comparison of the CEM to cross-section data shows impressive agreement (Fig.~\ref{fig:CEM_NLO}).

\subsection{Color-Singlet Model}\label{sec:Color_Singlet}

\begin{figure}[h!tb]
  \centering\includegraphics*[width=0.49\columnwidth]{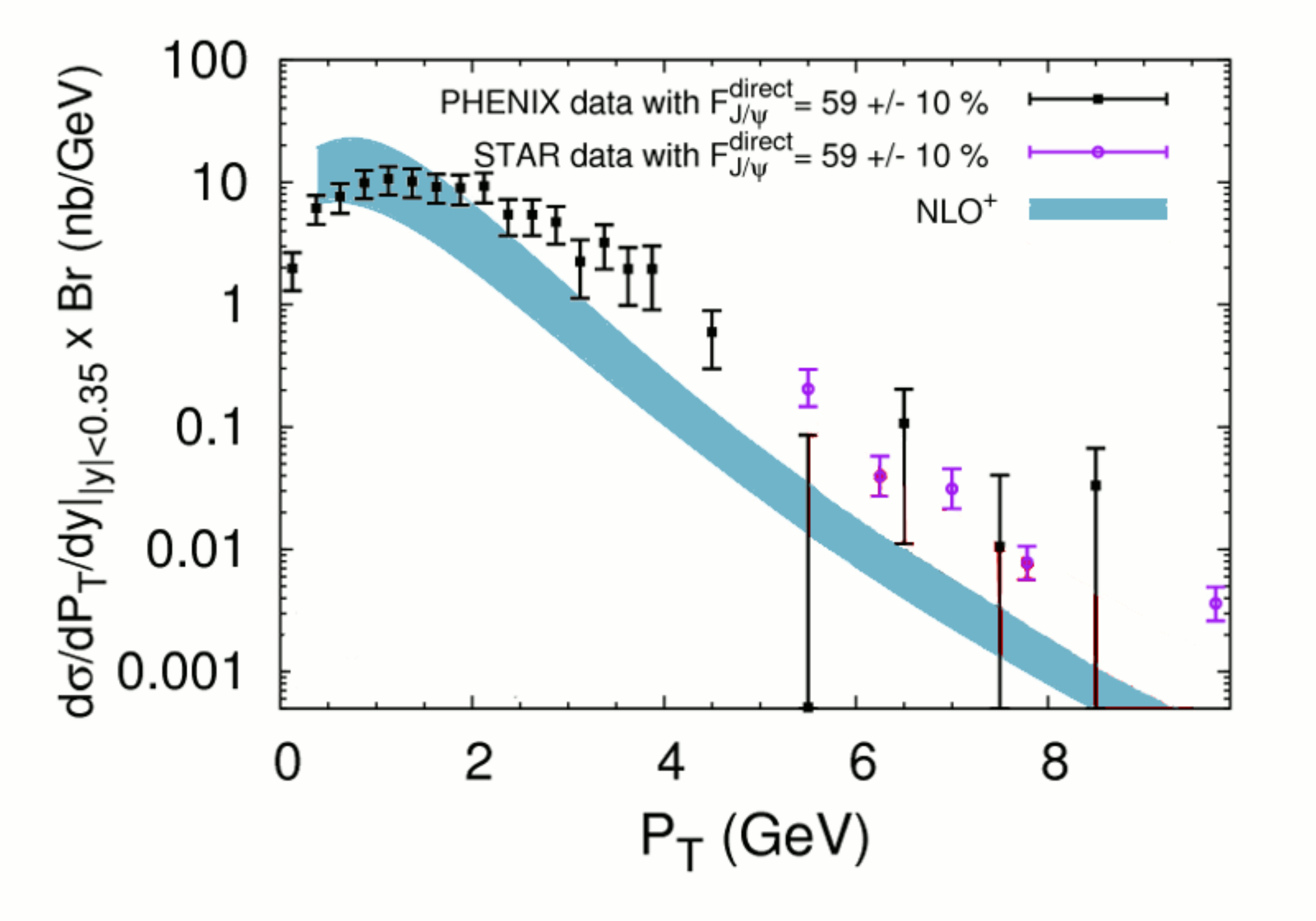} 
  \centering\includegraphics*[width=0.49\columnwidth]{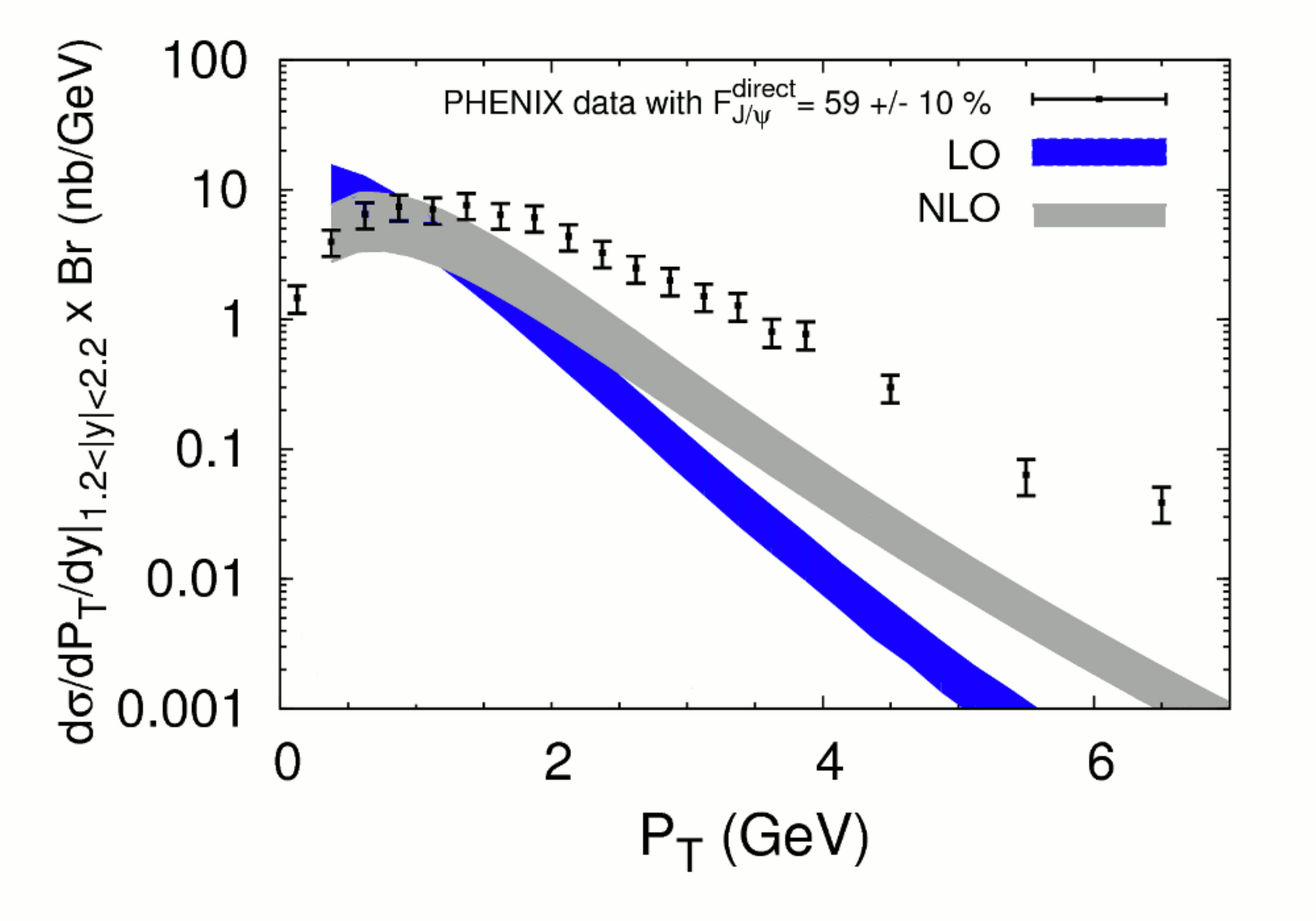} 
  \isucaption[Comparison of the CSM at NLO in $\alpha_{S}$ with $J/\psi$ cross-section data from RHIC.]{Comparison of the CSM at NLO in $\alpha_{S}$ with the inclusive $J/\psi$ cross-section from PHENIX and STAR at mid rapidity (left), and with the PHENIX inclusive $J/\psi$ cross-section at forward rapidity (right) (from~\cite{Lansberg:2010vq}).  Both plots assume that 59$\pm$10\% of the $J/\psi$ mesons are directly produced, and in both cases the model under-predicts the data.}\label{fig:CS_NLO_xsec}
\end{figure}

\begin{figure}[h!tb]
  \centering\includegraphics*[width=\columnwidth]{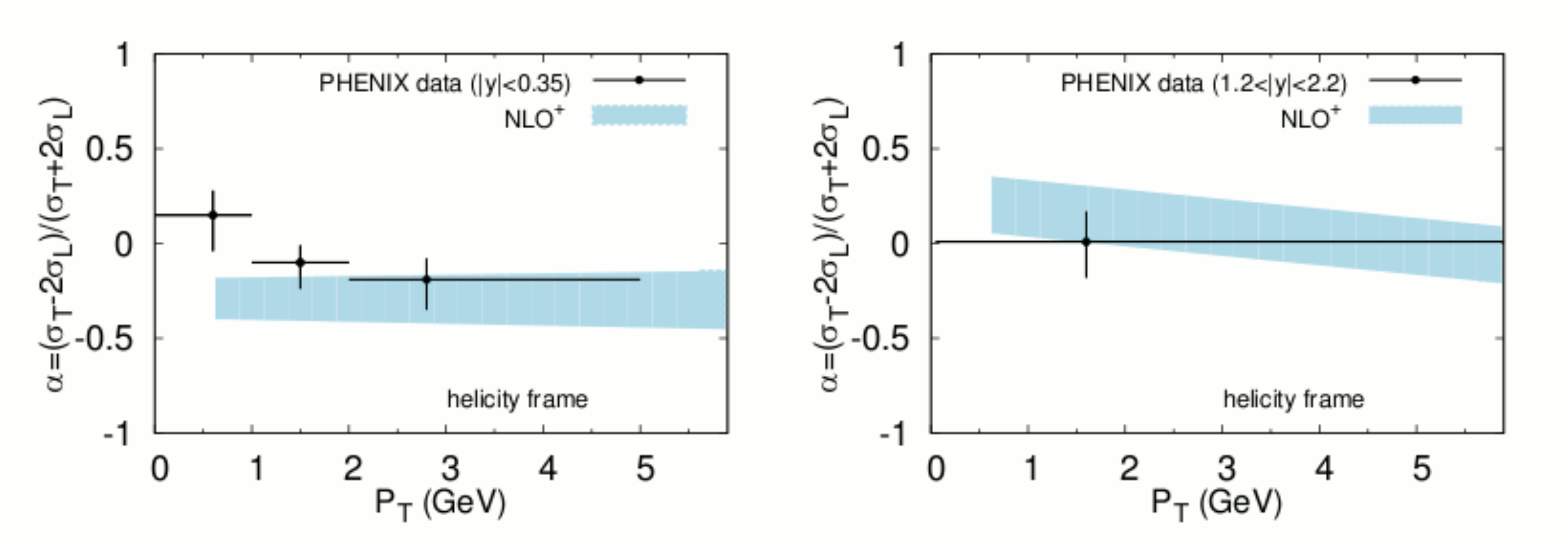} 
  \isucaption[Comparison of the CSM at NLO in $\alpha_{S}$ with $J/\psi$ spin-alignment data from PHENIX.]{Comparison of the CSM at NLO in $\alpha_{S}$ with the $\alpha$ spin-alignment coefficient (called `$\lambda_{\vartheta}$' in this document) for inclusive $J/\psi$ in the Helicity frame from PHENIX at mid rapidity (left) and forward rapidity (right) (from~\cite{Lansberg:2010vq}).  The data does not constrain the model very well.}\label{fig:CS_NLO_pol}
\end{figure}

The Color-Singlet Model (CSM)~\cite{Guberina:1980dc,Baier:1981uk} is an attempt to explicitly take into account the quantum state of the $c\bar{c}$ pair.  In this model, the $c\bar{c}$ pair emerging from the relativistic scattering diagram is assumed to be in the same quantum state as the produced $J/\psi$, and the non-relativistic amplitude is the real-space $J/\psi$ wavefunction evaluated at the origin:
\begin{equation}
d\sigma(J/\psi+X) = \int_{0}^{\infty} dM \frac{d\sigma_{c\bar{c}[^{3}S_{1}]+X}}{dM}\psi_{J/\psi}(r=0).
\end{equation}
The wavefunction $\psi_{J/\psi}$ is determined using hydrogen-like potential models for the hadronized $c\bar{c}$ pair~\cite{Gupta:1986xt}, many of which reproduce the charmonium spectroscopy quite well, and the model has no free parameters.  A comparison of a CSM calculation at next-to-leading order in $\alpha_{s}$ (NLO) with data from RHIC is shown for the differential cross section in Fig.~\ref{fig:CS_NLO_xsec} and the angular decay coefficient $\lambda_{\vartheta}$ in Fig.~\ref{fig:CS_NLO_pol}~\cite{Lansberg:2010vq}.   While the measurement of $\lambda_{\theta}$ does not constrain the model very well, the CSM clearly under-predicts the $J/\psi$ cross section at both mid and forward rapidities.  NLO diagrams increase the cross-section relative to leading order (LO), but the increase is not large enough to remove the discrepancy.  

Similar conclusions have been drawn about the CSM from comparisons of calculations at NLO with data from HERA and Zeus~\cite{Artoisenet:2009xh,Chang:2009uj}.  Recent calculations at NNLO, however, seem to accurately reproduce the measured $\Upsilon$ cross-section at CDF~\cite{Artoisenet:2008fc}.  The slow convergence of the perturbation theory seems to imply that the model does not properly order the contributing diagrams, and it is quite possible that important effects are either not yet included or are not included at the correct order~\cite{Lansberg:2005pc,Nayak:2007zb,Haberzettl:2007kj,Artoisenet:2009mk}.

\subsection{Color-Octet Model}\label{sec:Color_Octet}

\begin{figure}[h!tb]
        \centering\includegraphics*[width=0.25\columnwidth]{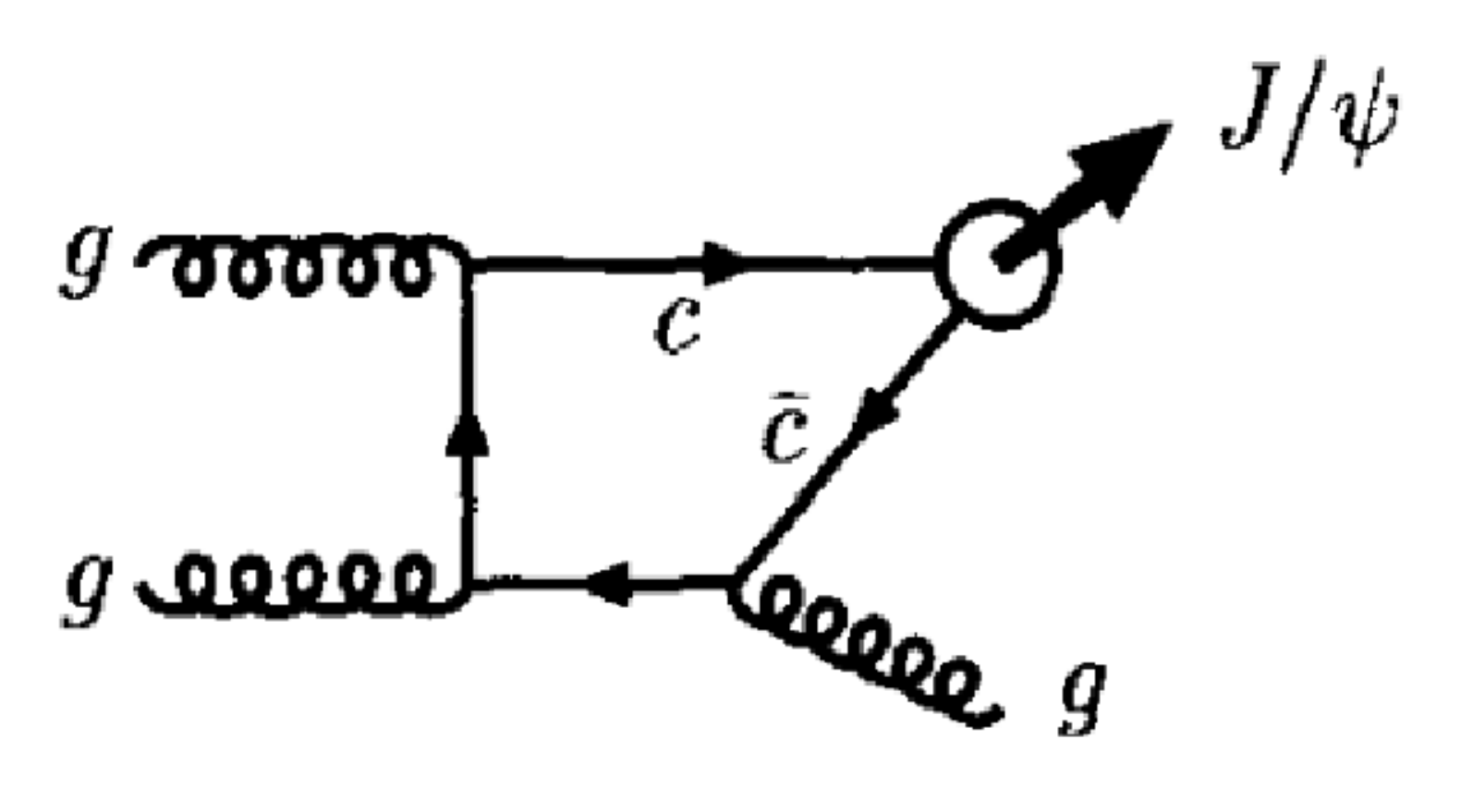} 
        \centering\includegraphics*[width=0.25\columnwidth]{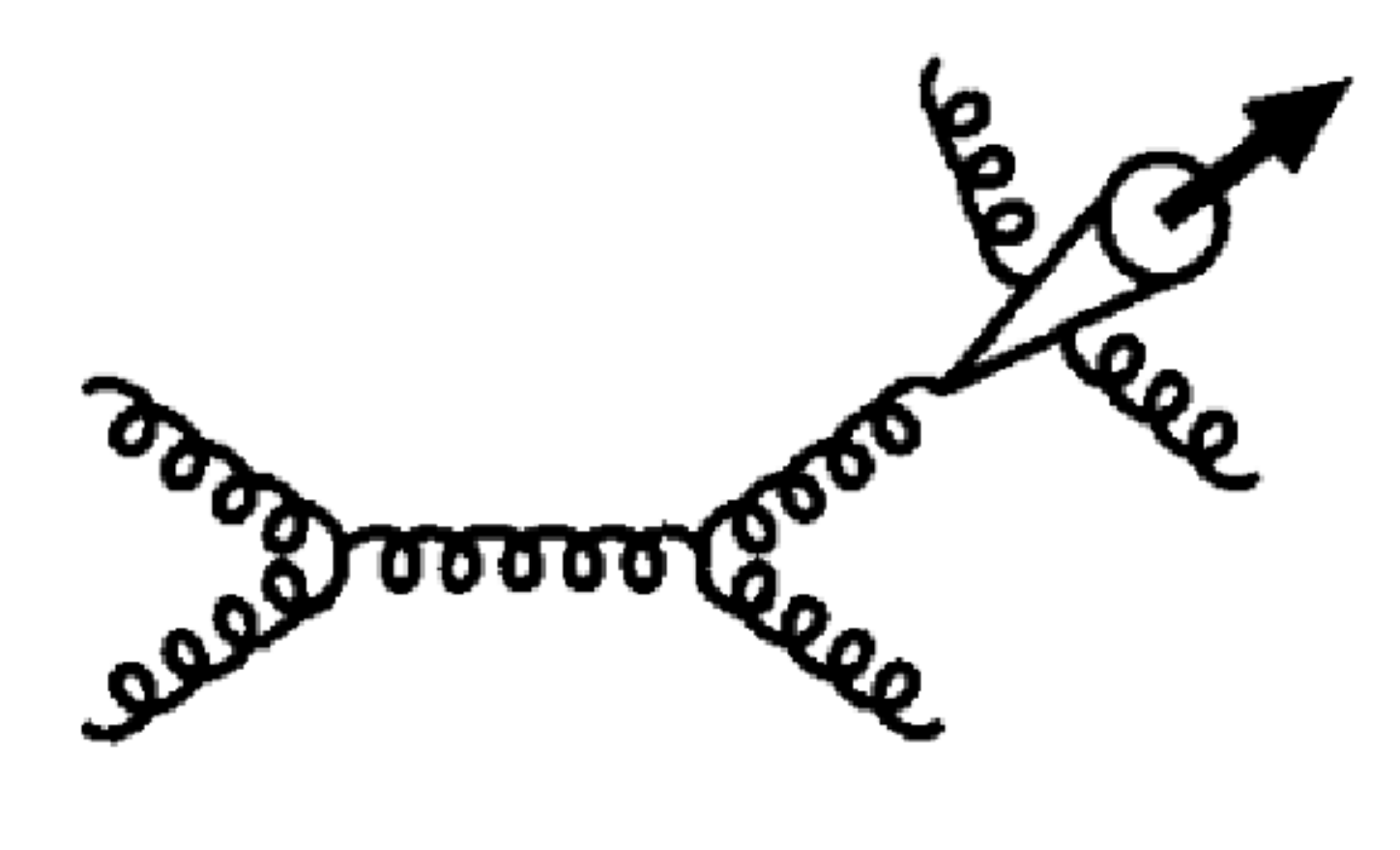} 
        \isucaption[Feynman Diagrams for color-singlet Charmonium Production.]{Feynman diagrams for color-singlet charmonium production at leading-order (left) and from color-singlet fragmentation (right).}\label{fig:CS_diagrams}
\end{figure}

\begin{figure}[h!tb]
        \centering\includegraphics*[width=0.25\columnwidth]{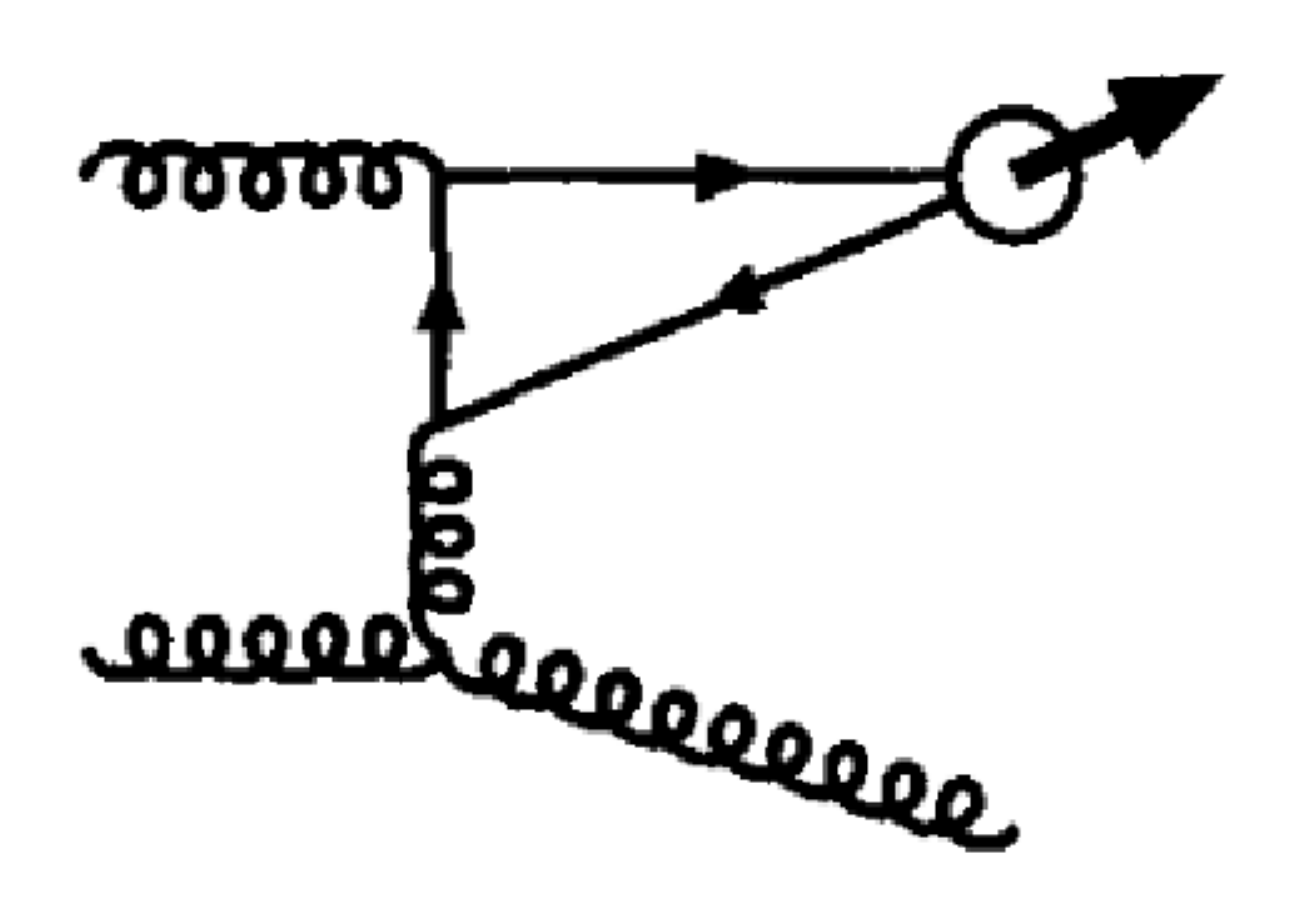} 
        \centering\includegraphics*[width=0.25\columnwidth]{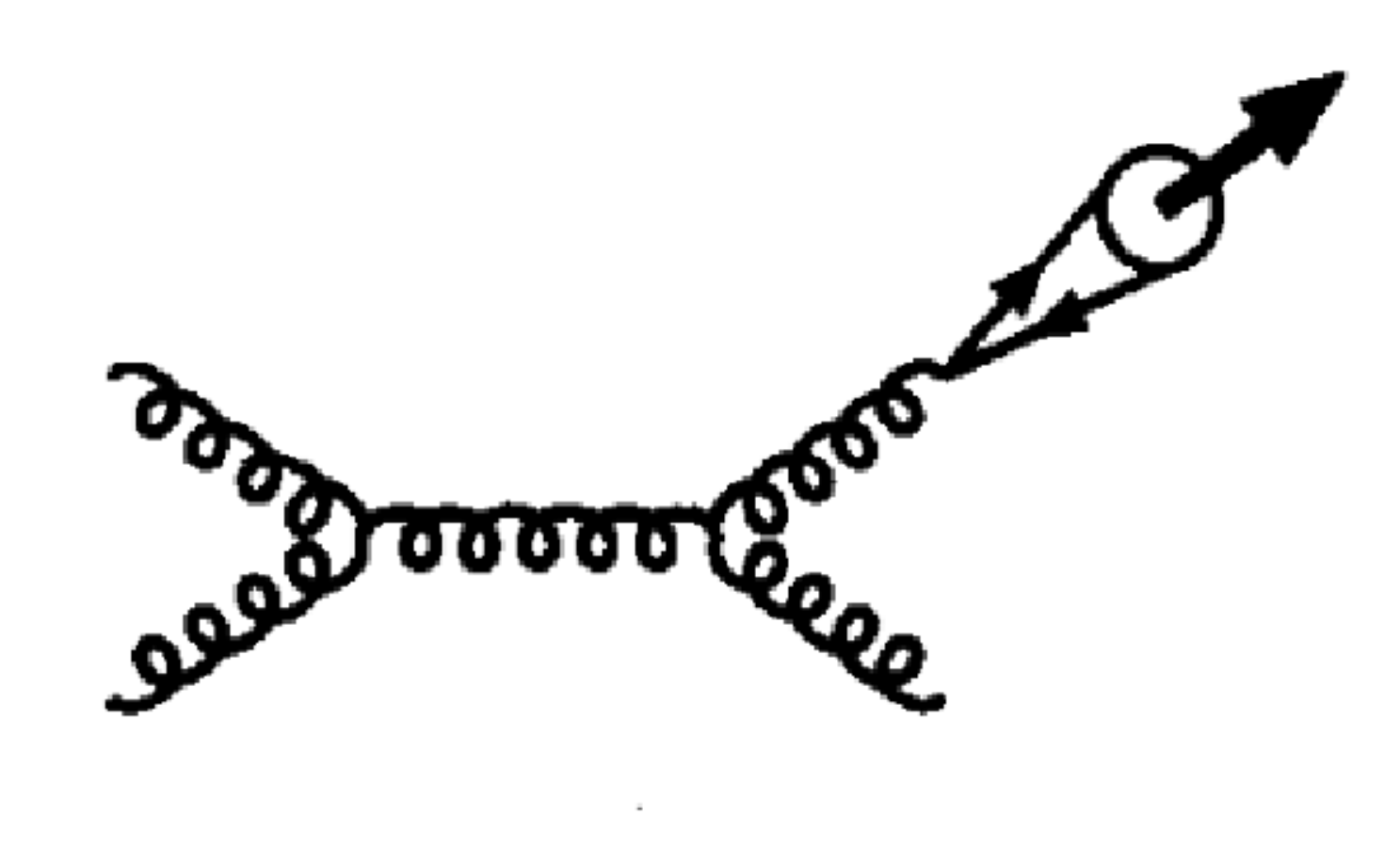} 
        \isucaption[Feynman Diagrams for color-singlet Charmonium Production.]{Feynman diagrams for color-octet charmonium production for leading-order production in the t-channel (left) and from color-octet fragmentation (right).}\label{fig:CO_diagrams}
\end{figure}

The Color-Octet Model (COM)~\cite{Bodwin:1994jh} attempts to formalize the factorization of relativistic and non-relativistic effects.  The model uses a generic expansion
\begin{equation}
d\sigma(J/\psi+X) = \displaystyle\sum_{n} \int_{0}^{\infty} dM \frac{d\sigma_{c\bar{c}[n]+X}}{dM}\langle O_{[n]}^{J/\psi} \rangle
\end{equation}
with parameters $\langle O_{[n]}^{J/\psi}\rangle$, non-relativistic matrix elements associated with the amplitude for producing a $J/\psi$ from a $c\bar{c}$ pair in state $[n]$.  Techniques developed in non-relativistic QCD~\cite{Caswell:1985ui} are then applied to determine the size of the $\langle O_{[n]}^{J/\psi} \rangle$ parameters in powers of $\textit{v}$, the relative velocity between the $c$ and $\bar{c}$, where $v^{2}$$\approx$0.3$c^{2}$ for the $J/\psi$~\cite{Kramer:1995nb}.  The model is thus a double expansion, about $v^{2}$ and $\alpha_{S}$.  The COM predicts that the leading color-singlet diagram is of order $\alpha_{s}^{3} (2 m_{c})^{4} / p_{T}^{8}$, while the color-octet is of order $\alpha_{s}^{3} (2 m_{c})^{2}\textit{v}^{4} / p_{T}^{6}$.  For $J/\psi$ mesons from fragmentation, the leading color-singlet diagram is of order $a_{s}^{5} / p_{T}^{4}$, and the color-octet is of order $\alpha_{s}^{3}\textit{v}^{4} / p_{T}^{4}$~\cite{Kramer:1998wn}.  These diagrams are shown schematically in Fig.~\ref{fig:CS_diagrams} and~\ref{fig:CO_diagrams}.  At large $p_{T}$, color-octet fragmentation becomes especially important, leading to the predicition of transverse spin-alignment at large $p_{T}$ in the Helicity reference frame for the COM (because the majority of the $J/\psi$ spin should come from a single transversely polarized gluon)~\cite{Beneke:1996yw,Braaten:1999qk}.  

\begin{figure}[h!tb]
  \centering\includegraphics*[width=0.45\columnwidth]{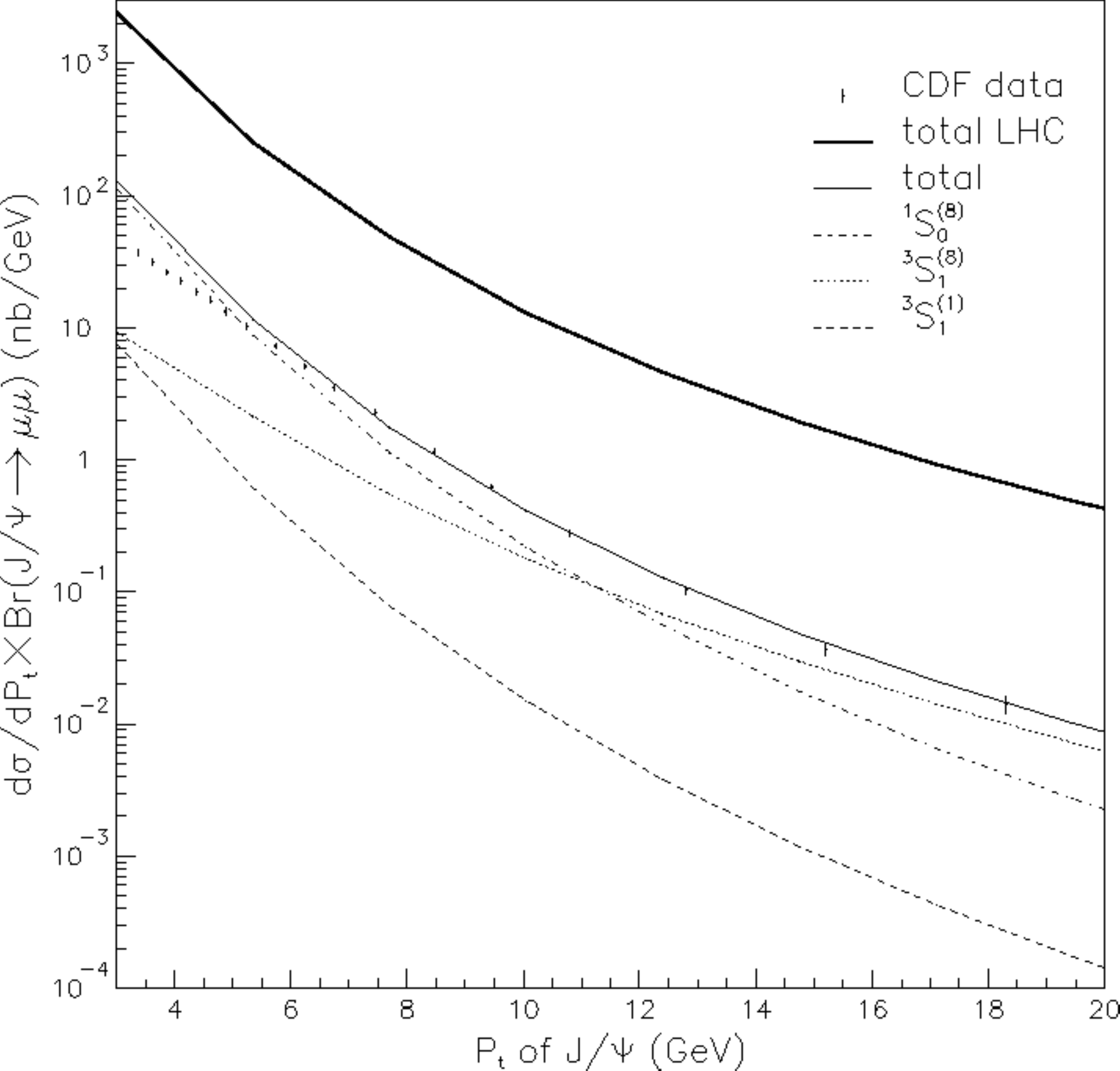} 
  \isucaption[Comparison of the COM at NLO in $\alpha_{S}$ with the $J/\psi$ cross-section from CDF.]{Comparison of the COM at NLO in $\alpha_{S}$ with the cross-section for directly produced $J/\psi$ mesons from CDF as well as predictions for the cross-section at the LHC (from~\cite{Gong:2008ft}).}\label{fig:CO_NLO_xsec}
\end{figure}

\begin{figure}[h!tb]
  \centering\includegraphics*[width=0.65\columnwidth]{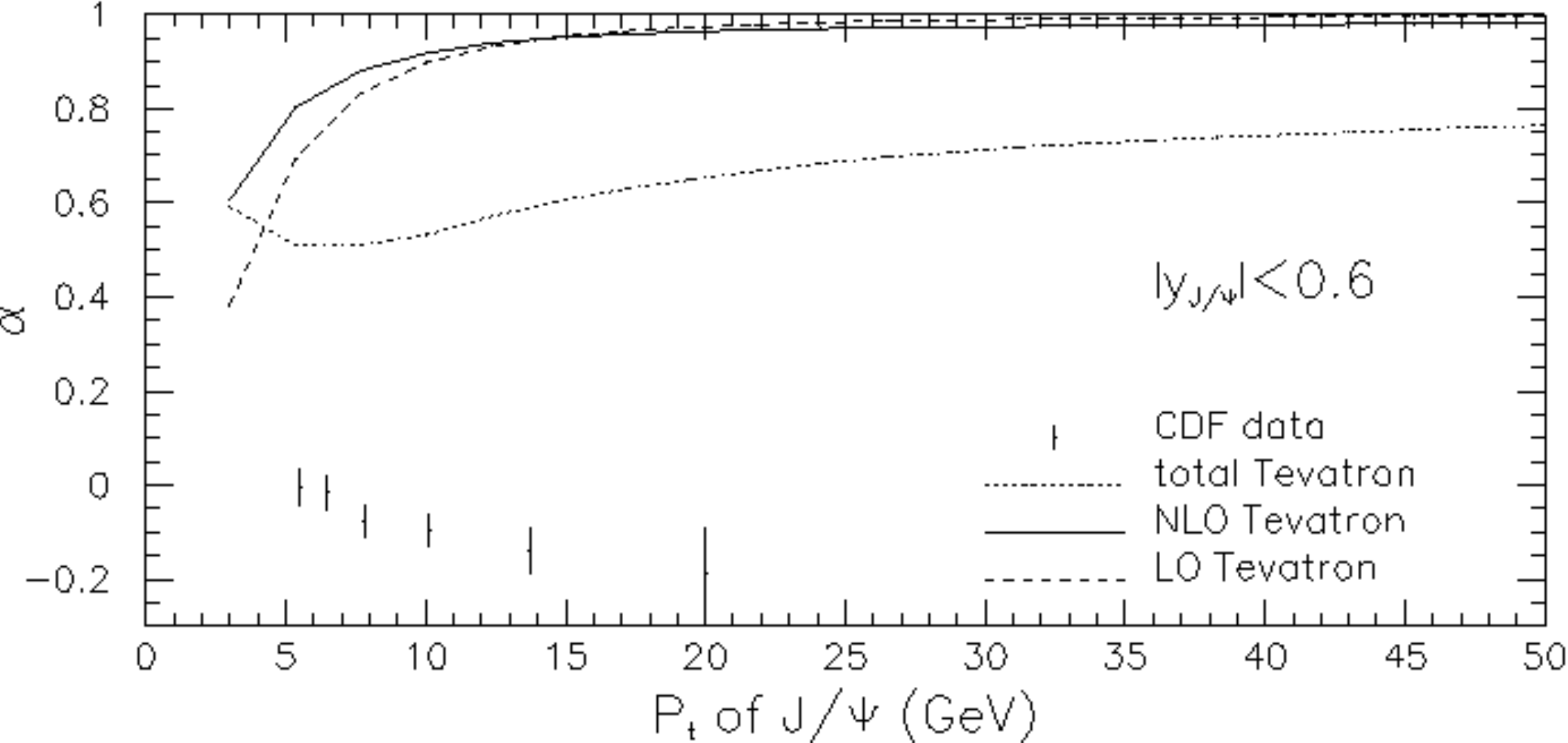} 
  \isucaption[Comparison of the COM at NLO in $\alpha_{S}$ with $J/\psi$ spin-alignment data from CDF.]{Comparison of the COM at NLO in $\alpha_{S}$ with the $\alpha$ spin-alignment coefficient (called `$\lambda_{\vartheta}$' in this document) for directly produced $J/\psi$ mesons from CDF (from~\cite{Gong:2008ft}).}\label{fig:CO_NLO_pol}
\end{figure}

The COM takes its name from the prediction that color-octet diagrams dominate for $J/\psi$ production at large $p_{T}$ for hadron colliders like the Tevatron, but it also predicts that color-singlet diagrams dominate for $J/\psi$ mesons with smaller $p_{T}$ from $e^{+}e^{-}$ collisions at B-factories like KEKB.   A comparison of a COM calculation at next-to-leading order in $\alpha_{s}$ (NLO) with data from CDF is shown for the cross-section in Fig.~\ref{fig:CO_NLO_xsec} and for the $\lambda_{\vartheta}$ angular decay coefficient in Fig.~\ref{fig:CO_NLO_pol}~\cite{Gong:2008ft}.  The model agrees very well with the cross-section but drastically disagrees with $\lambda_{\vartheta}$.

\section{Discussion}\label{sec:production_discussion}

None of the $J/\psi$ production mechanism models discussed here adequately describe the data.  Charmonium cross sections from hadronic collisions are well matched by the CEM, but the model does not make predictions for DIS or $e^{+}e^{-}$ collisions and has very little predictive power for angular decay coefficients.  The CSM makes predictions for the $\lambda_{\vartheta}$ angular decay coefficient which agree with data, but the model severely underpredicts measured $J/\psi$ cross sections.  The COM predictions, meanwhile, match measured cross sections but disagree in both sign and magnitude with the $\lambda_{\vartheta}$ angular decay coefficient.

Unfortunately, there is also some tension between separate measurements of $\lambda_{\vartheta}$.  CDF, the leading experiment reporting $\lambda_{\vartheta}$ at large $p_{T}$, found $\lambda_{\vartheta}$ consistent with the COM from Run-I data~\cite{Affolder:2000nn} but inconsistent with the model from Run-II~\cite{Abulencia:2007us}.  While the change of conclusion has been attributed to the identification of systematic effects which were not accounted for in the analysis of the Run-I data, further measurements are necessary to confirm disagreement with the COM.

Previous measurements of $\lambda_{\vartheta}$ at mid-rapidity from $p+p$ collisions at $\sqrt{s}$=200~GeV from PHENIX~\cite{Adare:2009js} are limited to small $p_{T}$, where predictions have been made from both the CSM~\cite{Lansberg:2010vq} and COM~\cite{Chung:2009xr}, but the COM predictions are questionable at small $p_{T}$ without further theoretical effort.  Furthermore, the data are not able to distinguish between the predicted $\lambda_{\vartheta}$ from the COM and CSM in the region where the calculations have been made.

A change in center of mass energy from $\sqrt{s}$=200~GeV to $\sqrt{s}$=500~GeV increases the $J/\psi$ cross section~\cite{Gavai:1994in}.  An increase in $\sqrt{s}$, along with the larger coverage of the PHENIX muon spectrometers relative to the central spectrometers (2 units of rapidity as opposed to 0.7), mean that a measurement at forward rapidity from $\sqrt{s}$=500~GeV $p+p$ collisions at PHENIX allows for significantly higher statistics.  Measurements with larger statistics at forward rapidity, which will be presented in Chapter~\ref{ch:jpsi_pol}, bring the $p_{T}$ coverage of PHENIX to a region where the COM is less questionable.

Furthermore, the full azimuthal coverage of the PHENIX muon spectrometers allows for measurements of the $\lambda_{\varphi}$ and $\lambda_{\vartheta\varphi}$ coefficients in addition to $\lambda_{\vartheta}$.  Measuring all three coefficients in several reference frames gives a much deeper understanding of $J/\psi$ production, providing several reference points for comparisons to production mechanism models.   Recent measurements from $p+N$ collisions at $\sqrt{s}$=41.6~GeV from HERA-B~\cite{Abt:2009nu} suggest that the Collins-Soper $\hat{z}$-axis is closer to the natural axis for $J/\psi$ spin alignment than the $\hat{z}$-axis of the Helicity frame, and theoretical attempts to determine the true natural axis have already begun~\cite{Braaten:2008xg}.  Measurements of all angular decay coefficients from PHENIX in several reference frames, as I will present in Chapter~\ref{ch:jpsi_pol}, can either confirm or contradict observations from HERA-B and provide input to theory, leading to a more comprehensive understanding of $J/\psi$ production.

%% file: chapters/detector/detector.tex
\chapter{RHIC and the PHENIX Experiment}\label{ch:detector_overview}

\section{RHIC}\label{sec:RHIC}

The Relativistic Heavy Ion Collider (RHIC) is a circular accelerator with 0.6~km radius in Upton, NY capable of colliding heavy ions (Au, Cu, Si, etc.) at beam energies up to 100~GeV and polarized protons at beam energies up to 250~GeV.  

For polarized proton collisions, an optically pumped polarized H$^{-}$ source provides protons to a Linear Accelerator (LINAC), which accelerates them to 200~MeV and injects into the Alternating Gradient Synchrotron (AGS).  The AGS further accelerates the protons to 1.5~GeV before injecting them into the RHIC accelerator ring, where they are brought to full energy and eventually into collision.

Beam polarizations are measured by two independent polarimeters, a fast carbon-target polarimeter~\cite{Nakagawa:2008zzb} for relative polarization measurements and a hydrogen-jet polarimeter~\cite{Okada:2005gu,Eyser:2006he} for an absolute measurement.  The carbon-target polarimeter measures the polarization for each store of beams in the RHIC accelerator, and the hydrogen-jet polarimeter is used over a longer time-period to obtain a calibration for the carbon-target measurements.  The stable polarization direction in the RHIC ring is transverse to the direction of the protons' motion, and the polarimeters take advantage of the transverse single-spin asymmetries (SSA) discussed in Section~\ref{ch:trans_protons} in order to measure the beam polarization.  

\begin{figure}[h!tb]
        \centering\includegraphics*[width=0.75\columnwidth]{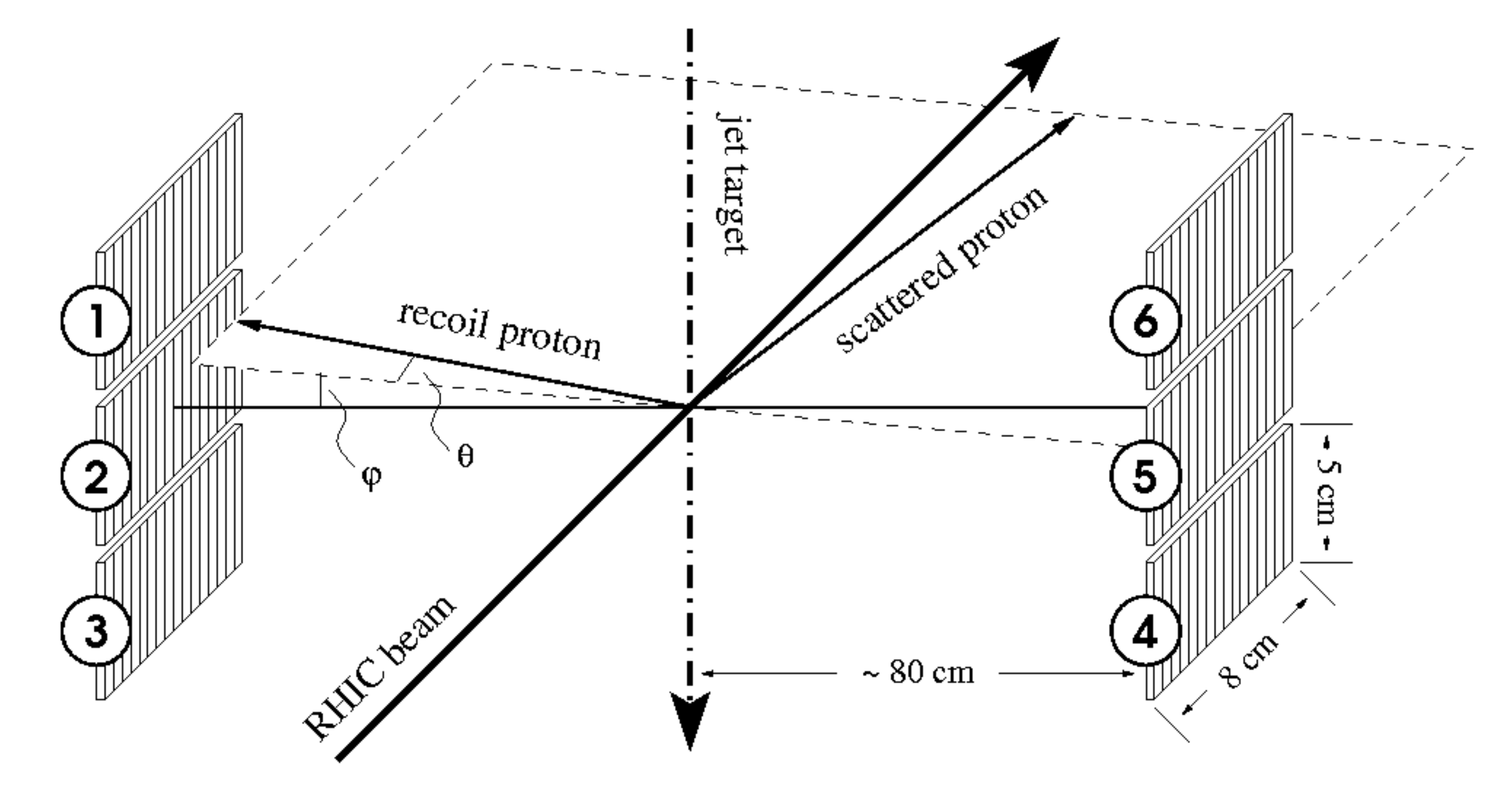} 
\isucaption[Setup of the RHIC hydrogen-jet polarimeter.]{Setup of the RHIC hydrogen-jet polarimeter.  Silicon detectors on the right and left measure the transverse single spin asymmetry of the recoil protons.}\label{fig:Hjet_polarimeter}
\end{figure}

For the hydrogen-jet polarimeter, a polarized hydrogen jet is brought into collision with each of the proton beams, and recoil protons are measured by silicon detectors (see Fig.~\ref{fig:Hjet_polarimeter}).  Data from the detectors is used to calculate $A_{N,target}$, the transverse SSA with an unpolarized beam and polarized target, and $A_{N,beam}$, the transverse SSA with a polarized beam and unpolarized target.\footnote{In reality both the beam and the target are always polarized, and statistics are combined in such a way that the asymmetry is only sensitive to one polarized particle.  The procedure and consequences for doing this will be discussed in Section~\ref{sec:asymmetry_formulae}.}  The beam polarization is then
\begin{equation}
\mathcal{P}_{beam} = -\frac{A_{N,beam}}{A_{N,target}}\mathcal{P}_{target}
\end{equation}
where the target polarization $\mathcal{P}_{target}$ is independently measured.

\begin{figure}[h!tb]
        \centering\includegraphics*[width=0.35\columnwidth]{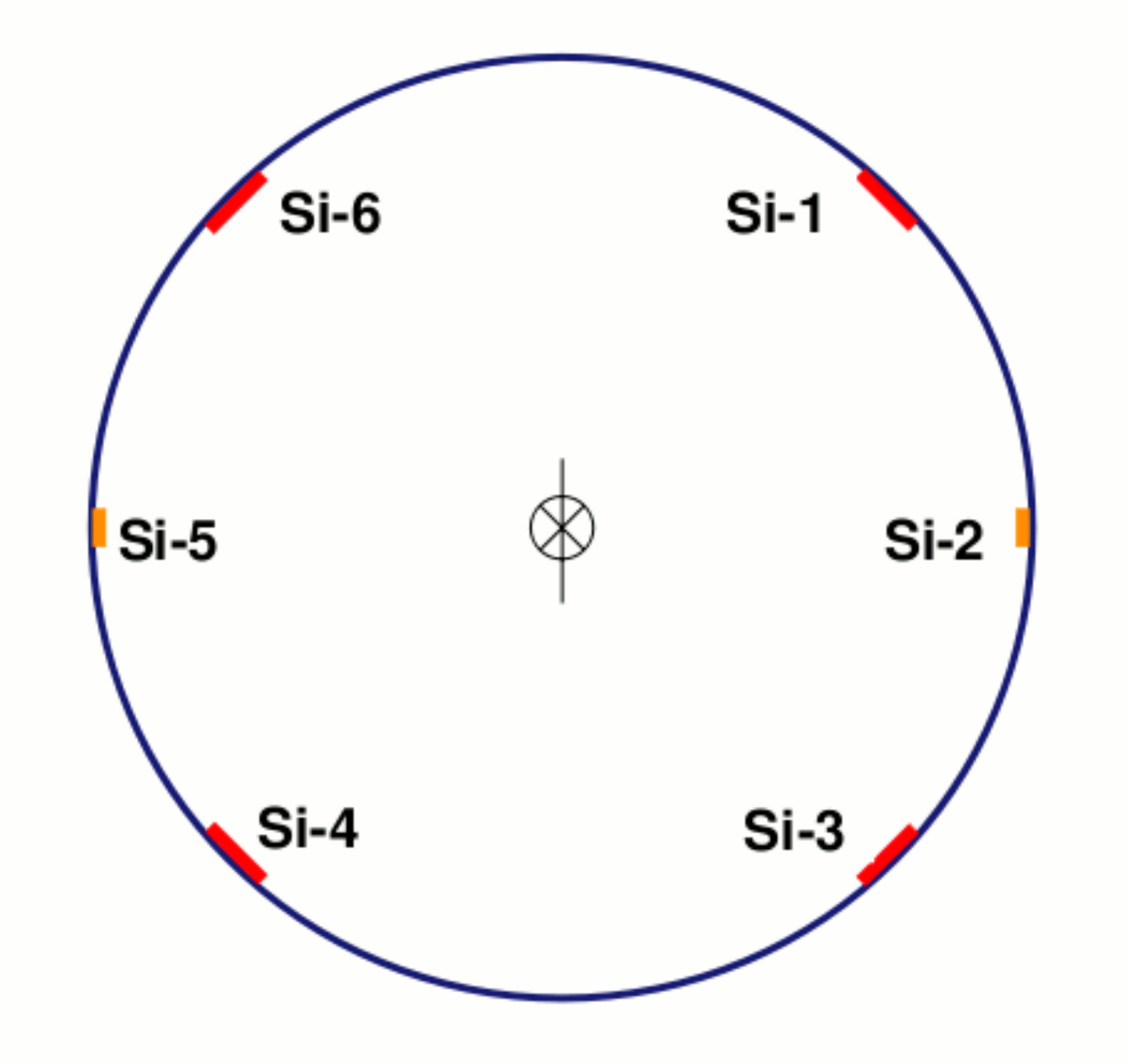} 
\isucaption[Setup of the RHIC carbon-target polarimeter.]{Setup of the RHIC carbon-target polarimeter.  Silicon detectors surround the target (represented by the cross-hairs in the center of the image).  The proton beam momentum direction is through the page.}\label{fig:pC_polarimeter}
\end{figure}

In the carbon-target polarimeter, an unpolarized ultra-thin carbon target is brought into collision with the polarized proton beams. The transverse single spin asymmetry $A_{N,beam}$ is then determined in each store by measuring the spin-dependent azimuthal distribution of scattered ions with silicon detectors surrounding the target~(Fig.~\ref{fig:pC_polarimeter}).  Because the transverse SSA scales with $\frac{1}{\mathcal{P}}$, only a single absolute measurement of $\mathcal{P}$ (from the hydrogen-jet polarimeter) is necessary to calibrate the carbon-target polarimeter and determine polarizations for all stores.

\section{PHENIX}

\begin{figure}[h!tb]
        \centering\includegraphics*[width=0.44\columnwidth]{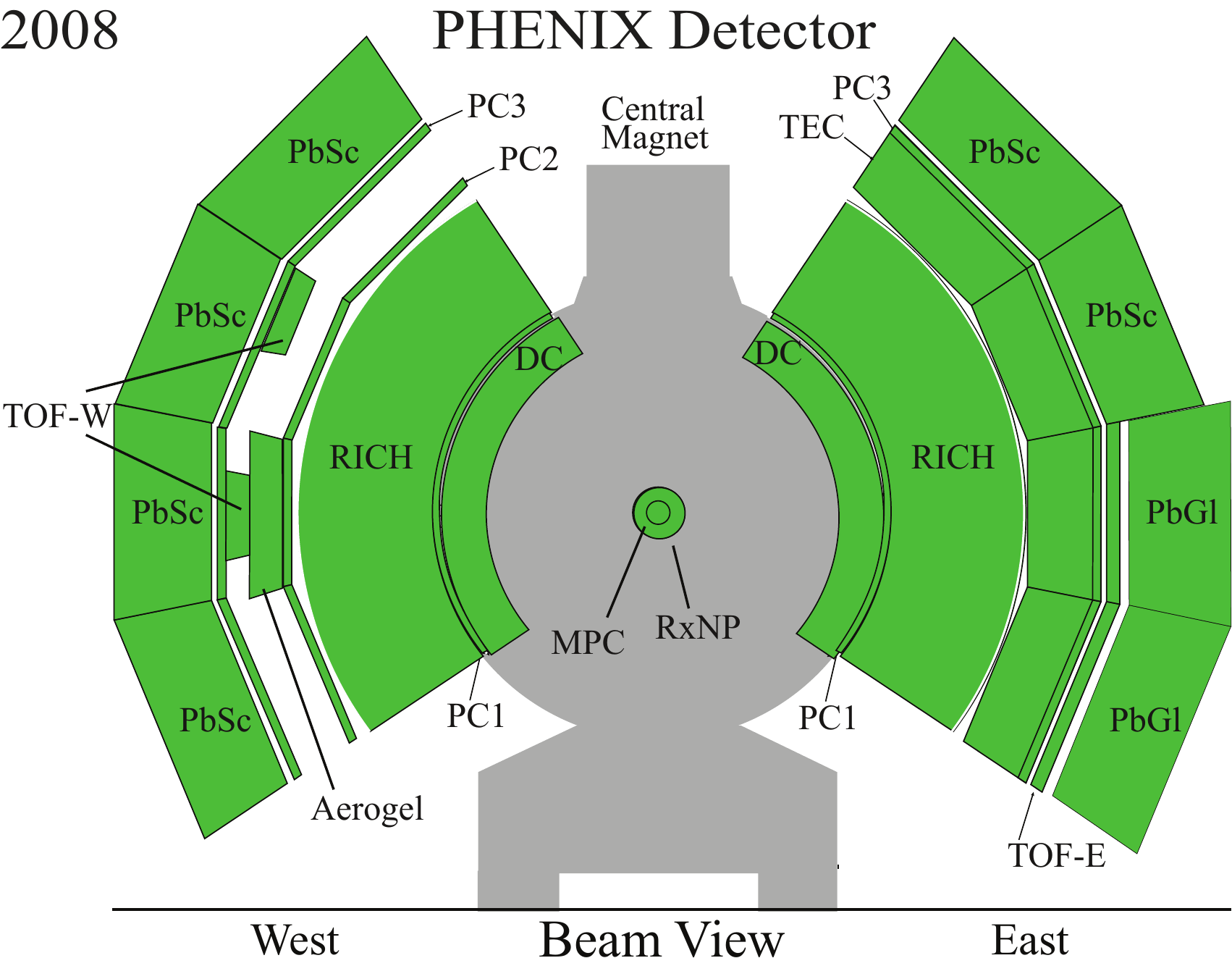} 
        \centering\includegraphics*[width=0.54\columnwidth]{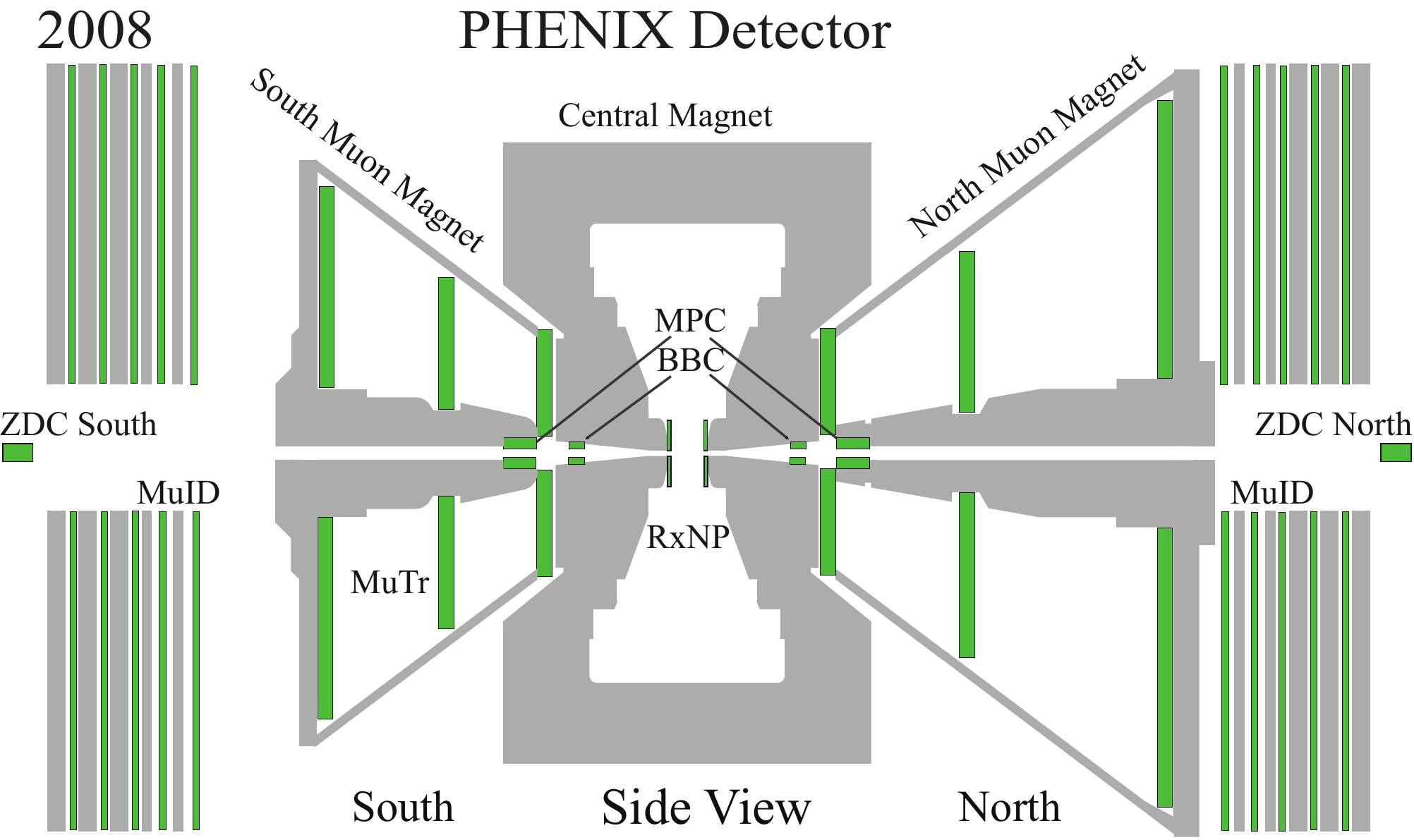}
\isucaption[PHENIX Detectors in beam and side views.]{Drawing of the PHENIX Detectors in beam view (left) and side view (right).}\label{fig:Detector_Drawings}
\end{figure}

The Pioneering High Energy Nuclear Interaction eXperiment (PHENIX) is a large, multi-purpose experiment on the RHIC accelerator ring.  It was designed with fine granularity, leading to good position and momentum resolution at the sacrifice of acceptance to produced particles~\cite{Adcox:2003zm}.  The experiment consists of four large spectrometers: two for tracking and calorimetry at central rapidity and two for muon tracking and identification at forward and backward rapidities.  The central detectors cover $|\eta|<$0.35 and $\Delta\phi=2\times\frac{\pi}{2}$, while the muon detectors cover approximately 1.2$<|\eta|<$2.2 and $\Delta\phi=2\pi$.  For global event characterization, there are beam-beam counters (BBC) to measure event time and position.  Additionally, a local polarimeter is used to monitor the proton polarization direction.  A schematic view of the various detector subsystems can be seen in Fig.~\ref{fig:Detector_Drawings}.  

\subsection{Detectors for Event Characterization}\label{sec:BBC}

In order to make a useful measurement with a high energy collider, it is typically necessary to determine both the time and position of each collision.  A determination of the time of the collision is necessary in order to tag the crossing in which the collision occurred, and the event vertex is necessary for tracking.  A measurement of the collision time is especially important for spin-dependent measurements, where the collision must be restricted to a single bunch in the accelerator in order to identify the polarization direction of the colliding protons.  PHENIX uses beam-beam counters (BBC) to measure the time and longitudinal position of each collision~\cite{Allen:2003zt}.\footnote{The transverse size of the colliding beams is less than 100~$\mu$m and has limited effects on reconstructed particle momenta.}

The BBC consists of two identical counters positioned 144~cm on either side of the center of the experiment.  Each counter consists of 64 photo-multiplier tubes (PMTs) mounted on a 3~cm thick quartz radiator in a cylinder of outer radius 30~cm and inner radius 10~cm  (corresponding to approximately $3.0<|\eta|<3.9$).  Raw signals from the PMTs are sent through time-to-voltage converters to flash analog-to-digital converters, which pass digitized timing and pulse-height information to a local level-1 trigger (LL1).  The LL1 makes a decision regarding whether or not the raw signal information should be sent to an event builder and recorded (triggering will be discussed in more detail in Section~\ref{sec:Triggering}).  The RHIC beam clock operates at approximately 9.6 MHz so that a crossing occurs every 104~ns.  Each BBC has an RMS timing resolution of 54$\pm$4~ps, which is clearly fine enough resolution to determine the crossing in which a collision has occurred.

\subsection{Local Polarimeter}\label{sec:local_pol}

While RHIC is responsible for measuring beam polarization using the polarimeters discussed in Section~\ref{sec:RHIC}, PHENIX is responsible for monitoring the direction of that polarization near the collision vertex.  This monitoring using measurements of neutrons from a zero-degree calorimeter (ZDC) \cite{Adler:2000bd} and shower maximum detector (SMD), collectively referred to as the local polarimeter.

The ZDC consists of three layers of hadronic calorimeter located approximately 18~m from the center of the interaction region.  Each layer is composed of a Cu-W alloy absorber and a Hamamatsu R329-2 photo-multiplier, which collects \v{C}erenkov light from optical fibers behind the absorber.\footnote{Since neutrons are not charged, they cannot directly produce \v{C}erenkov radiation.  Instead, the radiation is produced by fast electrons emitted as fission products from neutron interactions in the calorimeter.}  The entire layer corresponds to 1.7 interaction lengths of material, and the detector has an energy resolution of 21\% for a 100~GeV neutron.  An additional 3.3~mm layer of scintillator is placed in front of the ZDC for the identification and veto of charged particles.  Neutrons are typically selected by requiring an energy deposit between 20 and 120~GeV along the beam direction in the ZDC along with less than one minimum ionizing particle in the veto scintillator.

Interleaved between the first and second layer of the ZDC are position sensitive SMDs composed of 7 15~mm wide horizontal strips and 8 20~mm wide vertical strips.  The SMDs are tilted at an angle of 45$^\circ$ away from the interaction region with a position resolution of approximately 1~cm and an active area covering $\sim$0.3-1.4~mrad from the beam.

To determine a polarization direction, the transverse SSA of forward neutrons is measured as a function of azimuthal angle from the nominal spin direction and fit with 
\begin{equation}
A(\phi) = A_N \sin(\phi - \phi_0)
\end{equation}
to monitor a phase $\phi_0$ from the nominal spin orientation.

\subsection{Central Arm Detectors}\label{sec:Central_Arms}

The PHENIX central arm spectrometers are composed of separate subsystems for tracking, particle identification (PID), and calorimetry:

\begin{enumerate}
  \item
    Calorimetry is done using Pb-scintillator sampling calorimeters and a Pb-glass \v{C}erenkov calorimeter.~\cite{Aphecetche:2003zr}.
  \item 
    PID for electrons is handled by a ring-imaging \v{C}erenkov detector~\cite{Aizawa:2003zq}.
  \item 
    Tracking uses drift chambers and pad chambers~\cite{Adcox:2003zp}. 
\end{enumerate}
All contributing subsystems are used in coordination to make a meaningful measurement~\cite{Mitchell:2002wu}.

\subsubsection{Central Arm Calorimetry}\label{sec:Central_Calorimetry}

Two subsystems located at the outer radius of the PHENIX central arms make up the electromagnetic calorimeter (EMCAL).  Approximately 75$\%$ of the azimuthal acceptance of the detector uses a Pb-scintillator sampling calorimeter (PbSc), and the other 25$\%$ uses a Pb-glass \v{C}erenkov calorimeter (PbGl).  The PbGl has better granularity and energy resolution, while the PbSc has better timing resolution (Table~\ref{tab:PbSc_PbGl_properties}), making the two subsystems quite complimentary.

\begin{figure}[h!tb]
  \centering\includegraphics*[width=0.55\columnwidth]{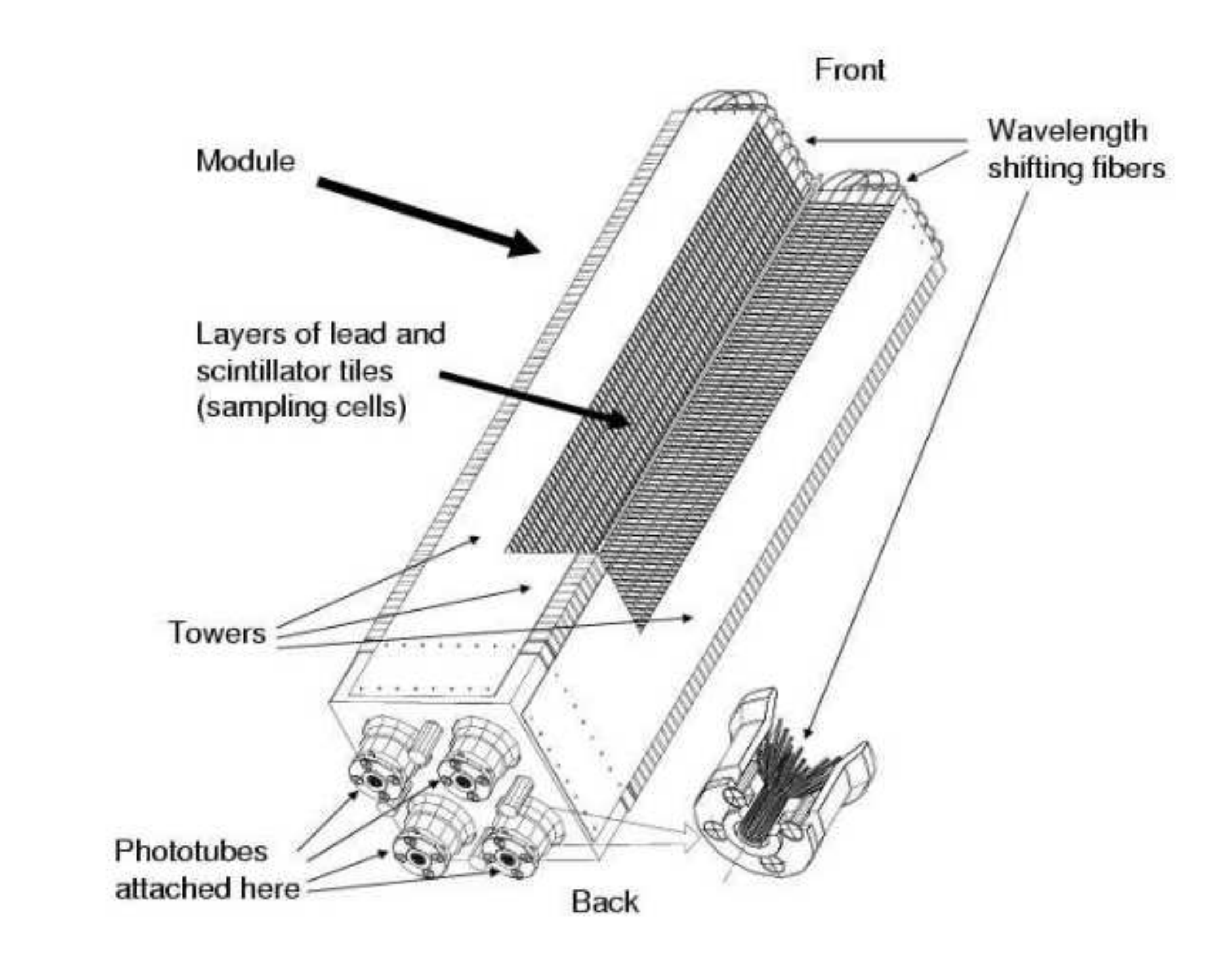}
  \isucaption{Diagrammatic view of a PHENIX Pb-scintillator calorimeter module.}\label{fig:PbSc_module}
\end{figure}

A diagrammatic drawing of a PbSc module is shown in Fig.~\ref{fig:PbSc_module}.  Each module consists of four towers composed of 66 5.535~cm$\times$5.535~cm sampling cells of alternating Pb and scintillator\footnote{The scintillator used in the sampling cells is p-bis[2-(5-Phenyloxazolyl)]-benzene (POPOP) with a fluorescent additive p-Terphenyl (PT)} with edges plated in Al.  The Pb acts as a passive absorber, while the scintillator samples incoming energy.  Light is collected by 36 fibers in each cell and read out by photo-tubes in the back of the tower.  One `super-module' is made up of 32 modules mounted on a single stainless steel skin.  Eighteen of these super-modules are combined to make one 2$\times$4 $m^{2}$ sector with its own steel support frame.

\begin{figure}[h!tb]
  \centering\includegraphics*[width=0.75\columnwidth]{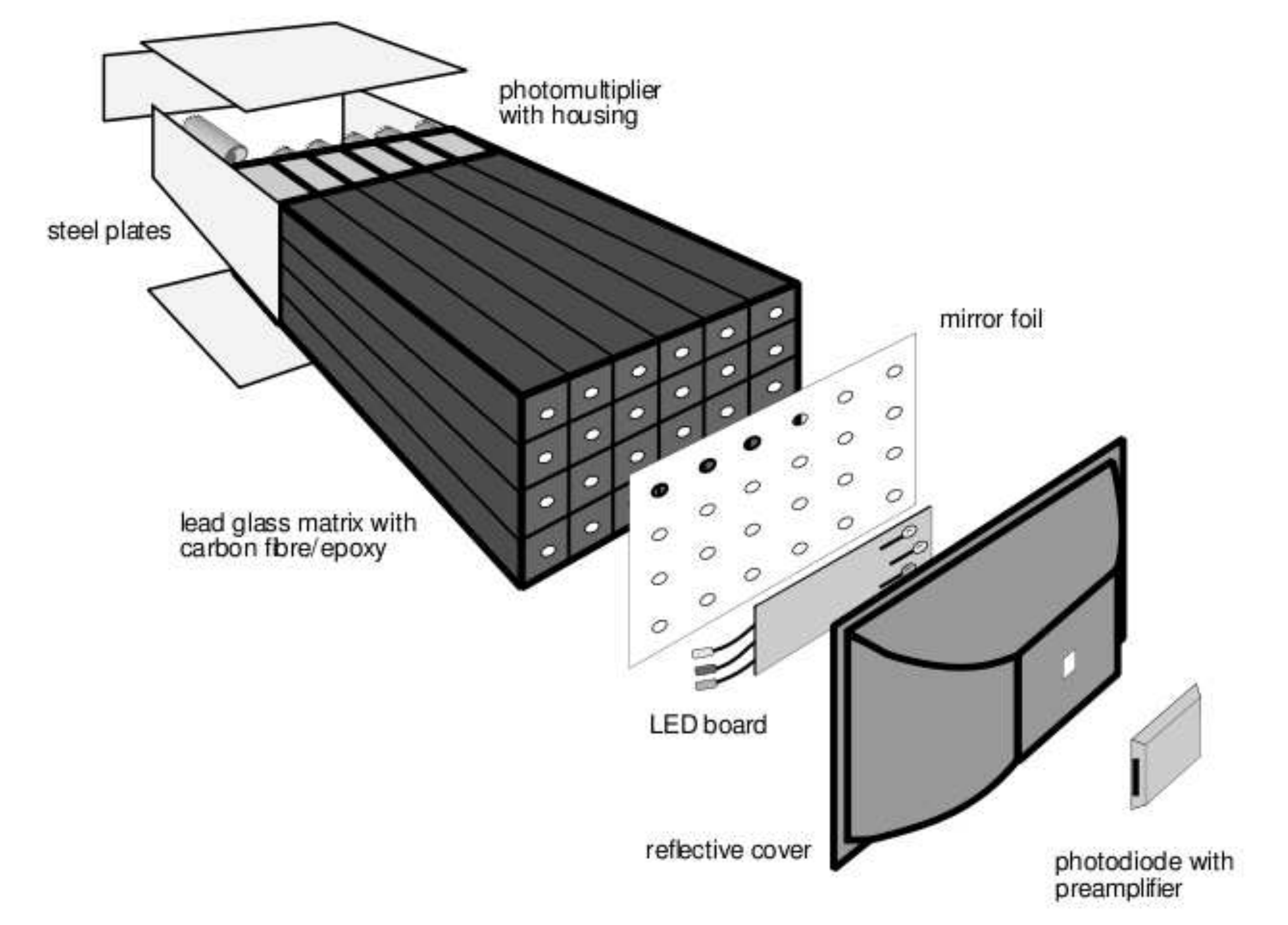}
  \isucaption{Exploded view of a PHENIX Pb-glass calorimeter super-module.}\label{fig:PbGl_Smodule}
\end{figure}

The PbGl section of the central calorimeter is composed of 192 super-modules, each with 24 separate Pb-glass modules.  The modules have a high index of refraction so that high energy charged particles moving through them radiate \v{C}erenkov light, and each module has a dedicated photo-multiplier tube for readout.  Fig.~\ref{fig:PbGl_Smodule} shows an exploded view of a PbGl super-module along with the LED board consisting of dedicated LEDs for calibrating each module.

\begin{table}[htbp]
\centering
\isucaption{Energy, spatial, and timing resolution of the PHENIX PbGl and PbSc calorimeters for electrons.}{\label{tab:PbSc_PbGl_properties}}
\begin{tabular}{cccc}
\hline
Detector & $\sigma_{E}(E) / E$                          & $\sigma_{x}(E)$~(mm)                 &  $\sigma_{t}(E)$~(ns)  \\ \hline
PbSc     & $2.1\%\oplus\frac{8.1\%}{\sqrt{E/GeV}}\quad$ & $1.55\oplus\frac{5.7}{\sqrt{E/GeV}}\quad$ &  $0.06\oplus\frac{0.03}{E/GeV - 0.01}$  \\ \hline 
PbGl     & $0.8\%\oplus\frac{5.9\%}{\sqrt{E/GeV}}\quad$ & $0.2\oplus\frac{8.4}{\sqrt{E/GeV}}\quad$  &  $0.075\oplus\frac{3.75}{\sqrt{500 \cdot E/GeV}}$  \\ \hline
\end{tabular}
\end{table}

\subsubsection{Central Arm Particle ID}

Located between the calorimeters and tracking detectors is the ring-imaging \v{C}erenkov (RICH) detector, used for discriminating between electrons and pions below the pion \v{C}erenkov threshold of 4~GeV/$c$.  The majority of the RICH volume consists of 40~m$^{3}$ of ethane held at 0.5'' of water above ambient between two glass plates.  Electrons moving through the ethane radiate a cone of \v{C}erenkov light, which is focused into $\sim$14.5~cm rings by spherical mirrors onto an array of photo-multiplier tubes.  Pions with momenta below 4~GeV/$c$ do not produce \v{C}erenkov light so that no ring is formed, and they can easily be rejected.  Imperfections in the focusing mirrors lead to a 2.5~mm inaccuracy in the ring position.

\subsubsection{Central Arm Tracking}

The Drift Chamber (DC) is the innermost detector of the central arm spectrometer and provides the primary momentum measurement and vector for tracking.  It is outside of the primary magnetic field, but a residual field of approximately 0.6~kGa accounts for $\sim$1$^{\circ}$ of track bending within the chamber.  The detector consists of 2m long cylindrically shaped detectors located 2-2.4~m from the beam axis.  Each chamber consists of 20 sectors, made up of 6 wire modules each.  There are a total of 6,500 anode wires, making up 13,000 readout channels, and leading to 165~$\mu$m  single wire resolution and approximately 2mm spatial resolution on each track. 

\begin{figure}[h!tb]
  \centering\includegraphics*[width=0.85\columnwidth]{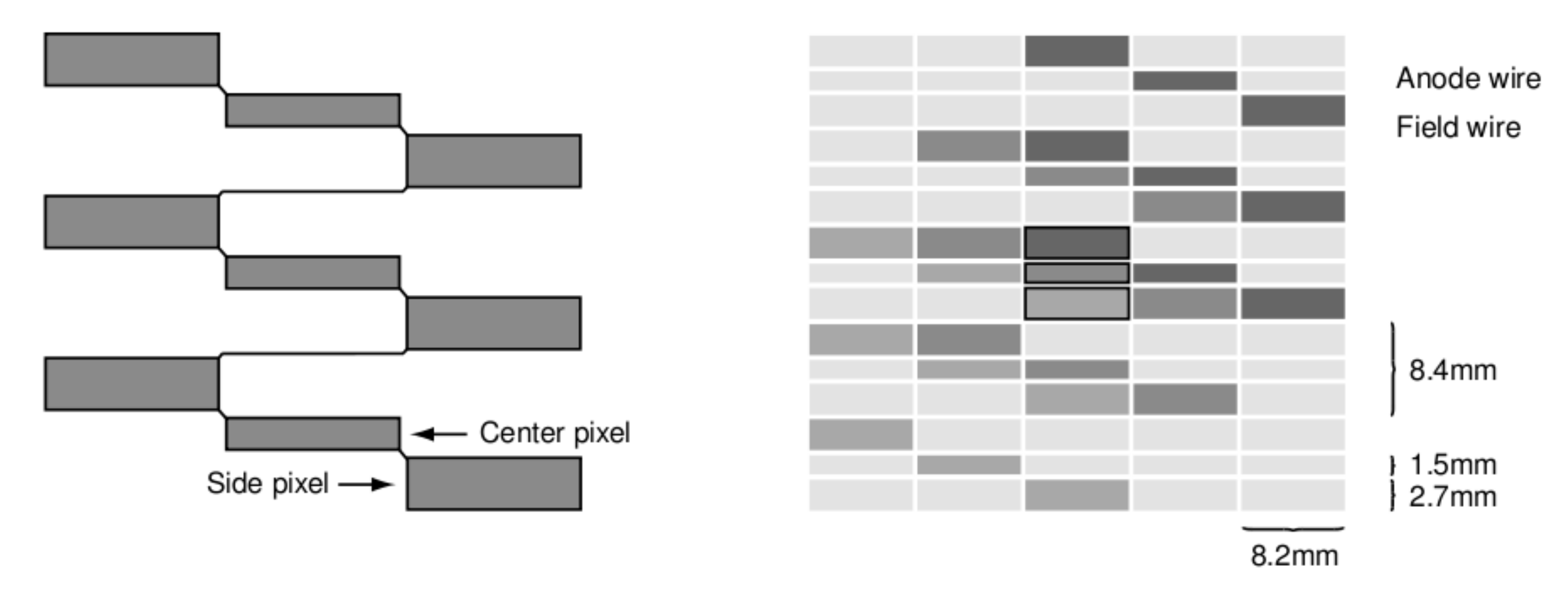}
  \isucaption[PHENIX Pad Chamber pad and cell configurations.]{PHENIX Pad Chamber pad and cell configurations.  On the left is one pad, composed of 9 pixels, and on the right is a number of interleaved pads.  The outlined boxes correspond to a hit in one cell (3 pads).\label{fig:PC_pads}}
\end{figure}

Outside of the DC are 2 layers of non-projective pad chamber (PC), one directly inside of the RICH and one inside of the EMCAL.  The West arm has an additional layer of PC directly outside of the RICH.  The inner layer of PC provides a z-coordinate at the exit of the DC and assists in projecting tracks through the RICH, while the larger outer layer provides matching to clusters in the EMCAL.  Each PC is segmented into pixels 8.2~mm in z by 8.4~mm in r-$\phi$.  In order to consolidate readout channels and minimize noise, 9 pixels are ganged into one `pad' and hits in 1 cell (3 pads) are required to register a single hit (see Fig.~\ref{fig:PC_pads}).

\subsection{Muon Arm Detectors}\label{sec:Muon_Arms}

\begin{figure}[h!tb]
  \centering\includegraphics*[width=0.45\columnwidth]{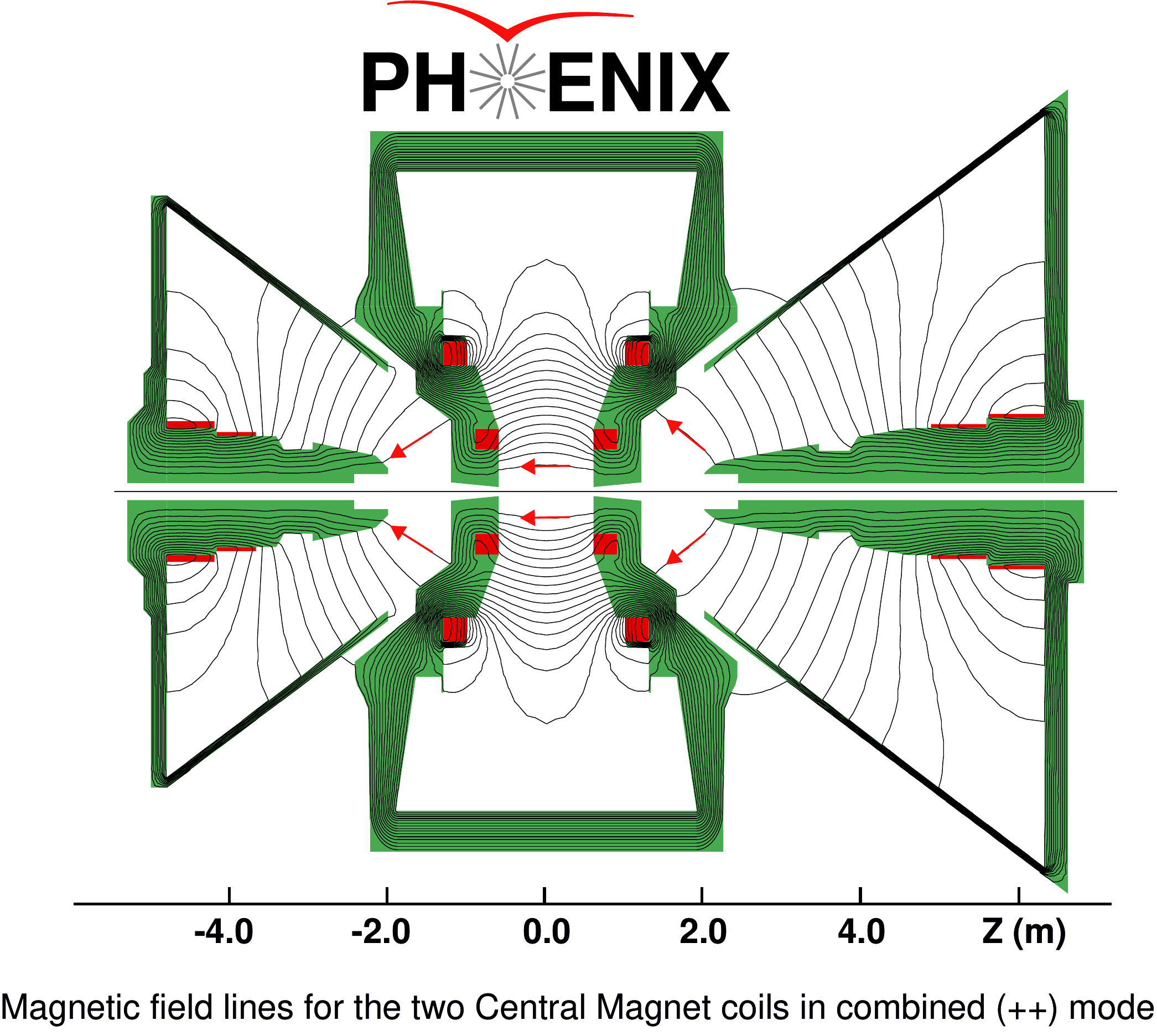}
  \isucaption[PHENIX magnetic field orientation during 2009 RHIC run.]{Drawing of the magnetic field orientation in the PHENIX detector during the 2009 RHIC Run.}\label{fig:B_Field_Drawing}
\end{figure}

At forward rapidities, spectrometry is devoted mainly to muon detection by two major subsystems, the Muon Tracker ($\mu$Tr), used for tracking, and the Muon Identifier ($\mu$Id) for muon identification~\cite{Akikawa:2003zs}.   The trajectory of a muon entering the detector is bent in azimuth by a radial magnetic field of several kGa, shown schematically in Fig.~\ref{fig:B_Field_Drawing} (magnets are described much more fully in~\cite{Aronson:2003zn}), so that their momenta can be measured by the finely segmented $\mu$Tr.  The majority of hadrons are absorbed by the thick steel plates of the $\mu$Id so that a muon trigger can be formed in the furthest gaps of the detector from the collision vertex. 

On each side of the collision vertex are 3 $\mu$Tr stations with 3 chambers in the closest two stations and two in the furthest station from the vertex.  Each chamber is constructed in octants with the 5 mm wide cathode strips positioned at 11.25$^{\circ}$ from radial.  The chambers are filled with a gas mixture of 50$\%$Ar + 30$\%$CO$_{2}$ + 20$\%$CF$_{4}$ and held at a voltage of approximately 1850~V.  As an ionizing particle moves through the gas, $\sim$80~fC of charge is deposited on the cathode strips.  


\begin{figure}[h!tb]
  \centering\includegraphics*[width=0.45\columnwidth]{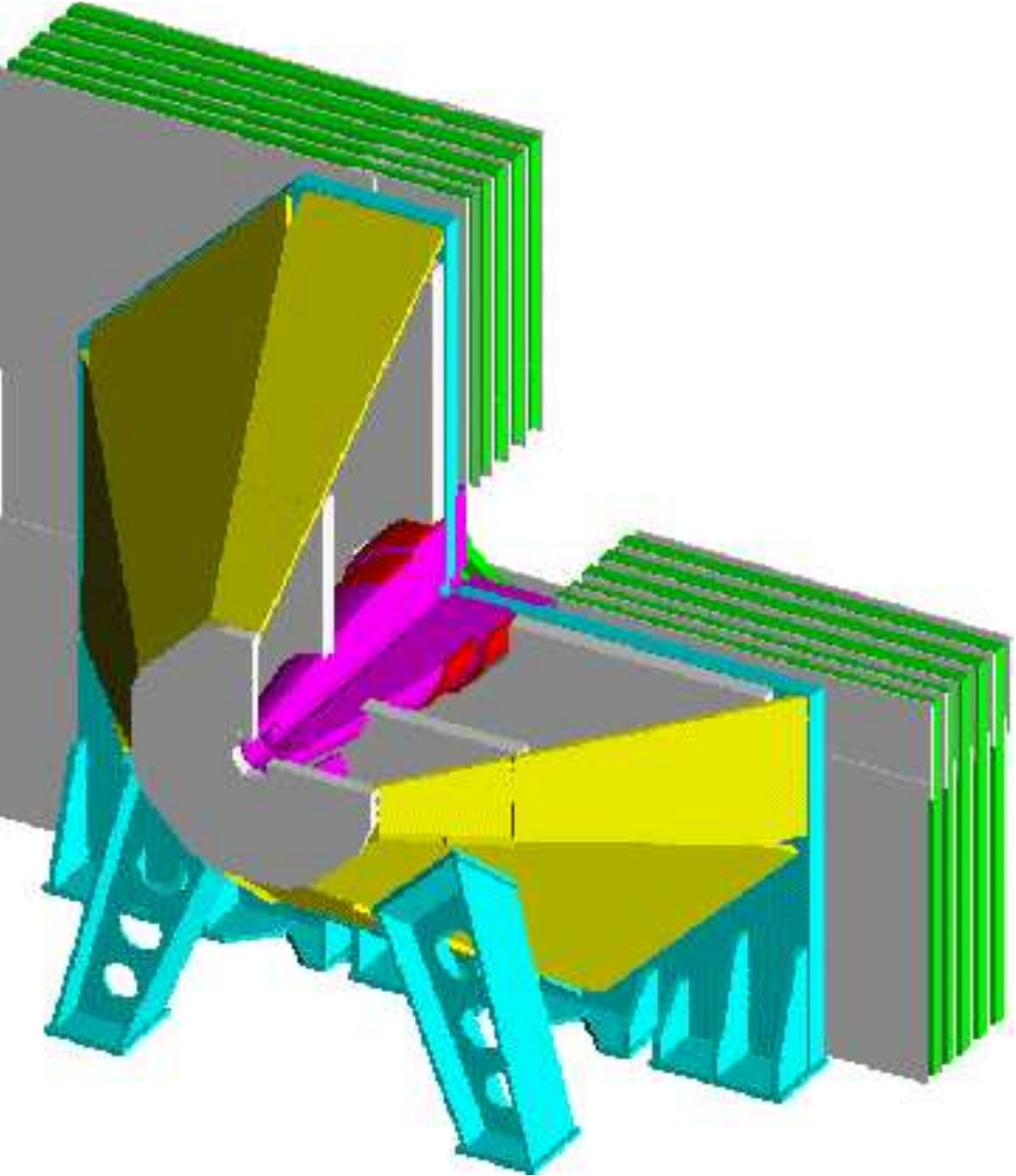}
  \isucaption[Cut-away view of the PHENIX muon detectors]{Cut-away view of the PHENIX muon detectors.  The three stations of the $\mu$Tr are located in the yellow structure while the $\mu$Id planes are drawn in green.}\label{fig:Muon_Detectors}
\end{figure}

Behind the $\mu$Tr on the north side of the experiment is a 30~cm thick steel backplate for the magnet, and a 20~cm thick backplate is behind the $\mu$Tr in the south.  The backplates are followed by 2 10~cm and 2 20~cm steel absorbers, creating a total of five gaps.  The gaps are instrumented with four planes of 8.4~cm wide Iarocci tubes each, two oriented vertically and two horizontally.  Signals between the two sets of tubes in the same orientation are OR'ed together creating a signal with higher efficiency and faster signal timing than would be produced by a single tube.  A muon must have an energy of 2.7~GeV in order to reach the final gap of the $\mu$Id, and the probability for a 4~GeV/$c$ pion to reach the final gap is 3$\%$ or less. 

\subsection{Triggering}\label{sec:Triggering}

The PHENIX data acquisition system is limited to $\sim$5 kHz of bandwidth for data throughput from all detectors, meaning that signals from subsystems cannot be constantly recorded.  Instead, data must be `triggered' using some minimal criteria for the detector signals.  In PHENIX, this is done with several local level-1 (LL1) trigger systems combined into a single global level-1 (GL1) trigger.  LL1 triggers are typically specific to a single subsystem, and the GL1 trigger combines trigger decisions from the separate LL1 triggers.  

In order to avoid writing useless data to disk, it is essential to determine that a collision has taken place.  The LL1 system responsible for determining a valid collision time and vertex is the BBC LL1 trigger, which takes raw input from the BBC and converts it into timing signals.  Dead or noisy channels are masked in both arms, and the number of hits in each arm are determined.  The trigger can then make a decision based either on the multiplicity in each BBC (at least one hit in each detector signifies a possible collision) or select events more strictly by determining whether the collision vertex occurred within some distance about the center of the experiment.

Triggering on electrons in the PHENIX central spectrometers is done using the EMCAL-RICH trigger circuit (ERT).  The electron trigger of the ERT takes raw signals from the central arm calorimeters and sums the energy in either 2x2 or 4x4 squares of towers.  If the energy summed within that region is above a specified threshold and there is corresponding activity in the RICH, the event is assumed to include a candidate electron and is triggered.\footnote{Note that a photon trigger can easily be made by requiring energy above a given threshold in a number of towers with no requirement on the RICH.}  In order to increase the rejection (number of events vetoed), the ERT is typically AND'ed with the BBC LL1 trigger (e.g. both triggers must accept the event in order for it to be recorded).

\begin{figure}[h!tb]
  \centering\includegraphics*[width=0.85\columnwidth]{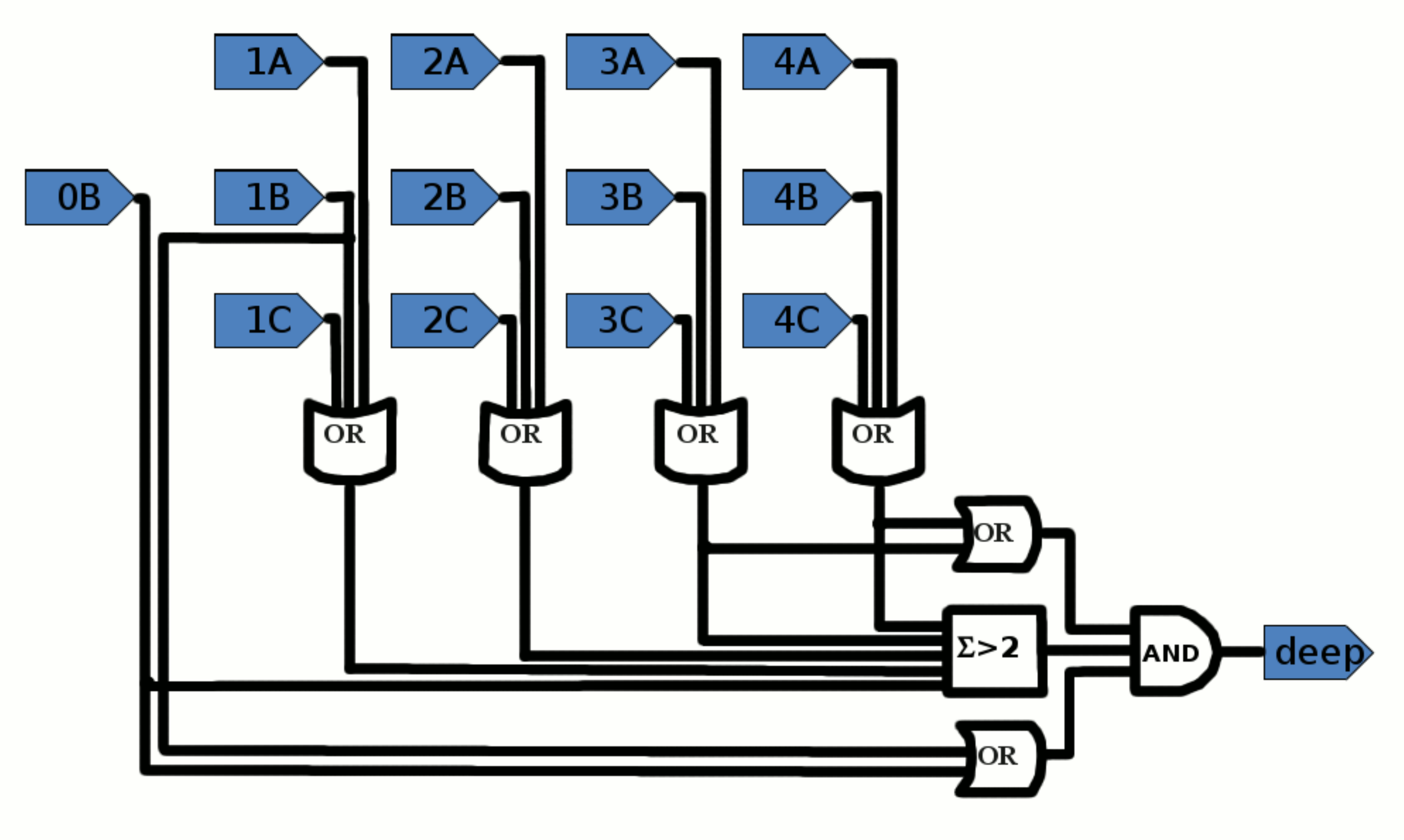}
  \isucaption[Logic for a `deep' $\mu$Id LL1 trigger.]{Logic for a `deep' trigger in the $\mu$Id LL1.  The A, B, and C represent separate groups of tubes contained in the logical unit for each gap.}\label{fig:MUID_Deep_logic}
\end{figure}

\begin{figure}[h!tb]
  \centering\includegraphics*[width=0.85\columnwidth]{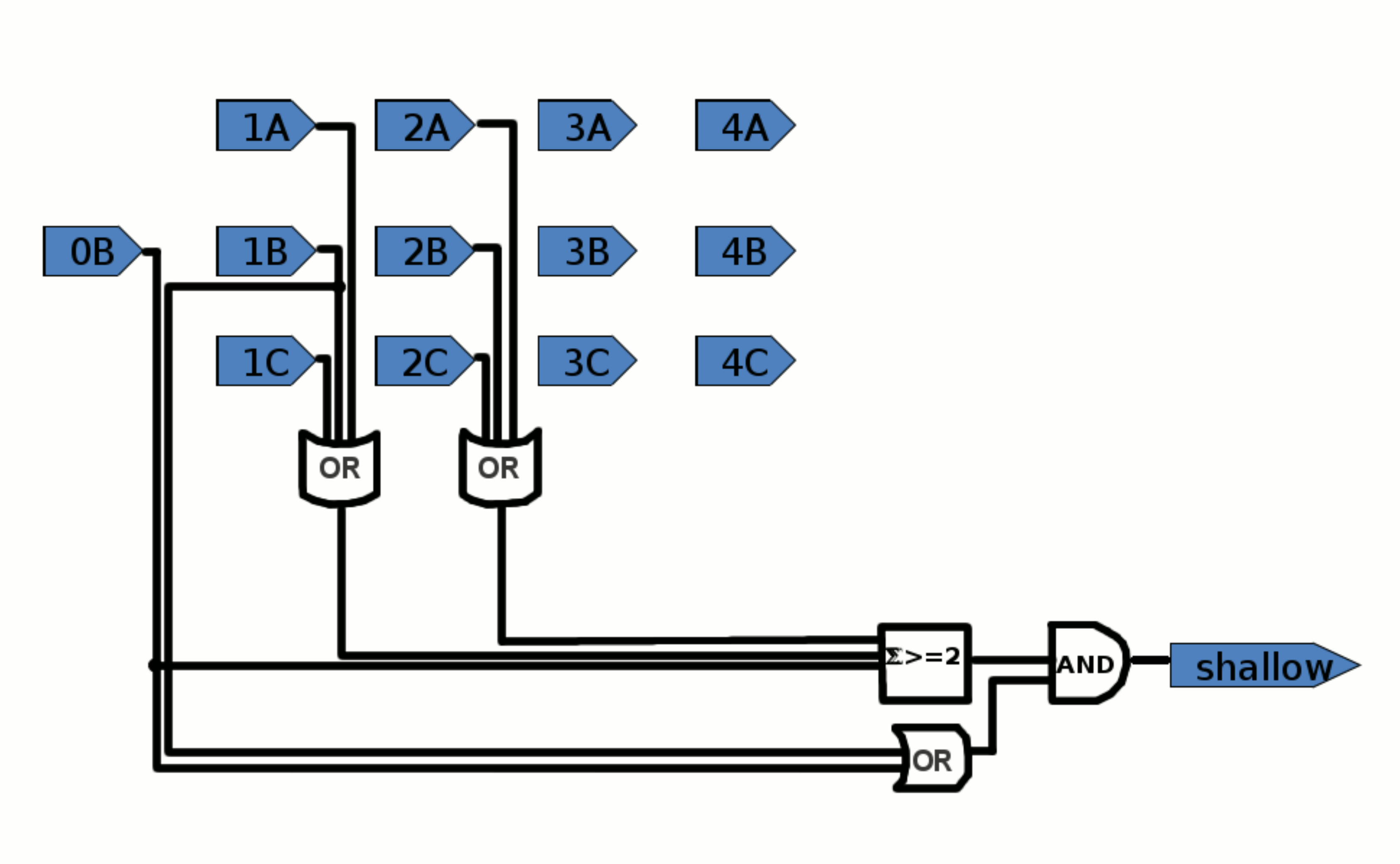}
  \isucaption[Logic for a `shallow' $\mu$Id LL1 trigger.]{Logic for a `shallow' trigger in the $\mu$Id LL1.  The A, B, and C represent separate groups of tubes contained in the logical unit for each gap.}\label{fig:MUID_Shallow_logic}
\end{figure}

For muons, the $\mu$Id is currently used as the primary triggering system, although upgrades are underway to implement a momentum-sensitive trigger using the $\mu$Tr.  The $\mu$Id LL1 trigger uses raw signals from logical combinations of tubes containing all tubes within a legitimate track projection in the detector (tracks should be oriented such that they are coming from the collision vertex).  Requirements are then made on the gaps hit within these projections.  The trigger for J/$\psi$ mesons, for instance, requires two separate logical units to have hits in 3 out of 5 gaps, one in either the first or second gap and another in either the fourth or fifth (furthest from the collision vertex).  A single event meeting these requirements is called a `deep' trigger, and a schematic for the logic is shown in Fig.~\ref{fig:MUID_Deep_logic}.  Occasionally, a similar `shallow' trigger (Fig.~\ref{fig:MUID_Shallow_logic}), which only makes requirements on the first three gaps, is used in coincidence with a deep trigger to select $J/\psi$ mesons.  Due to the poor timing resolution of the $\mu$Id, the $\mu$Id LL1 trigger is only used in coincidence with a BBC LL1 trigger.

\subsection{$J/\psi$ mesons at PHENIX}

The spectroscopy for electrons at mid-rapidity and muons at forward rapidity make PHENIX an ideal experiment for measuring $J/\psi$ meson decays to leptons over a large kinematic range.  Extensive measurements of $J/\psi$ mesons have been made at PHENIX from various collision systems, including:
\begin{itemize}
\item
Cross-section from $\sqrt{s}$=200~GeV $p$+$p$ collisions at mid and forward rapidity~\cite{Adare:2006kf}
\item
Angular decay coefficients from $\sqrt{s}$=200~GeV $p$+$p$ collisions at mid-rapidity~\cite{Adare:2009js}
\item
Cross-section from $\sqrt{s_{NN}}$=200~GeV Au+Au collisions at mid and forward rapidity~\cite{Adare:2006ns}
\item
Cross-section from ultra-peripheral $\sqrt{s_{NN}}$=200~GeV Au+Au collisions at mid-rapidity~\cite{Afanasiev:2009hy}
\item
Cross-section from $\sqrt{s_{NN}}$=200~GeV d+Au collisions at mid and forward rapidity~\cite{Adler:2005ph}
\item
Cross-section from $\sqrt{s_{NN}}$=200~GeV Cu+Cu collisions at mid and forward rapidity~\cite{Adare:2008sh}
\end{itemize}
In Chapter~\ref{ch:jpsi_AN}, I will present a measurement of the $J/\psi$ transverse SSA from $\sqrt{s}$=200~GeV $p$+$p$ collisions at both mid and forward rapidities, for the first time taking advantage of the polarized protons available to PHENIX in a $J/\psi$ measurement, and using nearly all of the subsystems discussed in this chapter.  The measurement of Chapter~\ref{ch:jpsi_pol} will present the first measurement of the $J/\psi$ angular decay coefficients using the muon spectrometers from $\sqrt{s}$=500~GeV $p$+$p$ collisions.

%% file: chapters/jpsi_AN/jpsi_AN.tex
\chapter{$J/\psi$ Transverse Single Spin Asymmetry}\label{ch:jpsi_AN}

\section{Motivation} \label{sec:jpsi_AN_motivation}

It was recently proposed~\cite{Yuan:2008vn} that the transverse single spin asymmetry (SSA) for direct $J/\psi$ production is sensitive to both the Sivers effect (Section~\ref{sec:Sivers}) and the $J/\psi$ production mechanism (Section~\ref{sec:production_mechanism}).   The direct production of $J/\psi$ mesons is dominated by processes involving only gluons in the initial state, and as I stated in Section~\ref{sec:Transversity}, gluons do not carry transversity.  This implies that neither transversity nor the Collins or Boer-Mulders effects can contribute, and only the Sivers effect can be responsible for any transverse SSA.

\begin{figure}[h!tb]
  \begin{center}
    \includegraphics*[width=\columnwidth]{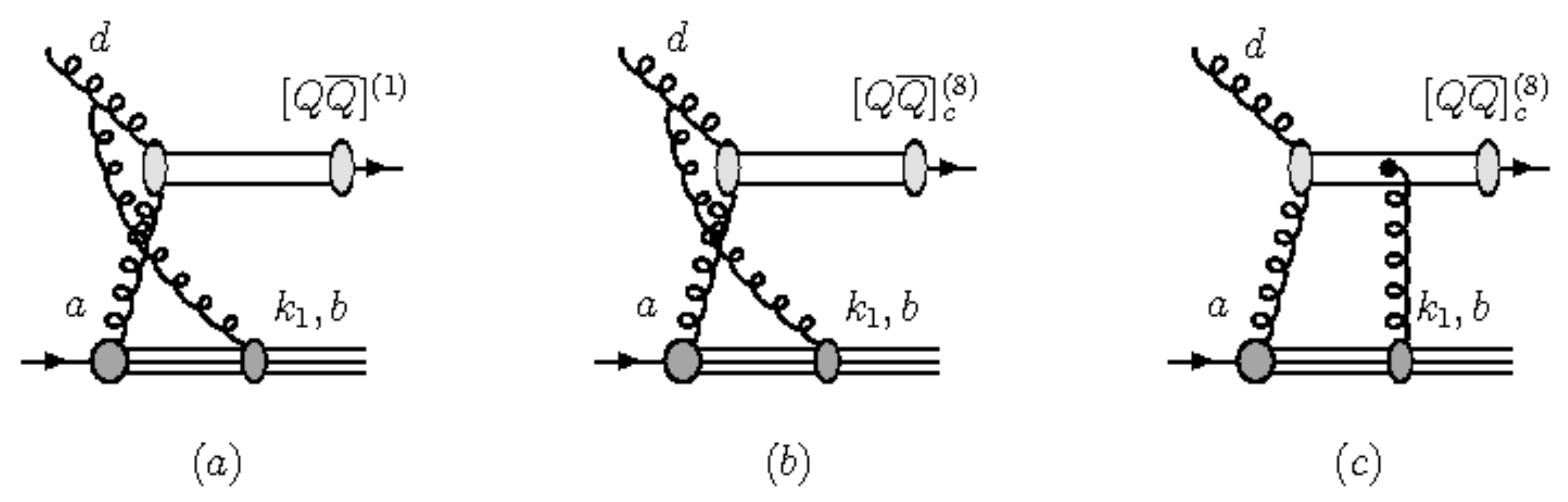} 
  \end{center}
  \isucaption[Transverse single-spin asymmetry generating diagrams for $J/\psi$ meson production]{\label{fig:yuan_diagrams}Diagrams which potentially produce an transverse SSA for $J/\psi$ meson production.  Only initial state interactions contribute to color-singlet production (a).  For color-octet production, interactions in the initial state (b) and final state (c) cancel in the TMD approach, causing the transverse SSA to vanish. (Fig. from~\cite{Yuan:2008vn})}
\end{figure}

In the TMD approach used in~\cite{Yuan:2008vn}, the transverse SSA is only non-zero if there is a non-zero gluon Sivers function and $J/\psi$ mesons are produced from $p$+$p$ collisions in the color-singlet configuration, where only initial state radiation contributes.  The transverse SSA cancels to all orders between the diagrams for initial and final state radiation for $J/\psi$ mesons produced in the color-octet configuration~\cite{Yuan:2008vn} (see Fig.~\ref{fig:yuan_diagrams}).  It is important to remember, however, that factorization is broken for the TMD approach to the Sivers effect in $p$+$p$ collisions, especially in the case of $J/\psi$ production at RHIC, where the transverse momentum of the $J/\psi$ is approximately equal to its mass.  The relationship between the transverse SSA and production mechanism is not as simple in the collinear approach, where there is partial but not full cancellation of transverse SSA generating terms for the color-octet configuration~\cite{Kang:2010pr}.

Transversely polarized protons were brought into collision for approximately 4 weeks at a center of mass collision energy of $\sqrt{s}$=200~GeV during both the 2006 and 2008 RHIC runs, and both of these data samples are used to determine the transverse SSA for inclusive $J/\psi$ production.\footnote{ Since the measurement is made for inclusive $J/\psi$ mesons and not only for those which are directly produced, the theoretical guidance given above will be helpful, but not complete. A more complete theoretical calculation will need to take into account $J/\psi$ mesons from $\chi_{c}$ decays.}  

In order to utilize the entire PHENIX acceptance, separate analyses are done for $J/\psi$$\rightarrow$$e^{+}e^{-}$ in the central spectrometers and  $J/\psi$$\rightarrow$$\mu^{+}\mu^{-}$ in the muon spectrometers.  The analysis for the central spectrometer is done using data from the 2006 run and for the muon spectrometer using data from the 2006 and 2008 runs.  Each analysis requires different considerations due to differences in the detectors and the number of $J/\psi$ mesons measured by each detector. I was primarily responsible for determining the asymmetry from the central spectrometers but also provided assistance in the final stages of the analysis in the muon spectrometers.  For the benefit of the reader and for a more comprehensive view of the result, I will present measurements from both detectors.  The result has recently been posted to the arXiv and submitted for publication~\cite{Adare:2010bd}.

\section{$J/\psi$ Selection}\label{sec:jpsi_AN_cuts}

To ensure that tracks used in the analysis are from electron and muon pairs from $J/\psi$ meson decays, cuts are applied to single tracks and to the pair.

For electron candidates, a coincidence is required between the BBC and the ERT triggers.\footnote{Descriptions of these triggers are given in Section~\ref{sec:Triggering}}  Electron tracks are required to have total momentum $>$0.5 GeV/$c$, to avoid background from electron conversions within the detector.  The energy measured in the EMCAL divided by momentum measured in the DC is required to be within 4 standard deviations ($\sigma$) of 1 to ensure that the track is made by an electron, and the position matching between the track in the DC and the energy cluster in the EMCAL is required to be $<4\sigma$ in both beam direction and azimuth.  Only events with a collision vertex measured by the BBC of $\pm$30~cm about the center of the detector are considered, because the tracking of the central spectrometer does not work very well for collision vertices outside of that range.

Muon candidates from the 2006 run are required to have a coincidence between the BBC, a $\mu$ID deep, and a $\mu$ID shallow trigger (discussed in Section~\ref{sec:Triggering}).  For the 2008 run, a coincidence is required between the BBC and two $\mu$ID deep triggers.  The $J/\psi$$\rightarrow$$\mu^{+}\mu^{-}$ analyses required the collision vertex to be within 35~cm of the center of the experiment along the beam direction.\footnote{A cut on the vertex is made at 30~cm by the BBC trigger, but the trigger does not perform a full vertex reconstruction, and the looser cut at 35~cm allows the inclusion of events with errors in the vertex determined by the trigger.}  Each track is required to have momentum along the beam direction $1.4 < p_{z} ($GeV/$c) < 20$ to avoid the trigger turn-on region, and the distance between the projection of the track in the $\mu$Id and the track in the $\mu$Tr to the first $\mu$Id gap is required to be $<$25(30)~cm for the spectrometer on the North~(South) side of the experiment.  The angle between the projected tracks is required to be $<$10$^{\circ}$ in both the North and South detector.  A fit of track pairs to the BBC vertex is required to have a $\chi^{2}/ndf<5$ with 4 degrees of freedom.

\begin{figure}[h!tb]
  \begin{center}
    \includegraphics*[width=0.65\columnwidth]{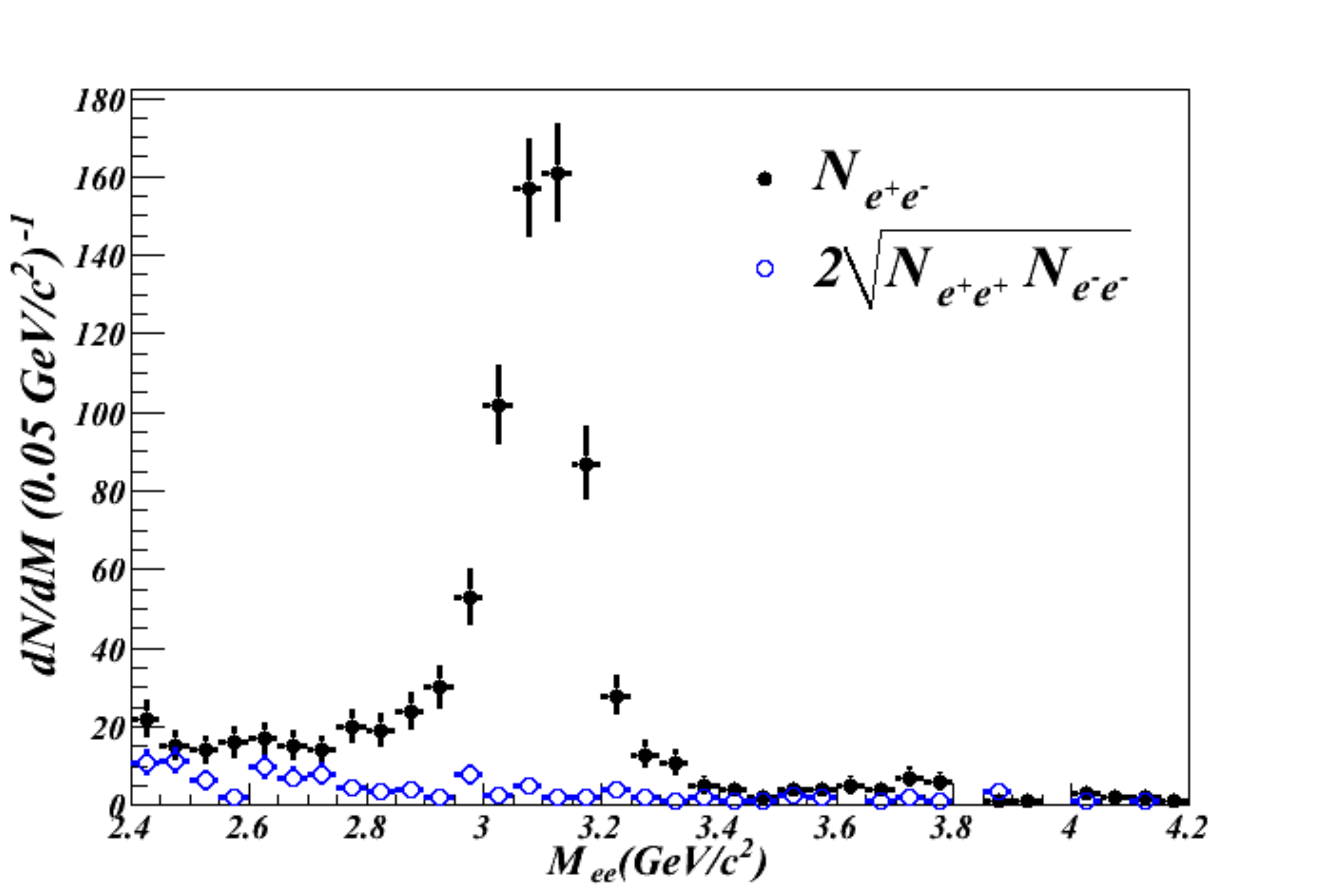}   
  \end{center}
  \isucaption[Invariant mass spectrum for electron candidate pairs (solid circles) and uncorrelated track pairs (open circles) in the central spectrometer.]{Invariant mass spectrum for electron candidate pairs and uncorrelated track pairs in the central spectrometer.\label{fig:central_mass}}
\end{figure}

\begin{figure}[h!tb]
  \begin{center}
    \includegraphics*[width=0.57\columnwidth]{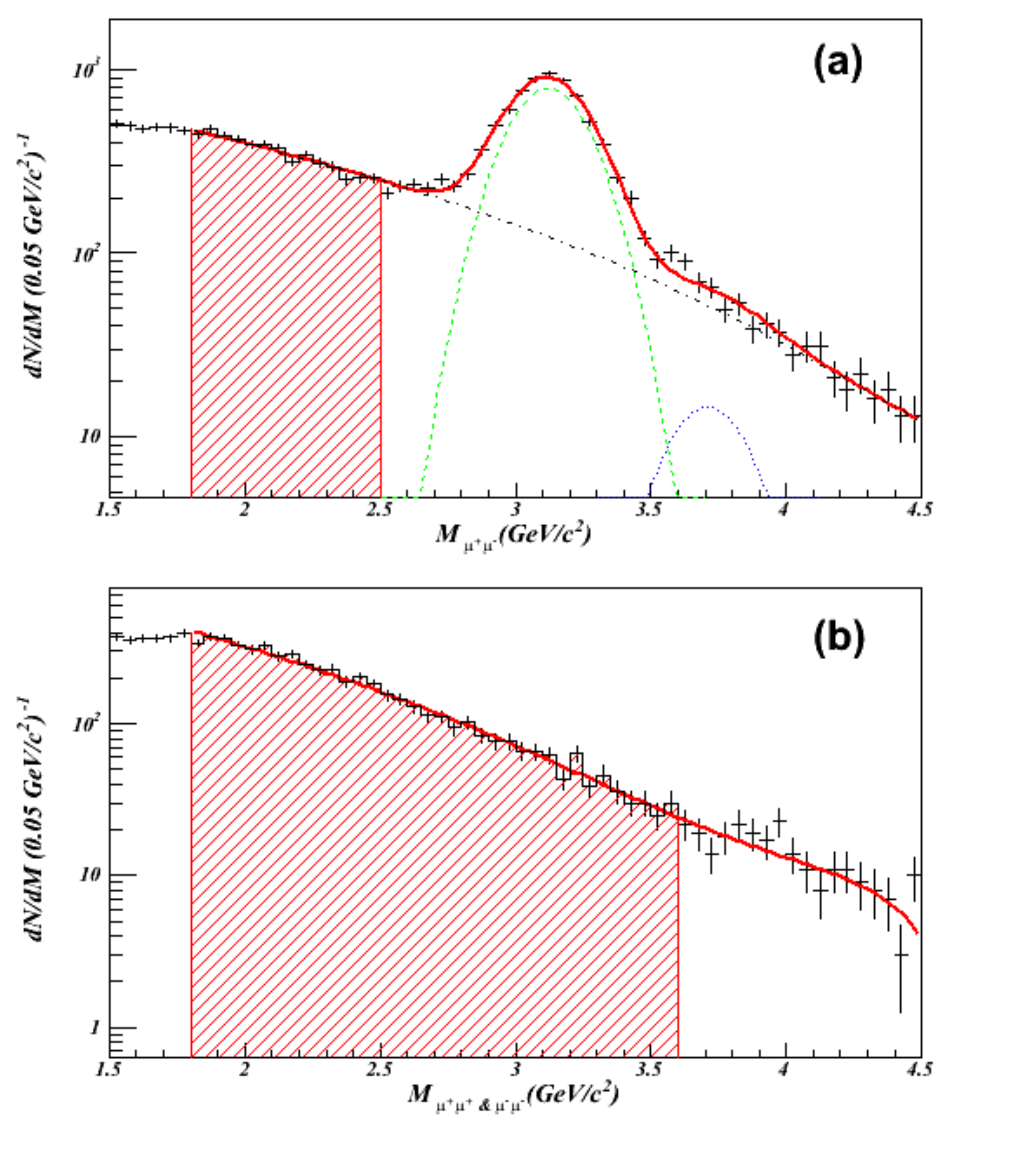}   
  \end{center}
  \isucaption[Invariant mass spectrum for oppositely-charged muon candidate pairs and charged candidate pairs with the same sign.]{\label{fig:muon_mass}Example invariant mass spectrum from the 2006 dataset in the North muon spectrometer  ($p_{z}>$0) for (a) oppositely-charged muon candidate pairs and (b) charged candidate pairs with the same sign.  The solid line in the figure is the sum of Gaussian distributions for the $J/\psi$ (dashed curve), $\psi^{\prime}$ (dotted curve), and a third-order polynomial background (dotted-dashed curve).  The shaded regions show the area used for determining the background $A_N$.}
\end{figure}

A plot of the invariant mass for electron pairs passing all cuts is shown in Fig.~\ref{fig:central_mass} and for muons in Fig.~\ref{fig:muon_mass}.  The resolution of the $J/\psi$ mass peak is found to be $61\pm2$MeV/$c^{2}$ in the central spectrometer and roughly $143\pm3$ ($154\pm3$) MeV/$c^{2}$ in the North~(South) muon spectrometer.  

\begin{figure}[h!tb]
  \begin{center}
    \includegraphics*[width=0.65\columnwidth]{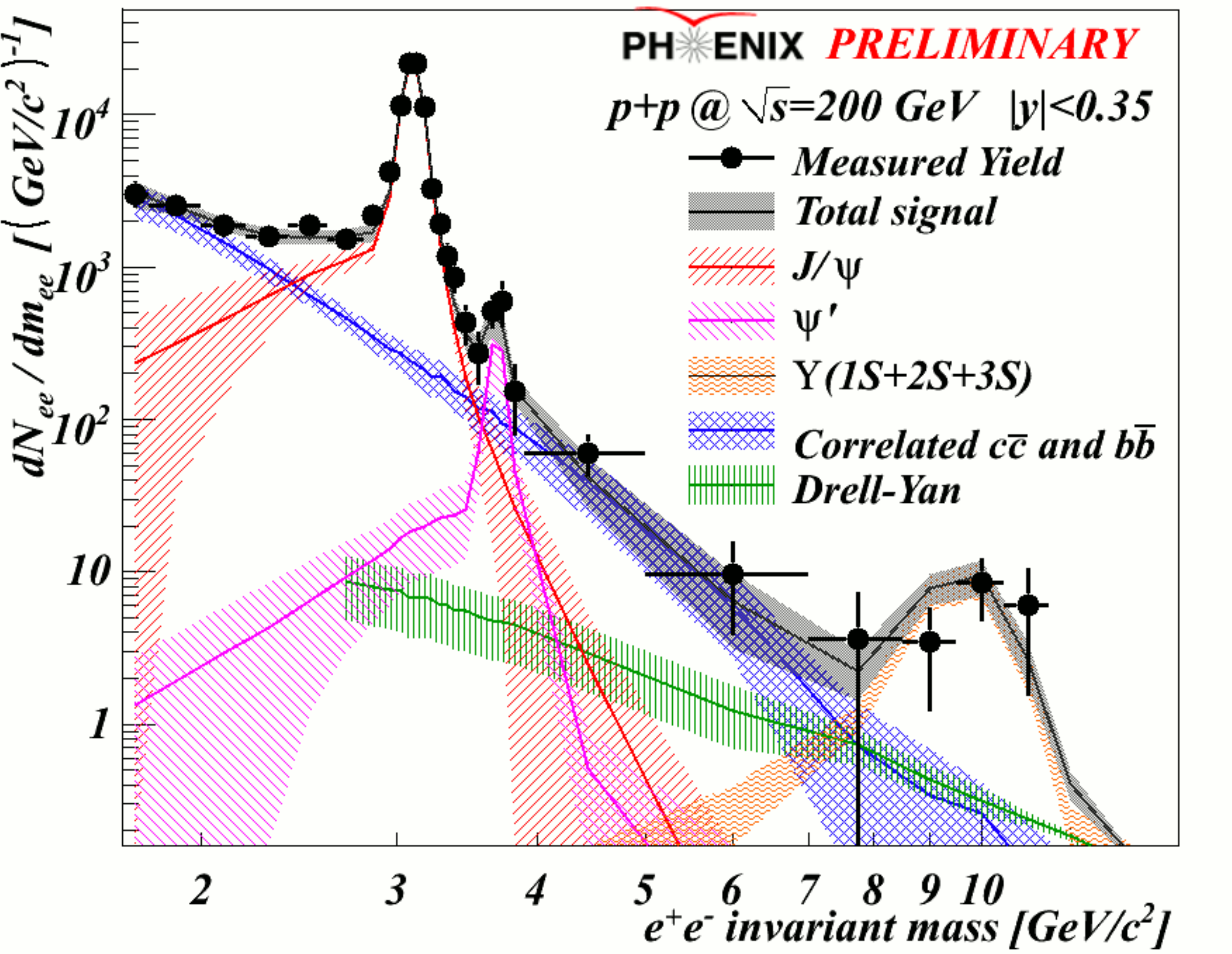}   
  \end{center}
  \isucaption {Invariant mass spectrum of electron pairs showing the contributions from all physical sources.\label{fig:invmass_fit}}
\end{figure}

In the central spectrometer, the contribution of pairs from random uncorrelated track combinations are estimated to be twice the geometric mean of the positive and negatively charged pairs, $2\sqrt{N_{e^{+}e^{+}}N_{e^{-}e^{-}}}$, and subtracted from the spectrum of oppositely charged pairs, leaving only those pairs from collision-related processes.  The fraction of remaining pairs coming from sources other than $J/\psi$ mesons is determined in a separate analysis using a full detector simulation of all physical processes contributing to the continuum (Fig.~\ref{fig:invmass_fit}).  Background fractions, defined as the ratio of correlated pairs from other physical sources to pairs from $J/\psi$ meson decays in a mass range from $2.7 < M ($GeV/$c) < 3.4$ were estimated with the same simulation (Table~\ref{tab:background_fractions}).

In the muon spectrometers, the number of background pairs is determined by fitting the invariant mass distribution of oppositely charged muon pairs with a function
\begin{eqnarray*}
\frac{dN}{dM} &=&(a_0+a_1M+a_2M^2+a_3M^3) \\
&+&
\frac{N_{J/\psi}}{2\pi\sqrt{\sigma}}e^{-\frac{(M-M_{J/\psi})^2}{2\sigma^2}}
+\frac{N_{\psi^\prime}}{2\pi\sqrt{\sigma^\prime}}e^{-\frac{(M-M_{\psi^\prime})^2}
{2{\sigma^\prime}^2}}.
\label{eq:muon_background_shape}
\end{eqnarray*}
where the first and second Gaussian distributions approximate the shape of the $J/\psi$ and $\psi^{\prime}$ peaks respectively, and the third order polynomial parameterizes the shape in invariant mass of both the uncorrelated pairs and the background from Drell-Yan production, $p$+$p$$\rightarrow$$\text{D}\overline{\rm D}X$$\rightarrow$$\mu^{+}\mu^{-}X$ (open charm), and $p$+$p$$\rightarrow$$\text{B}\overline{\rm B}X$$\rightarrow$$\mu^{+}\mu^{-}X$ (open bottom).   Background fractions from the muon spectrometers, defined as the total number of background pairs divided by the number of signal pairs in a mass range $2\sigma$ about the $J/\psi$ Gaussian were estimated from the fit (Table~\ref{tab:background_fractions}).  The background fractions are found to be larger for the 2006 dataset because the trigger requirement is not as restrictive, allowing for the reconstruction of more random track combinations.

\begin{table}[h!tb]
\begin{center}
\isucaption[Background fractions for $J/\psi$ mesons in the central and muon spectrometers at several transverse momenta.]{\label{tab:background_fractions}  Background fractions for $J/\psi$ mesons in the central and muon spectrometers at several transverse momenta.  The background fraction for the central spectrometer includes only physical processes, while the background fraction for the muon spectrometers also includes random track combinations.}
\begin{tabular}{ccccc}
\hline
$p_T$ (GeV/$c$)   &    dataset   & \ \ detector \ \       & background fraction \\ \hline
0--6              &     2006     &       South Muon       & 21.7$\pm$0.6\%    \\
                  &     2006     &       North Muon       & 19.1$\pm$0.4\%    \\
                  &     2008     &       South Muon       & 16.4$\pm$0.2\%    \\
                  &     2008     &       North Muon       & 14.2$\pm$0.2\%    \\
                  &     2006     &       Central          &  6.6$\pm$0.4\% \\ \hline
0--1.4            &     2006     &       South Muon       & 23.2$\pm$0.7\%    \\
                  &     2006     &       North Muon       & 22.0$\pm$0.7\%    \\
                  &     2008     &       South Muon       & 16.1$\pm$0.3\%    \\
                  &     2008     &       North Muon       & 15.5$\pm$0.3\%    \\
                  &     2006     &       Central          &  5.6$\pm$0.5\% \\ \hline
1.4--6            &     2006     &       South Muon       & 20.1$\pm$0.8\%    \\
                  &     2006     &       North Muon       & 14.1$\pm$0.5\%    \\
                  &     2008     &       South Muon       & 15.6$\pm$0.4\%    \\
                  &     2008     &       North Muon       & 10.5$\pm$0.2\%    \\
                  &     2006     &       Central          &  7.8$\pm$0.7\% \\
\hline
\end{tabular}
\end{center}
\end{table}

\begin{figure}[h!tb]
  \begin{center}
    \includegraphics*[width=0.5\columnwidth]{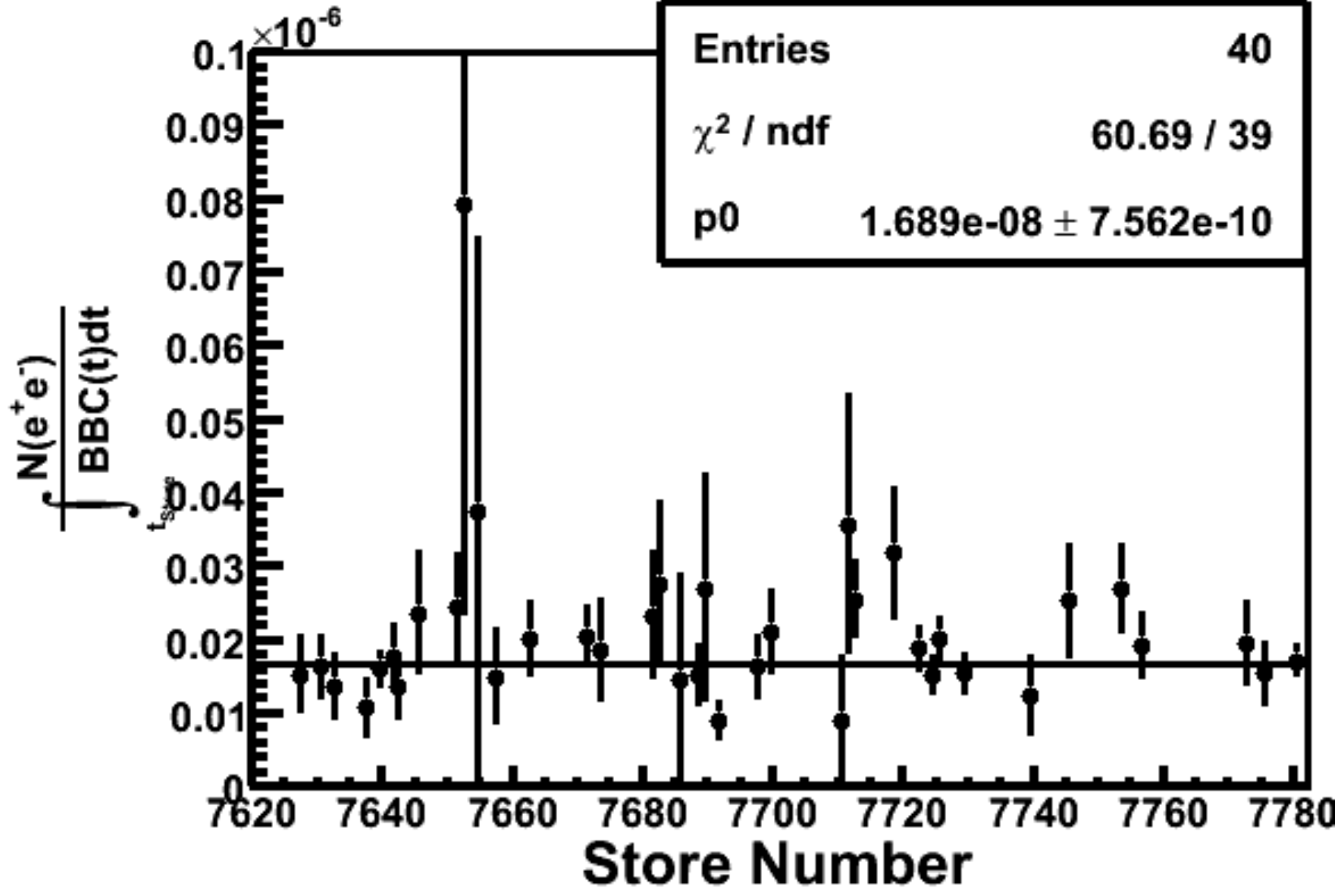}
  \end{center}
  \isucaption{\label{fig:ep_v_lummi}Number of $e^+e^-$ pairs with invariant masses $M$$\in$$\left[2.7,3.4\right]$~GeV/$c^{2}$ divided by number of counts in the BBC integrated over individual stores, plotted against store number, and fit with a constant.}
\end{figure}

\begin{figure}[h!tb]
  \begin{center}
    \includegraphics*[width=0.4\columnwidth]{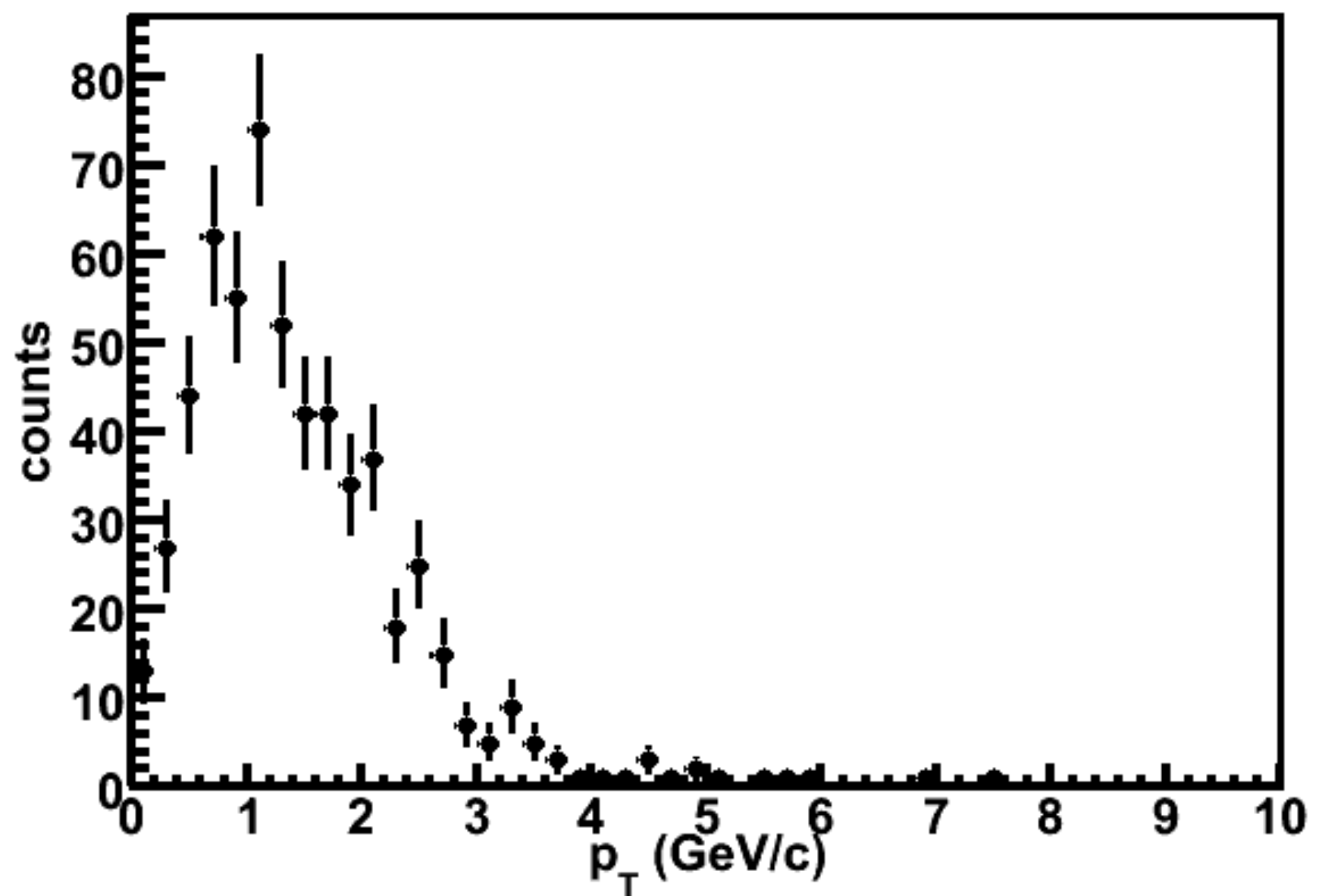}  \
    \includegraphics*[width=0.4\columnwidth]{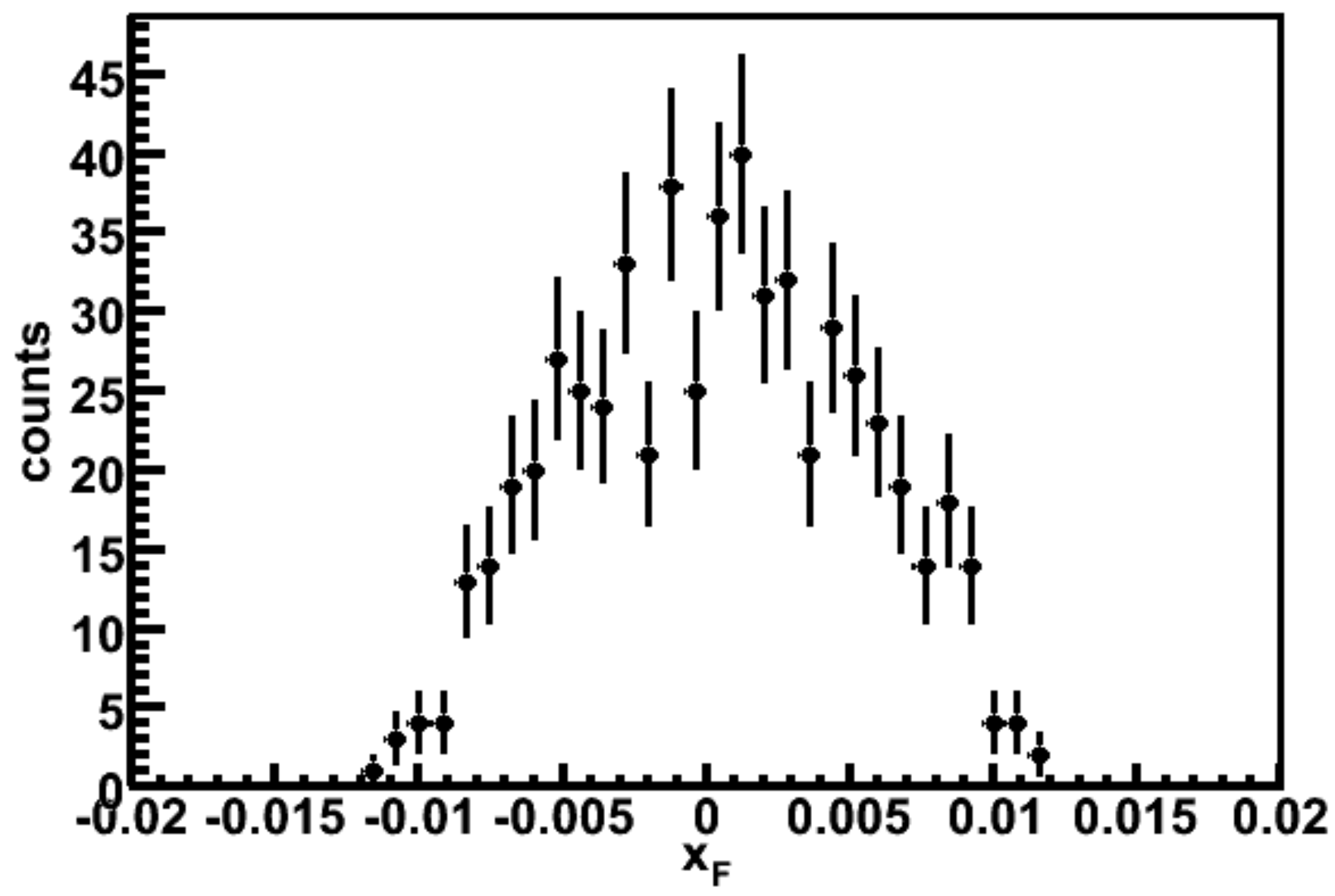}  \\
    \includegraphics*[width=0.4\columnwidth]{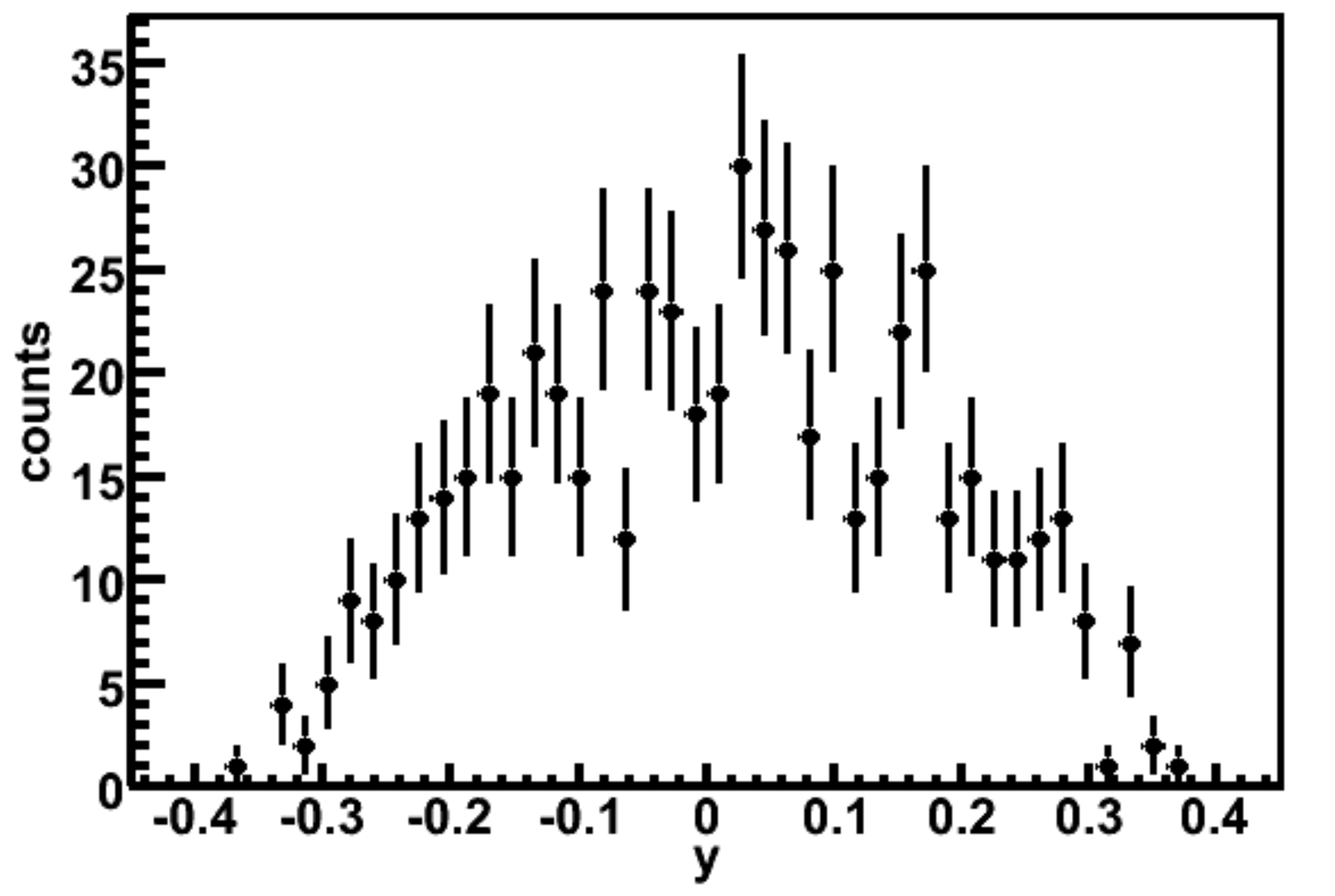}
  \end{center}
  \isucaption[Kinematics of accepted $e^{+}e^{-}$ pairs for $J/\psi$ $A_{N}$ analysis.]{\label{fig:pair_kinematics}On the top, transverse momentum (left) and
    Feynman-x (right), and on the bottom rapidity of accepted $e^{+}e^{-}$ pairs.}
\end{figure}

For the analysis in the central spectrometer, the number of $e^{+}e^{-}$ candidate pairs with invariant masses $2.7 < M ($GeV/$c^{2}) < 3.4$ divided by the number of counts in the BBC\footnote{The number of counts in the BBC is proportional to the provided luminosity, meaning that the number of $J/\psi$ mesons divided by the number of BBC counts should not change over time as long as the active area of the detector and the contributions from backgrounds do not change.} are found to be constant over the course of the running period, implying that both the backgrounds and acceptance are approximately constant over time. (Fig. \ref{fig:ep_v_lummi}).  Kinematic distributions for the accepted pairs are show in Fig. \ref{fig:pair_kinematics}.

\section{Asymmetry Formulae} \label{sec:asymmetry_formulae}

The general expression for a transverse single spin asymmetry (SSA) is given by Eq.~\ref{eq:AN_sigma}.  Since both proton beams are polarized at RHIC, however, we never measure a single spin asymmetry directly using this expression but instead average over the polarization of one beam, defining the single polarized cross section as
\begin{eqnarray}
\sigma^{\uparrow} \equiv \sigma^{\uparrow\uparrow} + \sigma^{\uparrow\downarrow} \nonumber \\
\sigma^{\downarrow} \equiv \sigma^{\downarrow\downarrow} + \sigma^{\downarrow\uparrow}
\label{eq:single_pol_sigma}
\end{eqnarray} 
where the $\uparrow$ and $\downarrow$ represent the polarization directions of the two beams.  In the following discussions, $\uparrow$ will be used for protons polarized radially outward from the RHIC ring and $\downarrow$ for those polarized radially inward for the analysis of 2006 data.  For the analysis of 2008 data, they denote up and down. Using Eq.\ref{eq:single_pol_sigma} we can perform statistically independent analyses for each polarized beam and combine the results to get a smaller statistical uncertainty in the final result.

Combining Eq.~\ref{eq:AN_sigma} and Eq.~\ref{eq:single_pol_sigma} gives
\begin{eqnarray}
A_N &=& \frac{f}{\mathcal{P}} \frac{(\sigma^{\uparrow\uparrow} + \sigma^{\uparrow\downarrow}) - (\sigma^{\downarrow\downarrow} + \sigma^{\downarrow\uparrow})}{(\sigma^{\uparrow\uparrow} + \sigma^{\uparrow\downarrow}) + (\sigma^{\downarrow\downarrow} + \sigma^{\uparrow\uparrow})} \nonumber \\
    &=& \frac{f}{\mathcal{P}} \frac{(N^{\uparrow\uparrow} + \mathcal{R}_{1}N^{\uparrow\downarrow}) - (\mathcal{R}_{2}N^{\downarrow\downarrow} + \mathcal{R}_{3}N^{\downarrow\uparrow})}{(N^{\uparrow\uparrow} + \mathcal{R}_{1}N^{\uparrow\downarrow}) + (\mathcal{R}_{2}N^{\downarrow\downarrow} + \mathcal{R}_{3}N^{\uparrow\uparrow})}
\label{eq:muon_an_formula}
\end{eqnarray}
where $N^{\uparrow\uparrow}$, $N^{\uparrow\downarrow}$,  $N^{\downarrow\uparrow}$,and  $N^{\downarrow\downarrow}$ are the experimental yields in each spin configuration, and $\mathcal{R}_{1}=\mathcal{L}^{\uparrow\uparrow}/\mathcal{L}^{\uparrow\downarrow}$,  $\mathcal{R}_{2}=\mathcal{L}^{\uparrow\uparrow}/\mathcal{L}^{\downarrow\uparrow}$, and  $\mathcal{R}_{1}=\mathcal{L}^{\uparrow\uparrow}/\mathcal{L}^{\downarrow\downarrow}$ are the ratios of provided luminosities.  By using a single polarization $\mathcal{P}$, we have implicitly assumed that the extent to which protons are polarized is the same in each spin configuration.

Eq.~\ref{eq:muon_an_formula} works in situations where there are enough statistics in $N^{\uparrow\uparrow}$, $N^{\uparrow\downarrow}$,  $N^{\downarrow\uparrow}$, and  $N^{\downarrow\downarrow}$  to determine an asymmetry for a time-period in which $R_{1}$, $R_{2}$, and $R_{3}$ are stable.  In the muon analyses this is possible, and each $R$ is determined for every (approximately eight hour long) store.  An asymmetry is then calculated by fitting a constant to the asymmetries from all stores.   In the central analysis there are significantly fewer $J/\psi$ counts; so we first assume that double spin asymmetries are small so that Eq.~\ref{eq:muon_an_formula} can be written as
\begin{equation}
A_N = \frac{f}{\mathcal{P}} \frac{N^{\uparrow} - \mathcal{R}N^{\downarrow}}{N^{\uparrow}+ \mathcal{R}N^{\downarrow}},
\label{eq:single_lumi_formula}
\end{equation}
where the spin orientation of the polarized beam is denoted by the arrow, and there is only one relative luminosity $\mathcal{R}=\frac{\mathcal{L}^{\uparrow}}{\mathcal{L}^{\downarrow}}$.  To further simplify, we can eliminate the explicit use of the relative luminosity by including both the asymmetry on the left and right\footnote{Recall from Section~\ref{ch:trans_protons} that Eq.~\ref{eq:AN_sigma} and hence all $A_N$ formulae above (including Eq.~\ref{eq:single_lumi_formula}) apply only to the left side of the polarized proton.  For the asymmetry on the right, an overall negative sign is required.} as
\begin{equation}
A_N = \frac{f^{\prime}}{\mathcal{P}} \frac{{\sqrt{N_L^{\uparrow}N_R^{\downarrow}}-\sqrt{N_L^{\downarrow} N_R^{\uparrow}}}}
{{\sqrt{N_L^{\uparrow} N_R^{\downarrow}}+\sqrt{N_L^{\downarrow}N_R^{\uparrow}}}}.
\label{eq:central_an_formula}
\end{equation}
The geometric scale factor
\begin{equation}
f^{\prime} = 2\left(\frac{\int^{\pi}_{0}  \varepsilon(\phi) \sin\phi d\phi}{\int^{\pi}_{0} \varepsilon(\phi) d\phi} - \frac{\int^{2\pi}_{\pi}  \varepsilon(\phi) \sin\phi d\phi}{\int^{2\pi}_{\pi} \varepsilon(\phi) d\phi} \right)^{-1}
\end{equation}
is slightly different from the factor $f$ in Eq.~\ref{eq:single_lumi_formula} (defined in Eq.~\ref{eq:f_factor}), because left and right are treated simultaneously, but this only leads to differences between Eq.~\ref{eq:single_lumi_formula} and Eq.~\ref{eq:central_an_formula} of order $\left(A_N\right)^{3}$.

Unfortunately, $J/\psi$ statistics in the central spectrometer are still limited to the extent that determining asymmetries with Eq.~\ref{eq:central_an_formula} from each store is not possible.  Instead, statistics are integrated across all stores, and a single $A_N$ is calculated for the entire running period.  The integration across $n$ time periods means that the asymmetry on the left is actually
\begin{equation}
A_{N} = \frac{1}{n} \displaystyle\sum_{i=1}^{n} \frac{f_{i}}{\mathcal{P}_{i}} \frac{N_{i}^{\uparrow} - \mathcal{R}_{i} N_{i}^{\downarrow}}{N_{i}^{\uparrow} + \mathcal{R}_{i} N_{i}^{\downarrow}} = \frac{1}{n} \displaystyle\sum_{i=1}^{n} A_{N,i}.
\label{eq:sum_over_stores}
\end{equation}
(using Eq.~\ref{eq:single_lumi_formula} for simplicity).  Since the measurement is statistically limited, we do not use individual $f_{i}$, $\mathcal{P}_{i}$, and $\mathcal{R}_{i}$ but instead assume that
\begin{equation}
\displaystyle A_{N} = \frac{\left<f\right>\displaystyle\sum_{i=1}^{n} \left( N_{i}^{\uparrow} - \left<\mathcal{R}\right> N_{i}^{\downarrow} \right)}{\left<\mathcal{P}\right>\displaystyle\sum_{i=1}^{n}\left( N_{i}^{\uparrow} + \left<\mathcal{R}\right> N_{i}^{\downarrow} \right)}
\end{equation}
is equivalent to Eq.~\ref{eq:sum_over_stores}, where the brackets denote luminosity weighted averages over the course of the running period.  In order for these two expressions to be truly equivalent it must be true that $\mathcal{R}_{i} = \left<\mathcal{R}\right>$, $f_{i}=\left<f\right>$, and $\mathcal{P}_{i}=\left<\mathcal{P}\right>$ for all $i$.

The acceptance of the central spectrometer is stable enough in the 2006 data that the assumption is valid for $f$.  Likewise, variations in the polarization are consistent with systematic errors on the measurement of the polarization, meaning that our inclusion of polarization uncertainties will take care of any difference between $\mathcal{P}_{i}$ and $\left<\mathcal{P}\right>$.   Unlike variations in $f$ and $\mathcal{P}$, which are under control and can only affect the overall scale of the asymmetry, variations in $\mathcal{R}$ are significant over the course of the measurement and can lead to false asymmetries.

In order to stabilize $\mathcal{R}$ for the measurement in the central spectrometer, a procedure was developed wherein several bunches of colliding protons are removed from the analysis so that each $\mathcal{R}_{i}$ (the relative luminosity in a given store) is brought as close to unity as possible.  In principle, any constant value for the relative luminosity will work, but unity is chosen for convenience.

First, all bunches with luminosities greater than 2$\sigma$ away from the mean bunch luminosity of a store are removed.  A bunch is then chosen at random, and if removing that bunch from the analysis brings the relative luminosity closer to unity, it is removed.  Otherwise it is kept, and another bunch is chosen at random.  The process continues until the relative luminosity is within 1$\%$ of unity or as close to unity as possible given the finite number of bunches.  The corrected relative luminosities are distributed with an RMS of approximately 1.5$\%$ away from unity, and the entire procedure removed approximately 5$\%$ of the provided luminosity from the data sample used in the analysis.

Since removed bunches are chosen at random, the result of the analysis is not unique, and a systematic uncertainty is introduced.  In order to determine this systematic uncertainty, the analysis is run 5k times, and the resulting asymmetries and statistical uncertainties are histogrammed.  The mean values of these two histograms are taken as the central value and statistical uncertainty in the data point, and the RMS of the histogrammed asymmetry is taken as a source of systematic uncertainty.

Statistics in the muon analyses are plentiful enough to allow for the measurement of a background asymmetry, $A_N^{BG}$, which makes possible a calculation of the signal asymmetry, $A_N^{J/\psi}$, which is not diluted by background.  The signal asymmetry is found with
\begin{equation}
A_N^{J/\psi}=\frac{A_N^{Incl}-r\cdot A_N^{BG}}{1-r},
\label{eq:AN-Phy}
\end{equation}
where $r$ is the background fraction (Table~\ref{tab:background_fractions}).  The background asymmetries are determined for the analysis of the 2006 
dataset using oppositely-charged muon pairs in the invariant mass range $1.8 < m($GeV$/c^{2}) < 2.5$ along with charged pairs of the same sign in invariant mass range $1.8 < m ($GeV$/c^{2}) < 3.6$ (shaded areas in Fig.~\ref{fig:muon_mass}(a) and (b) respectively).  For the analysis of the 2008 dataset the lower limit of the mass range is 2.0~GeV/$c^{2}$.  Background asymmetries from this study are shown in Table~\ref{tab:Bg-asymmetries}.  In the central spectrometer, statistics do not allow for the measurement of a background asymmetry, and $A_N^{BG}$ is assumed to be zero in Eq.~\ref{eq:AN-Phy}.

\begin{table}[h!tb]
\begin{center}
\isucaption{\label{tab:Bg-asymmetries}  Background asymmetries as a function of $p_{T}$ for the PHENIX muon spectrometers.}
\begin{tabular}{cccc}
\hline
$p_T$ (GeV/$c$)&  $<x_{F}>$  &   dataset   &  $A_N^{BG}$  \\ \hline

0--6           &   -0.07     &     2006    & -0.003$\pm$0.028  \\
               &   -0.07     &     2008    & -0.072$\pm$0.034  \\
               &    0.08     &     2006    & -0.008$\pm$0.028  \\ 
               &    0.08     &     2008    & -0.003$\pm$0.035  \\ \hline
0--1.4         &   -0.07     &     2006    & -0.002$\pm$0.031  \\
               &   -0.07     &     2008    & -0.043$\pm$0.039  \\
               &    0.08     &     2006    & -0.021$\pm$0.038  \\
               &    0.08     &     2008    & -0.060$\pm$0.046  \\ \hline
1.4--6         &   -0.08     &     2006    & -0.066$\pm$0.050  \\
               &   -0.08     &     2008    &  0.047$\pm$0.064  \\
               &    0.08     &     2006    &  0.039$\pm$0.056  \\ 
               &    0.08     &     2008    & -0.072$\pm$0.070  \\ \hline
\end{tabular}
\end{center}
\end{table}

A final additional complication is that even and odd numbered crossings are triggered by separate circuits in the ERT with slightly different gains.  This means that detector efficiencies can vary depending on the spin orientation and potentially lead to false asymmetries in the central spectrometer.   We restore the equality in efficiency by measuring asymmetries separately for even and odd crossings and combining the resulting asymmetries.

\section{Polarization and Geometric Scale Factors}

 Average beam polarizations are measured using the RHIC polarimeters discussed in Section~\ref{sec:RHIC} and found for the 2006 run to be
\begin{eqnarray*}
& 0.53\pm 0.02^\text{syst} & \text{(clockwise)} \\
& 0.52\pm 0.02^\text{syst} & \text{(counterclockwise)}
\end{eqnarray*}
with standard deviations of 0.03 and 0.04 respectively.  The labels represent the direction of the beam circulation, and the systematic uncertainties are uncorrelated between beams.  There is an additional systematic uncertainty of $3.5\%$ correlated between the two beams.

During the 2008 run, the average beam polarizations were
\begin{eqnarray*}
& 0.48\pm 0.02^\text{syst} & \text{(clockwise)} \\
& 0.41\pm 0.02^\text{syst} & \text{(counterclockwise)}
\end{eqnarray*}
each with a standard deviation of 0.04 and an additional systematic uncertainty of $3.9\%$ correlated between the beams.

The deviation of the polarization from nominal at the PHENIX interaction region is measured using the local polarimeter discussed in Section~\ref{sec:local_pol} and is found, in radians, to be
\begin{eqnarray*}
& 0.064 \pm 0.040^\text{stat} \pm 0.086^\text{syst} & \text{(clockwise)} \\
& 0.109 \pm 0.038^\text{stat} \pm 0.036^\text{syst} & \text{(counterclockwise)}
\end{eqnarray*}
for the 2006 data and
\begin{eqnarray*}
& 0.263 \pm 0.030^\text{stat} \pm 0.090^\text{syst} & \text{(clockwise)} \\
& 0.019 \pm 0.048^\text{stat} \pm 0.103^\text{syst} & \text{(couterclockwise)}
\end{eqnarray*}
for the 2008 data.  The polarization directions of the counterclockwise-going beam in 2006 and clockwise-going beam in 2008 are considered significant and used in the calculation of $A_N$.

The geometric scale factor from 2006 data in the muon spectrometers of $f=1.57\pm0.04$ is determined from $J/\psi$ azimuthal distributions in data and is found to be independent of $p_{T}$ within statistical uncertainties.  For the 2008 data, the factors are $f=1.64\pm0.01$ for the clockwise circulating beam and $f=1.56\pm0.01$ for the counter-clockwise circulating beam.  In the central spectrometer, the factor $f^{\prime}$ is determined from a GEANT~\cite{GEANT} Monte Carlo simulation of single $J/\psi$ decays with a full geometric description of the detector including all known inefficiencies, and the resulting geometric scale factors are listed in Table~\ref{tab:f_factors}.

\begin{table}[h!tb]
\centering
\isucaption[Geometric scale factors for the central spectrometer]{\label{tab:f_factors}Geometric scale factors for the central spectrometer determined with a Monte Carlo simulation of the detector.}
\begin{tabular}{cc}
\hline
$p_{T}$ (GeV/$c$)    & $f^{\prime}$    \\ \hline
0--6                 & 1.62$\pm$0.01 \\ \hline
0--1.4               & 1.61$\pm$0.01 \\ \hline
1.4--6               & 1.70$\pm$0.02 \\ \hline
\end{tabular}
\end{table}

\section{Systematic Uncertainties}

Systematic uncertainties can generally be classified as one of three types: uncertainties which are uncorrelated between data points (Type A), uncertainties which are correlated between data points, usually with an unknown correlation matrix (Type B), and uncertainties in the vertical scale which move all data points in the same direction and by the same magnitude while maintaining the statistical significance from zero (Type C).

As discussed in Section~\ref{sec:asymmetry_formulae}, a Type A systematic uncertainty exists in the central measurement due to the stabilization of the relative luminosity.  There is also Type A systematic uncertainty in the muon measurement due to the fit which determines the background fraction $r$ in Eq.~\ref{eq:AN-Phy}. This uncertainty is estimated by calculating $r$ using a slightly wider mass window in the fit to Eq.~\ref{eq:muon_background_shape}.  The difference between the resulting $A_N$ and the nominal value is taken as the magnitude of the uncertainty.

Type B systematic uncertainties arise from uncertainties on the polarization which are either uncorrelated between beams or uncorrelated between year and are all 0.003 or smaller (numerical values can be found in Appendix~\ref{sec:data_tables}).\footnote{These uncertainties would na\"ively be Type C, but the different statistical weights used to average the uncertainty between beams or years for each data point mean that the magnitude of the uncertainty is not necessarily uniform.}  Unlike typical Type B systematic uncertainties, the uncertainties in this analysis maintain the significance of the data points from zero.

Type C systematic uncertainties come from sources of polarization uncertainty which are fully correlated between beams and years (e.g. systematic uncertainty in the hydrogen jet measurement due to contamination of the jet).  This leads to a scale uncertainty of 3.4\% for 2006, 3.0\% for 2008, and 2.4\% for the combined 2006 and 2008 datasets.

A summary of all uncertainties can be found in Table~\ref{tab:AN} of Appendix~\ref{sec:data_tables}.

\section{Results and Discussion}
The final asymmetries included 539$\pm$25 $J/\psi$ mesons in the central spectrometer, 3507$\pm$59 (3354$\pm$58) in the North (South) muon spectrometer from the 2006 run, and 8540$\pm$92 (8105$\pm$90) in the North (South) muon spectrometer from the 2008 run.  The values of $A_N$ obtained by analyses from both the central and muon spectrometers are shown in Fig.~\ref{fig:AN_v_pT} and~\ref{fig:AN_v_xF}, and the data points are given in Table~\ref{tab:AN} of Appendix~\ref{sec:data_tables}.  It is quite exciting to note that the $p_{T}$ integrated data point at $\left<x_{F}\right>$=0.08 is 3.3$\sigma$ less than zero, meaning that there is $\sim$99.9\% probability that $A_{N}$ is negative at that $x_{F}$.  The central and backward data points in $x_{F}$ are consistent with zero, and there is no observable $p_{T}$ dependence.

\begin{figure}[h!tb]
  \begin{center}
    \includegraphics*[width=0.75\columnwidth]{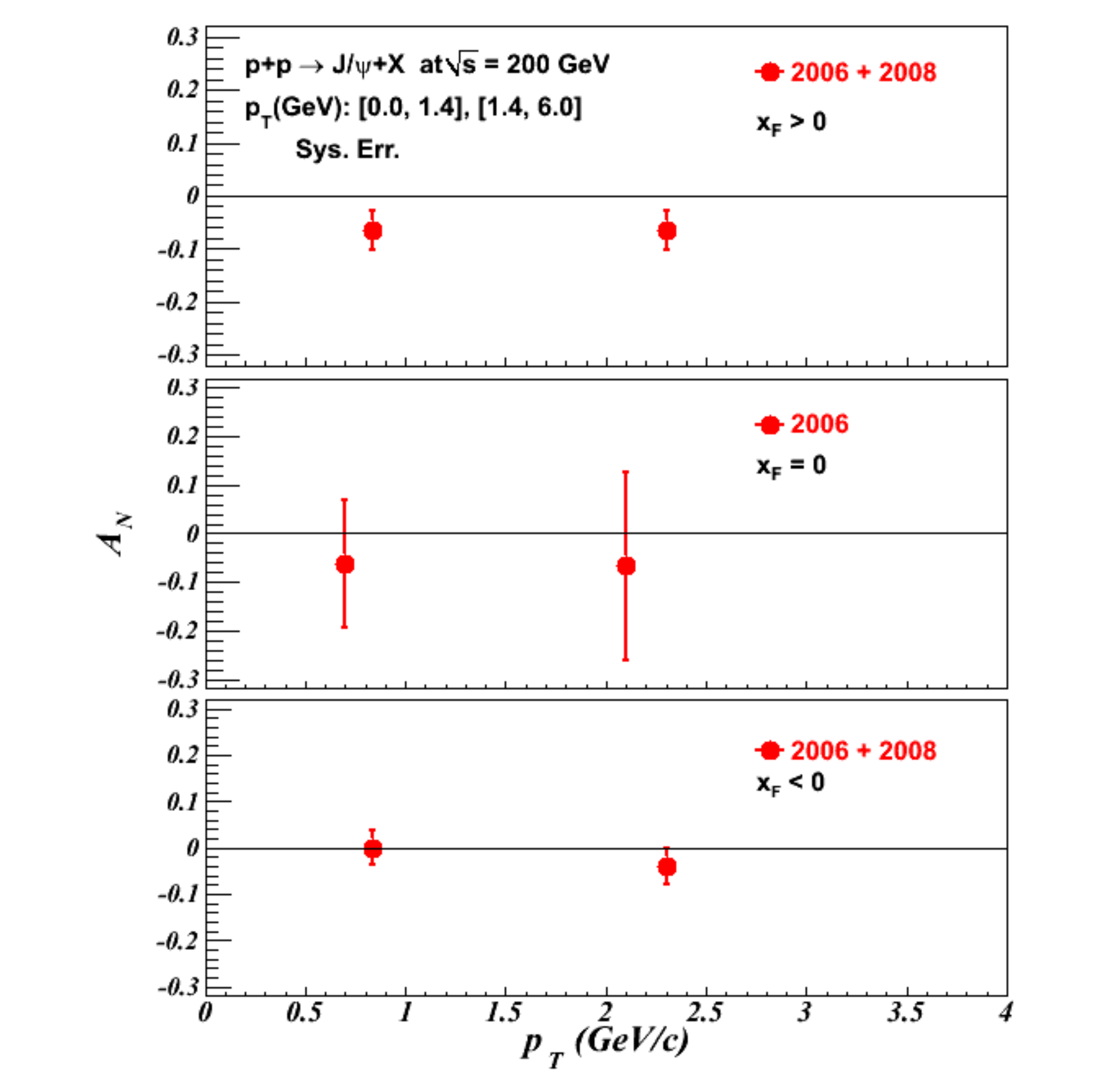} 
  \end{center}
\isucaption[$J/\psi$ transverse SSA measured at PHENIX from $\sqrt{s}=$200GeV $p$+$p$ collisions plotted against transverse momentum.]{$J/\psi$ transverse SSA measured at PHENIX from $\sqrt{s}=$200GeV $p$+$p$ collisions plotted against transverse momentum. Type A systematic uncertainties are added in quadrature.  Type B systematic uncertainties are all less than 0.003 and are not shown.  Type C systematic uncertainties are also not shown but are 3.4\%~(3.0\%) for the 2006~(2008) dataset, and 2.4\% for the 2006+2008 dataset.\label{fig:AN_v_pT}}
\end{figure}

\begin{figure}[h!tb]
\begin{center}
    \includegraphics*[width=0.49\columnwidth]{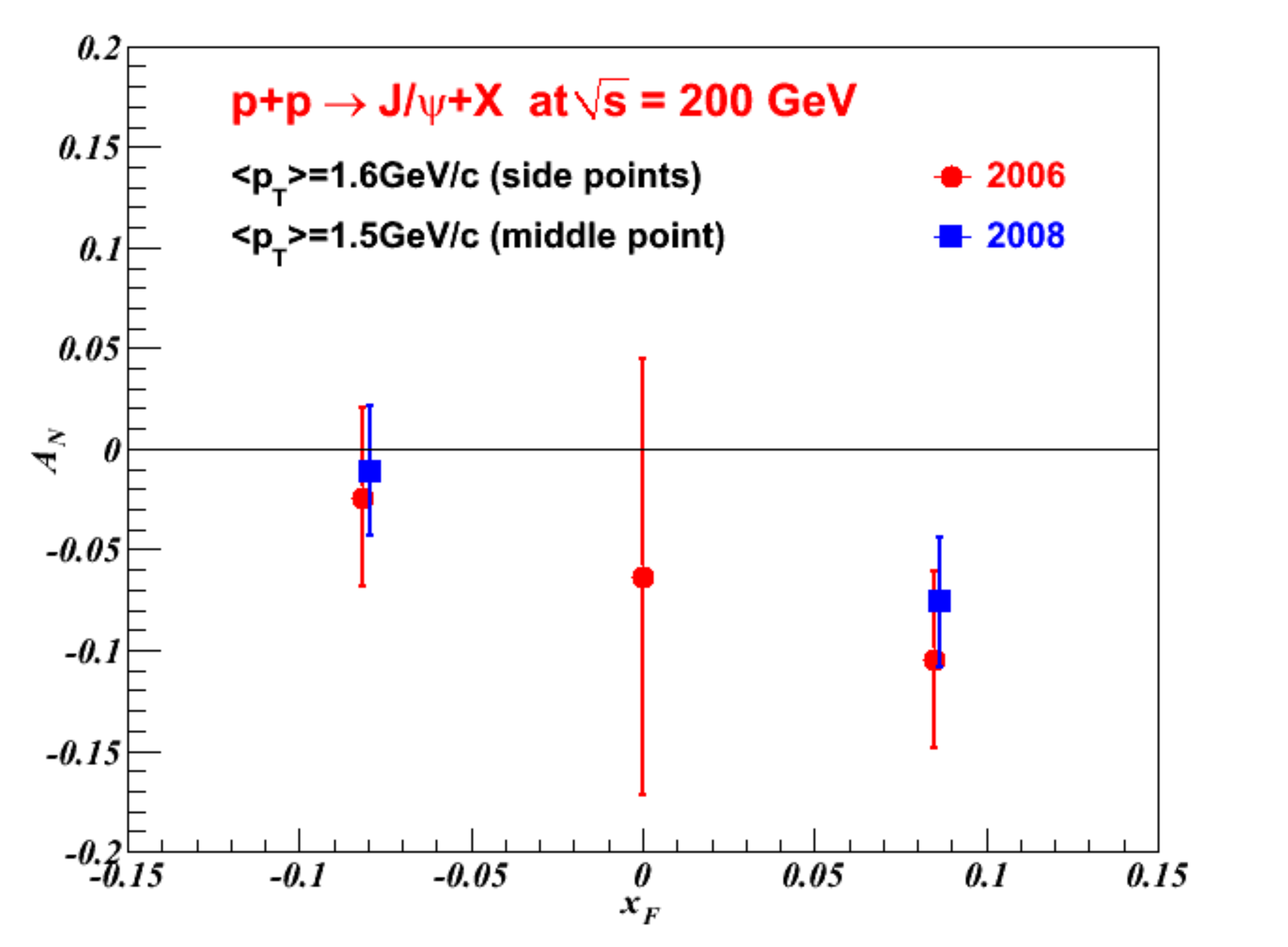} 
    \includegraphics*[width=0.48\columnwidth]{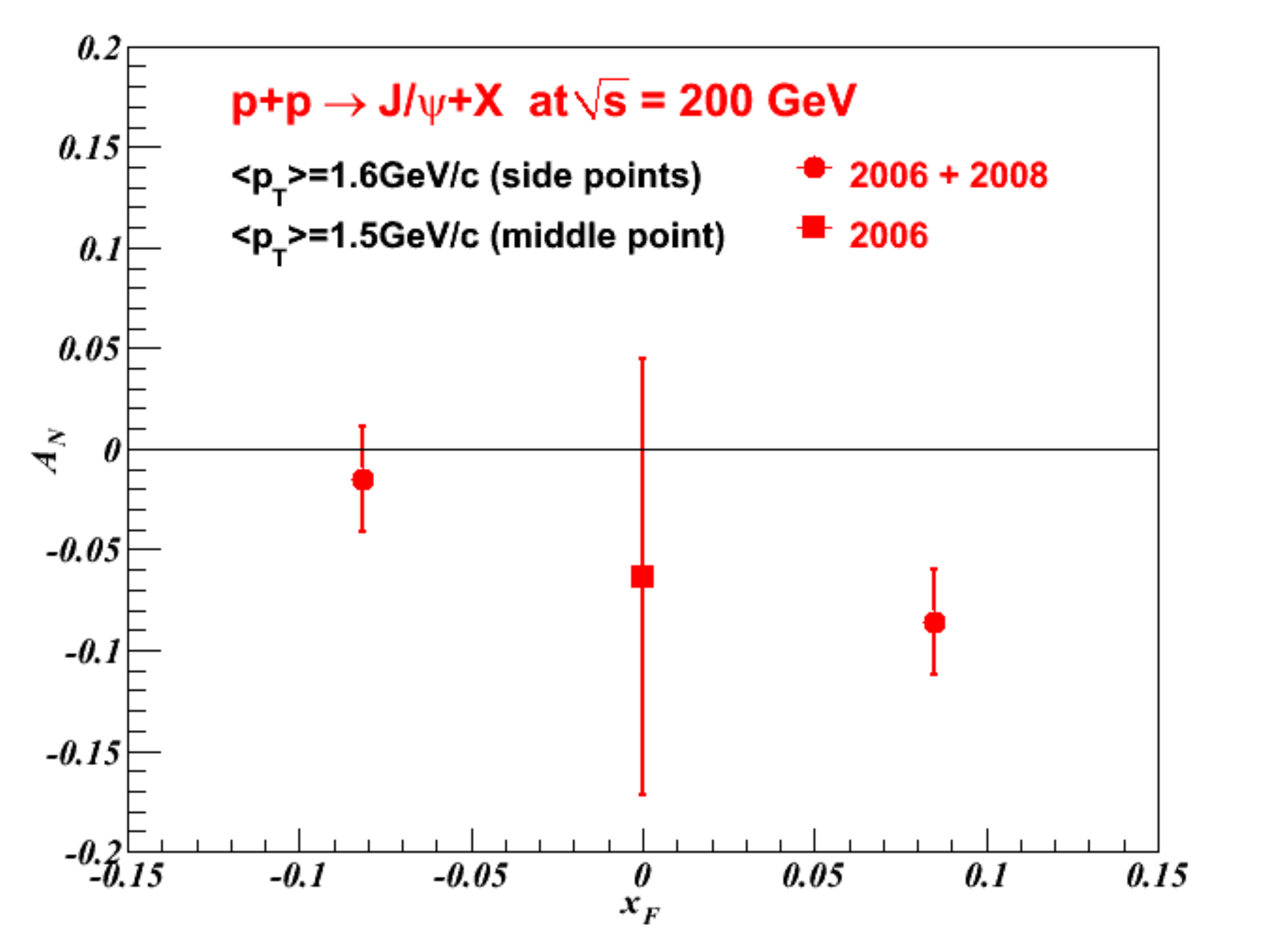}
\end{center}
\isucaption [$J/\psi$ transverse SSA measured at PHENIX from $\sqrt{s}=$200GeV $p$+$p$ collisions plotted against Feynman-x.]{ $J/\psi$ transverse SSA measured at PHENIX from $\sqrt{s}=$200GeV $p$+$p$ collisions plotted against Feynman-x.  Type A systematic uncertainties are added in quadrature.  Type B systematic uncertainties are all less than 0.003 and are not shown.  Type C systematic uncertainties are also not shown but are 3.4\%~(3.0\%) for the 2006~(2008) dataset, and 2.4\% for the 2006+2008 dataset.  On the left, $x_{F}>0$ data points from the 2006 and 2008 runs are shown separately, and on the right they are combined.\label{fig:AN_v_xF}}
\end{figure}



\begin{figure}[h!tb]
  \begin{center}
    \includegraphics*[width=0.6\columnwidth]{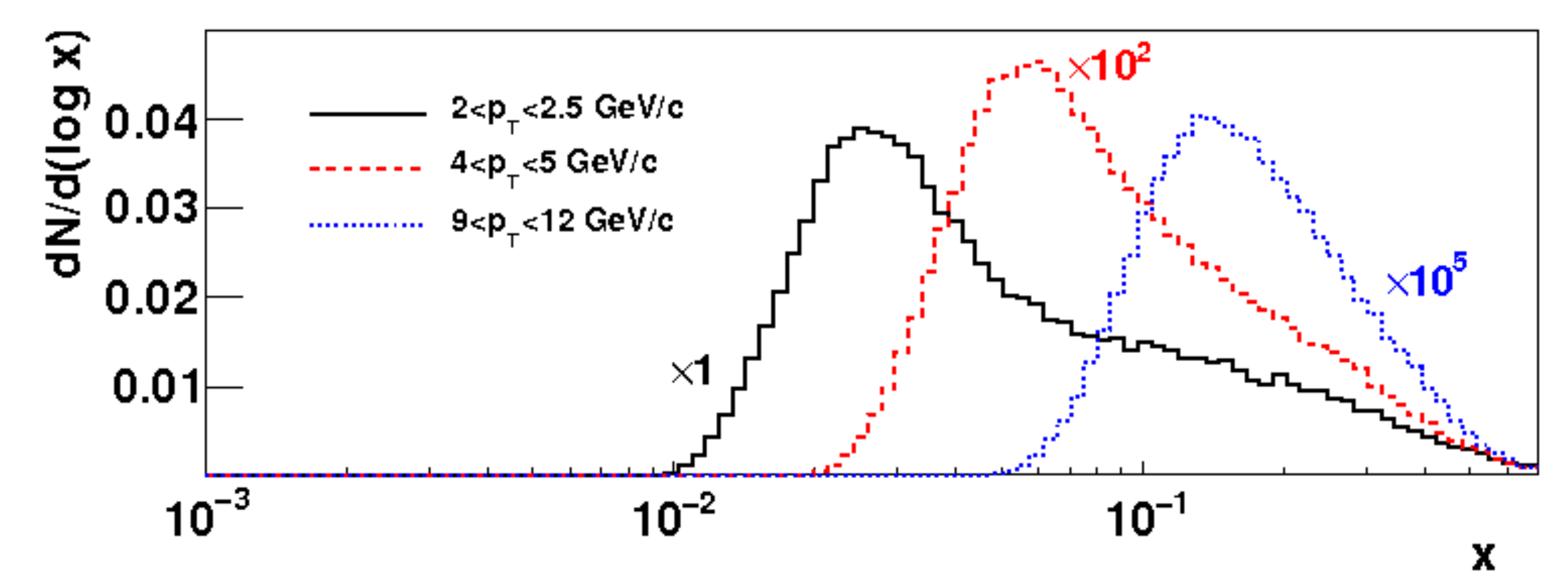}  \\
    \includegraphics*[width=0.6\columnwidth]{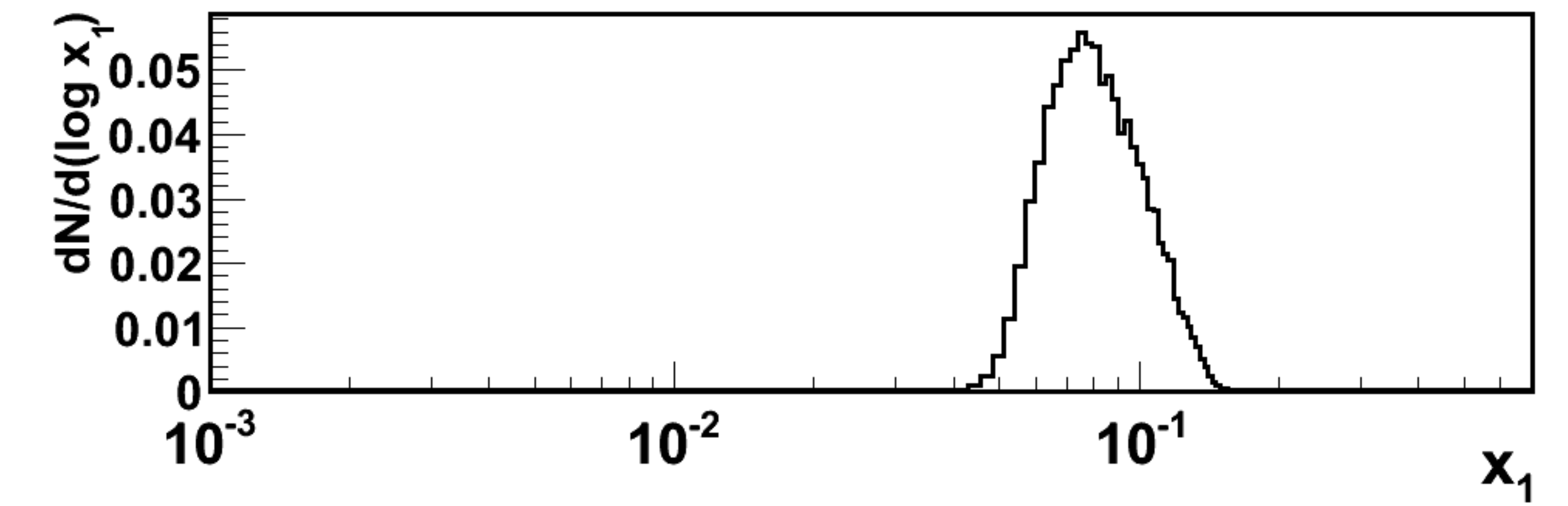} 
  \end{center}
  \isucaption[$x$ range covered for various $p_{T}$ ranges of $\pi^{0}$ production at mid-rapidity in the PHENIX central spectrometers and for the parton in the polarized proton from $J/\psi$ production in the PHENIX muon spectrometers at $x_F$$>$0 from $p$+$p$ collisions at $\sqrt{s}$=200~GeV.]{\label{fig:xcoverage}$x$ range covered for various $p_{T}$ ranges of $\pi^{0}$ production at mid-rapidity in the PHENIX central spectrometers (top, from~\cite{Adare:2008px}) and for the parton in the polarized proton from $J/\psi$ production in the PHENIX muon spectrometers at $x_F$$>$0 from $p$+$p$ collisions at $\sqrt{s}$=200~GeV (bottom).}
\end{figure}

Because there appears to be a non-zero asymmetry at forward $x_F$, it is interesting to compare this result with other analyses from PHENIX.  Preliminary measurements of the transverse SSA for mid-rapidity $\pi^{0}$ production from PHENIX shown back in Fig.~\ref{fig:PHENIX_pi0} cover an $x$ range of approximately [0.02,0.3] and show no sign of a non-zero Sivers or multi-parton correlation function.  The $x$ range covered for the parton in the transversely polarized proton for the $x_F$$>$0 data point in this analysis is shown along with the $x$ coverage for $\pi^{0}$ production as a function of $p_{T}$ in Fig.~\ref{fig:xcoverage}.  For forward $J/\psi$ production, the PHENIX muon arms cover approximately $x_{1}\in$[0.05,0.15].

While the overlap in coverage for forward $J/\psi$ production and central $\pi^{0}$ production at $p_{T}\gtrsim$5~GeV/$c$ mean that the two results appear to disagree, the processes contributing to the production of the two particles are also important.  Looking back at Fig.~\ref{fig:pi0_processes}, we can see that at $p_{T}\gtrsim$5~GeV/$c$, $\pi^{0}$ production at central rapidity is dominated not by gluon-gluon scattering but instead by quark-gluon scattering.  This means that any effect from a non-zero gluon Sivers or trigluon correlation function for mid-rapidity $\pi^{0}$ production will be diluted by collisions where a quark, not a gluon, enters the scattering from the polarized proton.   For $J/\psi$ production, on the other hand, gluon-gluon scattering dominates for the entire $p_{T}$ range so that there is no dilution.  Likewise, color factors from the contributing diagrams have been shown to have a large effect on the resulting transverse SSA~\cite{Bacchetta:2007sz} so that a full comparison between the $J/\psi$ and $\pi^{0}$ transverse SSA will require further theoretical activity.

At the current state of theoretical understanding it is unclear whether the sign of $A_N$ is be determined by the Sivers effect or the production mechanism.  It is also unclear how the magnitude compares to that expected by theory, as~\cite{Yuan:2008vn} gives a possible source for the asymmetry but no numerical guidance.\footnote{Admittedly, numerical guidance is difficult to obtain, because neither the size of the gluon Sivers function nor the $J/\psi$ production mechanism have been determined.}  However, the fact that the $J/\psi$ transverse SSA is non-zero in any kinematic region is quite exciting because of the implication that either the Sivers or trigluon correlation functions are non-zero.  In the TMD approach, this would additionally imply that color-singlet diagrams contribute substantially to the $J/\psi$ cross section for $p+p$ collisions at $\sqrt{s}=200$~GeV.

A great deal of theoretical development and further measurements are necessary to untangle both the gluon Sivers effect and the $J/\psi$ production mechanism in transverse SSAs (this will be discussed further in Chapter~\ref{ch:conclusions}).  Nevertheless, the non-zero effect at forward $x_F$ which is implied by Fig.~\ref{fig:AN_v_xF} is an exciting and important step. 

%% file: chapters/jpsi_pol/jpsi_pol.tex
\chapter{$J/\psi$ Angular Decay Coefficients}\label{ch:jpsi_pol}

\section{Introduction}

In 2009 the PHENIX experiment collected data from $p$+$p$ collisions at $\sqrt{s}$=500~GeV, providing access for the first time to $J/\psi$ angular decay coefficients from $p$+$p$ collisions at that energy.  In this chapter, I will discuss a measurement of the angular decay coefficients $\lambda_{\vartheta}$, $\lambda_{\varphi}$, and $\lambda_{\vartheta\varphi}$ using the PHENIX muon spectrometers.

\subsection{Measuring $\lambda_{\vartheta}$, $\lambda_{\varphi}$, and $\lambda_{\vartheta\varphi}$}\label{sec:measurement_outline}

If we require the $\hat{z}$-axis of our reference frame to be in the production plane, the full angular distribution of leptons from $J/\psi$ decays is given by Eq.~\ref{eq:angle_dist}.  It is common, however, to use distributions integrated of $\varphi$ and $\cos\vartheta$ written, respectively, as
\begin{equation}
\frac{dN}{d(\cos\vartheta)} \propto 1 + \lambda_{\vartheta} \cos^{2}\vartheta
\label{eq:lambda_theta}
\end{equation}
and 
\begin{equation}
\frac{dN}{d\varphi} \propto 1 + \frac{2\lambda_{\varphi}}{3+\lambda_{\vartheta}} \cos2\varphi
\label{eq:lambda_phi}
\end{equation}
as expressions for determining $\lambda_{\vartheta}$ and $\lambda_{\varphi}$.
To determine $\lambda_{\vartheta\varphi}$ one can then define an angle $\varphi_\vartheta$, as proposed in~\cite{Faccioli:2010kd}, with
\begin{equation}
\varphi_{\vartheta} =
\begin{cases}
\varphi - \frac{\pi}{4}  \text{  for $\cos\vartheta >$ 0} \\
\varphi - \frac{3\pi}{4} \text{  for $\cos\vartheta <$ 0}  
\end{cases}
\end{equation}
in order to get a distribution
\begin{equation}
\frac{dN}{d\varphi_{\vartheta}} \propto 1 + \frac{\sqrt{2}\lambda_{\vartheta\varphi}}{3+\lambda_{\vartheta}} \cos\varphi_{\vartheta}
\label{eq:lambda_thetaphi}
\end{equation}
for $\lambda_{\vartheta\varphi}$.  In order to determine the angular coefficients, each of these distributions is measured in real data and a simulation with an isotropic decay distribution ($\lambda_{\vartheta}$=$\lambda_{\varphi}$=$\lambda_{\vartheta\varphi}$=0).  To correct for the acceptance, the distributions from data are divided by those from simulation.  These acceptance-corrected distributions are then fit with Eq.~\ref{eq:lambda_theta}, \ref{eq:lambda_phi}, and \ref{eq:lambda_thetaphi} to determine each of the coefficients.

One must be very careful, however, to estimate the effect of acceptance on these integrated distributions.  If the acceptance creates `holes' in the $cos\vartheta$-$\varphi$ phase-space, the acceptance of these `hole' regions cannot be corrected through division by a simulated distribution.  The relative acceptance of the PHENIX muon spectrometers can be found in Appendix~\ref{sec:angular_acceptance} and indeed creates large holes.  To avoid the effects of holes in the muon acceptance and to include the effects of correlations between the various coefficients, we will instead perform a two-dimensional fit with Eq.~\ref{eq:angle_dist}.  To correct for acceptance, we perform a GEANT~\cite{GEANT} Monte Carlo simulation using a full description of the PHENIX detector to determine the shape of the angular distributions in our detectors with an isotropic decay distribution.  The angular distributions from real data are fit with a convolution of Eq.~\ref{eq:angle_dist} and the simulated acceptance. 

Avoiding muons with momenta near the trigger threshold is quite important, even if the simulation exactly describes the trigger.  Because the angular distribution is very different for $\lambda_{\vartheta}$=1 than for $\lambda_{\vartheta}$=0 around the trigger threshold, the physical acceptance of the detector in $\cos\vartheta$ can be different for these two situations such that dividing the data by a simulation with $\lambda_{\vartheta}$=0 does not properly correct the data for acceptance.  An extreme example would be as follows:  

Consider two mesons with equal momenta.  For the first meson, decay muons are produced along the meson's momentum, while, for the second, muons are produced transverse to the meson's momentum.   One decay muon from the first meson will get a boost along the meson's momentum direction, while the other gets a boost in the opposite direction.  For the second meson, both muons will get a boost in a direction transverse to the meson's momentum.  It is less likely that both muons from the first meson will make it through the $\mu$ID, because one of them is being boosted in the wrong direction.  Because we require that both muons make it through the detector, the trigger turn-on curve will be shifted to higher $p_{z}$ for the first meson.  Our acceptance correction will not account for the difference because it uses an isotropic simulated distribution.

\begin{figure}[h!tb]
        \centering\includegraphics*[width=0.49\columnwidth]{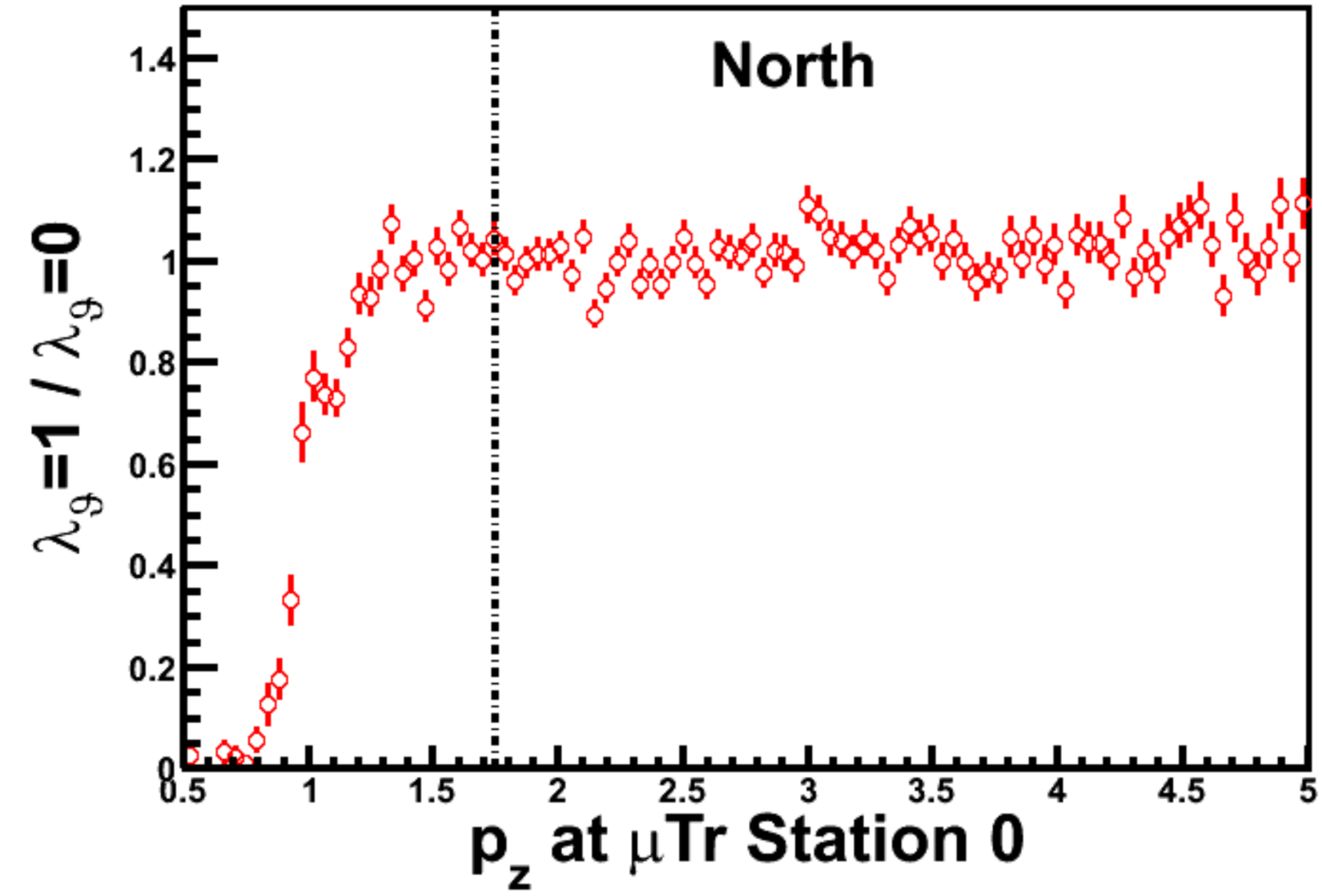} \
        \centering\includegraphics*[width=0.49\columnwidth]{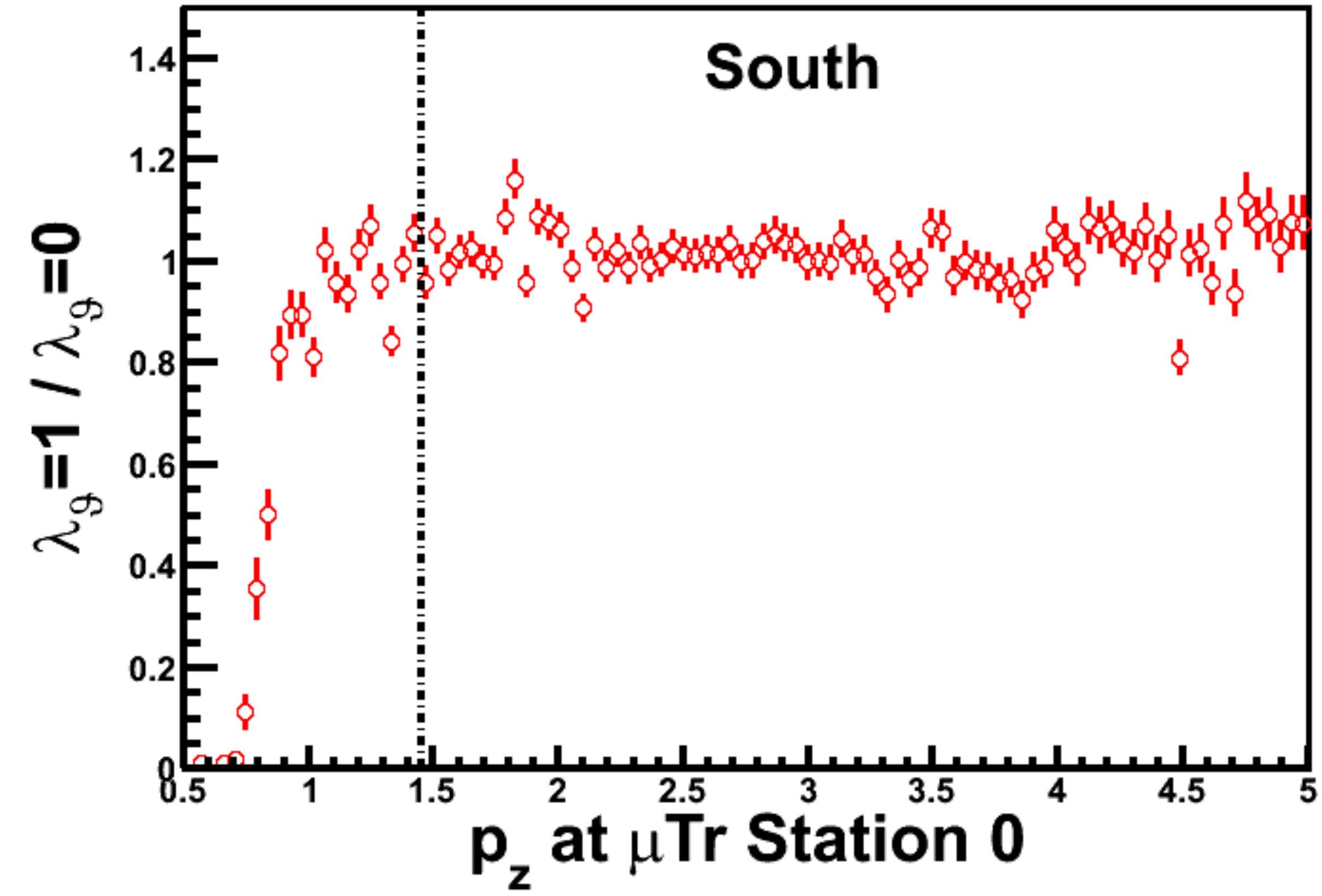}
        \isucaption{$p_{z}$ at the first $\mu$Tr station from a full simulation with $\lambda_{\vartheta}$=1 in the Helicity frame divided by a full simulation with $\lambda_{\vartheta}$=0 for the North (left) and South (right).  The $\lambda_{\vartheta}$=1 distribution is clearly shifted to larger $p_{z}$ Dashed lines show the cut placed on this quantity in the analysis.}\label{fig:st0_pz_pol}
\end{figure}

The effect can be seen for realistic angular distributions from $\lambda_{\vartheta}$=0 and $\lambda_{\vartheta}$=1 for the Helicity frame in Fig.~\ref{fig:st0_pz_pol} where the trigger turn-on curve is clearly shifted to larger $p_{z}$ for $\lambda_{\vartheta}$=1.  The cut placed on the trigger threshold, $p_{z}$ measured at the first $\mu$Tr station, has been made tight enough in this analysis to exclude any data around the trigger turn-on.

\subsection{Simulation Procedure}\label{sec:procedure}

\begin{figure}[h!tb]
        \centering\includegraphics*[width=0.49\columnwidth]{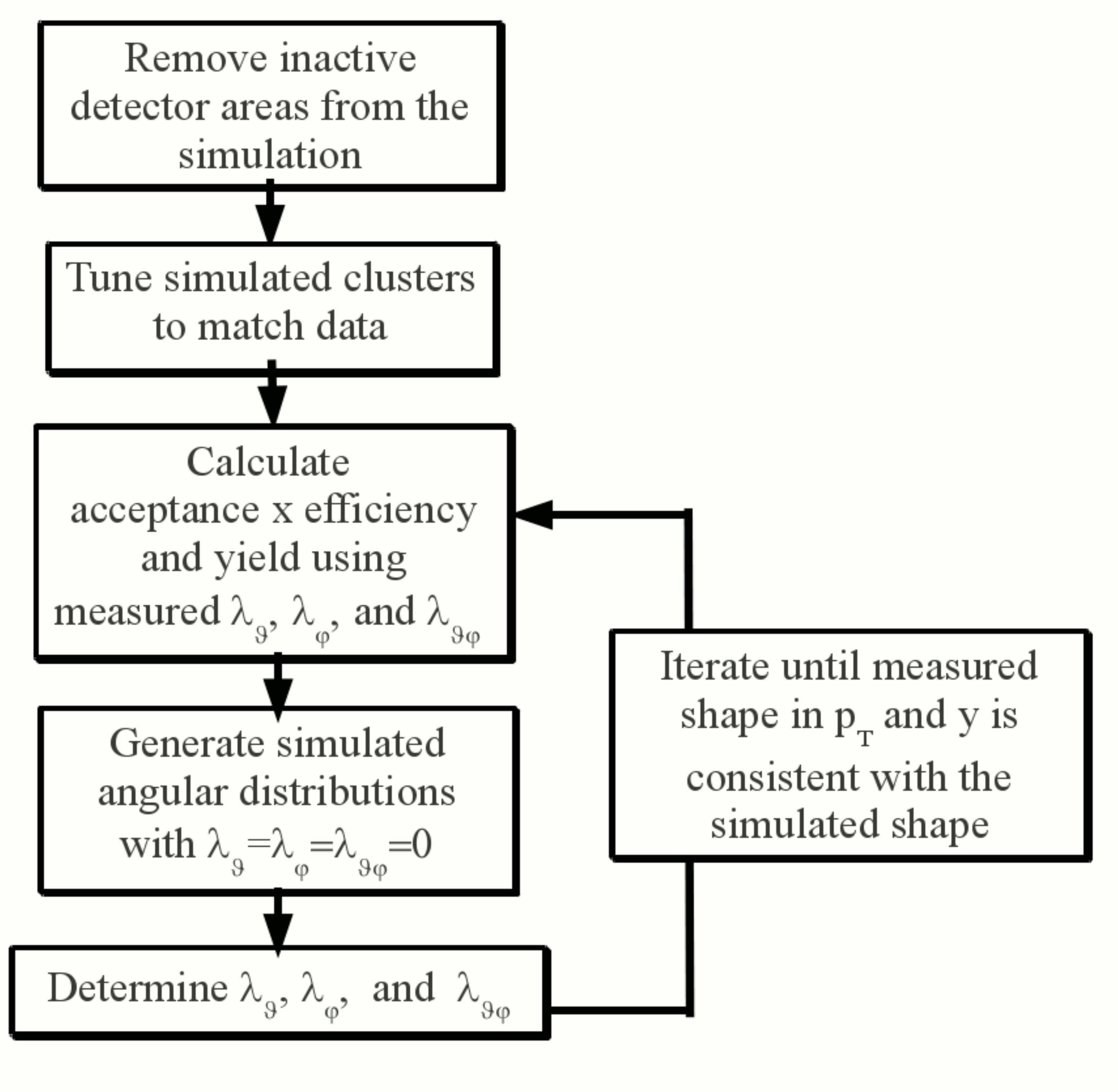} 
        \isucaption{Flowchart of the iterative procedure for tuning the simulation to real data.}\label{fig:flowchart}
\end{figure}

Because we are using a simulation to obtain the acceptance, the simulation itself needs to be well tuned to reproduce both low-level cluster-related quantities and high level kinematic distributions.  In this section, I will give a general outline of the procedure used to tune the simulation to data.  More detailed information and plots from the simulation will be given throughout the rest of this document.  The procedure will be as follows (a flowchart can also be found in Fig.~\ref{fig:flowchart}):
\begin{enumerate}
  \item 
    Introduce inactive areas into the simulation. 
  \item
    Tune the total cluster charge, tracking resolution, and cluster shape to match those found in real data. 
  \item
    Use the tuned simulation to determine the acceptance $\times$ efficiency and yield for $J/\psi$ mesons as a function of transverse momentum~($p_{T}$) and rapidity~($y$).
  \item
    Generate a second set of events using the shape of the measured yield in $p_{T}$ and $y$ from Step 3 and an isotropic decay distribution.  Use this simulation to determine $\lambda_{\vartheta}$, $\lambda_{\varphi}$, and $\lambda_{\vartheta\varphi}$.
  \item
    Repeat Step 3 using the measured $\lambda_{\vartheta}$, $\lambda_{\varphi}$, and $\lambda_{\vartheta\varphi}$ from Step 4.  Continue iterating until the shape used in Step 3 is consistent with the shape of the measured yield.
\end{enumerate}
 
\section{Quality Assurance}

Before the data can be taken seriously, we need to place criteria on both the subset of the data used in the analysis and the tracks in the detector considered to be muons.  A determination of the angular decay coefficients is especially sensitive to large changes in detector acceptance; so we will use criteria which require that the acceptance stay relatively constant over the course of the analysis.  Requirements will also be placed on tracks to increase the number of muons from the $J/\psi$ signal relative to background from random track combinations in the detector.

\subsection{Run Selection}

\begin{figure}[h!tb]
        \centering\includegraphics*[width=0.4\columnwidth]{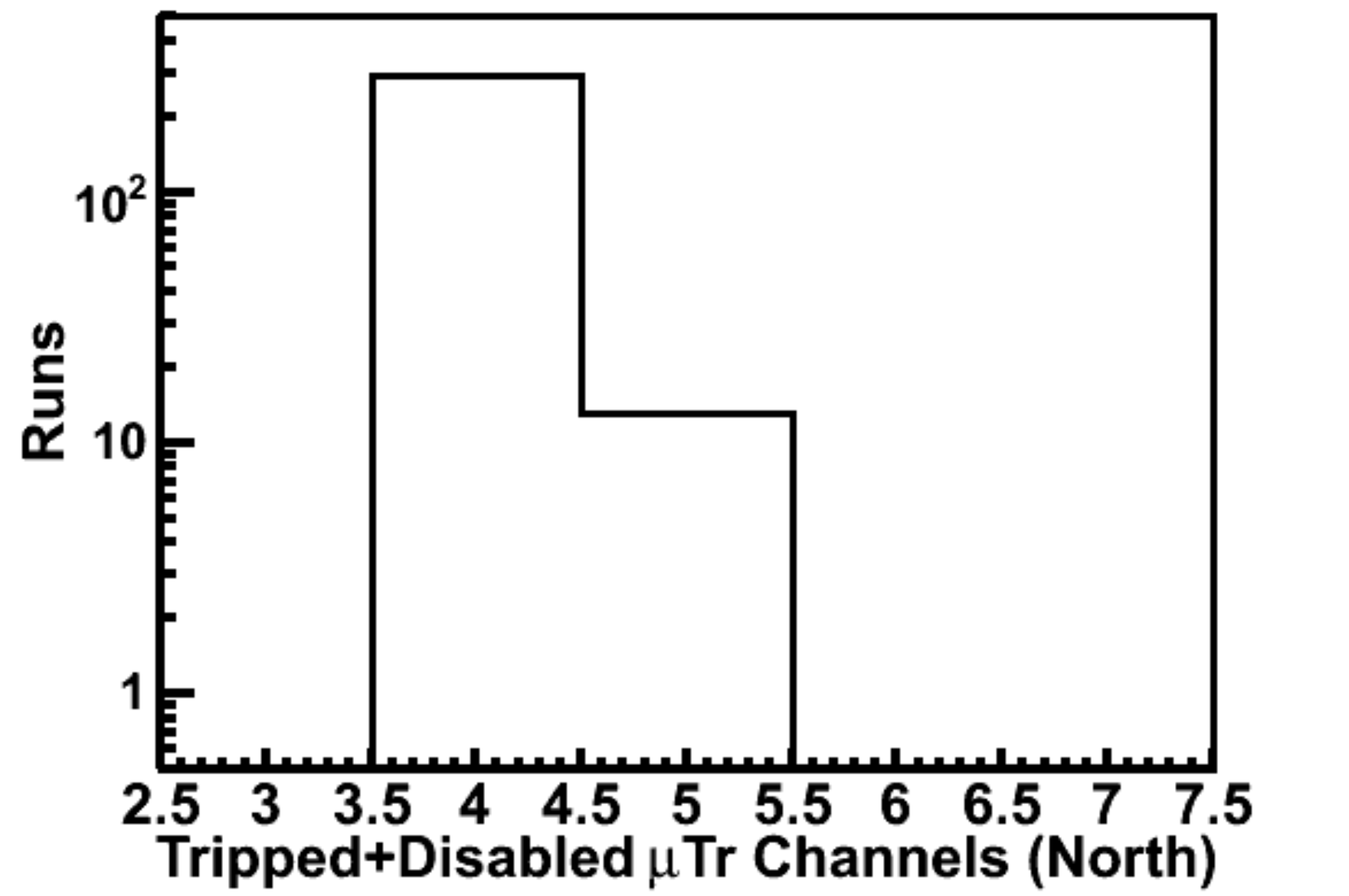} 
        \centering\includegraphics*[width=0.4\columnwidth]{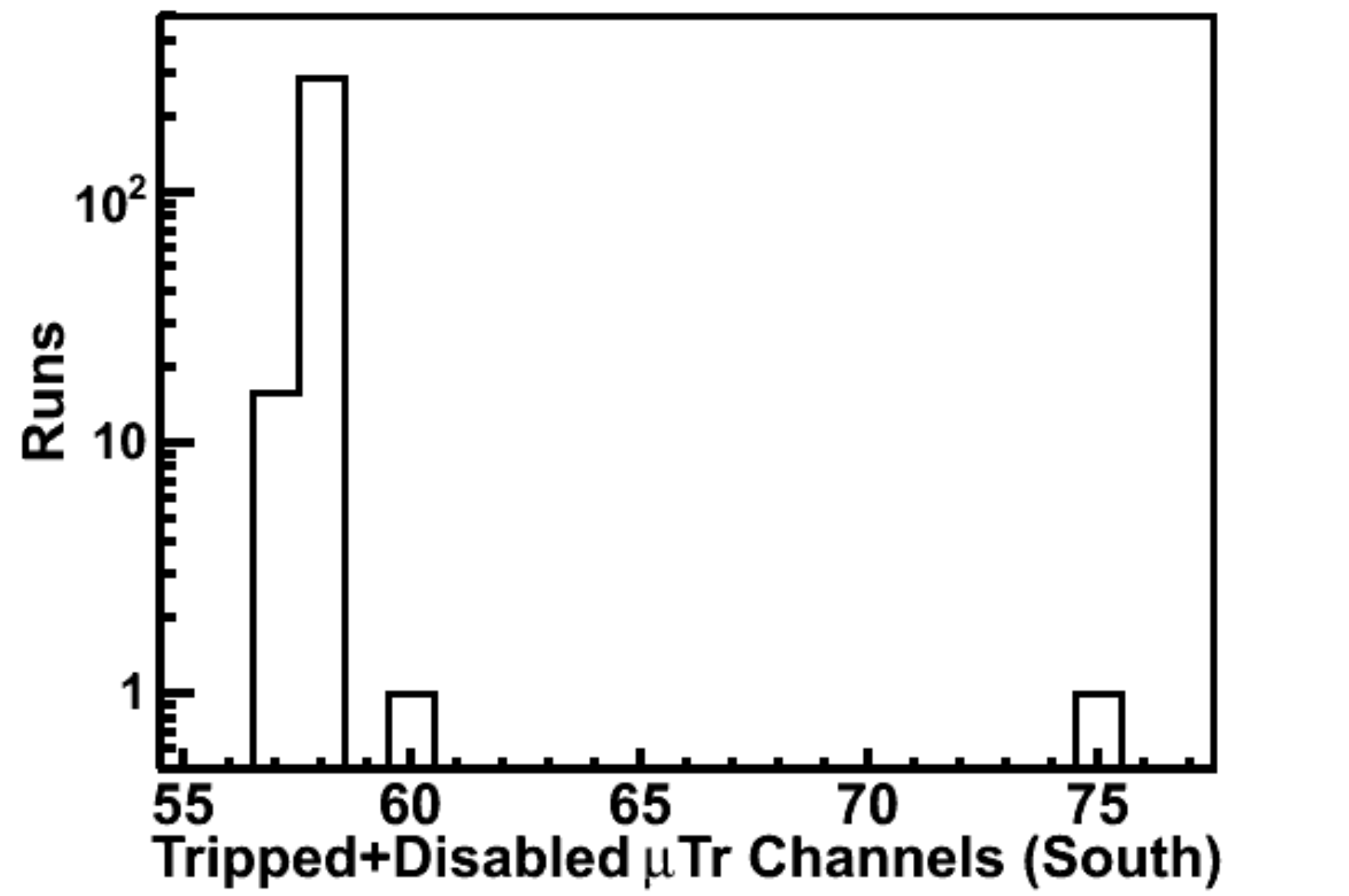} 
        \isucaption[Number of tripped or disabled channels in the $\mu$Tr for the angular decay coefficient analysis.]{Number of tripped or disabled channels in the $\mu$Tr histogrammed by run for the arm on the North (South) side of PHENIX on the top (bottom).}\label{fig:MuTr_trips}
\end{figure}

\begin{figure}[h!tb]
        \centering\includegraphics*[width=0.4\columnwidth]{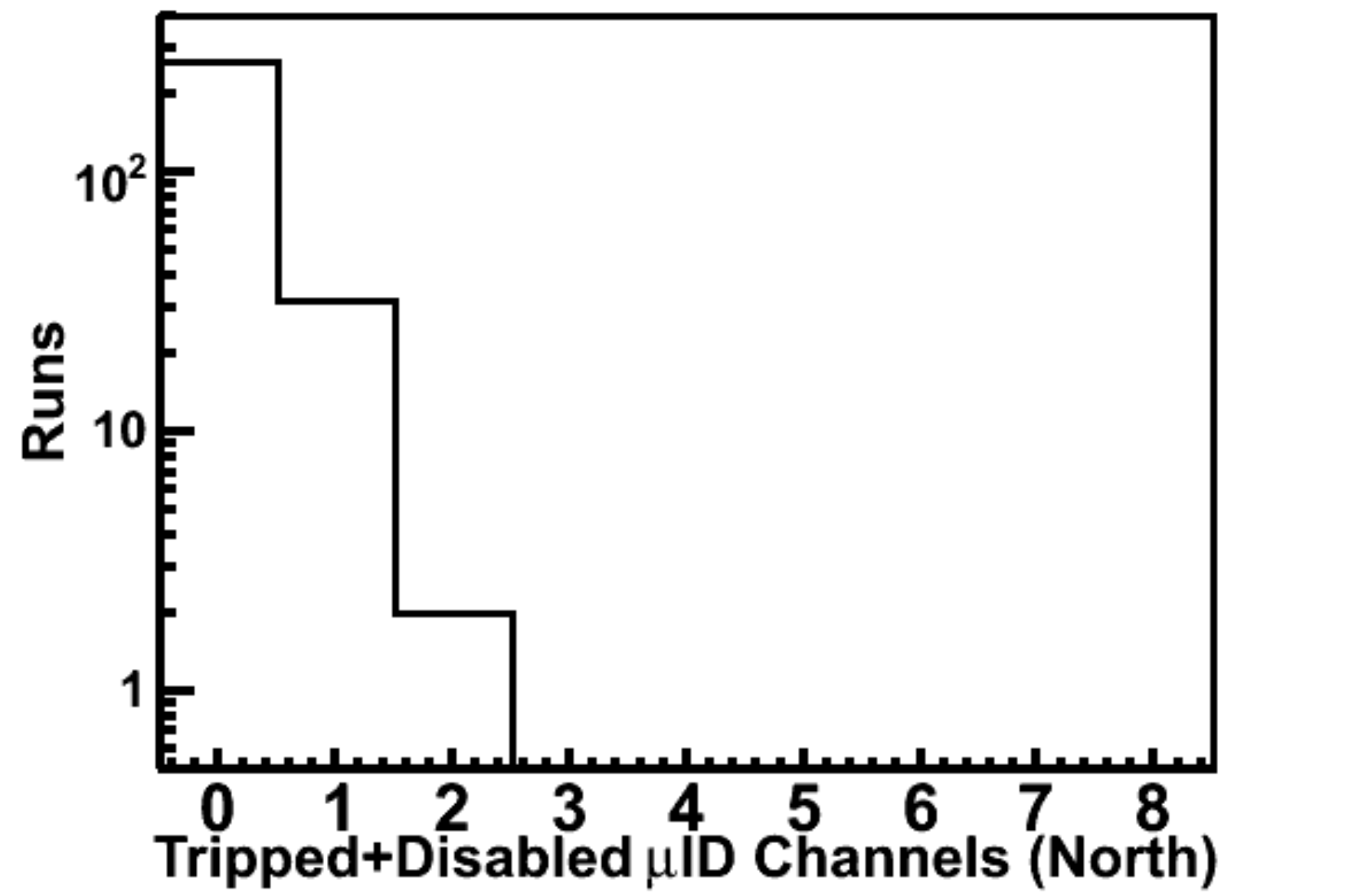} 
        \centering\includegraphics*[width=0.4\columnwidth]{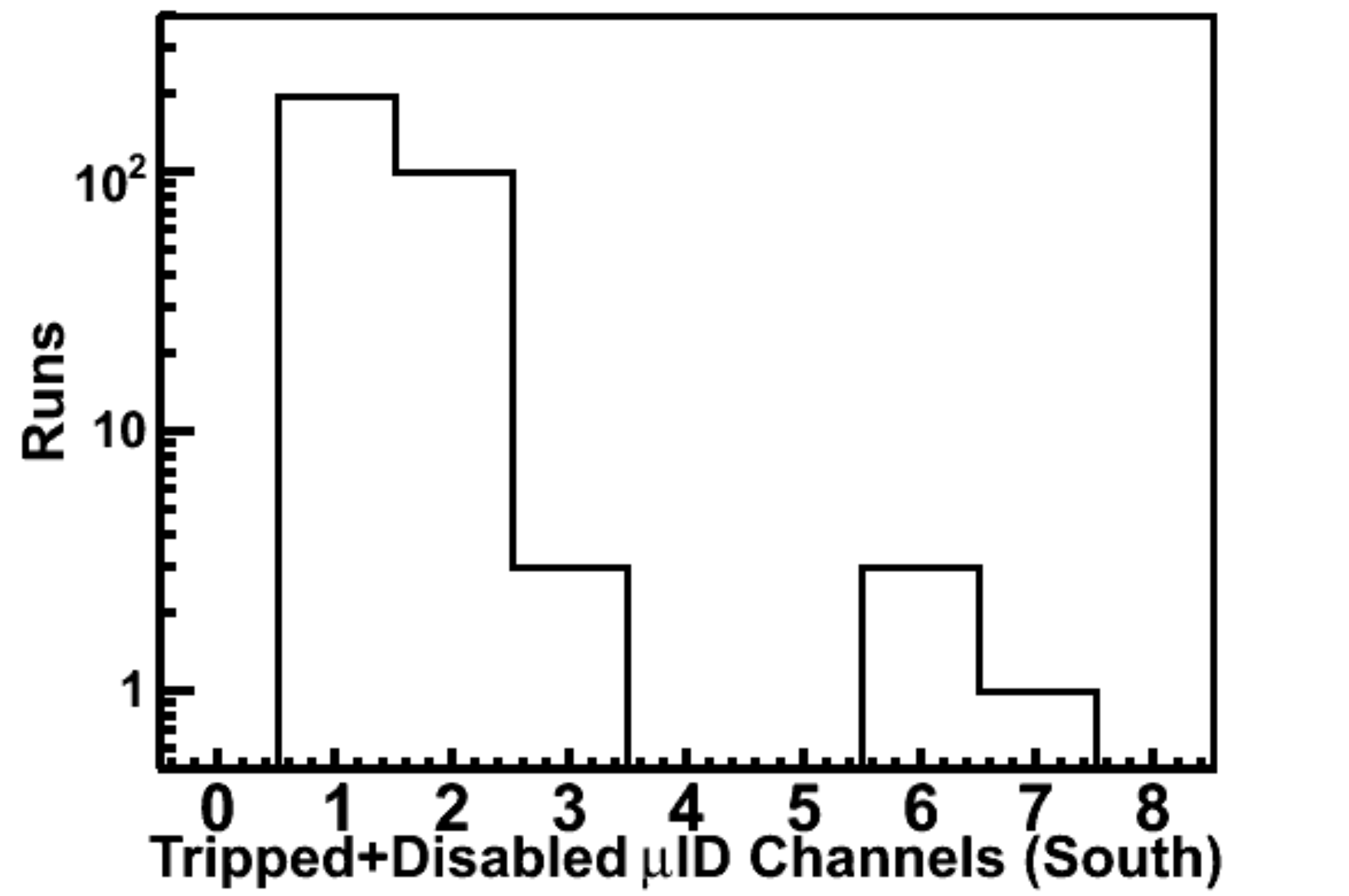} 
        \isucaption[Number of tripped or disabled channels in the $\mu$ID for angular decay coefficient analysis.]{Number of tripped or disabled channels in the $\mu$ID histogrammed by run for the arm on the North (South) side of PHENIX on the top (bottom).}\label{fig:MUID_trips}
\end{figure}

Data is taken by the PHENIX experiment in approximately one hour segments (called runs) to allow for the rejection of data which are taken with less than ideal conditions.  To ensure that the acceptance is approximately constant across all runs in the analysis, runs are required to meet the following criteria:
\begin{itemize}
  \item $<$ 4(59) tripped or disabled high voltage channels in the North~(South) $\mu$Tr (see Fig.~\ref{fig:MuTr_trips}).  Each channel corresponds to approximately $1/4$ of one octant in a gap at one of the three stations.\footnote{There are significantly more disabled channels in the South $\mu$Tr than the North because of high backgrounds in the inner channels of the first station.}
  \item $<$ 2(3) tripped high voltage channels in the North~(South) $\mu$ID (see Fig.~\ref{fig:MUID_trips}).  Each channel corresponds to approximately $1/3$ of one of the 6 panels which make up either the horizontal or vertical plane of one gap.
  \item $<$ 1\% of the data packets lost from the $\mu$Tr during transmission to the data acquisition system.  Losing one packet of data corresponds to a loss of that event for approximately one octant at one station in the $\mu$Tr.\footnote{There are only two data packets for the each $\mu$ID; so we require that all $\mu$ID data packets be received.}
  \item runs should not have been ended due to problems with any of the muon detectors
  \item the data acquisition system should indicate that the run ended cleanly
\end{itemize}
Of the 259 runs applicable to the analysis, the run selection criteria left 236 (238) runs for analysis in the North (South) muon spectrometer.  The total integrated luminosity was 17.6~pb$^{-1}$, and after run selection we are left with 16.0~(16.5)~pb$^{-1}$ in the North (South) spectrometer.

\subsection{$J/\psi$ Selection}

Criteria for selecting $J/\psi$ mesons must simultaneously reduce background and maintain signal.  In order to determine whether or not a pair of tracks come from a $J/\psi$, we need to ensure that both tracks behave like muons and originate from a real collision.

For an event to be considered, the following conditions must be met in either the North or South muon spectrometer:
\begin{description}
\item[$\mu$ID 2-deep and no vertex BBC Trigger]
  Two deep requirements must be met in the either the North or South $\mu$ID LL1, discussed in Section~\ref{sec:Triggering}, along with at least one tube fired in both the North and South BBC. 
\item[BBC z-vertex]
  The position of the event vertex along the beam direction must be within 40~cm of the center of the central spectrometer as measured by the BBC in order to avoid steel from the magnets located at $\sim$45~cm. 
\item[DDG0.]
  The angle between the track in the $\mu$Tr and the track in the $\mu$Id projected to the closest $\mu$Id gap to the interaction point must be less than 9$^{\circ}$ to ensure that tracks in the two detectors are associated with each other. 
\item[Vertex $\chi^{2}/ndf$]
  A fit of the two tracks in the $\mu$Tr to the vertex measured by the BBC must have a $\chi^{2}/ndf<$4 for 4 degrees of freedom.  Such a requirement removes accidental combinations of random tracks when constructing the pair.
\item[Minimum $p_{z}$ at the first $\mu$Tr station]
  The magnitude of the track momentum along the beam direction measured at the first $\mu$Tr Station must be greater than 1.75~(1.45) GeV/$c$ in the North~(South) spectrometer to avoid the turn-on curve associated with the trigger (Fig.~\ref{fig:st0_pz_pol}).
\end{description}

\section{Low Level Simulation Tuning}\label{sec:lowlevel_tuning}

To ensure that our simulation correctly reproduces data, we need to first check basic quantities like the size and shape of charge clusters in the $\mu$Tr as well as dead areas in both the $\mu$Tr and $\mu$ID.

\begin{figure}
        \centering\includegraphics*[width=0.75\columnwidth]{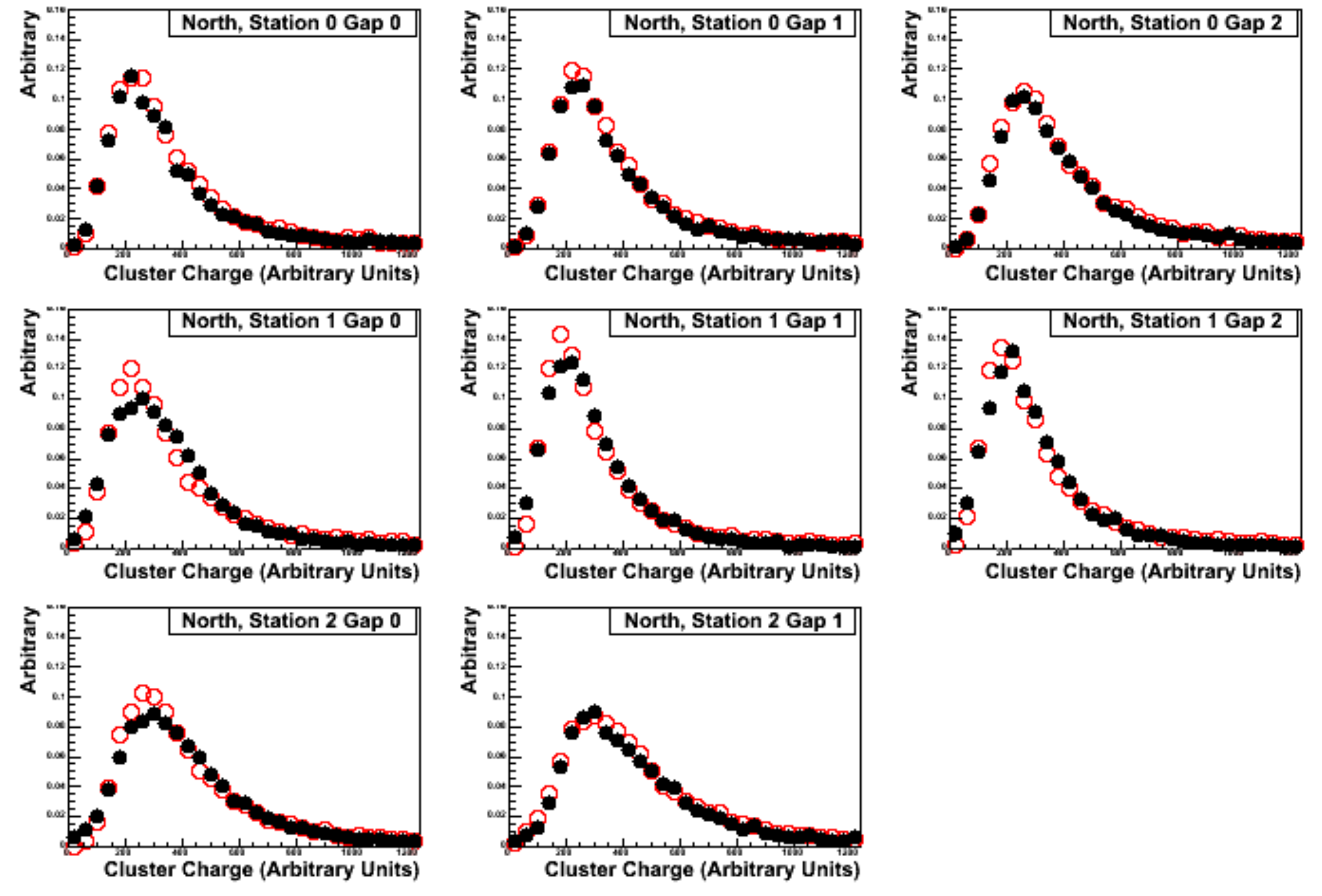}
        \isucaption[Comparison of total cluster charge in the North $\mu$Tr from simulation and data.]{Comparison of total cluster charge in the North $\mu$Tr in simulation (open circles) and data (closed circles).}\label{fig:N_cluster_charge}
\end{figure}

\begin{figure}
        \centering\includegraphics*[width=0.75\columnwidth]{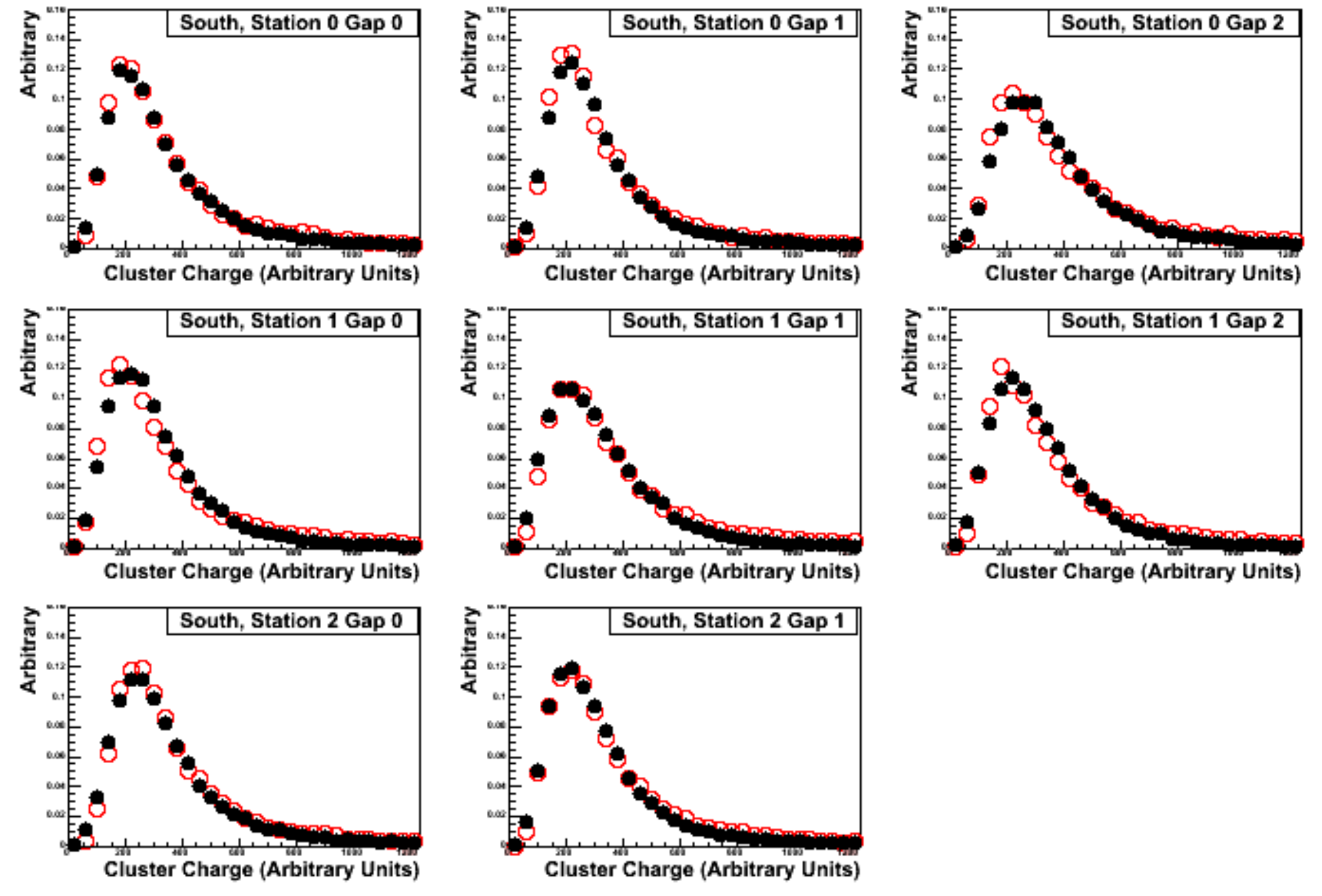}
        \isucaption[Comparison of total cluster charge in the South $\mu$Tr from simulation and data.]{Comparison of total cluster charge in the South $\mu$Tr in simulation (open circles) and data (closed circles).}\label{fig:S_cluster_charge}
\end{figure}

Charge is deposited in a cluster of strips by particles traversing the $\mu$Tr, and the total charge in those clusters is simulated by a Landau distribution
\begin{equation}
P(q) = q_{0} + \frac{q_{1}}{\pi}\displaystyle\int_{0}^{\infty}e^{-t \log t - q t} sin(\pi t) dt 
\end{equation}
where the parameters $q_{0}$ and $q_{1}$ are chosen octant-by-octant for each gap in the detector such that the distribution of total cluster charge matches the data.  The total cluster charge from a simulation of single muons is compared with cluster charges from data in Fig.~\ref{fig:N_cluster_charge} and~\ref{fig:S_cluster_charge} and agrees quite well after the tuning of $q_{0}$ and $q_{1}$.


\begin{figure}
        \centering\includegraphics*[width=0.75\columnwidth]{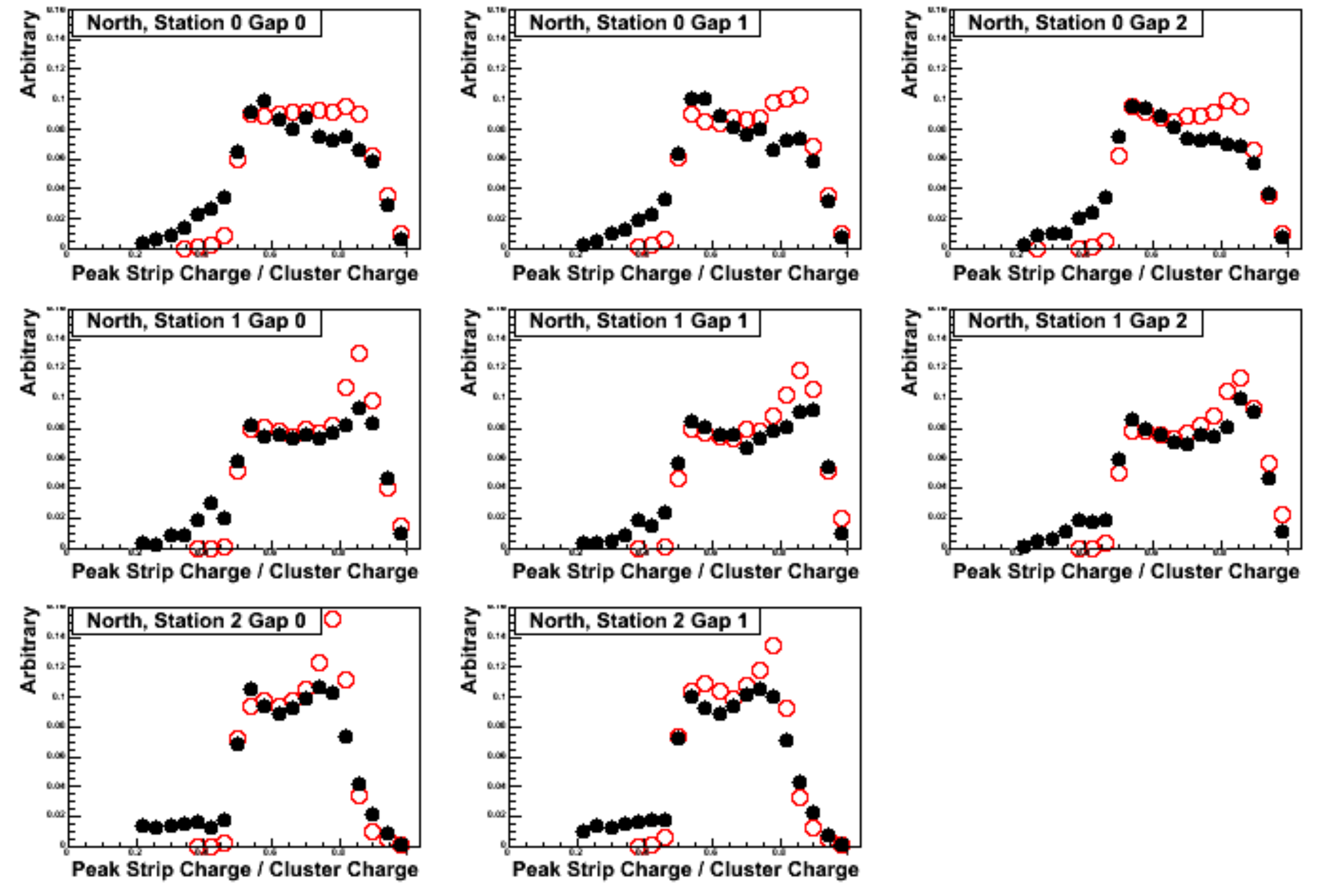}
        \isucaption[Comparison of peak strip charge over total cluster charge in the North $\mu$Tr from simulation and data.]{Comparison of peak strip charge over total cluster charge in the North $\mu$Tr in simulation (open circles) and data (closed circles).}\label{fig:N_cluster_width}
\end{figure}

\begin{figure}
        \centering\includegraphics*[width=0.75\columnwidth]{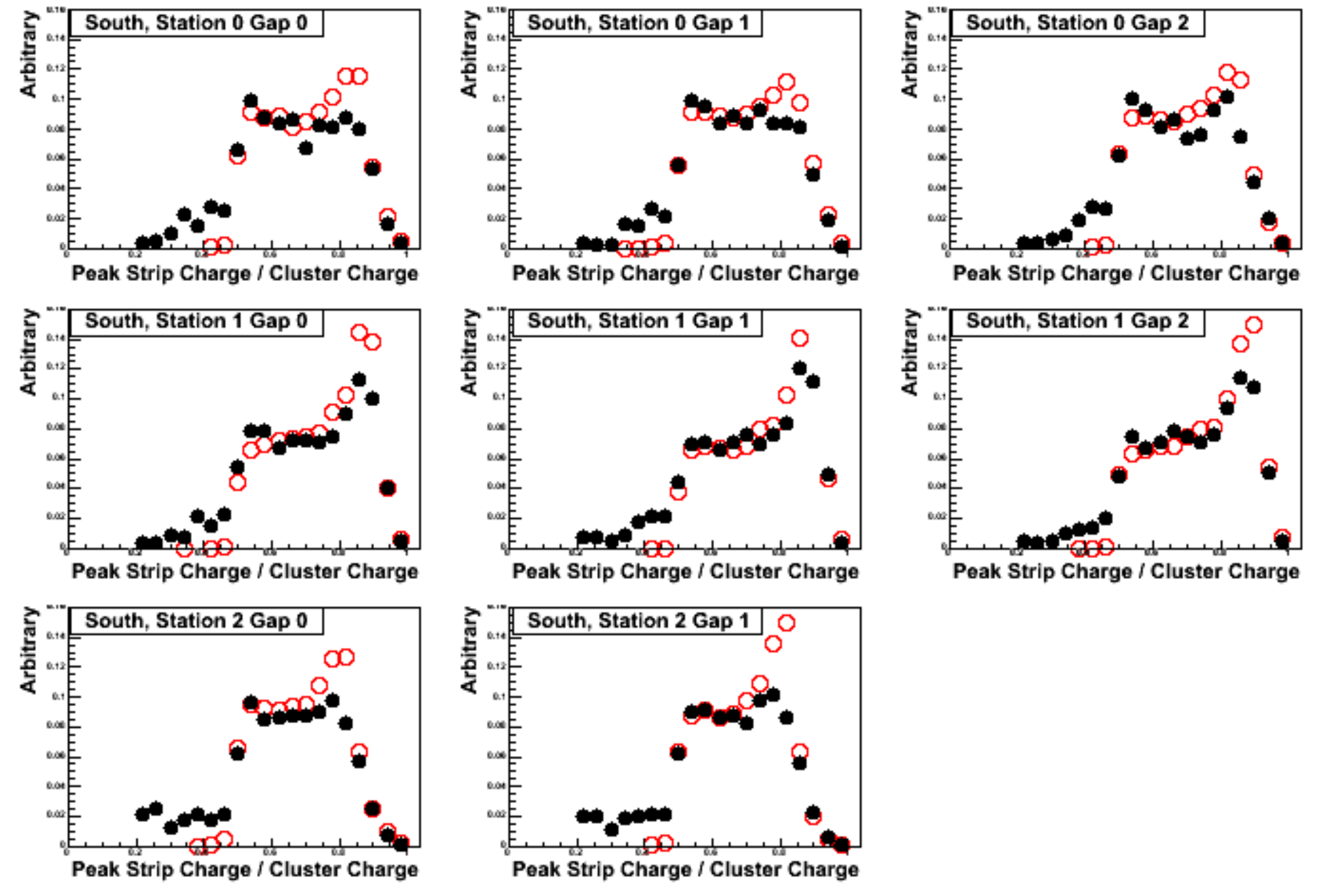}
        \isucaption[Comparison of peak strip charge over total cluster charge in the South $\mu$Tr from simulation and data.]{Comparison of peak strip charge over total cluster charge in the South $\mu$Tr in simulation (open circles) and data (closed circles).}\label{fig:S_cluster_width}
\end{figure}

The simulated shape of clusters in the $\mu$Tr is determined by a Mathieson distribution~\cite{Gatti:1979sn,Mathieson:1985uz}.  The capacitive coupling between neighboring cathodes and the spacing between cathode and anode are tuned for each octant in each gap so that the fraction of the total cluster charge carried by the peak strip matches between simulation and data, and a comparison between the simulated and measured distributions is shown in Fig.~\ref{fig:N_cluster_width} and~\ref{fig:S_cluster_width}.  While these distributions do not match perfectly, the effect of the cluster shape on acceptance should be minimal compared to differences in the total charge, which couple directly to the efficiency of the detector.


Previous PHENIX analyses have assumed that both the total cluster charge and cluster shape are constant across a given gap in the $\mu$Tr.  Through the course of this analysis, I found that there are significant variations from octant-to-octant within a gap (due to differences in the electronics which measure gain).  The analysis presented here is the first to use simulated cluster charges and shapes which are not only tuned gap-by-gap but also octant-by-octant.

\begin{figure}
        \centering\includegraphics*[width=0.75\columnwidth]{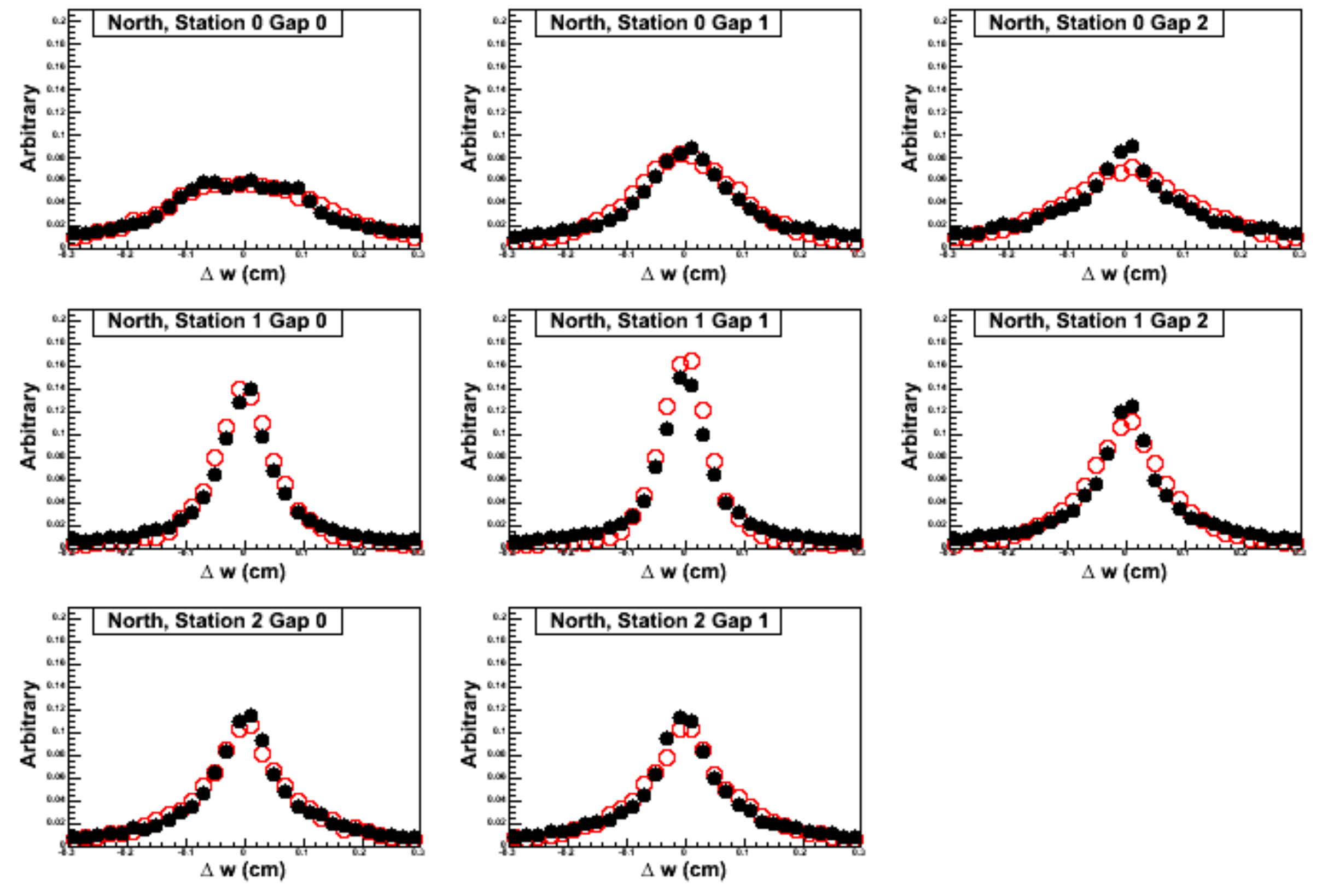}
        \isucaption[Comparison of tracking residuals in the North $\mu$Tr from simulation and data.]{Comparison of the tracking residual $\Delta w$, the difference between the cluster and track w coordinate in the North $\mu$Tr between simulation (open circles) and data (closed circles).}\label{fig:N_deltaw}
\end{figure}

\begin{figure}
        \centering\includegraphics*[width=0.75\columnwidth]{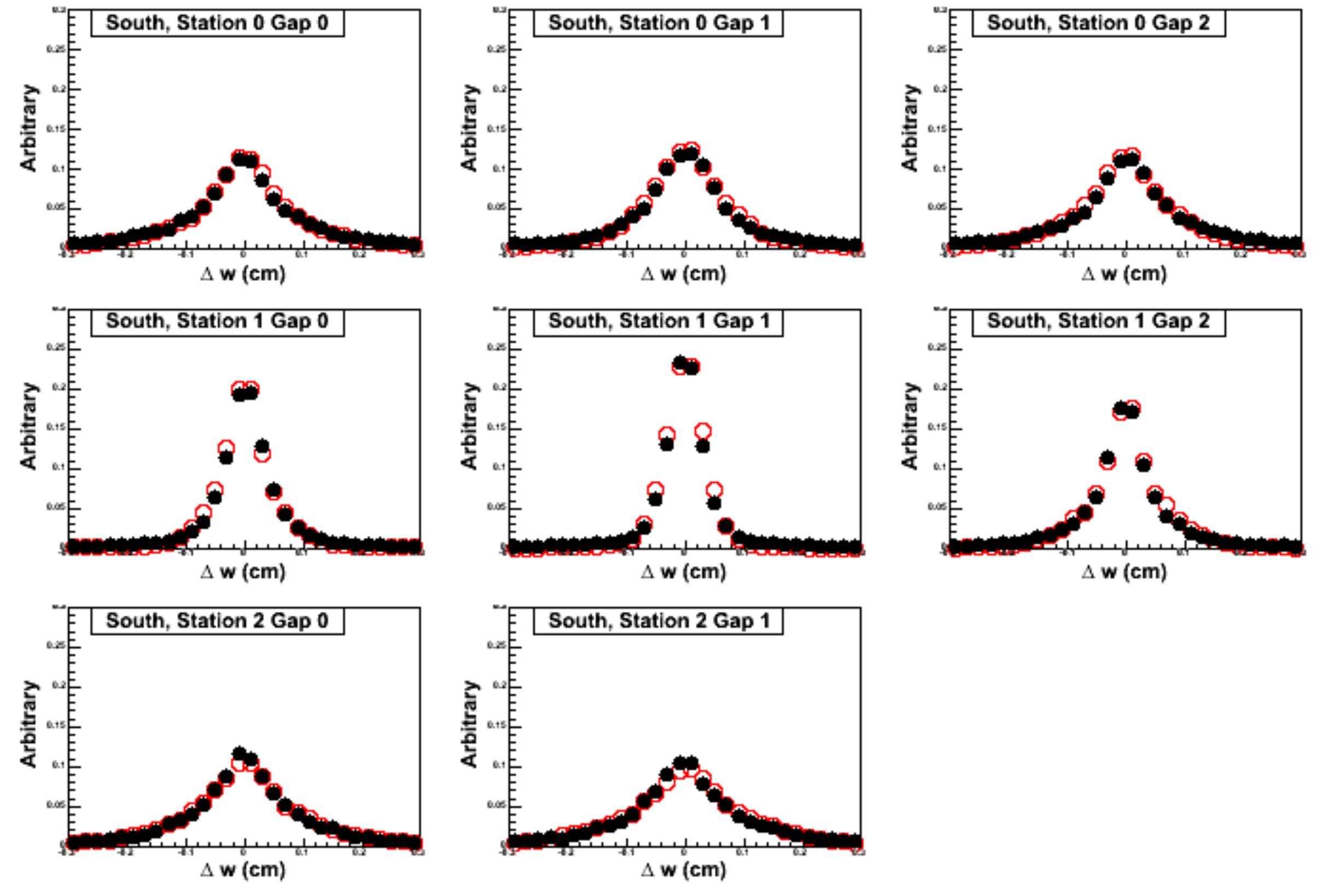}
        \isucaption[Comparison of tracking residuals in the South $\mu$Tr from simulation and data.]{Comparison of the tracking residual $\Delta w$, the difference between the cluster and track w coordinate in the North $\mu$Tr between simulation (open circles) and data (closed circles).}\label{fig:S_deltaw}
\end{figure}

In addition to the size and shape of clusters, it is important that the tracking resolution of the simulated $\mu$Tr match the resolution of the data.  This resolution is measured by the `w' coordinate, defined as the distance along a direction perpendicular to the strips in the chamber.  Unfortunately, the resolution of the simulated $\mu$Tr is typically much better than the real detector,\footnote{The difference is likely due to misalignments in the real detector which are not properly accounted for in the reconstruction or simulation.  For example, the large width of the residual in the North Arm Station 0, Gap 0 seen in Fig.~\ref{fig:N_deltaw} is entirely due to a 0.1~cm misalignment found in that gap after the data was produced.} and in order to correct for this difference, the simulated charge distributions are smeared by an additional amount $A$
\begin{eqnarray}
       q_{smeared} &=& q_{generated} + A P \frac{r}{g} \\
\Delta q_{smeared} &=& A \frac{r}{g}
\end{eqnarray}
where $P$ is a random number chosen from a normal distribution centered about the origin with $\sigma$=1, $r$ and $g$ are the root mean square ADC counts (rms) and gain, respectively, measured from the calibration of the strip with charge $q$.  The parameters $A$ for each $\mu$Tr gap are tuned so that the simulated difference between the w coordinate of the cluster and the track (the residual) match the measured distributions.  The simulated and measured tracking residuals are shown in Fig.~\ref{fig:N_deltaw} and~\ref{fig:S_deltaw}.


\begin{figure}[h!tb]
        \centering\includegraphics*[width=0.49\columnwidth]{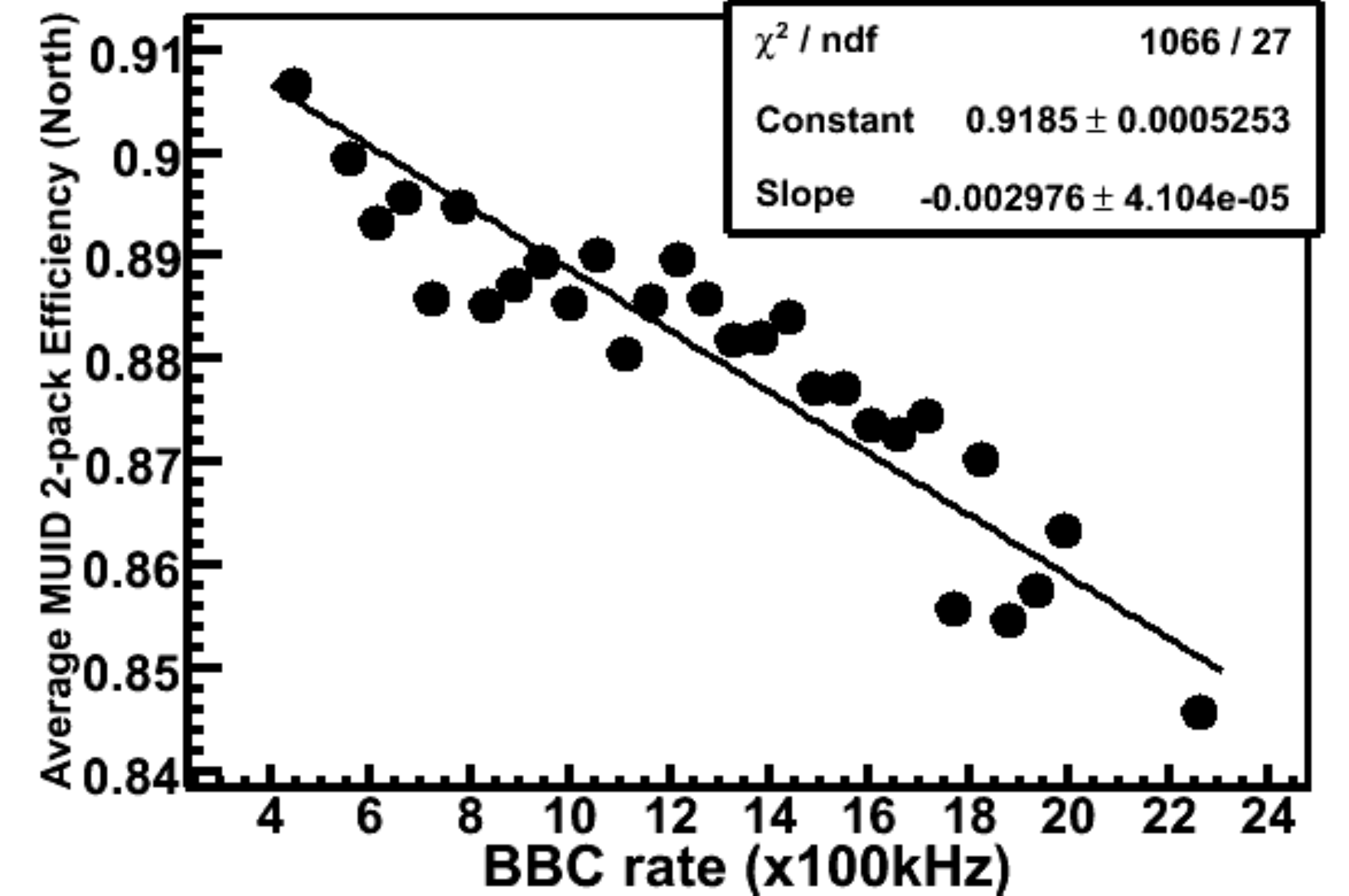} \
        \centering\includegraphics*[width=0.49\columnwidth]{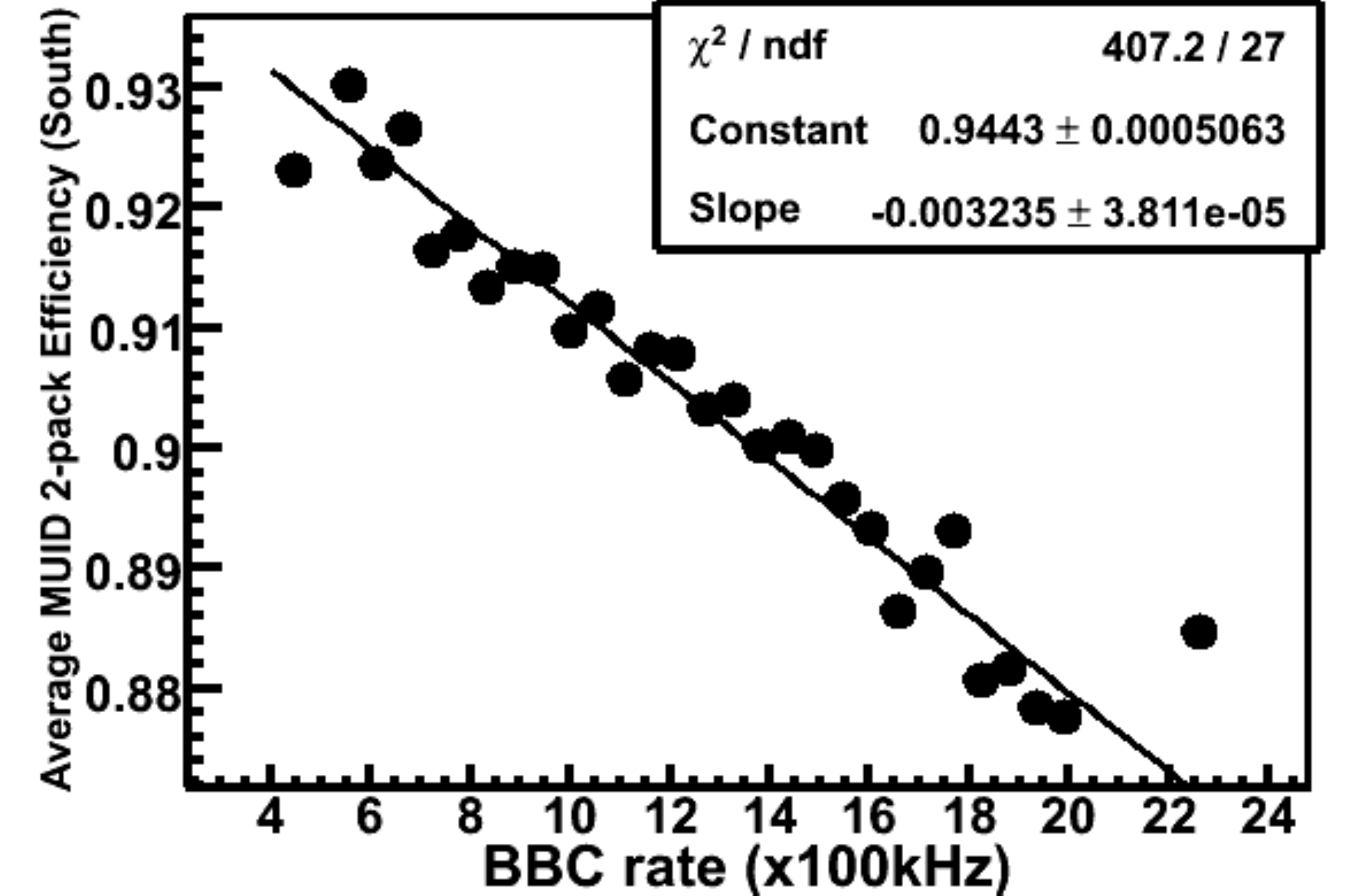}
        \isucaption[Average efficiency for pairs of Iarocci tubes in the $\mu$ID plotted against BBC rate.]{Average efficiency for pairs of Iarocci tubes (called `2-packs') in the $\mu$ID plotted against BBC rate in the North (South) arm on the left (right).}\label{fig:MUID_effic_rate}
\end{figure}

In addition to the cluster charge distributions and residuals, it is important that the physical acceptance of both the $\mu$Tr and $\mu$ID match between simulation and data.  For the $\mu$ID this matching is accomplished by measuring the efficiency of each set of two Iarocci tubes in both the North and South detectors and applying those efficiencies to the simulation.  The $\mu$ID efficiency was determined from data using tracks with hits in nearly every 2-pack (pair of tubes).  The efficiency is calculated as
\begin{equation}
\epsilon_{\text{plane of interest}} =\frac{\text{Tracks with hits in all 10 planes}}{\text{Tracks with hits in at least 9 planes, excluding the plane of interest}}
\label{eq:MUID_effic}
\end{equation}
and is projected to 2-packs in the $\mu$ID to determine an efficiency for each 2-pack.  The average efficiency is found to decrease with interaction rate (Fig.~\ref{fig:MUID_effic_rate}),\footnote{The decrease in efficiency with interaction rate is due mostly to an associated increase in beam-related backgrounds.  As the number of hits in the $\mu$ID increases, the average current in each tube is increased, decreasing the voltage across the tube (which decreases the efficiency).} but we simply use the mean value of the efficiency over the course of the running period for each 2-pack, as a uniform change in efficiency should have little effect on the acceptance to angular distributions.  The mean is calculated by integrating both the numerator and denominator of Eq.~\ref{eq:MUID_effic} over the course of the running period.

In the $\mu$Tr, channels are removed from the simulation if the high voltage for that channel was off for the majority of the running period.  Fiducial cuts are then made to the dead regions in order to ensure that no tracks reconstructing into those regions are used.  

\begin{figure}[h!tb]
        \centering\includegraphics*[width=0.49\columnwidth]{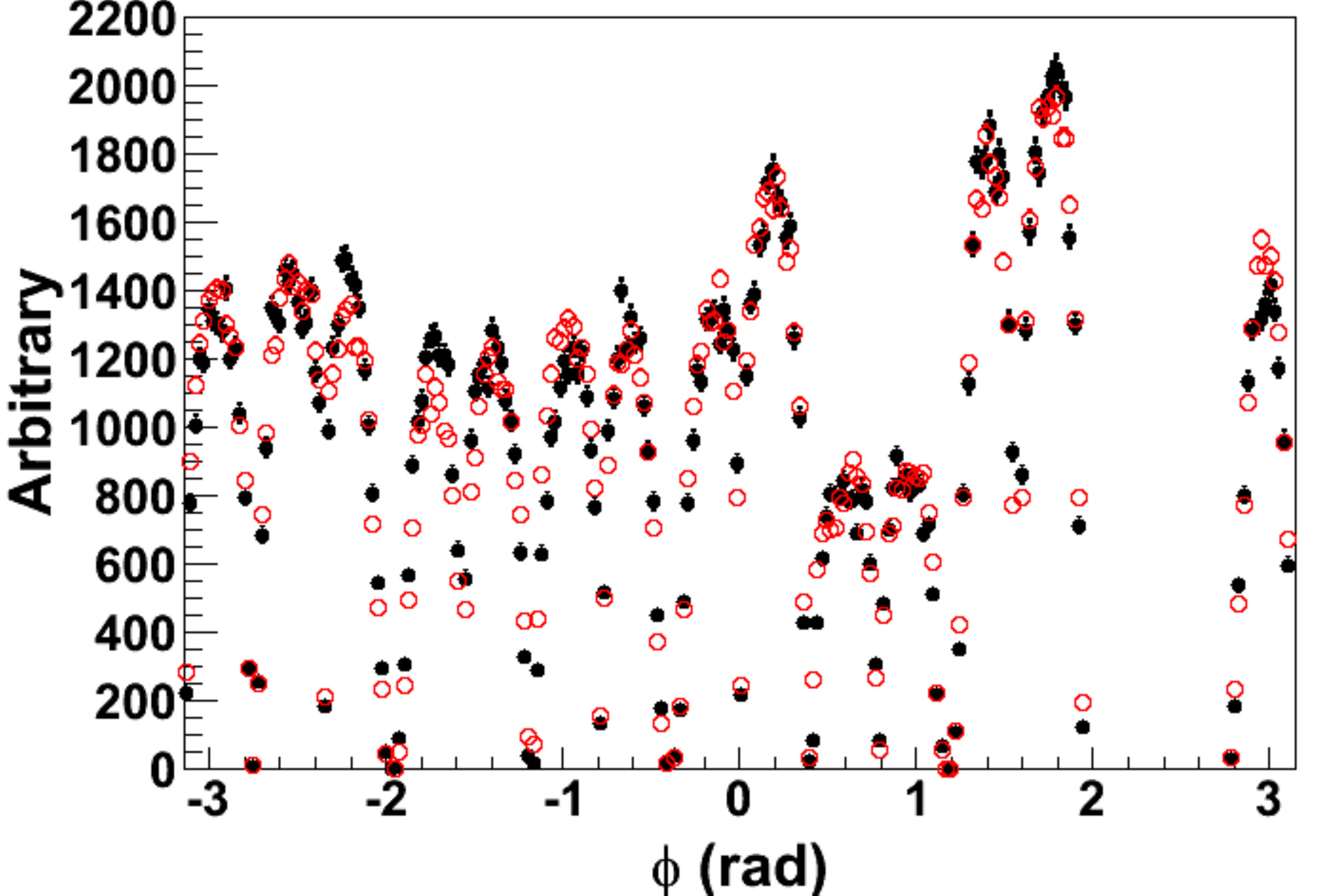}  \
        \centering\includegraphics*[width=0.49\columnwidth]{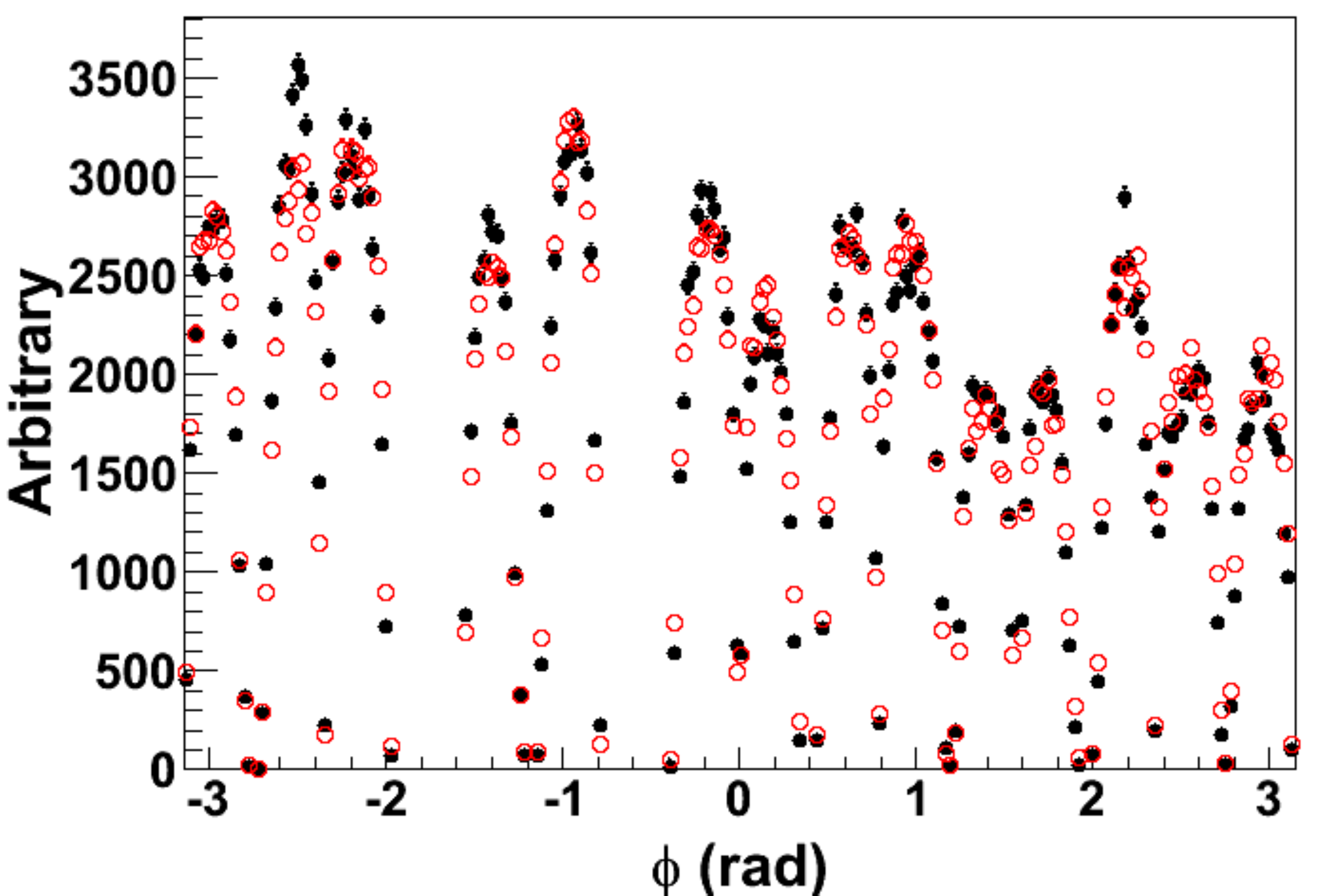}  
        \isucaption[Comparison of azimuthal track projections between simulation and data.]{Comparison of azimuthal track projections between simulation (open circles) and data (closed circles) in the North (left) and South (right).}\label{fig:phi_compare}
\end{figure}

\begin{figure}[h!tb]
        \centering\includegraphics*[width=0.45\columnwidth]{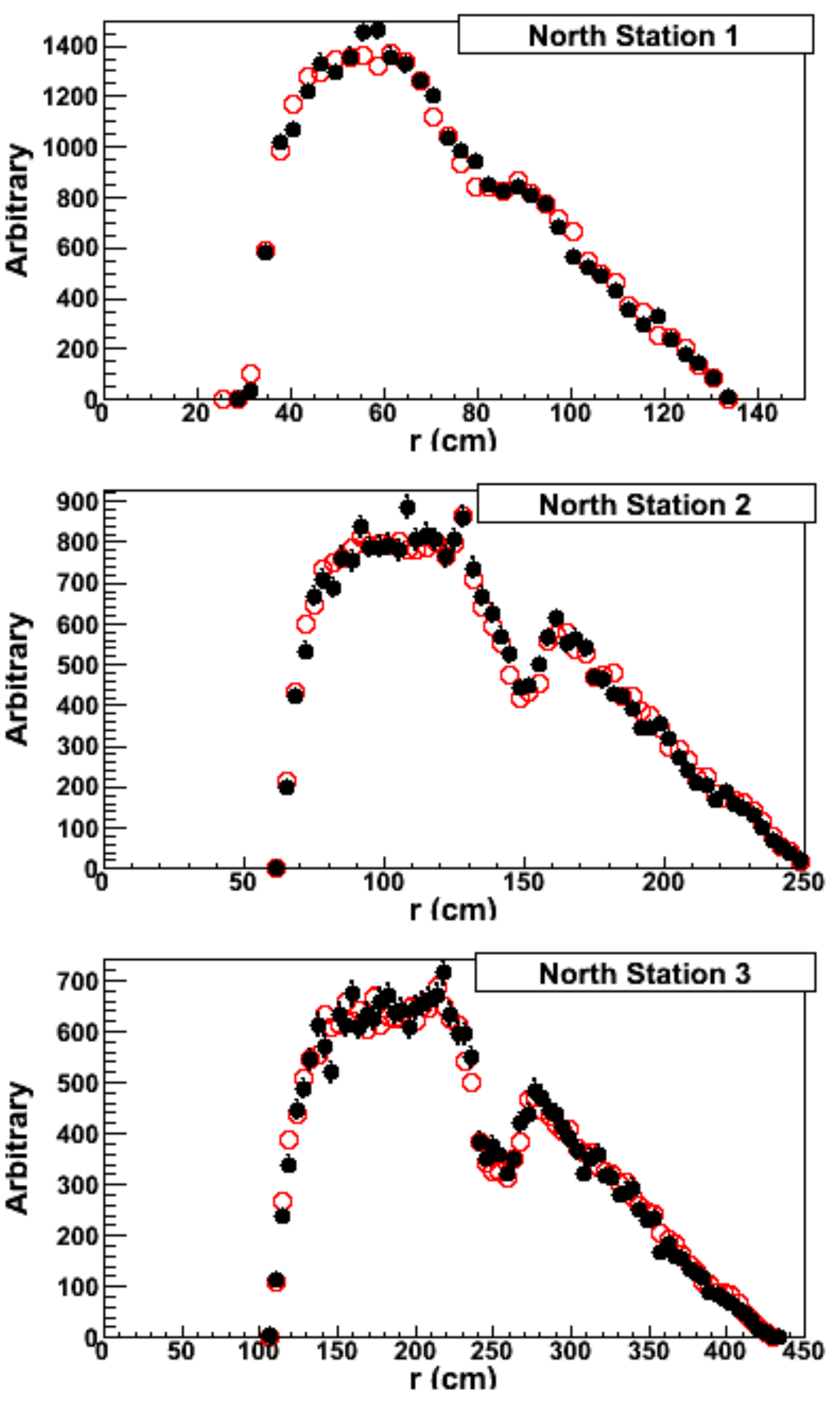}  \
        \centering\includegraphics*[width=0.45\columnwidth]{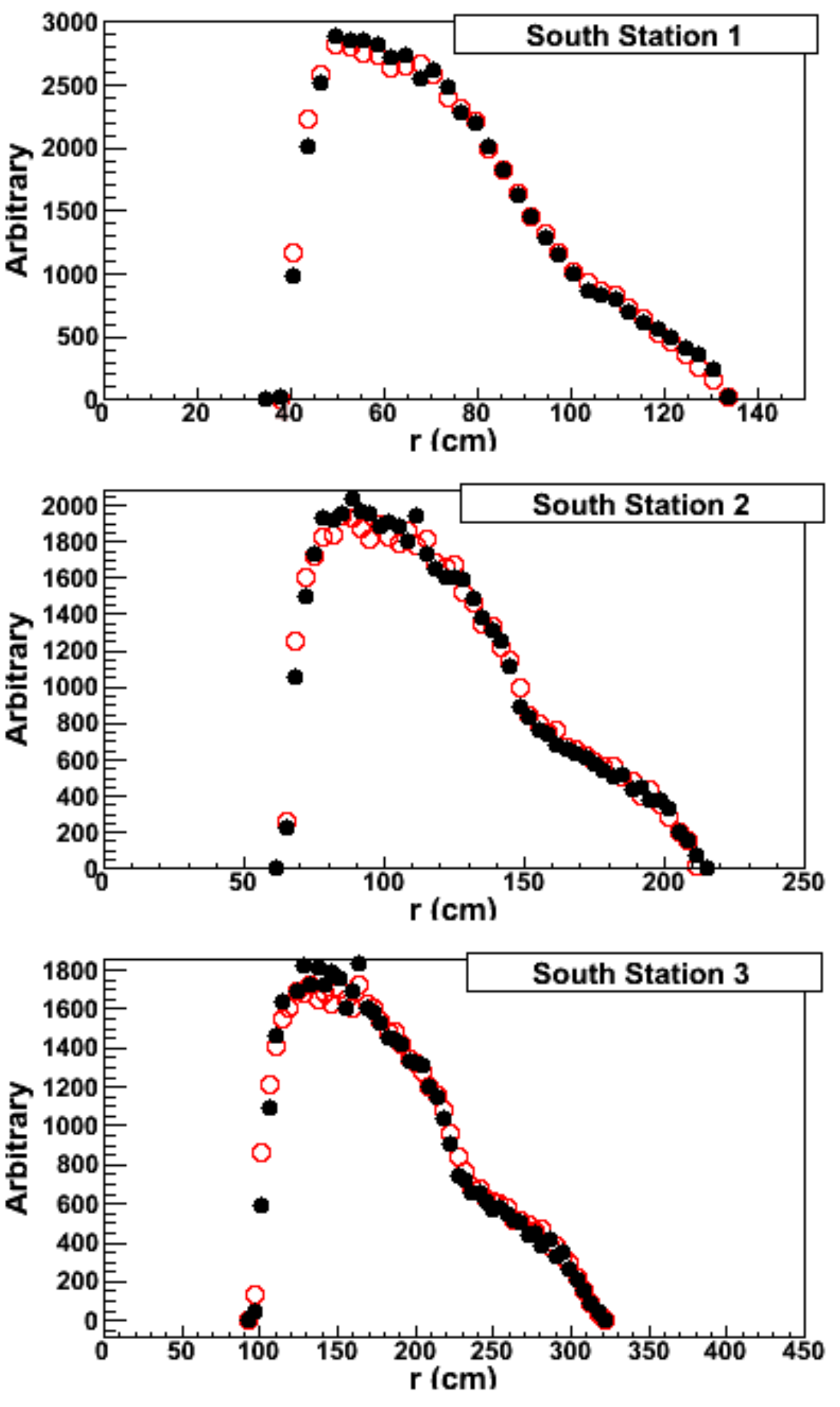}  
        \isucaption[Comparison of radial track projections between simulation and data.]{Comparison of radial track projections between simulation (open circles) and data (closed circles) for each $\mu$Tr station in the North (left) and South (right).}\label{fig:r_compare}
\end{figure}

After low-level tuning and fiducial cuts are applied, we can look at the distributions of the azimuthal and radial track projections on each $\mu$Tr station to see that the simulation adequately reproduces the data.  These distributions are shown in Fig.~\ref{fig:phi_compare} and~\ref{fig:r_compare} respectively and show good agreement between the simulated distributions and those from data.

\section{$J/\psi$ Yield}\label{sec:jpsi_yield}

A necessary step in measuring the angular decay coefficients for $J/\psi$ mesons is to determine the shape of the $J/\psi$ yield in $p_{T}$ and rapidity in order to have a proper kinematic shape for the simulation.  Since no previous measurements have been made of the $J/\psi$ yield from $p$+$p$ collisions at $\sqrt{s}$=500~GeV in 1.2$<|y|<$2.2, we will need to determine this shape and the uncertainty associated with it from data.

In our data, the kinematic distributions of $\mu^{+}\mu^{-}$ pairs in the $J/\psi$ mass region contain some background, and it is important to remove the contributions from this background.  Two types of background contribute to the number of oppositely charged muon pairs in the $J/\psi$ mass region:  (1) random combinations of tracks in the detector (called the combinatorial background), and (2) physical processes like $p$+$p$$\rightarrow$$\text{D}\overline{\rm D}X$$\rightarrow$$\mu^{+}\mu^{-}X$ (open charm), $p$+$p$$\rightarrow$$\text{B}\overline{\rm B}X$$\rightarrow$$\mu^{+}\mu^{-}X$ (open bottom), and Drell-Yan production, which all produce a continuum of correlated $\mu^{+}\mu^{-}$ pairs (called the continuum background).

To estimate the combinatorial background we will us the same method discussed in Section~\ref{sec:jpsi_AN_cuts}, which is to take twice the geometric mean between the number of positive and negatively charged muon pairs, $2\sqrt{N_{\mu^{+}\mu^{+}}N_{\mu^{-}\mu^{-}}}$.   Physical background is more difficult to estimate than combinatorial background, but we are helped by the fact that the physical backgrounds form a continuum which can be approximated with a falling exponential under the $J/\psi$ peak.\footnote{While there is no clear theoretical reason that an exponential should describe the combination of the Drell-Yan and open-heavy flavor backgrounds in the continuum, such a shape has historically parameterized the background very well.}  The following functional form is used to describe the dimuon continuum in mass:
\begin{equation}
\frac{d N }{d M} = A e^{-b M} + \frac{N_{J/\psi}}{\sqrt{2\pi\sigma_{J/\psi,1}^{2}}} e^{-\frac{(M-M_{J/\psi})^{2}}{2\sigma_{J/\psi,2}^{2}}} + \frac{f_{2} N_{J/\psi}}{\sqrt{2\pi\sigma_{J/\psi,2}^{2}}} e^{-\frac{(M-M_{J/\psi})^{2}}{2\sigma_{J/\psi,2}^{2}}}.
\label{eq:mass_func}
\end{equation}
The second normal distribution accounts for large tails in the $J/\psi$ mass peak found in simulation, and the $f_2$ is fixed by simulation.  The $J/\psi$ resonance is quite wide in data because of the resolution of the spectrometer.  Because of this width, the $\psi^{\prime}$ resonance is not very prominent, especially in regions of little acceptance, and is not included in the fit.

The exponential shape of the background is modified at low mass by our trigger requirement.  In order to take this into account and obtain a better fit, a simple Monte Carlo simulation is used to determine the effects of the acceptance and trigger on an exponentially falling background.  Lepton pairs are generated with a normal distribution in rapidity and an exponentially falling distribution in mass and are put through a simple model of the detector acceptance and trigger requirement.  The output distributions from simulation are then applied as a weight to the exponential used in the fit in order to obtain a more realistic distribution.\footnote{Another way of taking the effect of the trigger into account is to use a polynomial to describe the background shape, as we did in Section~\ref{sec:jpsi_AN_cuts}.}

\begin{figure}[h!tb]
        \centering\includegraphics*[width=0.49\columnwidth]{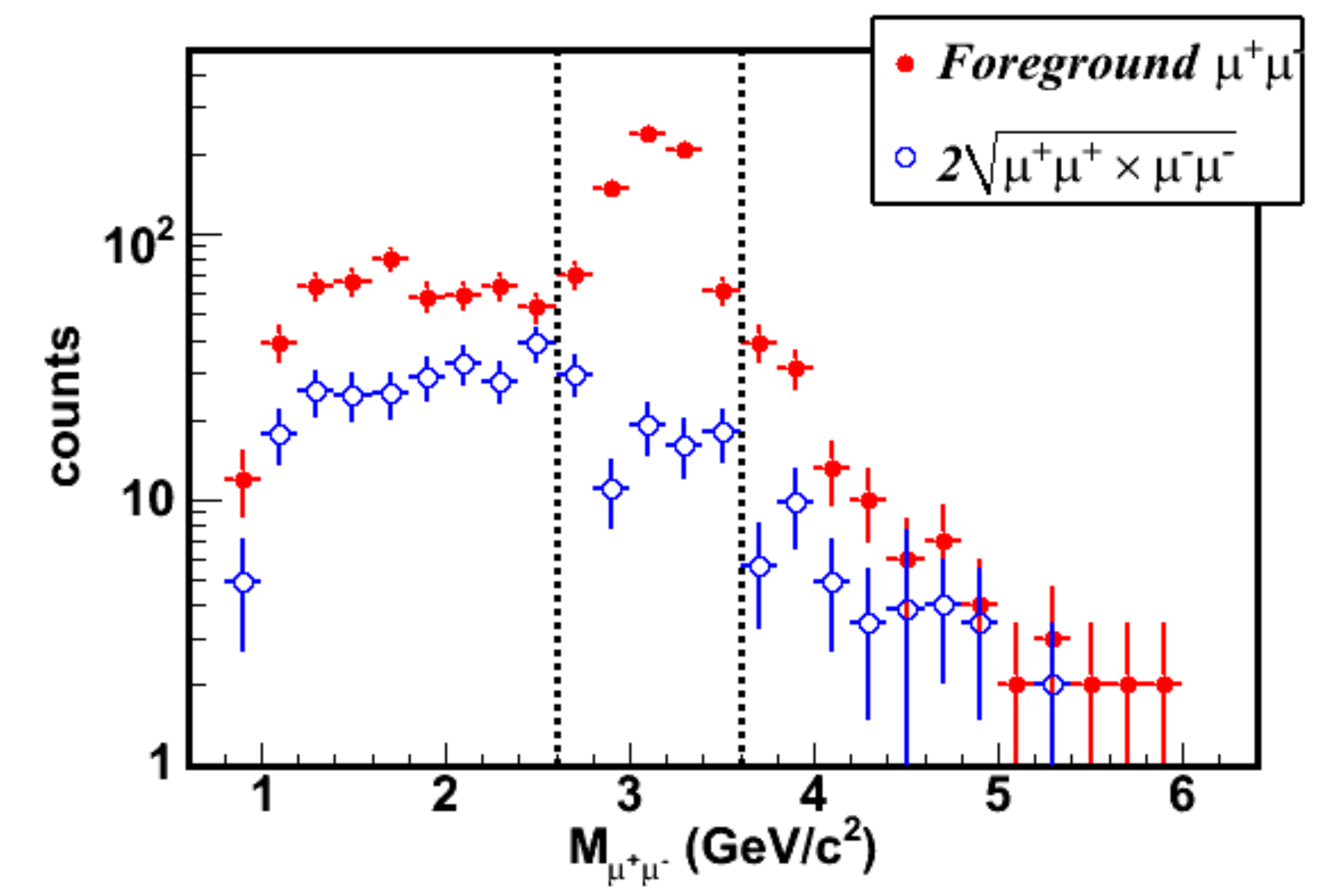} \
        \centering\includegraphics*[width=0.49\columnwidth]{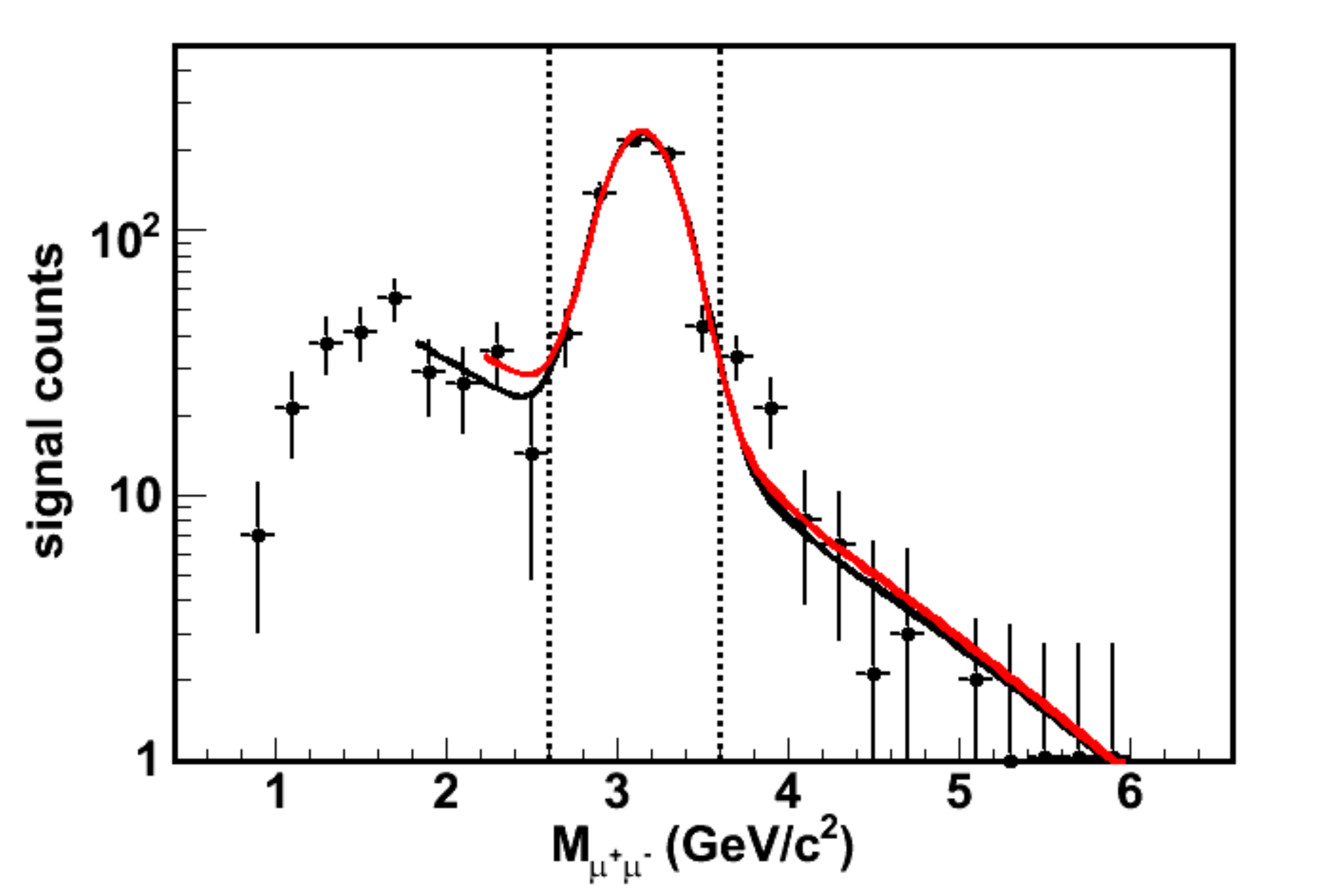}
        \isucaption[Total counts and combinatorial background for an example $p_{T}$ range and fits to the signal.]{Example plots of the total counts of oppositely charged muons from one kinematic bin along with the associated combinatorial background estimation (left).  Example fits to the background subtracted distribution (right).  Separate fit ranges are show in red and black, and dashed lines show the mass range in used to determine the signal.}\label{fig:example_fits}
\end{figure}

Varying the range for the fit to Eq.~\ref{eq:mass_func} can potentially change the number of $J/\psi$ measured.  In order to account for this variation, fits are repeated with two separate fit ranges: M$\in$[1.8,7.0]~GeV$/c^{2}$ and M$\in$[2.2,6.0]~GeV$/c^{2}$.  Separate fits are also done with $f_{2}$ increased and decreased by 25\%.

 The separate fit ranges and 3 scenarios for $f_2$ account for 6 fits, and they are plotted together in Fig.~\ref{fig:example_fits}, where the 3 separate scenarios for $f_2$ are indistinguishable.  The mean of the integrals of the normal $J/\psi$ distributions from the 6 signal extractions is taken as the central value of the data point for each bin in $p_{T}$ and rapidity, and the RMS about the mean number of $J/\psi$ from all signal extractions is taken as a source of systematic uncertainty which is mostly uncorrelated between bins.  We will assume that this uncertainty is entirely uncorrelated between bins and add it in quadrature with the statistical uncertainty.

When fitting the yields in $p_{T}$ and rapidity, it is important that the data points be located at the mean $p_{T}$ or rapidity for that bin (depending on which is being histogrammed).  To determine the mean $p_{T}$ and rapidity for each bin, we generate a sample of simulated $J/\psi$ mesons and put them through a GEANT~\cite{GEANT} model of the detector, keeping track of both the generated and reconstructed kinematics.  For each bin in reconstructed $p_{T}$ and rapidity we histogram the generated quantities, taking the mean value as the centroid of the bin in our final yield and the RMS as the uncertainty for that bin in abscissa.

After the data points have been determined, we need to parameterize the $J/\psi$ kinematics for use in simulation.  A Kaplan function~\cite{Kaplan:1977kr,Jostlein:1978vp} is used to parameterize the $p_{T}$ distribution 
\begin{equation}
\frac{1}{2\pi p_{T}}\frac{dN}{d p_{T}} = A \left(1 + \left(\frac{p_{T}}{p_{0}}\right)^{2}\right)^{-n},
\label{eq:Kaplan}
\end{equation}
and a normal distribution centered about zero parameterizes the shape in rapidity.  To get a reasonable distribution of event vertices in our simulation, vertices are taken from a sample of events for which at least one tube fired in both the North and South BBC.

\begin{figure}[h!tb]
        \centering\includegraphics*[width=0.49\columnwidth]{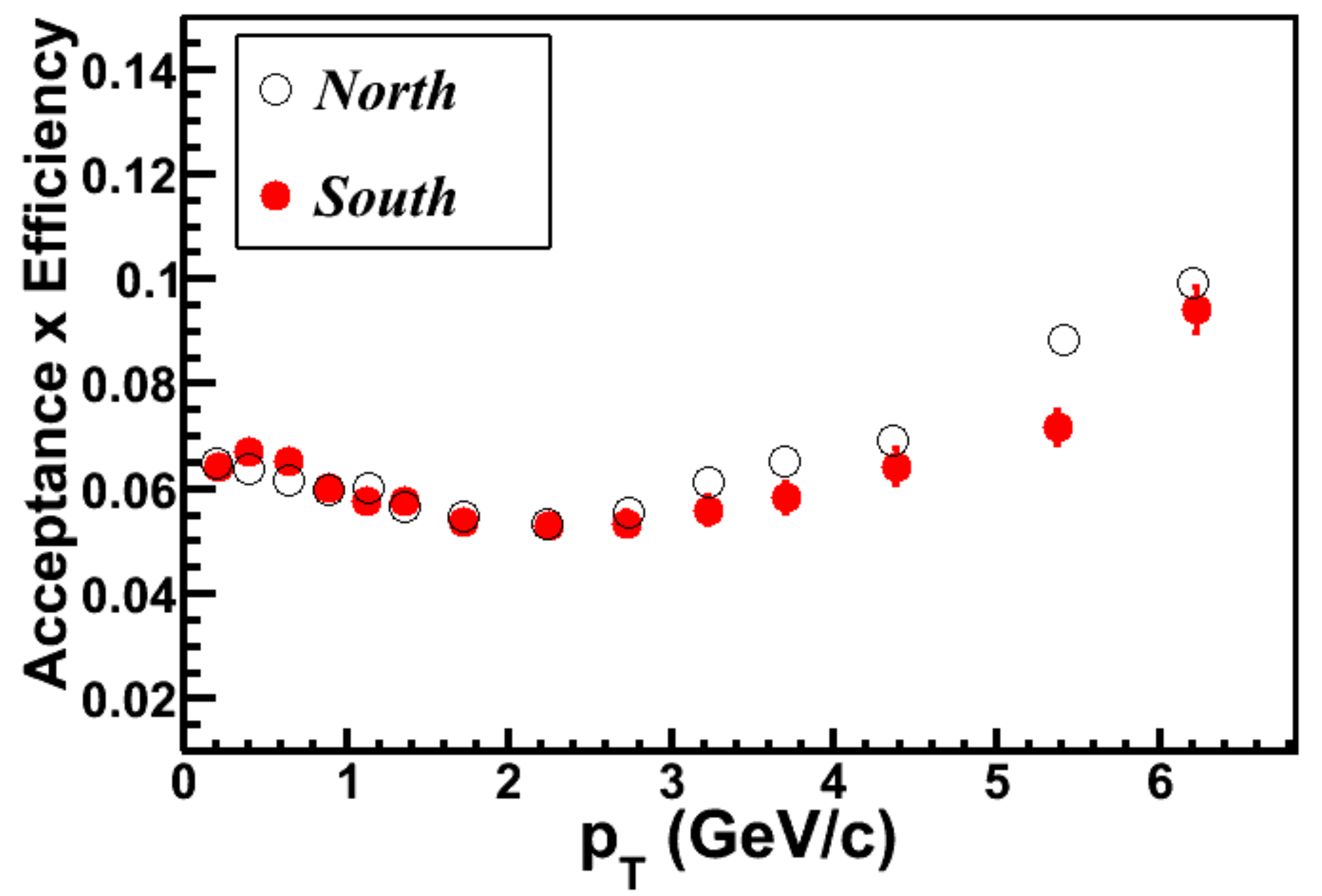} \
        \centering\includegraphics*[width=0.49\columnwidth]{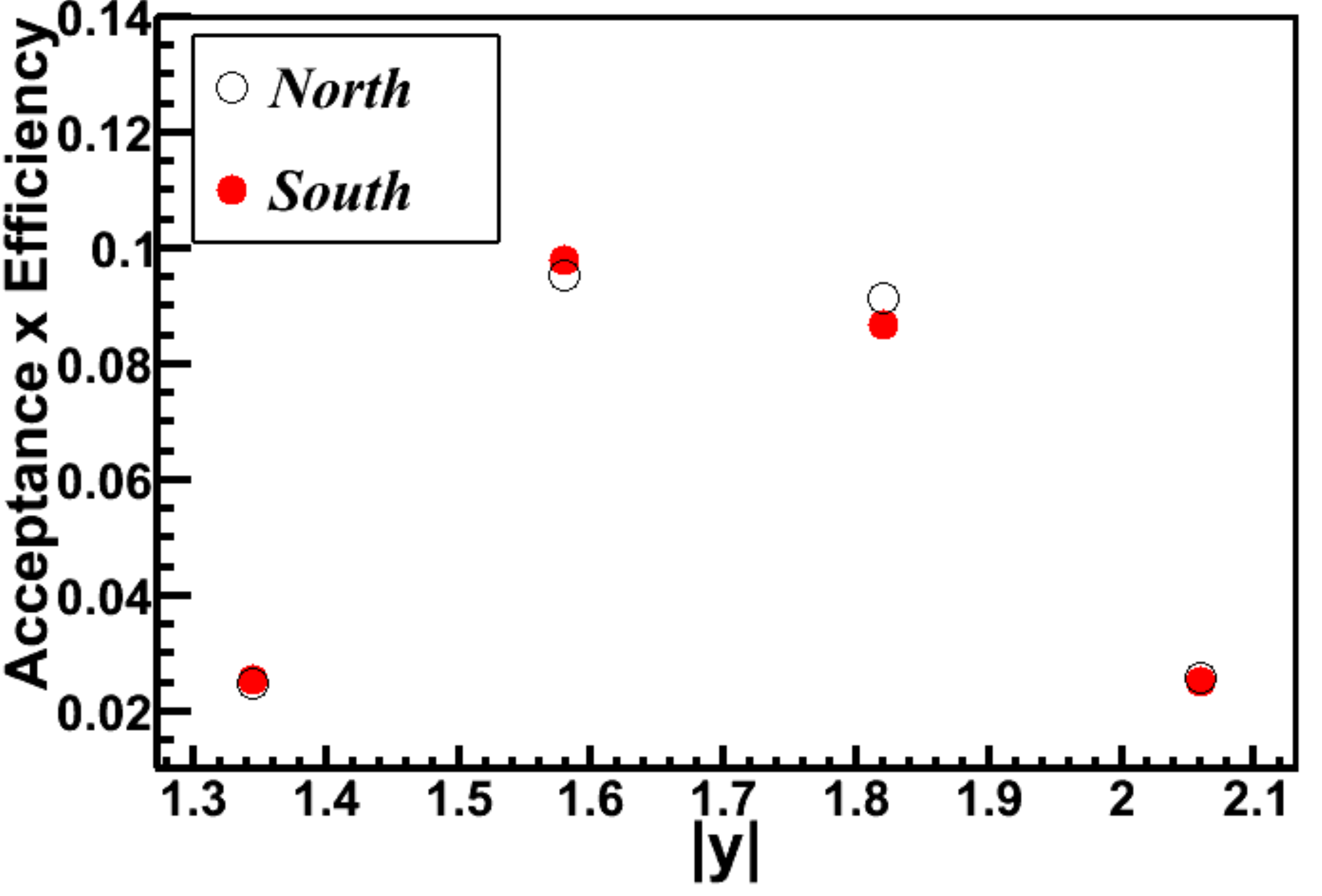}
        \isucaption[Acceptance $\times$ efficiency plotted against transverse momentum and rapidity.]{Acceptance $\times$ efficiency plotted against transverse momentum on the left and rapidity on the right for the North (open circles) and South (closed circles) muon spectrometer.}\label{fig:acceff}
\end{figure}

\begin{figure}[h!tb]
        \centering\includegraphics*[width=0.49\columnwidth]{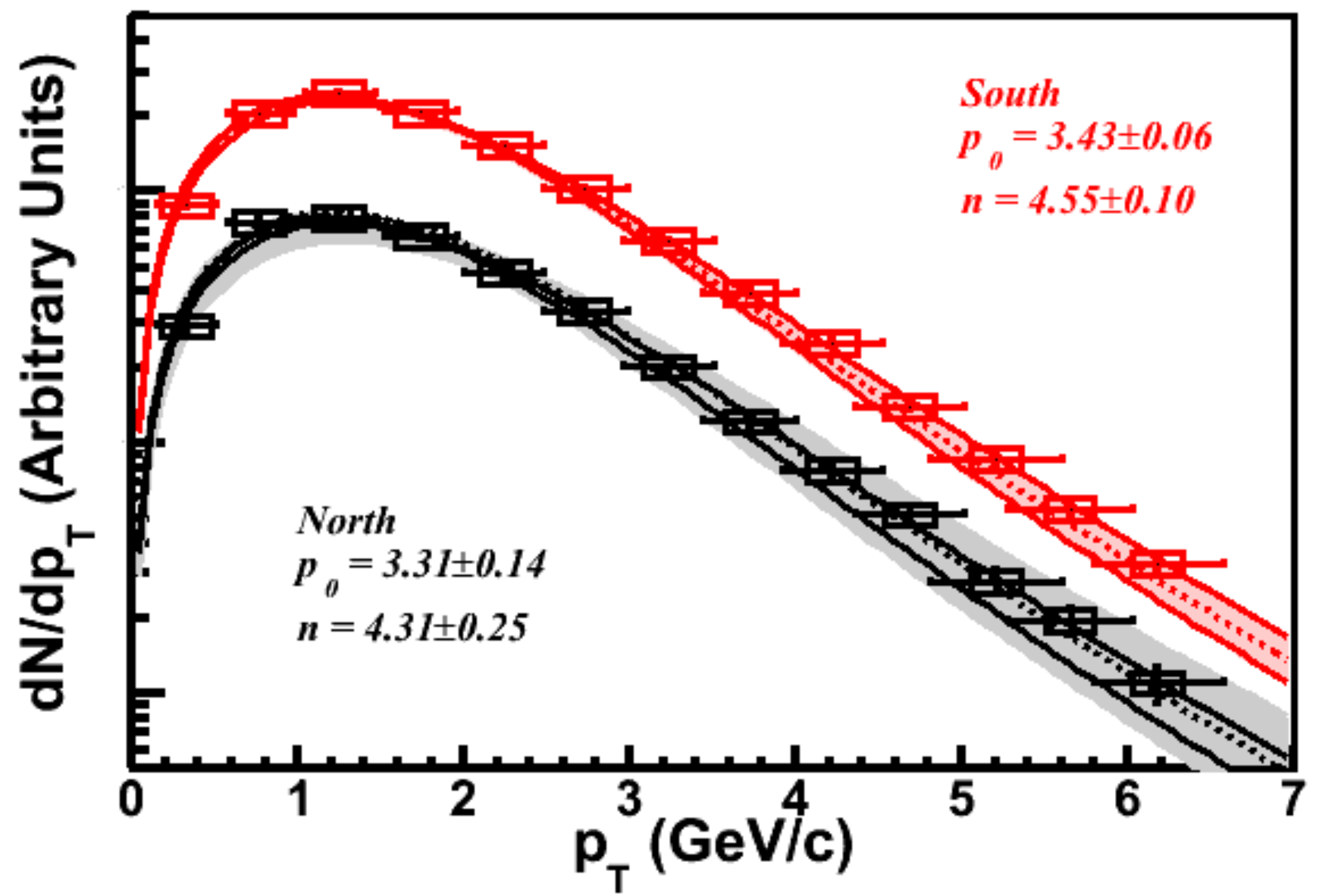} \
        \centering\includegraphics*[width=0.49\columnwidth]{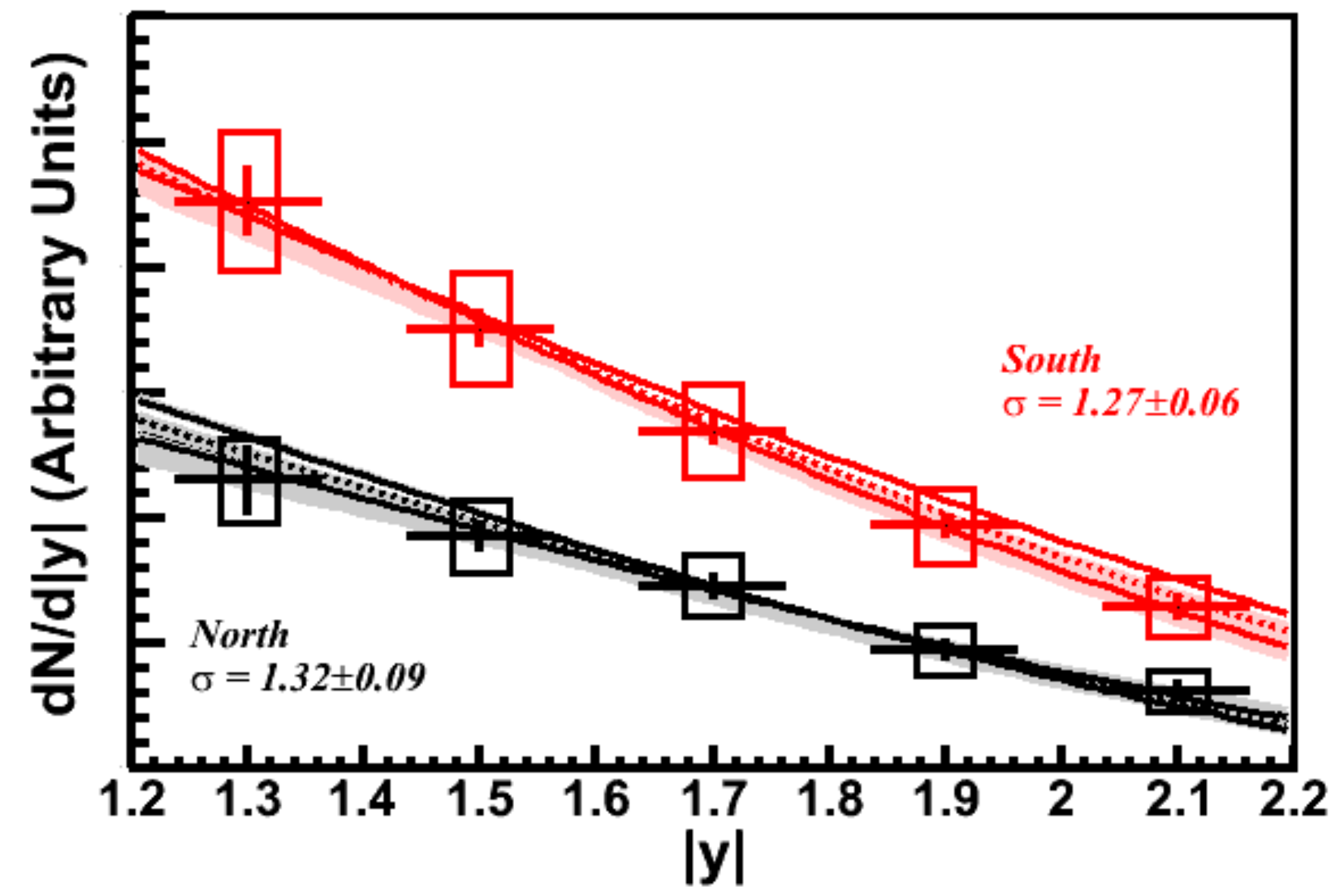} 
        \isucaption[Acceptance corrected $J/\psi$ yield plotted against transverse momentum and rapidity.]{Acceptance corrected $J/\psi$ yield plotted against transverse momentum on the left and rapidity on the right for the North (bottom) and South (top) muon spectrometer.  Dashed lines show the average fits to a Kaplan function on the left and normal distribution on the right, and solid lines show the fits at $\pm$1$\sigma$.}\label{fig:yield_plots}
\end{figure}

As discussed in Section~\ref{sec:measurement_outline}, the method for determining the $J/\psi$ yield is iterative.  First we measure the yield using the acceptance $\times$ efficiency from a given simulation, then we repeat using the measured yield shape as input to the simulation.  The final iteration used to determine the acceptance $\times$ efficiency had $n$=4.5 and $p_{0}$=3.4 in the Kaplan function and $\sigma$=1.27 for the normal distribution in rapidity.  For the angular decay coefficients we use a linear interpolation between the measured angular decay coefficients of the Helicity frame as a function of $p_{T}$ (consistent with the final measured values).  Fig.~\ref{fig:acceff} shows the acceptance $\times$ efficiency from the simulation plotted against $p_{T}$ and rapidity, and Fig.~\ref{fig:yield_plots} shows the yields.  From the fit parameters shown in the latter figure, it is clear that the input shape in $p_{T}$ and rapidity is consistent with the measured shape in both spectrometers.

Because the parameterizations of the yield shape in $p_{T}$ and rapidity are not perfectly determined, we will need to introduce systematic uncertainties to account for the full range of fits in both the North and South muon spectrometers.  When determining the uncertainty, a 10\% systematic uncertainty, correlated between data points, is attributed to the acceptance $\times$ efficiency (because our simulation is not an exact description of the detector).  Yields are fit with a modified likelihood (described in~\cite{Adare:2008cg}) which takes these correlated systematic uncertainties into account.  The resulting parameters from the fit are then averaged, and the values at $\pm1\sigma$ are taken as the extreme shapes of the yield, used to determine the systematic uncertainty in Section~\ref{sec:angular_coefficients}.  The nominal Kaplan function in $p_T$ was found to have $n=4.52$ and $p_0=3.41$, while the steep function at the $1\sigma$ limit had $n=4.43$ and $p_0=3.46$ and the shallow function had $n=4.61$ and $p_0=3.35$.  The nominal normal distribution in rapidity had $\sigma=1.29$, while the steepest and shallowest distributions had $\sigma=1.25$ and $\sigma=1.33$ respectively.

\section{$J/\psi$ Angular Decay Coefficients}\label{sec:angular_coefficients}

To determine angular decay coefficients, two dimensional histograms of 5 bins in $\cos\vartheta$ by 8 bins in $\varphi$ are filled from data and an isotropic simulation (plots of the acceptance in all frames are shown in Appendix~\ref{sec:angular_acceptance}). The combinatorial background is then estimated and subtracted, and the mass distribution is fit with Eq.~\ref{eq:mass_func} as discussed in Section~\ref{sec:jpsi_yield} to extract the number of $J/\psi$ mesons in each kinematic bin.  Finally, two dimensional histograms of the number of $J/\psi$ in $\cos\vartheta$-$\varphi$ from data are fit with the full angular distribution of Eq.~\ref{eq:angle_dist} convoluted with the histogram from simulation.

We quantify the uncertainty due to the input shape of the yield by using the extreme shapes in $p_T$ and rapidity described in Section~\ref{sec:jpsi_yield}.  A $\cos\vartheta$-$\varphi$ surface is drawn for the nominal shape in $p_T$ and the two extreme shapes in rapidity as well as the nominal shape in rapidity and the two extreme shapes in $p_T$.  The surface is then varied 15k times with a Gaussian probability about the nominal shape with the extreme shapes at $\pm1\sigma$.  The resulting coefficients are histogrammed to determine the systematic uncertainty separately for the $p_T$ and rapidity shapes.  These uncertainties are summed in quadrature and considered correlated between data points.\footnote{Note that the correlation matrix of this uncertainty between data points is different for angular decay coefficients plotted against $p_T$ than it is for coefficients plotted against $x_F$.}

The contribution of background to the angular decay coefficient $\lambda_{\vartheta}$ is
\begin{equation}
\lambda_{\vartheta,\text S} = \frac{\lambda_{\vartheta,\text fit}}{R_{\text S}} - \lambda_{\vartheta,B} \frac{(1-R_{\text S})}{R_{\text S}}
\label{eq:theta_BG}
\end{equation}
where $R_{\text S}$ is the fraction of pairs which come from $J/\psi$ (a derivation can be found in Appendix~\ref{sec:bg_contributions}).  Likewise, for $\lambda_{\varphi}$ and $\lambda_{\vartheta\varphi}$ we have
\begin{equation}
\lambda_{\varphi,\text S} = \frac{\lambda_{\varphi,\text{fit}}}{R_{\text S}}\frac{3+\lambda_{\vartheta,\text S}}{3+\lambda_{\vartheta,\text{fit}}} - \lambda_{\varphi,\text B} \frac{(1-R_{\text S})}{R_{\text S}} \frac{3+\lambda_{\vartheta,\text S}}{3+\lambda_{\vartheta,\text B}}
\label{eq:phi_BG}
\end{equation}
and
\begin{equation}
\lambda_{\vartheta\varphi,\text S} = \frac{\lambda_{\vartheta\varphi,\text{fit}}}{R_{\text S}}\frac{3+\lambda_{\vartheta,\text S}}{3+\lambda_{\vartheta,\text{fit}}} - \lambda_{\vartheta\varphi,\text B} \frac{(1-R_{\text S})}{R_{\text S}} \frac{3+\lambda_{\vartheta,\text S}}{3+\lambda_{\vartheta,\text B}}
\end{equation}
(Derivations of these expressions are in Appendix~\ref{sec:bg_contributions}).  In practice, the background coefficients $\lambda_{\vartheta,B}$, etc. are very difficult to quantify.  Even if the combinatoric background is subtracted and the decay coefficients of the continuum are assumed to be negligible, the difference in $p_T$ and rapidity shape between the continuum background and signal can cause large false asymmetries to be measured.  To make matters worse, the Drell-Yan background, which makes up a small part of the continuum,\footnote{The Drell-Yan contribution is estimated in the $J/\psi$ mass region to be approximately 2\% of the $J/\psi$ yield from a PHENIX analysis of $p$+$p$ collisions at $\sqrt{s}$=200~GeV~\cite{Gadrat:2005se}} is known to have a large $\lambda_{\vartheta}$ in the Collins-Soper frame~\cite{Zhu:2008sj}.  To remove background contributions, fits are performed in each two-dimensional bin of the $\cos\vartheta-\varphi$ distribution using the distribution in Eq.~\ref{eq:mass_func} as described in Section~\ref{sec:jpsi_yield}.

\begin{figure}
        \centering\includegraphics*[width=0.48\columnwidth]{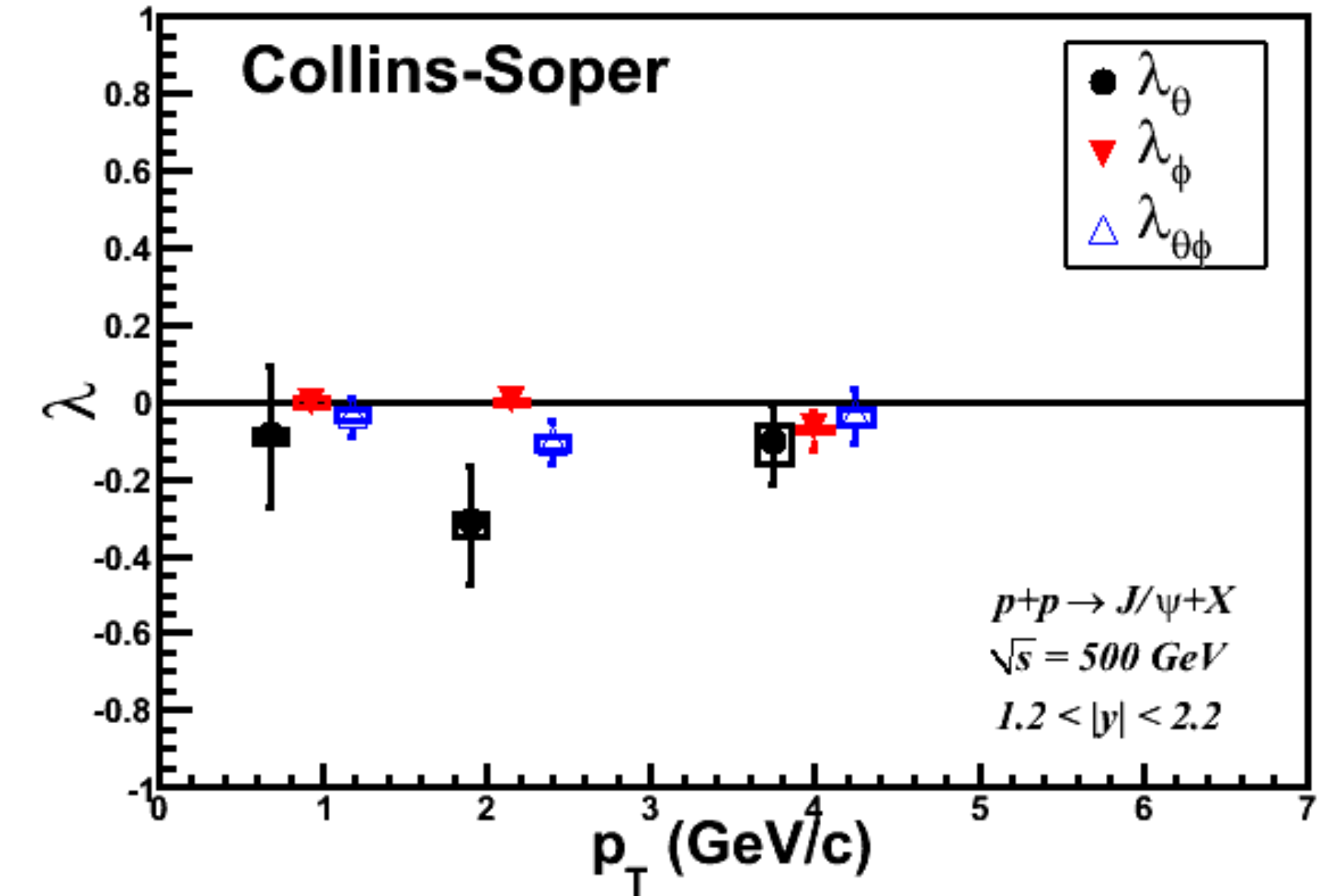} \
        \centering\includegraphics*[width=0.48\columnwidth]{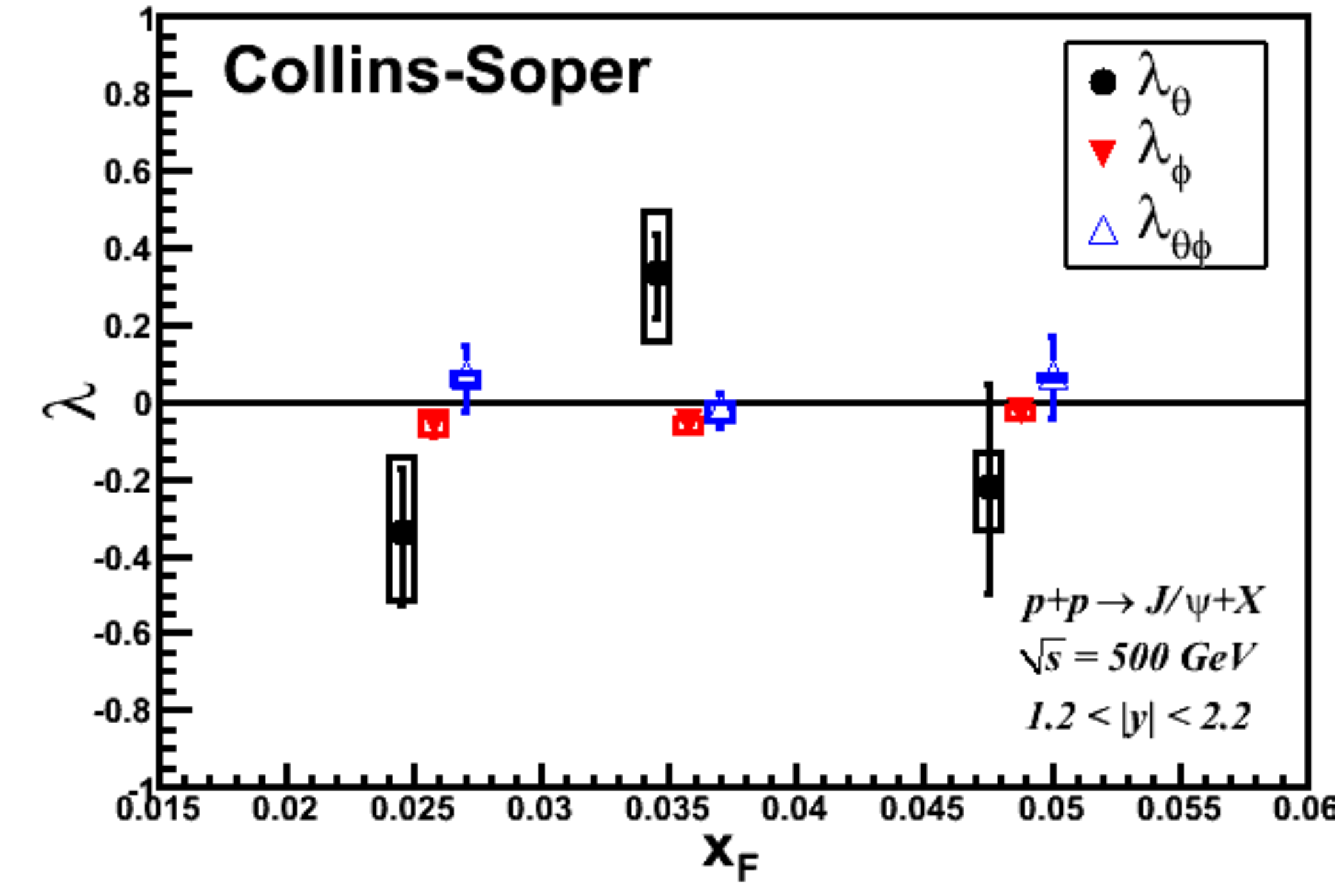} \\
        \centering\includegraphics*[width=0.48\columnwidth]{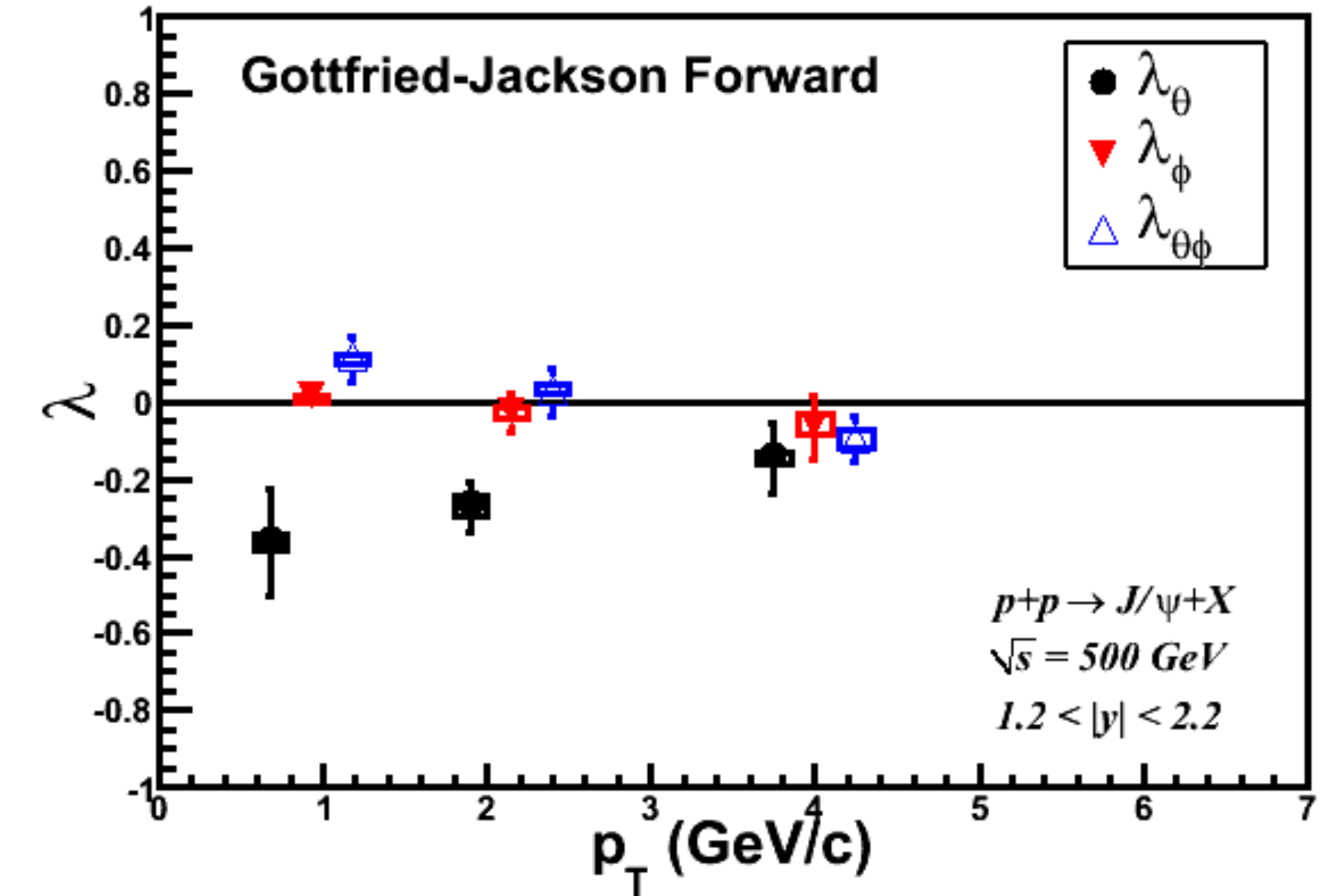} \
        \centering\includegraphics*[width=0.48\columnwidth]{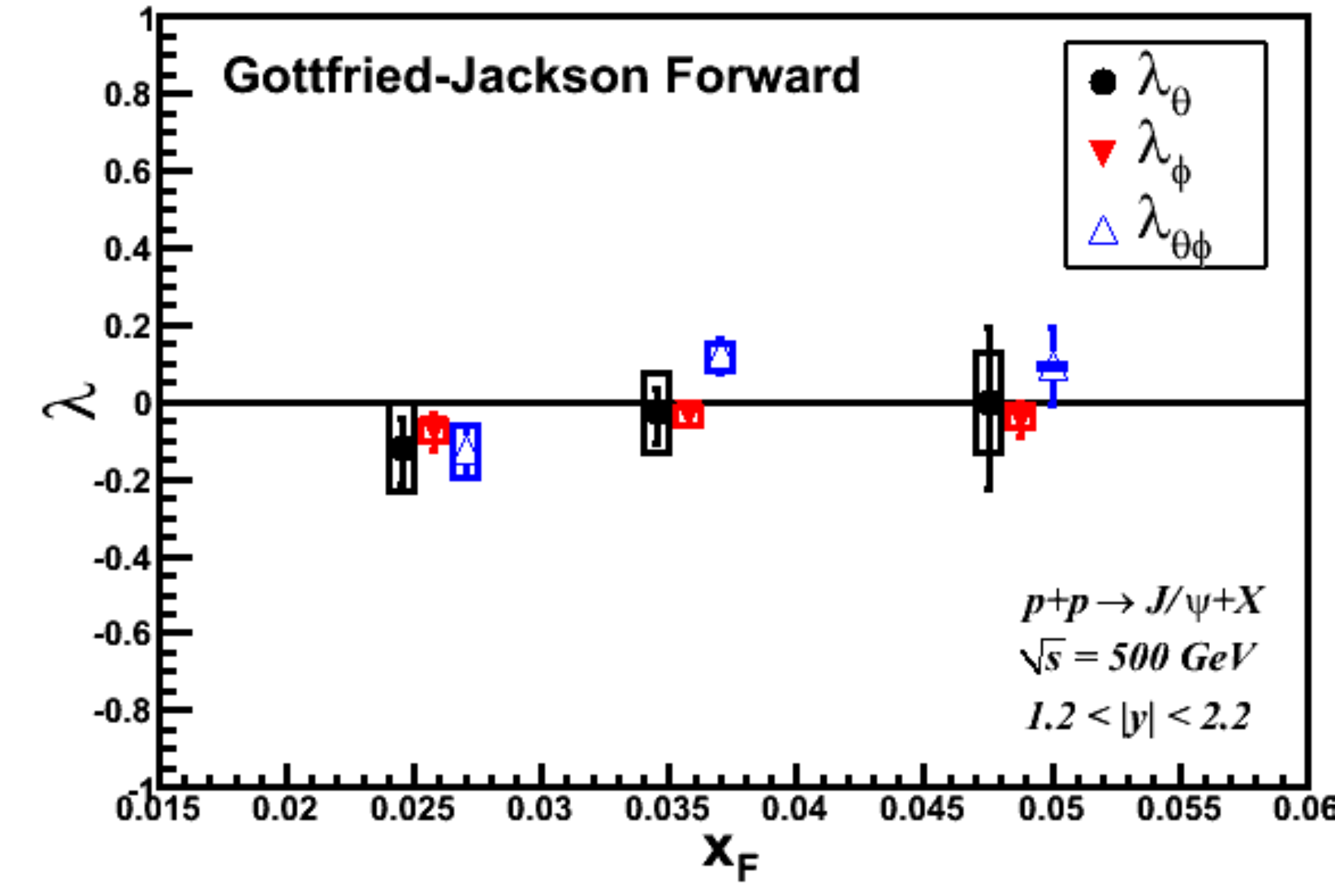} \\
        \centering\includegraphics*[width=0.48\columnwidth]{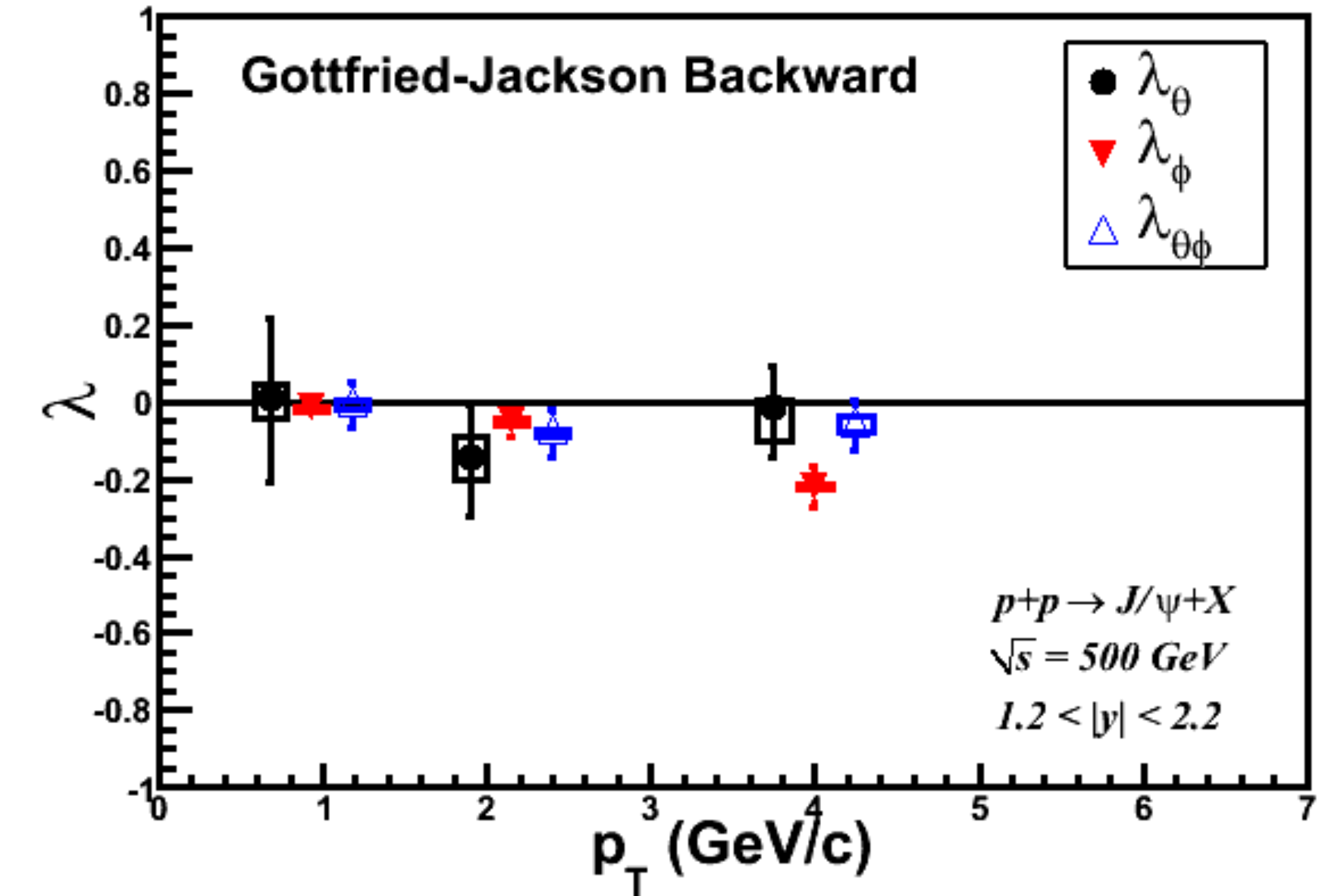} \
        \centering\includegraphics*[width=0.48\columnwidth]{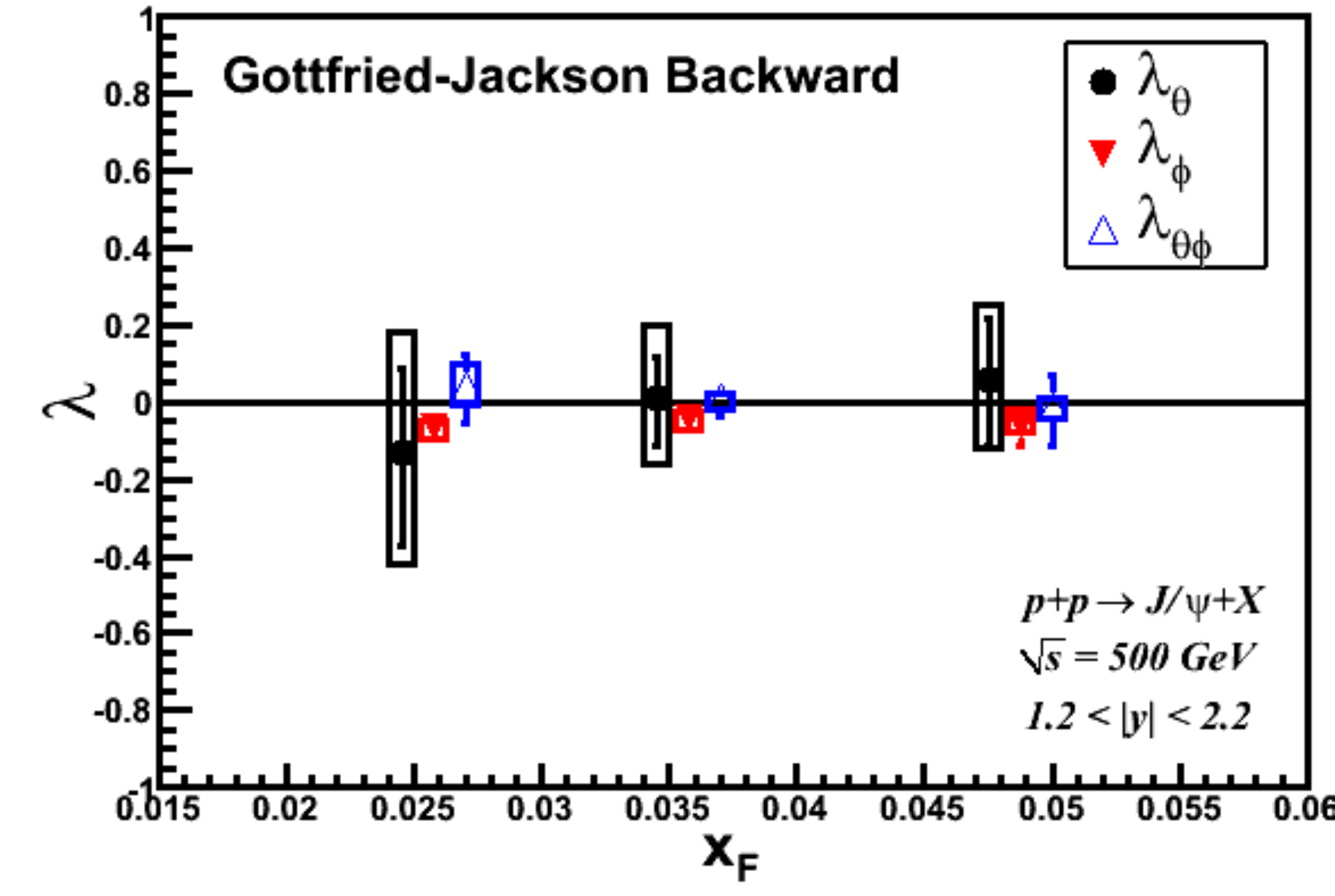} \\
        \centering\includegraphics*[width=0.48\columnwidth]{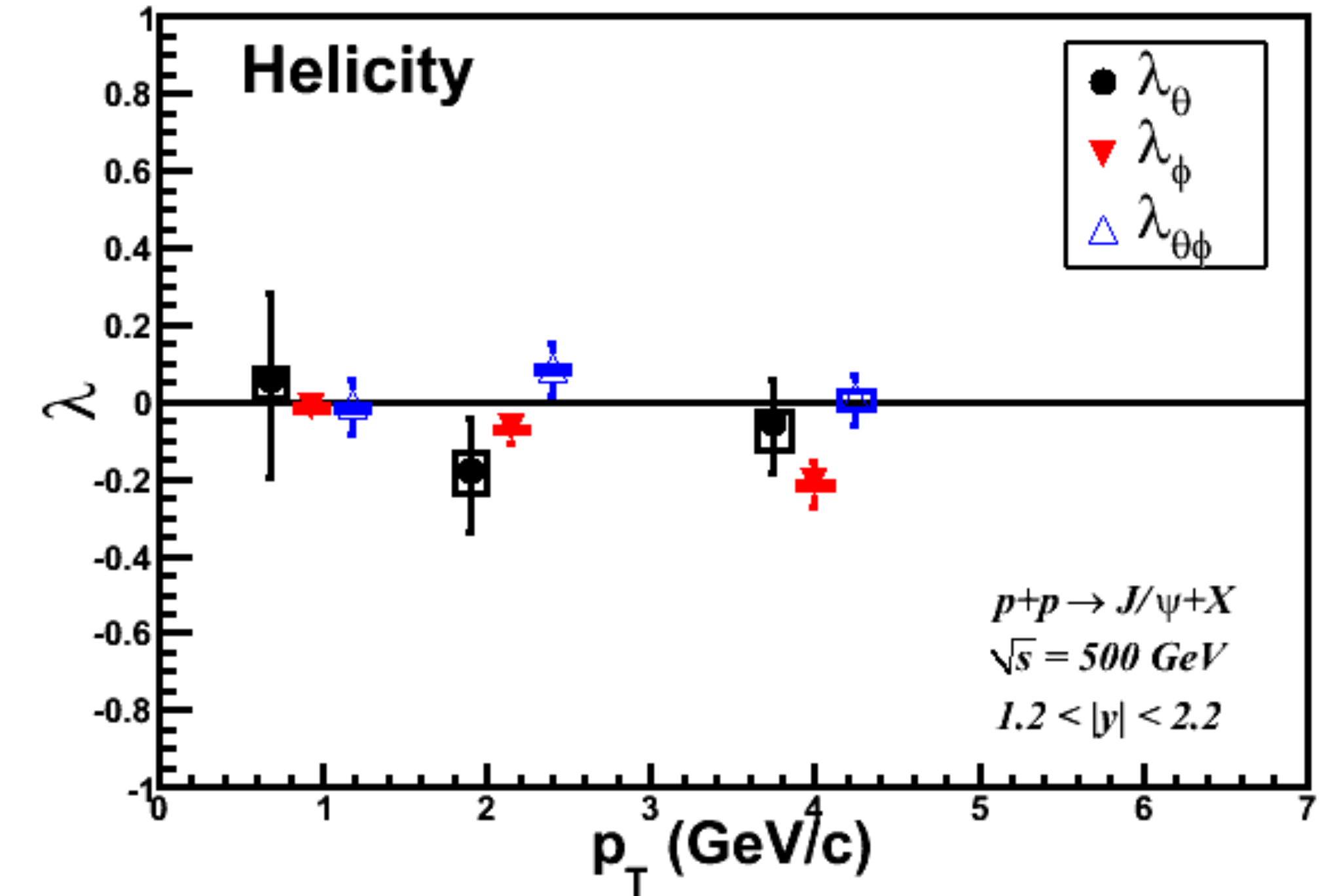} \ 
        \centering\includegraphics*[width=0.48\columnwidth]{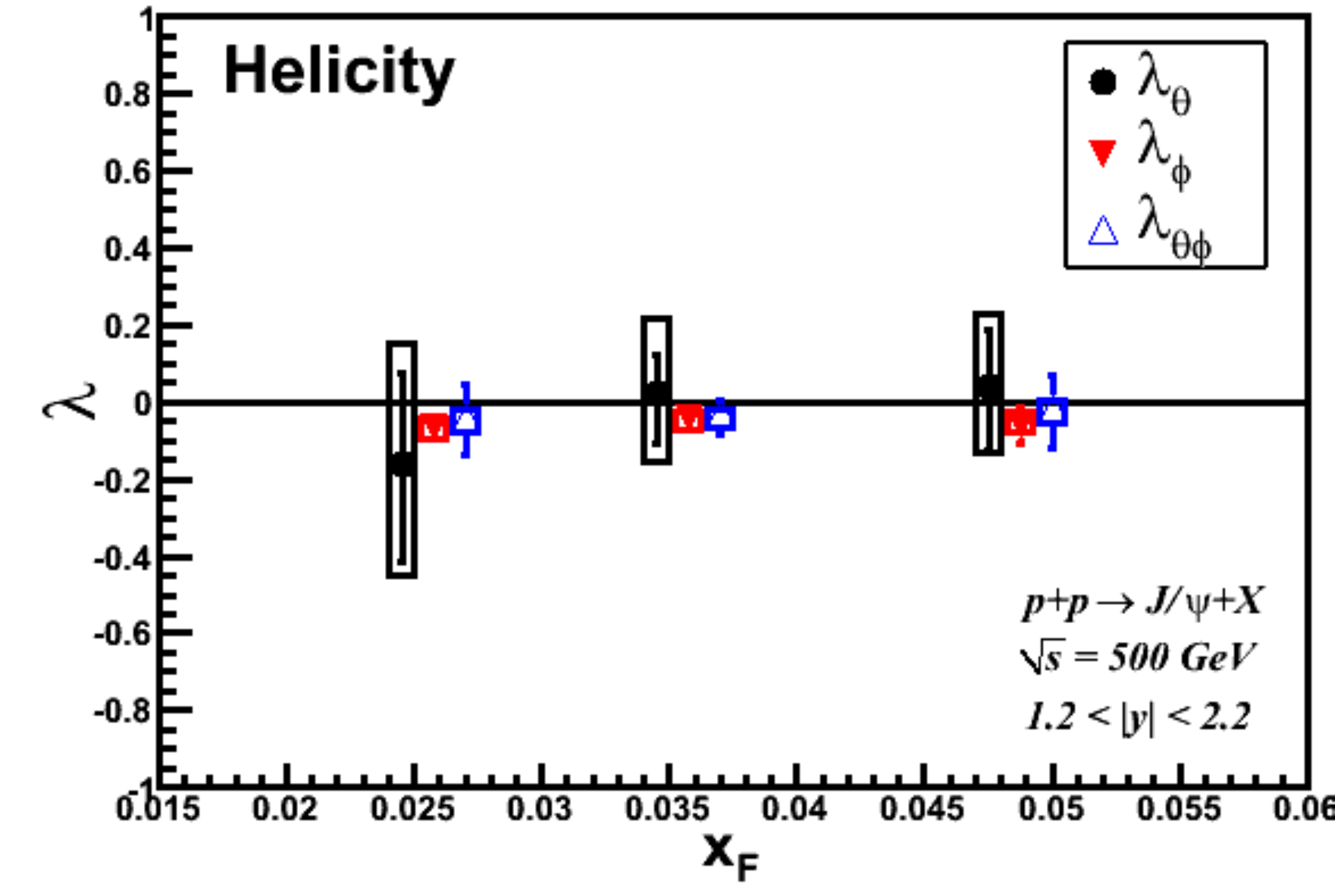} \\
        \isucaption[Angular decay coefficients for $J/\psi\rightarrow\mu^{+}\mu^{-}$]{From top to bottom:  Angular decay coefficients from the Collins-Soper, Gottfried-Jackson forward, and Gottfried-Jackson backward, and Helicity frames plotted against $p_{T}$ on the left and $x_F$ on the right.  Point-to-point uncorrelated uncertainties are shown with error bands, and point-to-point correlated uncertainties with boxes.}\label{fig:lambdas}
\end{figure}

Measured angular decay coefficients included approximately 13167 $J/\psi$ mesons in the North and 25390 in the South muon spectrometer.  After all fits are performed and systematics included for the input kinematic shapes, the resulting angular decay coefficients are plotted against $p_T$ and $x_F$ (Fig.~\ref{fig:lambdas}).  A single $p_T$ and $x_F$ integrated data point is shown in Fig.~\ref{fig:integrated_lambdas}, and numerical values for each data point can be found in Appendix~\ref{sec:data_tables}.

\section{Discussion}

The integrated $\lambda_{\vartheta}$ coefficient is consistent with zero in all reference frames, but the coefficient appears to be negative at low $p_T$ in the Gottfried-Jackson forward frame.  Unfortunately, predictions do not exist yet for $J/\psi$ decay coefficients from $p$+$p$ collisions at $\sqrt{s}$=500~GeV, but the prediction from the CSM at NLO shown back in Fig.~\ref{fig:CS_NLO_pol} for $p$+$p$ collisions at $\sqrt{s}$=200~GeV appears to be consistent with the measured $\lambda_\vartheta$ in the Helicity frame, and the angular decay coefficients are not expected to change drastically with $\sqrt{s}$.  Predictions from the COM are typically valid for a minimum $p_{T}$ of 5~GeV/$c$, larger than the $p_{T}$ of this measurement.

\begin{figure}[h!tb]
        \centering\includegraphics*[width=\columnwidth]{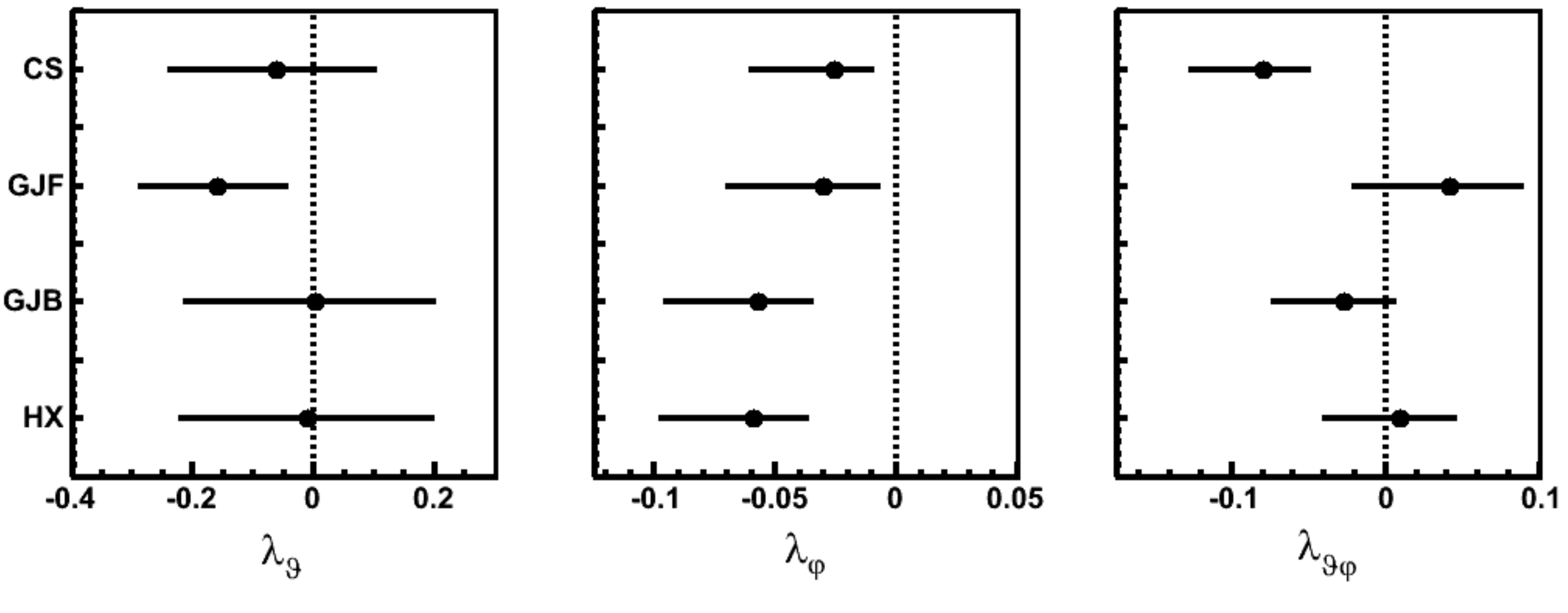}
        \isucaption[Magnitude of the angular decay coefficients in four reference frames integrated across $p_{T}$ and $x_{F}$]{Magnitude of the angular decay coefficients in the Collins-Soper (CS), Helicity (HX), Gottfried-Jackson Forward and Backward (GJF and GJB) integrated across $p_{T}$ and $x_{F}$.  Statistical and systematic uncertainties are summed in quadrature.}\label{fig:integrated_lambdas}
\end{figure}

From a phenomenological point of view, the Gottfried-Jackson Forward frame is slightly preferred as closest to the `natural' reference frame for the $J/\psi$  spin alignment (the frame where only $\lambda_{\vartheta}$ is non-zero) as $\lambda_{\vartheta}$ is largest in magnitude in that frame, while $\lambda_{\varphi}$ and $\lambda_{\vartheta\varphi}$ are smallest (Fig.~\ref{fig:integrated_lambdas}).  Measurements from the HERA-B experiment for $p$+N collisions~\cite{Abt:2009nu} found the Collins-Soper frame to be the closest to natural, but those experiments were carried out in a fixed-target environment where only a single Gottfried-Jackson frame is relevant.

\begin{figure}[h!tb]
        \centering\includegraphics*[width=0.49\columnwidth]{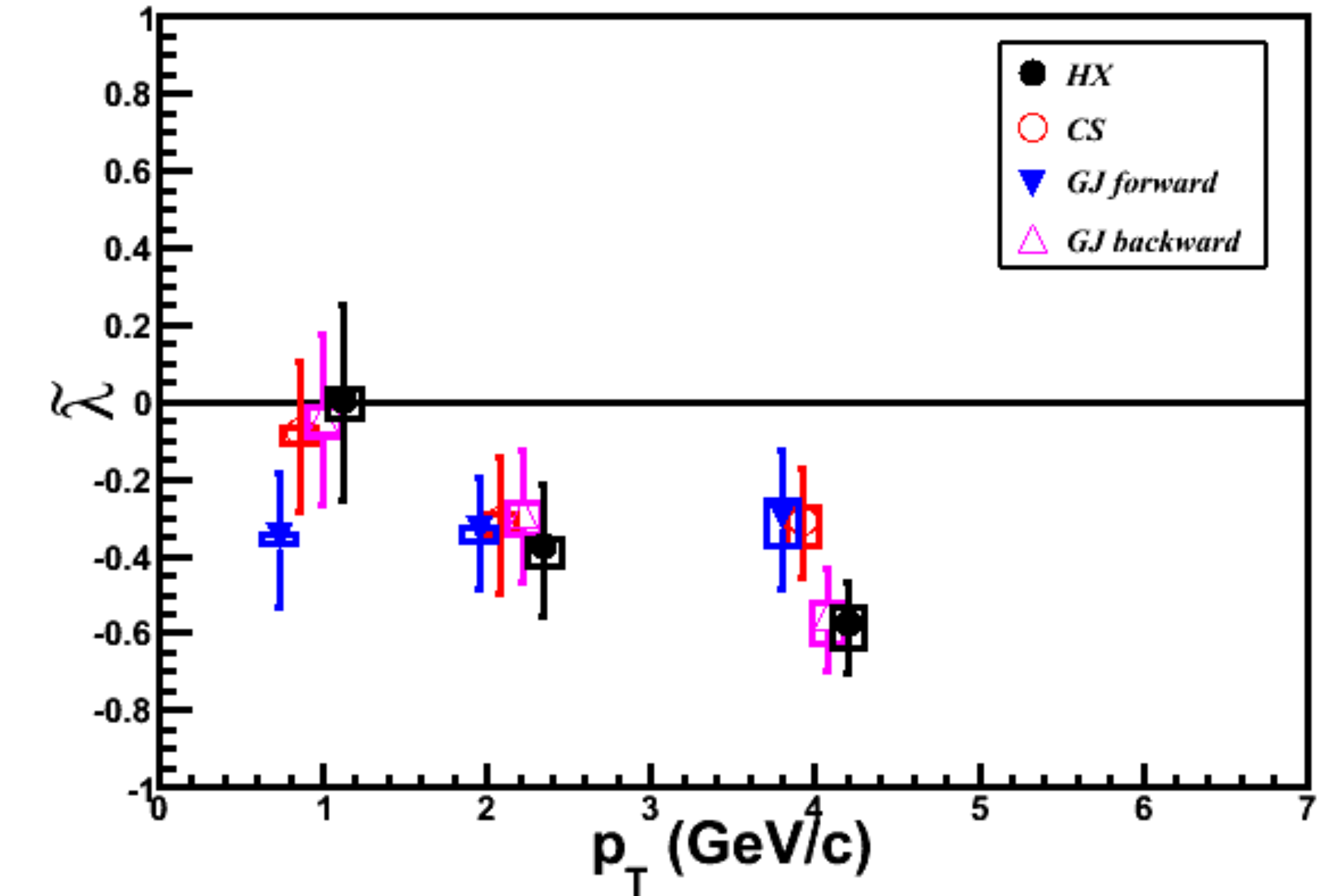} \
        \centering\includegraphics*[width=0.49\columnwidth]{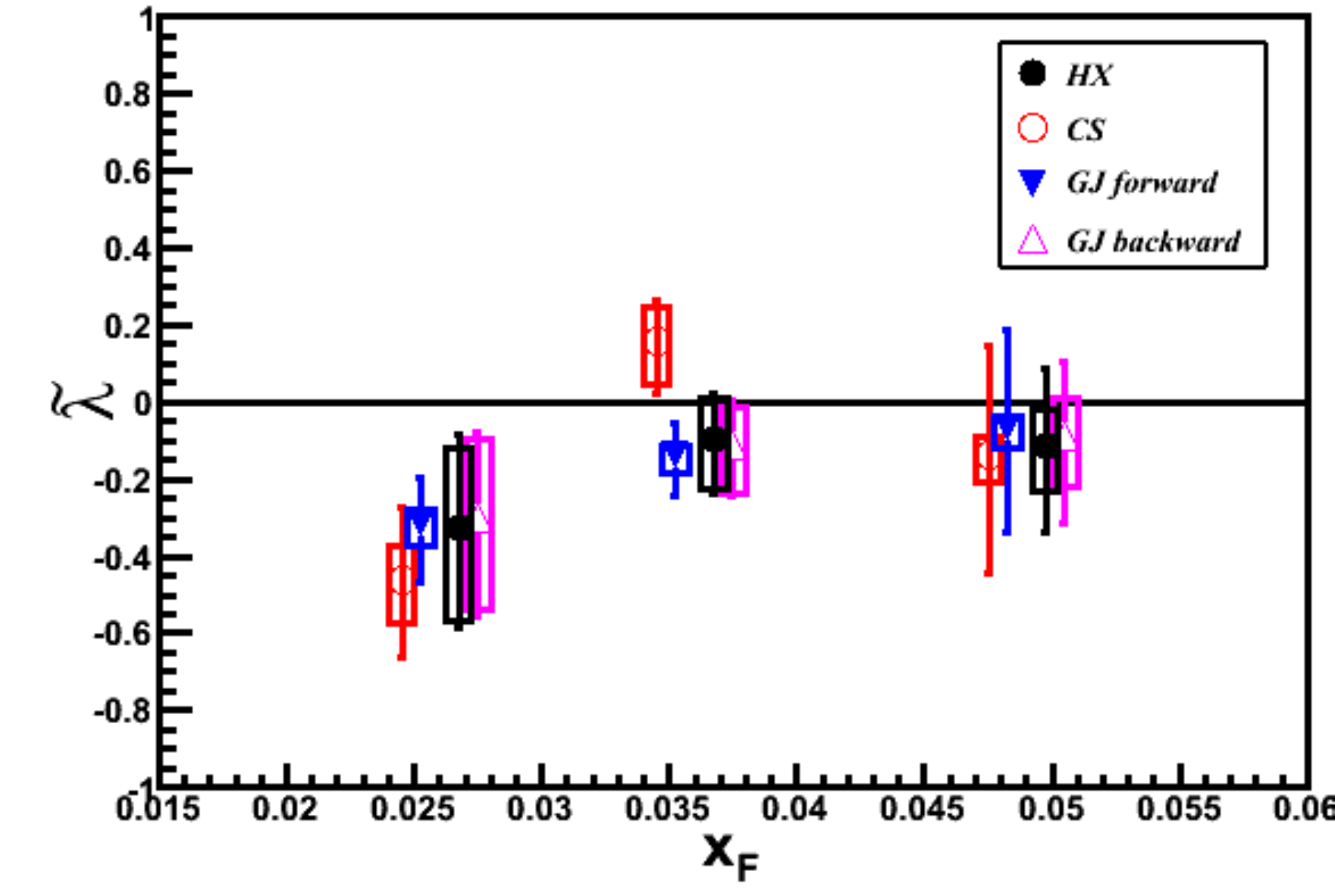} \\
        \centering\includegraphics*[width=0.49\columnwidth]{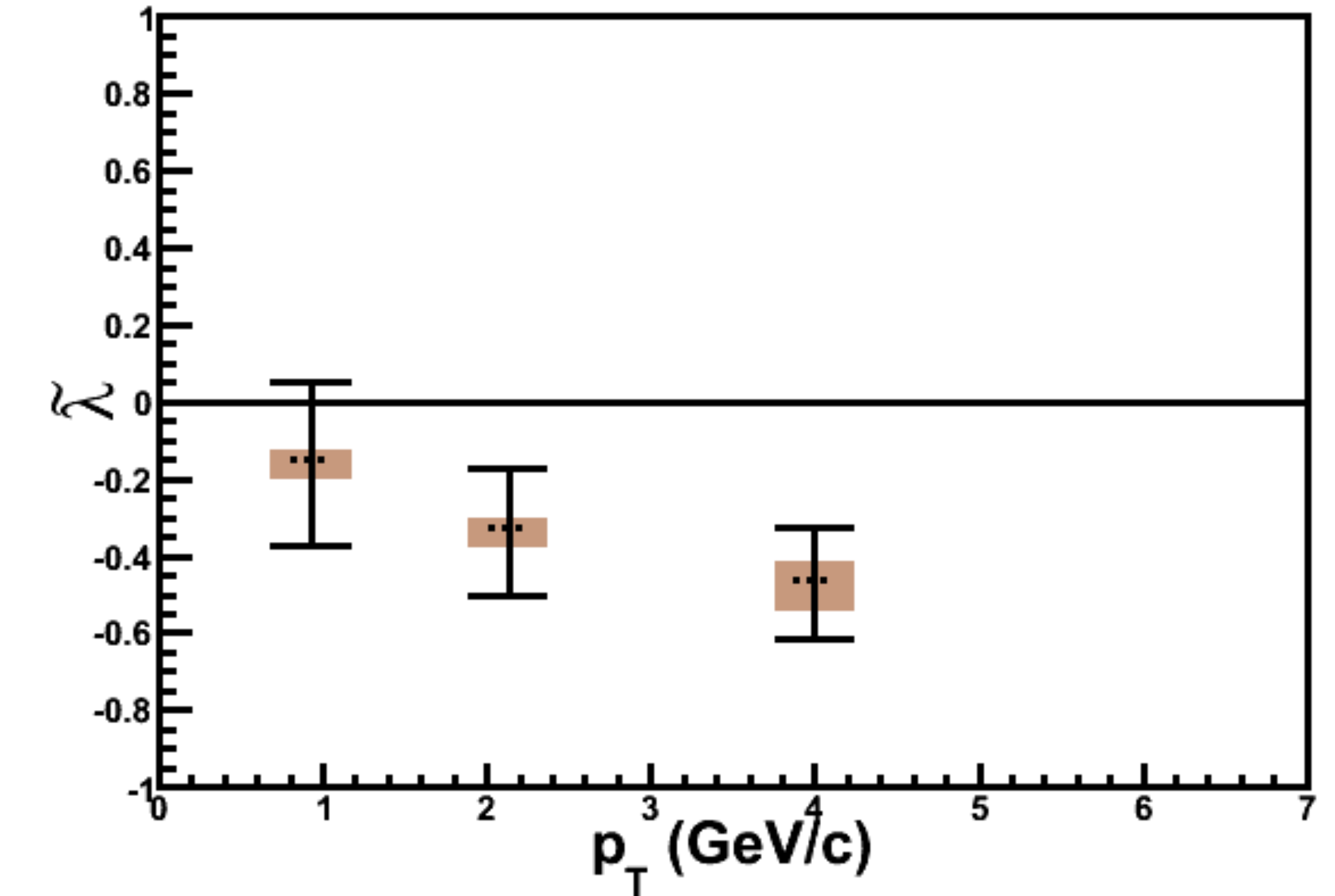} \
        \centering\includegraphics*[width=0.49\columnwidth]{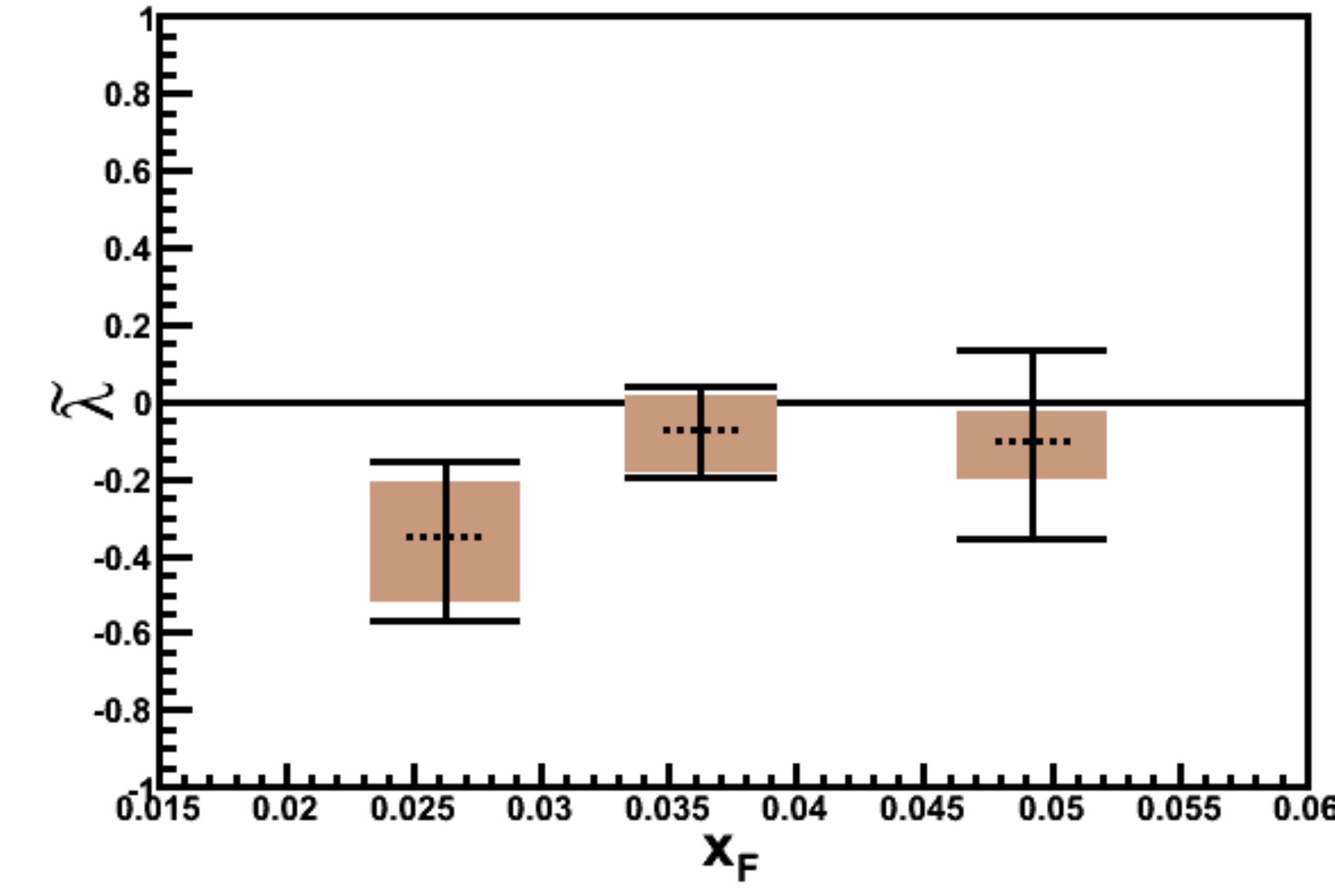} 
        \isucaption[Frame-invariant angular decay coefficient, $\widetilde{\lambda}$, from measurements in four different reference frames.]{Frame-invariant angular decay coefficient, $\widetilde{\lambda}$, from measurements in the Collins-Soper (CS), Helicity (HX), Gottfried-Jackson Forward and Backward (GJF and GJB) reference frames plotted against $p_{T}$ and $x_{F}$ on the top.  On the bottom, the average of the four measuremnets where wide lines correspond to uncertainty uncorrelated between points and wide boxes correspond to correlated uncertainty.}\label{fig:tilde_plots}
\end{figure}

It is also interesting to look at the frame-invariant $\widetilde{\lambda}$ directly using Eq.~\ref{eq:lambda_tilde}.   Values of $\widetilde{\lambda}$ along with their uncertainties are plotted for each reference frame against $p_{T}$ and $x_{F}$ in Fig.~\ref{fig:tilde_plots}.  Also plotted in that figure are the mean values of $\widetilde{\lambda}$ assuming that uncertainties are completely correlated between frames.  The measured $\widetilde{\lambda}$ are consistent between frames within their systematic uncertainties, and it is quite interesting to note that the mean $\widetilde{\lambda}$ is consistent with a non-zero longitudinal spin-alignment which increases in magnitude with increasing $p_T$.  This suggest that there exists a natural reference frame wherein the $J/\psi$ has a longitudinal spin-alignment and that none of the measured frames are optimal for maximizing $|\lambda_{\vartheta}|$.

\begin{table}[h!tb]
\centering
\renewcommand\arraystretch{1.5}
\isucaption[Frame-invariant angular decay coefficient, $\widetilde{\lambda}$, calculated for a single integrated data point.]{Frame-invariant angular decay coefficient, $\widetilde{\lambda}$, calculated for a single data point integrated over $p_{T}\in[0,7]$~GeV/$c$ and $|x_{F}|\in[0.015,0.06]$\label{tab:lambda_tilde}}
\begin{tabular}{cccc}
\hline
Reference Frame    & $\widetilde{\lambda}$  &          uncorr.      &  corr.                \\ \hline
Collins-Soper      &      -0.14             &    $+0.08\atop-0.11$  &  $+0.07\atop-0.10$  \\ \hline
Gottfried-Jackson  &                        &                       &                       \\ 
Forward            &      -0.25             &    $+0.06\atop-0.08$  &  $+0.02\atop-0.04$  \\ \hline 
Gottfried-Jackson  &                        &                       &                       \\ 
Backward           &      -0.16             &    $+0.08\atop-0.11$  &  $+0.10\atop-0.12$  \\ \hline
Helicity           &      -0.17             &    $+0.09\atop-0.11$  &  $+0.10\atop-0.13$  \\ \hline
\end{tabular}
\end{table}
To summarize, we can make several strong statements regarding the $J/\psi$ angular decay coefficients from these measurements:
\begin{itemize}
\item
  $\lambda_\vartheta$ is small in each of the four measured reference frames.
\item
  The Gottfried-Jackson Forward frame appears to be the closest to the natural reference frame, with the smallest $|\lambda_{\varphi}|$ and $|\lambda_{\vartheta\varphi}|$ and the largest $|\lambda_{\vartheta}|$.
\item
  Using the frame invariant approach proposed in~\cite{Faccioli:2010kd}, we find a $\widetilde{\lambda}$ which is likely negative and increasing in magnitude with increasing $p_T$.
\end{itemize}
All of these conclusions give useful information about the $J/\psi$ production mechanism and can be used in correlation with future measurements from other collision systems and energies to determine how the $J/\psi$ is produced.

%% file: chapters/conclusions/conclusions.tex
\chapter{Conclusions and Future Measurements}\label{ch:conclusions}

Two measurements have been presented in this document:  a transverse single spin asymmetry (SSA) of $J/\psi$ mesons and a comprehensive measurement of all relevant $J/\psi$ angular decay coefficients in various reference frames.  Both measurements show information which is new and relevant. 

The transverse SSA of $J/\psi$ mesons is 3.3$\sigma$ less than zero at $<$$x_{F}$$>$=0.8 and consistent with zero for mid and backward $x_{F}$.  The trend of having a non-zero transverse SSA at forward $x_{F}$ is consistent with transverse SSA measurements of other particles but is unexpected in the case of the $J/\psi$.  The fact that the transverse SSA is non-zero potentially implies a non-zero gluon Sivers or trigluon correlation function in the proton.  It might also mean that forward $J/\psi$ production from $p$+$p$ collisions at $\sqrt{s}$=200~GeV is dominated not by color-octet production but more likely by color-singlet production.  The calculation of trigluon correlation functions for $J/\psi$ production would clarify the relationship between the transverse SSA and production mechanism.

Measurements of the $J/\psi$ transverse SSA with other center of mass energies and collisions systems would help to confirm the existence of a transverse SSA at forward $x_{F}$ and to determine the mechanism for such a transverse SSA.  While the evolution in $Q^{2}$ of the gluon Sivers and trigluon correlation functions are unknown, transverse SSAs for other particles have persisted through a wide range in $\sqrt{s}$ in various collision systems.  A measurement of the $J/\psi$ transverse SSA in SIDIS would be especially useful, as~\cite{Yuan:2008vn} predicts a vanishing asymmetry for the color-singlet model in SIDIS but non-zero asymmetry for the color-octet model.  Mapping the dependence of the effect on collision system, therefore, could reveal considerably more information about the $J/\psi$ production mechanism.

The $J/\psi$ spin-alignment from $p$+$p$ collisions at $\sqrt{s}$=500~GeV for 1.2$<$$|$$y$$|$$<$2.2 is consistent with zero in all measured reference frames when integrated over $p_T$ and $x_F$.  The Gottfried-Jackson forward frame appears to have a negative $\lambda_\vartheta$ at low $p_T$ and is the closest to the natural reference frame.  The frame invariant $\widetilde{\lambda}$ implies that there exists a natural frame wherein $\lambda_\vartheta$ becomes increasingly negative as $p_T$ increases.

Measurements of all $J/\psi$ angular decay coefficients (not just $\lambda_{\vartheta}$) using various collisions species and energies would lead to a much better understanding of the production mechanism, as the kinematic dependencies of the color-singlet and color-octet diagrams are quite different.  In particular, a measurement of this kind at larger $p_T$ could definitively show whether or not the Color-Octet Model describes data, and if not, in what respects it disagrees (i.e. is the Helicity frame really the natural frame for $J/\psi$ spin-alignment?).  

The $J/\psi$ production mechanism provides access to basic QCD dynamics, but the road to understanding it has not always been smooth, and we have not yet reached the end.  The measurements presented in this document represent the application of novel ways of thinking about the production mechanism, and I hope that these measurements, along with similar measurements from other experiments, will lead to a better understanding of the $J/\psi$ in particular and QCD in general.

%% file: chapters/bg_contributions/bg_contributions.tex
\chapter{Background Contributions to Angular Distributions}\label{sec:bg_contributions}

This appendix presents a derivation of the background contribution to the angular decay coefficients discussed in Section~\ref{sec:angular_coefficients}.  If we explicitly include background in Eq.~\ref{eq:lambda_theta}, the differential yield of inclusive pairs $N_I$ becomes 
\begin{equation}
\frac{dN_{\text I}}{d\cos\vartheta} \propto f_{\text S} \left(1 + \lambda_{\vartheta,\text S} \cos^{2}\vartheta\right) A\epsilon_{S}(\cos\vartheta) +  f_{\text B} \left(1 + \lambda_{\vartheta,\text B} \cos^{2}\vartheta\right) A\epsilon_{B}(\cos\vartheta)
\end{equation}
where $f_{\text S}$ and $f_{\text B}$ are the fraction of pairs in the signal and background respectively and $\lambda_{\vartheta,\text S}$, $\lambda_{\vartheta,\text B}$ are the signal and background coefficients.  To simplify the expression, we will assume that the acceptance times efficiency $A\epsilon(\cos\vartheta)$ is the same for the signal and background:
\begin{equation}
\frac{dN_{\text I}}{d\cos\vartheta} \propto \left(1 + \frac{f_{\text S} \lambda_{\vartheta, \text S} + f_{\text B} \lambda_{\vartheta, \text B}}{f_{\text S} + f_{\text B}} \cos^{2}\vartheta\right) A\epsilon(\cos\vartheta)
\end{equation}
If we define $R_{\text S}\equiv\frac{f_{\text S}}{f_{\text S} +  f_{\text B}}$, we can then see that $\lambda_{\vartheta}$ determined by our fit to the inclusive distributions is
\begin{equation}
\lambda_{\vartheta,\text{fit}} = R_{\text S} \lambda_{\vartheta, \text S} + (1 - R_{\text S}) \lambda_{\vartheta, \text B},
\end{equation}
which can be rearranged to give Eq.~\ref{eq:theta_BG}.

The derivation for $\lambda_{\varphi}$ and $\lambda_{\vartheta\varphi}$ are very similar, and we will only present the derivation for $\lambda_{\varphi}$.  Explicitly including background in Eq~\ref{eq:lambda_phi}, we have
\begin{equation}
\frac{dN}{d(\varphi)} \propto f_{\text S} \left(1 + \frac{2 \lambda_{\varphi, \text S}}{3 + \lambda_{\vartheta, \text S}} \cos{2\varphi}\right) A\epsilon_{\text S}(\varphi) + f_{\text B}\left(1 + \frac{2 \lambda_{\varphi, \text B}}{3 + \lambda_{\vartheta, \text B}} \cos{2\varphi}\right) A\epsilon_{\text B}(\varphi).
\end{equation}
Assuming that $A\epsilon_{\text S}(\varphi)$=$A\epsilon_{\text B}(\varphi)$, this expression simplifies to
\begin{equation}
\frac{dN}{d(\varphi)} \propto \left( 1 + \frac{2 R_{\text S}\lambda_{\varphi,\text S}}{3+\lambda_{\vartheta,\text S}} \cos 2\varphi + \frac{2(1-R_{\text S})\lambda_{\varphi,\text B}}{(3 + \lambda_{\vartheta,\text B})} \cos 2\varphi\right) A\epsilon(\varphi).
\end{equation}
from which we can identify
\begin{equation}
\frac{\lambda_{\varphi,\text{fit}}}{3 + \lambda_{\vartheta,\text{fit}}} =  \frac{R_{\text S}\lambda_{\varphi,\text S}}{3+\lambda_{\vartheta,\text S}} + \frac{(1-R_{\text S})\lambda_{\varphi,\text B}}{(3 + \lambda_{\vartheta,\text B})},
\end{equation}
which can be rearranged to give Eq.~\ref{eq:phi_BG}.

%% file: chapters/angular_acceptance/angular_acceptance.tex
\chapter{Relative Acceptance for Decay Angles}\label{sec:angular_acceptance}

\begin{figure}[h!tb]
        \centering\includegraphics*[width=\columnwidth]{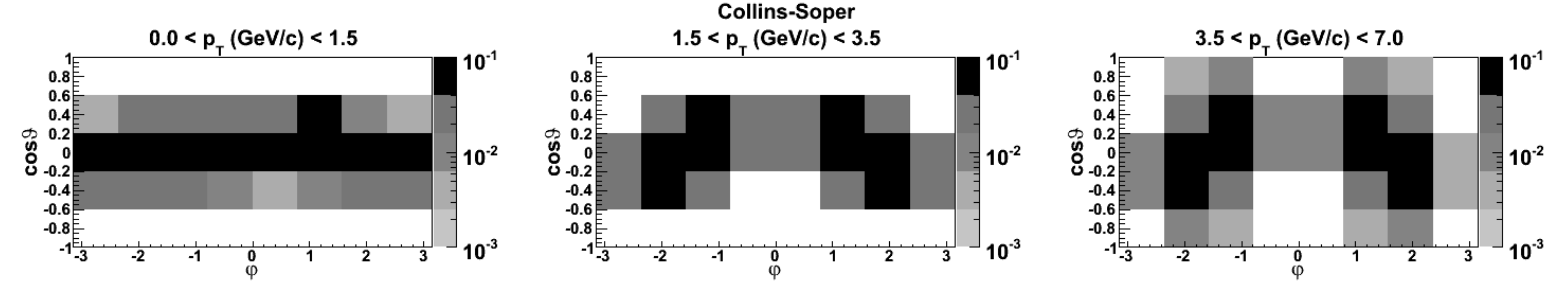} \\
        \centering\includegraphics*[width=\columnwidth]{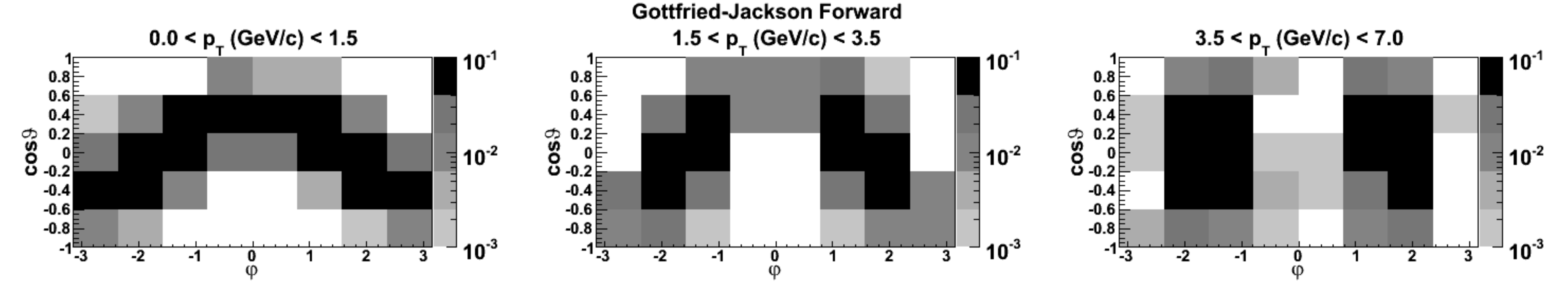} \\
        \centering\includegraphics*[width=\columnwidth]{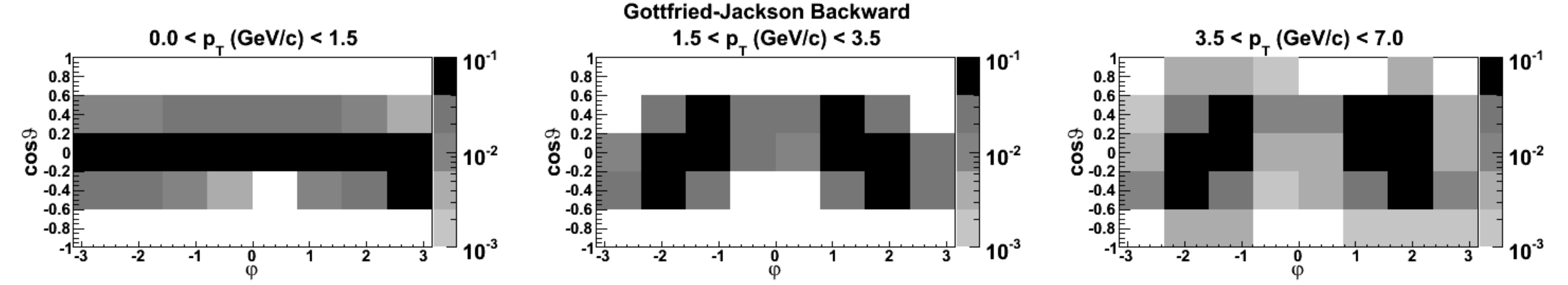} \\
        \centering\includegraphics*[width=\columnwidth]{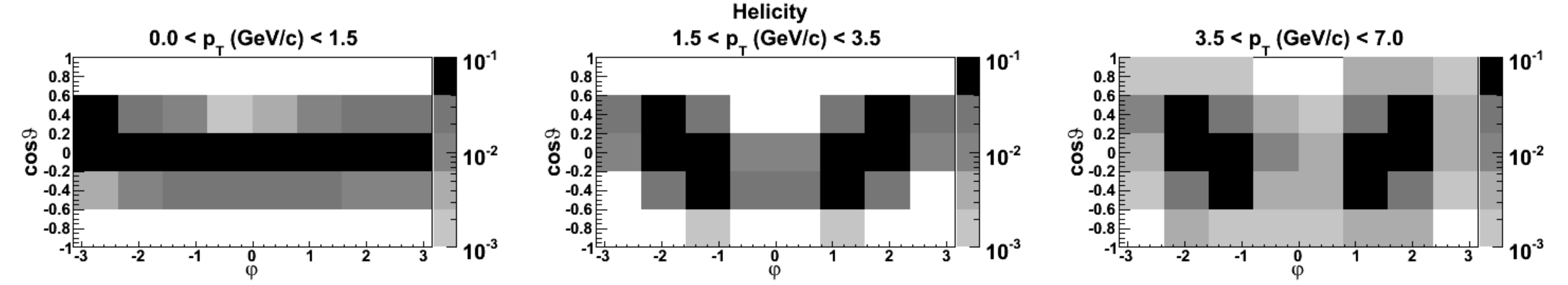} 
        \isucaption[Relative acceptance for $\cos\vartheta$-$\varphi$ for various ranges in $p_{T}$.]{Relative acceptance for $\cos\vartheta$-$\varphi$ in (from top to bottom) the Collins-Soper, Gottfried-Jackson Forward, Gottfried-Jackson Backward, and Helicity frames for increasing $p_{T}$ from left to right.  Plots are drawn from real data for the North muon spectrometer, and the acceptance is nearly identical for the spectrometer in the South. }\label{fig:theta_pt_acc}
\end{figure}

\begin{figure}[h!tb]
        \centering\includegraphics*[width=\columnwidth]{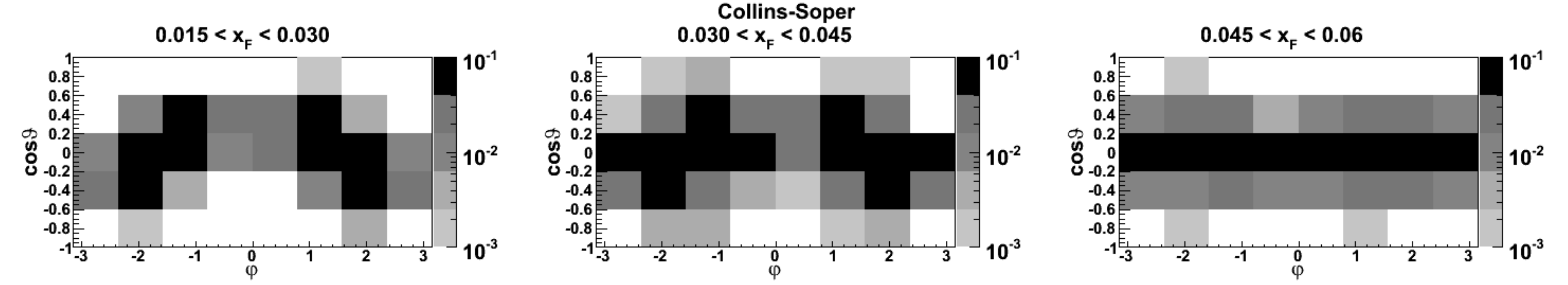} \\
        \centering\includegraphics*[width=\columnwidth]{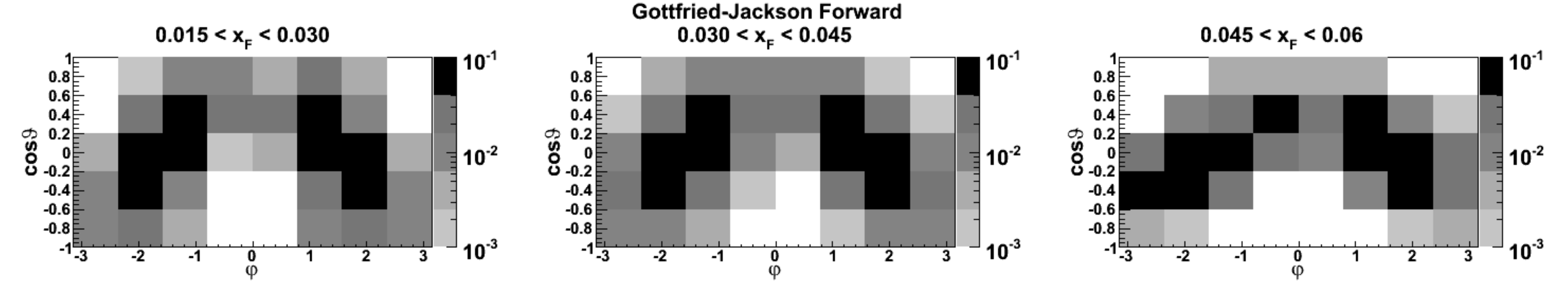} \\
        \centering\includegraphics*[width=\columnwidth]{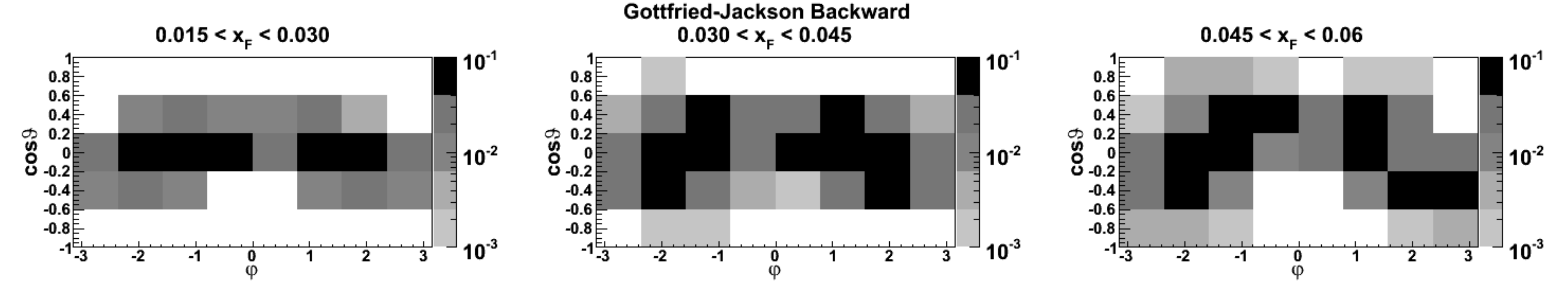} \\
        \centering\includegraphics*[width=\columnwidth]{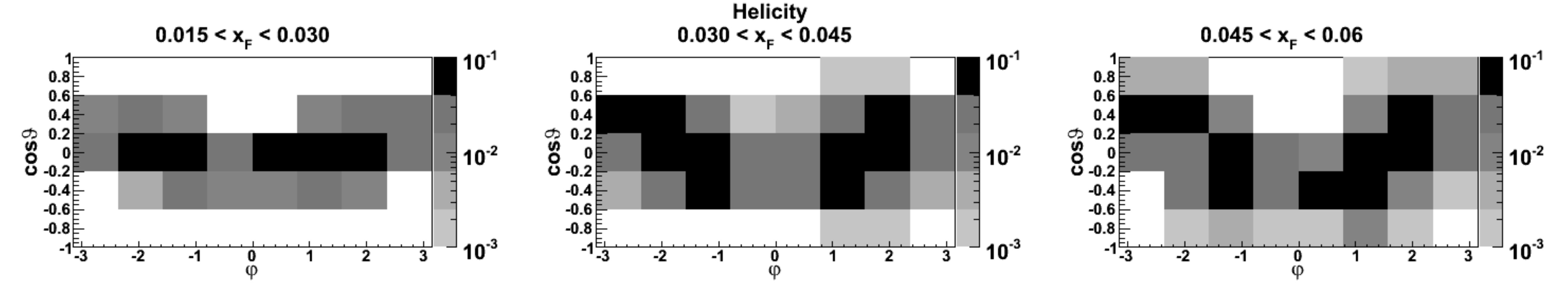} 
        \isucaption[Relative acceptance for $\cos\vartheta$-$\varphi$ for various ranges in $x_{F}$.]{Relative acceptance for $\cos\vartheta$-$\varphi$ in (from top to bottom) the Collins-Soper, Gottfried-Jackson Forward, Gottfried-Jackson Backward, and Helicity frames for increasing $x_{F}$ from left to right.  Plots are drawn from real data for the North muon spectrometer, and the acceptance is nearly identical for the spectrometer in the South. }\label{fig:theta_xf_acc}
\end{figure}

%% file: chapters/data_tables/data_tables.tex
\chapter{Data Tables}\label{sec:data_tables}

This appendix contains a collection of data tables for the $J/\psi$ SSA presented in Chapter~\ref{ch:jpsi_AN} and the $J/\psi$ angular decay coefficients presented in Chapter~\ref{ch:jpsi_pol}.

\begin{sidewaystable}[htbp]
\isucaption[$J/\psi$ SSA vs. $p_T$ in forward, backward and central rapidities]{\label{tab:AN}$J/\psi$ SSA vs. $p_T$ in forward, backward and central rapidities.  Systematic uncertainties in the last two columns are due to the geometric scale factor and the polarization, respectively.  There are additional Type C uncertainties due to the polarization of 3.4\%, 3.0\%, and 2.4\% for the 2006, 2008, and combined 2006 and 2008 results.}

\centering
\begin{tabular}{cccccccc}
\hline\hline
 $p_T$      & Data Sample& $<x_F>$ & $A_N$& $\delta A_N$ & $\delta A_N$ & $\delta A_N^f$ (\%) & $\delta A_N^P$ (\%) \\
(GeV/c)     &            &          &        & (stat.)  & (Type A syst.)& (Type B syst.)      &  (Type B syst.)     \\ \hline
           & 2006        & -0.081   & -0.024  & 0.044   & 0.003         & 0.6                 & 2.3 \\
           & 2008        & -0.082   & -0.010  & 0.032   & 0.004         & 0.4                 & 3.4 \\
           & 2006 + 2008 & -0.081   & -0.015  & 0.026   & 0.002         & 0.4                 & 2.8 \\
           \hline

0--6       & 2006        &  0.000   & -0.064  & 0.106   & 0.026         & 0.6                 & 2.3 \\
\hline

           & 2006        &  0.084   & -0.105  & 0.044   & 0.005         & 0.6                 & 2.3 \\
           & 2008        &  0.086   & -0.075  & 0.032   & 0.003         & 0.4                 & 3.3 \\
           & 2006 + 2008 &  0.085   & -0.086  & 0.026   & 0.003         & 0.4                 & 2.7 \\
           \hline\hline

            & 2006        & -0.081   &  0.050  & 0.067  & 0.007         & 0.6                 & 2.3 \\
            & 2008        & -0.081   & -0.025  & 0.046  & 0.008         & 0.4                 & 3.4 \\
            & 2006 + 2008 & -0.081   & -0.001  & 0.038  & 0.005         & 0.4                 & 2.8 \\
            \hline

0--1.4      & 2006        & 0.000    & -0.063  & 0.128  & 0.031         & 0.6                 & 2.3\\
\hline

            & 2006        & 0.085   & -0.065  & 0.066   & 0.005         & 0.6                 & 2.3 \\
            & 2008        & 0.087   & -0.064  & 0.045   & 0.003         & 0.4                 & 3.4 \\
            & 2006 + 2008 & 0.086   & -0.064  & 0.037   & 0.003         & 0.4                 & 2.7 \\
            \hline\hline

            & 2006        & -0.081   & -0.073  & 0.065 & 0.002          & 0.6                 & 2.3 \\
            & 2008        & -0.082   & -0.023  & 0.046 & 0.010          & 0.4                 & 3.5 \\
            & 2006 + 2008 & -0.082   & -0.039  & 0.038 & 0.002          & 0.4                 & 2.8 \\
            \hline

1.4--6      & 2006        & 0.000   & -0.068  & 0.188  & 0.045          & 1.2                 & 2.3 \\
\hline

            & 2006        & 0.084   & -0.046  & 0.064  & 0.005          & 0.6                 & 2.3 \\
            & 2008        & 0.086   & -0.073  & 0.046  & 0.007          & 0.4                 & 3.3 \\
            & 2006 + 2008 & 0.085   & -0.064  & 0.037  & 0.004          & 0.4                 & 2.7 \\
            \hline\hline
\end{tabular}
\end{sidewaystable}

\begin{sidewaystable}[htbp]
\renewcommand\arraystretch{1.5}
\isucaption[$J/\psi$ meson angular decay coefficients and uncertainties binned in $p_{T}$]{\label{tab:pT_lambdas} $J/\psi$ meson angular decay coefficients, point-to-point uncorrelated uncertainties, and point-to-point correlated uncertainties binned in $p_{T}$}
\begin{tabular}{cccccccccccccccc}
\cline{3-14}
                &     &    \multicolumn{3}{|c|}{$p_{T} \in [0,1.5]$~GeV/$c$}
                &         \multicolumn{3}{|c|}{$p_{T} \in [1.5,3.0]$~GeV/$c$}
                &         \multicolumn{3}{|c|}{$p_{T} \in [3.0,7.0]$~GeV/$c$}
                &         \multicolumn{3}{|c|}{$p_{T} \in [0,7.0]$~GeV/$c$}    \\ 
                &     &    \multicolumn{3}{|c|}{$<p_{T}>=0.93$ GeV/$c$}
                &         \multicolumn{3}{|c|}{$<p_{T}>=2.15$ GeV/$c$}
                &         \multicolumn{3}{|c|}{$<p_{T}>=4.00$ GeV/$c$}
                &         \multicolumn{3}{|c|}{$<p_{T}>=1.95$ GeV/$c$}     \\ \hline
Frame           &    Coeff.    &     \multicolumn{1}{|c}{\small point}     &    \small uncorr.       &  \multicolumn{1}{c|}{\small corr.}
                               &     \multicolumn{1}{|c}{\small point}     &    \small uncorr.       &  \multicolumn{1}{c|}{\small corr.}
                               &     \multicolumn{1}{|c}{\small point}     &    \small uncorr.       &  \multicolumn{1}{c|}{\small corr.}  
                               &     \multicolumn{1}{|c}{\small point}     &    \small uncorr.       &  \multicolumn{1}{c|}{\small corr.}  \\ \hline \hline
                &  $\lambda_{\vartheta}$         &   -0.081   &  $+0.176\atop-0.194$  &  $+0.007\atop-0.024$    
                                                 &   -0.309   &  $+0.144\atop-0.166$  &  $+0.019\atop-0.039$ 
                                                 &   -0.099   &  $+0.094\atop-0.116$  &  $+0.037\atop-0.063$  
                                                 &   -0.062   &  $+0.077\atop-0.093$  &  $+0.142\atop-0.149$  \\ 
                &  $\lambda_{\varphi}$           &    0.001   &  $+0.016\atop-0.024$  &  $+0.007\atop-0.012$ 
                                                 &    0.003   &  $+0.020\atop-0.025$  &  $+0.002\atop-0.013$ 
                                                 &   -0.068   &  $+0.041\atop-0.054$  &  $+0.004\atop-0.010$  
                                                 &   -0.026   &  $+0.008\atop-0.022$  &  $+0.013\atop-0.026$  \\
 Collins-Soper   &  $\lambda_{\vartheta\varphi}$ &   -0.031   &  $+0.043\atop-0.057$  &  $+0.010\atop-0.018$ 
                                                 &   -0.101   &  $+0.053\atop-0.062$  &  $+0.013\atop-0.025$ 
                                                 &   -0.033   &  $+0.066\atop-0.074$  &  $+0.015\atop-0.025$ 
                                                 &   -0.080   &  $+0.027\atop-0.038$  &  $+0.011\atop-0.027$   \\  
                 &  $\widetilde{\lambda}$        &   -0.078   &  $+0.183\atop-0.207$  &  $+0.012\atop-0.031$ 
                                                 &   -0.305   &  $+0.160\atop-0.190$  &  $+0.014\atop-0.039$ 
                                                 &   -0.309   &  $+0.134\atop-0.146$  &  $+0.038\atop-0.067$ 
                                                 &   -0.137   &  $+0.083\atop-0.107$  &  $+0.072\atop-0.098$   \\  \hline
                  &  $\lambda_{\vartheta}$       &   -0.354   &  $+0.129\atop-0.151$  &  $+0.011\atop-0.030$  
                                                 &   -0.259   &  $+0.054\atop-0.076$  &  $+0.019\atop-0.037$                    
                                                 &   -0.136   &  $+0.081\atop-0.099$  &  $+0.006\atop-0.028$                   
                                                 &   -0.160   &  $+0.045\atop-0.065$  &  $+0.105\atop-0.109$     \\ 
 Gottfried-Jackson &  $\lambda_{\varphi}$        &    0.016   &  $+0.019\atop-0.031$  &  $+0.002\atop-0.016$ 
                                                 &   -0.016   &  $+0.045\atop-0.055$  &  $+0.010\atop-0.018$                  
                                                 &   -0.054   &  $+0.077\atop-0.088$  &  $+0.026\atop-0.027$                 
                                                 &   -0.030   &  $+0.012\atop-0.023$  &  $+0.019\atop-0.033$    \\ 
 Forward         &  $\lambda_{\vartheta\varphi}$ &    0.116   &  $+0.057\atop-0.063$  &  $+0.005\atop-0.018$                  
                                                 &    0.029   &  $+0.059\atop-0.066$  &  $+0.019\atop-0.006$                 
                                                 &   -0.094   &  $+0.055\atop-0.065$  &  $+0.018\atop-0.029$                
                                                 &    0.042   &  $+0.031\atop-0.039$  &  $+0.035\atop-0.048$    \\
                 &  $\widetilde{\lambda}$        &   -0.349   &  $+0.164\atop-0.186$  &  $+0.008\atop-0.021$ 
                                                 &   -0.333   &  $+0.138\atop-0.152$  &  $+0.007\atop-0.030$ 
                                                 &   -0.298   &  $+0.173\atop-0.187$  &  $+0.045\atop-0.078$ 
                                                 &   -0.246   &  $+0.061\atop-0.079$  &  $+0.022\atop-0.044$   \\  \hline
                &  $\lambda_{\vartheta}$         &    0.012   &  $+0.203\atop-0.217$  &  $+0.032\atop-0.053$ 
                                                 &   -0.140   &  $+0.135\atop-0.155$  &  $+0.048\atop-0.063$   
                                                 &   -0.014   &  $+0.109\atop-0.131$  &  $+0.021\atop-0.086$    
                                                 &    0.001   &  $+0.074\atop-0.096$  &  $+0.182\atop-0.188$   \\ 
 Gottfried-Jackson &  $\lambda_{\varphi}$        &   -0.016   &  $+0.013\atop-0.027$  &  $+0.004\atop-0.011$                  
                                                 &   -0.051   &  $+0.023\atop-0.037$  &  $+0.007\atop-0.009$                  
                                                 &   -0.217   &  $+0.049\atop-0.056$  &  $+0.004\atop-0.011$                  
                                                 &   -0.057   &  $+0.010\atop-0.020$  &  $+0.020\atop-0.032$    \\ 
 Backward        &  $\lambda_{\vartheta\varphi}$ &   -0.003   &  $+0.055\atop-0.065$  &  $+0.005\atop-0.013$                   
                                                 &   -0.073   &  $+0.056\atop-0.069$  &  $+0.004\atop-0.015$                  
                                                 &   -0.054   &  $+0.062\atop-0.068$  &  $+0.016\atop-0.025$                  
                                                 &   -0.028   &  $+0.030\atop-0.040$  &  $+0.012\atop-0.020$    \\ 
                 &  $\widetilde{\lambda}$        &   -0.040   &  $+0.215\atop-0.225$  &  $+0.028\atop-0.048$ 
                                                 &   -0.288   &  $+0.163\atop-0.177$  &  $+0.025\atop-0.054$ 
                                                 &   -0.557   &  $+0.122\atop-0.138$  &  $+0.039\atop-0.068$ 
                                                 &   -0.158   &  $+0.083\atop-0.107$  &  $+0.098\atop-0.125$   \\  \hline
                &  $\lambda_{\vartheta}$         &    0.055   &  $+0.230\atop-0.250$  &  $+0.035\atop-0.046$           
                                                 &   -0.180   &  $+0.135\atop-0.155$  &  $+0.048\atop-0.058$ 
                                                 &   -0.056   &  $+0.111\atop-0.129$  &  $+0.032\atop-0.071$ 
                                                 &   -0.010   &  $+0.075\atop-0.095$  &  $+0.190\atop-0.185$   \\  
                &  $\lambda_{\varphi}$           &   -0.011   &  $+0.019\atop-0.026$  &  $+0.006\atop-0.012$ 
                                                 &   -0.068   &  $+0.030\atop-0.040$  &  $+0.004\atop-0.011$   
                                                 &   -0.210   &  $+0.053\atop-0.062$  &  $+0.003\atop-0.016$  
                                                 &   -0.059   &  $+0.012\atop-0.023$  &  $+0.018\atop-0.030$   \\ 
 Helicity        &  $\lambda_{\vartheta\varphi}$ &   -0.009   &  $+0.066\atop-0.074$  &  $+0.003\atop-0.014$ 
                                                 &    0.090   &  $+0.063\atop-0.072$  &  $+0.006\atop-0.016$ 
                                                 &    0.012   &  $+0.061\atop-0.069$  &  $+0.016\atop-0.028$
                                                 &    0.010   &  $+0.033\atop-0.042$  &  $+0.013\atop-0.024$    \\ 
                 &  $\widetilde{\lambda}$        &    0.006   &  $+0.249\atop-0.261$  &  $+0.028\atop-0.052$ 
                                                 &   -0.375   &  $+0.160\atop-0.180$  &  $+0.021\atop-0.051$ 
                                                 &   -0.573   &  $+0.108\atop-0.132$  &  $+0.038\atop-0.065$ 
                                                 &   -0.174   &  $+0.089\atop-0.111$  &  $+0.097\atop-0.128$   \\  \hline
\end{tabular}
\end{sidewaystable}

\begin{sidewaystable}[htbp]
\renewcommand\arraystretch{1.5}
\isucaption{\label{tab:pT_chi2ndf} Average $\chi^{2}$/$ndf$ for two-dimensional fits in each arm to determine angular decay coefficients as a function of $p_T$.  Each fit has 36 degrees of freedom.}
\begin{tabular}{ccccccc}
\cline{2-7}
                    &    \multicolumn{2}{|c|}{$p_{T} \in [0,1.5]$~GeV/$c$}                     
                    &    \multicolumn{2}{|c|}{$p_{T} \in [1.5,3.0]$~GeV/$c$}        
                    &    \multicolumn{2}{|c|}{$p_{T} \in [3.0,7.0]$~GeV/$c$}    \\ 
                    &    \multicolumn{2}{|c|}{$<p_{T}>=0.93$ GeV/$c$}
                    &    \multicolumn{2}{|c|}{$<p_{T}>=2.15$ GeV/$c$}
                    &    \multicolumn{2}{|c|}{$<p_{T}>=4.00$ GeV/$c$}    \\ \hline
                    &    North   &  South 
                    &    North   &  South 
                    &    North   &  South  \\ \hline
\small Frame        &    \multicolumn{1}{|c}{\tiny $<\chi^{2}$/$ndf$$>$}   
                                             &   \multicolumn{1}{|c}{\tiny $<\chi^{2}$/$ndf$$>$} 
                                             &   \multicolumn{1}{|c}{\tiny $<\chi^{2}$/$ndf$$>$} 
                                             &   \multicolumn{1}{|c}{\tiny $<\chi^{2}$/$ndf$$>$} 
                                             &   \multicolumn{1}{|c}{\tiny $<\chi^{2}$/$ndf$$>$} 
                                             &   \multicolumn{1}{|c}{\tiny $<\chi^{2}$/$ndf$$>$}    \\ \hline \hline
 \small Collins-Soper   & 1.7 &  1.2  
                        & 1.9 &  1.5  
                        & 1.5 &  2.1  \\ \hline
 \small Gottfried-Jackson &     & 
                          &     &  
                          &     &      \\
 \small Forward           & 2.3 & 1.7
                          & 1.6 & 3.2
                          & 1.8 & 2.5  \\ \hline
 \small Gottfried-Jackson &     &
                          &     &
                          &     &      \\
 \small Backward          & 1.8 & 1.2
                          & 1.7 & 1.7
                          & 2.2 & 2.0  \\ \hline
 \small Helicity        & 1.6 & 1.1 
                        & 1.4 & 1.6 
                        & 2.5 & 1.9 \\ \hline
\end{tabular}
\end{sidewaystable}

\begin{sidewaystable}[htbp]
\renewcommand\arraystretch{1.5}
\isucaption[$J/\psi$ meson angular decay coefficients and uncertainties binned in $x_{F}$]{\label{tab:xF_lambdas} $J/\psi$ meson angular decay coefficients, point-to-point uncorrelated uncertainties, and point-to-point correlated uncertainties binned in $x_{F}$}
\begin{tabular}{ccccccccccccc}
\cline{3-11}
                &     &    \multicolumn{3}{|c|}{$x_{F} \in [0.015,0.030]$}                     
                &          \multicolumn{3}{|c|}{$x_{F} \in [0.030,0.045]$}                 
                &          \multicolumn{3}{|c|}{$x_{F} \in [0.040,0.060]$}    \\ 
                &     &    \multicolumn{3}{|c|}{$<x_{F}>=0.026$}
                &          \multicolumn{3}{|c|}{$<x_{F}>=0.036$}
                &          \multicolumn{3}{|c|}{$<x_{F}>=0.049$}    \\ \hline
Frame           &    Coeff.    &     \multicolumn{1}{|c}{\small point}     &    \small uncorr.       &  \multicolumn{1}{c|}{\small corr.}      
                               &     \multicolumn{1}{|c}{\small point}     &    \small uncorr.       &  \multicolumn{1}{c|}{\small corr.}      
                               &     \multicolumn{1}{|c}{\small point}     &    \small uncorr.       &  \multicolumn{1}{c|}{\small corr.}   \\ \hline \hline
                & $\lambda_{\vartheta}$        &    -0.337     &  $+0.162\atop-0.188$    &  $+0.195\atop-0.180$
                                               &     0.337     &  $+0.098\atop-0.122$    &  $+0.160\atop-0.176$             
                                               &    -0.217     &  $+0.262\atop-0.278$    &  $+0.083\atop-0.114$  \\ 
                & $\lambda_{\varphi}$          &    -0.052     &  $+0.029\atop-0.041$    &  $+0.026\atop-0.035$                 
                                               &    -0.055     &  $+0.012\atop-0.023$    &  $+0.012\atop-0.025$                  
                                               &    -0.023     &  $+0.026\atop-0.034$    &  $+0.010\atop-0.019$  \\ 
Collins-Soper   & $\lambda_{\vartheta\varphi}$ &     0.068     &  $+0.079\atop-0.091$    &  $+0.009\atop-0.024$    
                                               &    -0.019     &  $+0.041\atop-0.049$    &  $+0.016\atop-0.029$
                                               &     0.067     &  $+0.110\atop-0.110$    &  $+0.004\atop-0.009$  \\ 
                & $\widetilde{\lambda}$        &    -0.459     &  $+0.184\atop-0.206$    &  $+0.088\atop-0.116$    
                                               &     0.156     &  $+0.109\atop-0.131$    &  $+0.093\atop-0.112$
                                               &    -0.137     &  $+0.282\atop-0.308$    &  $+0.048\atop-0.068$  \\  \hline
                  & $\lambda_{\vartheta}$      &    -0.119     &  $+0.074\atop-0.096$    &  $+0.116\atop-0.111$ 
                                               &    -0.025     &  $+0.060\atop-0.080$    &  $+0.102\atop-0.105$        
                                               &    -0.004     &  $+0.199\atop-0.221$    &  $+0.134\atop-0.126$  \\ 
Gottfried-Jackson & $\lambda_{\varphi}$        &    -0.071     &  $+0.038\atop-0.052$    &  $+0.021\atop-0.033$
                                               &    -0.033     &  $+0.021\atop-0.029$    &  $+0.022\atop-0.027$ 
                                               &    -0.040     &  $+0.037\atop-0.048$    &  $+0.036\atop-0.027$  \\ 
Forward         &$\lambda_{\vartheta\varphi}$  &    -0.124     &  $+0.062\atop-0.073$    &  $+0.066\atop-0.075$    
                                               &     0.125     &  $+0.038\atop-0.047$    &  $+0.029\atop-0.040$ 
                                               &     0.095     &  $+0.097\atop-0.103$    &  $+0.002\atop-0.010$  \\
                &$\widetilde{\lambda}$         &    -0.317     &  $+0.122\atop-0.148$    &  $+0.036\atop-0.057$    
                                               &    -0.138     &  $+0.083\atop-0.107$    &  $+0.027\atop-0.046$ 
                                               &    -0.067     &  $+0.252\atop-0.268$    &  $+0.030\atop-0.054$  \\ \hline
                & $\lambda_{\vartheta}$        &     -0.130    &  $+0.215\atop-0.245$    &  $+0.313\atop-0.293$    
                                               &      0.011    &  $+0.104\atop-0.126$    &  $+0.186\atop-0.173$  
                                               &      0.056    &  $+0.159\atop-0.171$    &  $+0.199\atop-0.178$  \\ 
Gottfried-Jackson & $\lambda_{\varphi}$        &     -0.065    &  $+0.023\atop-0.032$    &  $+0.019\atop-0.032$                 
                                               &     -0.042    &  $+0.015\atop-0.025$    &  $+0.015\atop-0.030$                 
                                               &     -0.060    &  $+0.037\atop-0.053$    &  $+0.044\atop-0.017$  \\ 
Backward        &$\lambda_{\vartheta\varphi}$  &      0.040    &  $+0.083\atop-0.092$    &  $+0.057\atop-0.045$ 
                                               &      0.009    &  $+0.034\atop-0.046$    &  $+0.013\atop-0.026$ 
                                               &     -0.014    &  $+0.086\atop-0.099$    &  $+0.024\atop-0.032$  \\ 
                &$\widetilde{\lambda}$         &     -0.303    &  $+0.228\atop-0.252$    &  $+0.206\atop-0.235$    
                                               &     -0.112    &  $+0.117\atop-0.133$    &  $+0.098\atop-0.126$ 
                                               &     -0.090    &  $+0.195\atop-0.225$    &  $+0.100\atop-0.127$  \\ \hline
                & $\lambda_{\vartheta}$        &    -0.158      &  $+0.233\atop-0.257$   &  $+0.312\atop-0.293$
                                               &     0.020      &  $+0.105\atop-0.125$   &  $+0.195\atop-0.174$                 
                                               &     0.040      &  $+0.145\atop-0.165$   &  $+0.188\atop-0.170$  \\ 
                & $\lambda_{\varphi}$          &    -0.063      &  $+0.026\atop-0.034$   &  $+0.022\atop-0.032$                  
                                               &    -0.040      &  $+0.017\atop-0.028$   &  $+0.015\atop-0.030$                  
                                               &    -0.058      &  $+0.046\atop-0.049$   &  $+0.032\atop-0.019$  \\ 
Helicity        & $\lambda_{\vartheta\varphi}$ &    -0.040      &  $+0.088\atop-0.097$   &  $+0.021\atop-0.041$                 
                                               &    -0.033      &  $+0.041\atop-0.049$   &  $+0.016\atop-0.032$                  
                                               &    -0.016      &  $+0.089\atop-0.101$   &  $+0.022\atop-0.037$  \\
                & $\widetilde{\lambda}$        &    -0.327      &  $+0.242\atop-0.258$   &  $+0.210\atop-0.239$                 
                                               &    -0.097      &  $+0.122\atop-0.138$   &  $+0.108\atop-0.127$                  
                                               &    -0.116      &  $+0.201\atop-0.219$   &  $+0.095\atop-0.114$  \\   \hline
\end{tabular}
\end{sidewaystable}

\begin{sidewaystable}[htbp]
\renewcommand\arraystretch{1.5}
\isucaption{\label{tab:xF_chi2ndf} Average $\chi^{2}$/$ndf$ for two-dimensional fits in each arm to determine angular decay coefficients as a function of $x_F$.  Each fit has 36 degrees of freedom.}
\begin{tabular}{ccccccc}
\cline{2-7}
                &          \multicolumn{2}{|c|}{$x_{F} \in [0.015,0.030]$}
                &          \multicolumn{2}{|c|}{$x_{F} \in [0.030,0.040]$}                
                &          \multicolumn{2}{|c|}{$x_{F} \in [0.040,0.060]$}    \\ 
                &          \multicolumn{2}{|c|}{$<x_{F}>=0.026$}                     
                &          \multicolumn{2}{|c|}{$<x_{F}>=0.035$}                 
                &          \multicolumn{2}{|c|}{$<x_{F}>=0.045$}    \\ \hline
                &          North   &  South 
                &          North   &  South 
                &          North   &  South  \\ \hline
\small Frame               &   \multicolumn{1}{|c}{\tiny $<\chi^{2}$/$ndf$$>$}
                           &   \multicolumn{1}{|c}{\tiny $<\chi^{2}$/$ndf$$>$}
                           &   \multicolumn{1}{|c}{\tiny $<\chi^{2}$/$ndf$$>$}
                           &   \multicolumn{1}{|c}{\tiny $<\chi^{2}$/$ndf$$>$}
                           &   \multicolumn{1}{|c}{\tiny $<\chi^{2}$/$ndf$$>$}
                           &   \multicolumn{1}{|c}{\tiny $<\chi^{2}$/$ndf$$>$} \\ \hline \hline
 \small Collins-Soper     &  1.6  &  1.7
                          &  2.2  &  14.9
                          &  1.3  &  1.7  \\ \hline
 \small Gottfried-Jackson &     &
                          &     &
                          &     &     \\
 \small Forward           & 2.1 & 4.9
                          & 3.5 & 2.2
                          & 1.7 & 1.9 \\ \hline
 \small Gottfried-Jackson  &     &
                           &     &
                           &     &     \\
 \small Backward           & 1.3 & 1.5
                           & 2.5 & 2.4
                           & 2.4 & 2.0 \\ \hline
 \small Helicity          & 1.2 & 2.1
                          & 2.7 & 2.2
                          & 2.2 & 1.8 \\ \hline
\end{tabular}
\end{sidewaystable}